\documentclass[aps,twocolumn,prx,floatfix,showpacs,lengthcheck,citeautoscript,superscriptaddress,longbibliography]{revtex4-2}
\usepackage{amsmath}
\usepackage{amssymb}
\usepackage{graphicx}
\usepackage{bm}
\usepackage{color,soul}
\usepackage{braket}
\usepackage[dvipsnames,svgnames,table]{xcolor}
\usepackage{times}
\usepackage{MnSymbol}

\usepackage{bbold}
\usepackage{babel}
\usepackage{hyperref}
\usepackage{orcidlink}

\definecolor{Nathanpink}{rgb}{0.94,0.317,0.9607}

\definecolor{titles}{rgb}{0.,0.24,0.51}
\newcommand{\titles}[1]{{\color{titles}#1}}
\hypersetup{
pdfstartview={FitH},
colorlinks=true,    
linkcolor=NavyBlue, 
citecolor=NavyBlue,   
filecolor=NavyBlue, 
urlcolor=NavyBlue   
}

\newcommand{\bk}{\bm{k}}

\newcommand{\bea}{\begin{eqnarray}}
\newcommand{\eea}{\end{eqnarray}}
\newcommand{\be}{\begin{equation}}
\newcommand{\ee}{\end{equation}}

\renewcommand{\Ket}[1]{| #1 \rangle\rangle}
\renewcommand{\Bra}[1]{\langle\langle #1 |}

\newcommand{\HFk}{\hat{\bm{H}}_{\bm{k}}^F}
\newcommand{\UFk}{\hat{\bm{R}}_{\bm{k}}}
\newcommand{\tUFk}{\tilde{\bm{R}}_{\bm{k}}}
\newcommand{\Uk}{\hat{R}_{\bm{k}}}
\newcommand{\Pk}{\hat{P}_{\bm{k}}}

\begin{document}
\title{The St\v{r}eda Formula for Floquet Systems:\\ Topological Invariants and Quantized Anomalies from Ces\`aro Summation}
\author{Lucila {Peralta Gavensky}\,\orcidlink{0000-0002-5598-7303}}
\email{lucila.peralta.gavensky@ulb.be}
\affiliation{Center for Nonlinear Phenomena and Complex Systems, Universit\'e Libre de Bruxelles, CP 231, Campus Plaine, B-1050 Brussels, Belgium}
\affiliation{International Solvay Institutes, 1050 Brussels, Belgium}
\author{Gonzalo {Usaj}\,\orcidlink{0000-0002-3044-5778}}
\affiliation{Center for Nonlinear Phenomena and Complex Systems, Universit\'e Libre de Bruxelles, CP 231, Campus Plaine, B-1050 Brussels, Belgium}
\affiliation{Centro At\'omico Bariloche and Instituto Balseiro, Comisi\'on Nacional de Energ\'ia
Atomica (CNEA)- Universidad Nacional de Cuyo (UNCUYO), 8400 Bariloche, Argentina.}
\affiliation{Instituto de Nanociencia y Nanotecnolog\'ia (INN-Bariloche), Consejo Nacional de Investigaciones Cient\'ificas y Tecnicas (CONICET), Argentina.}
\author{Nathan Goldman\,\orcidlink{0000-0002-0757-7289}}
\affiliation{Laboratoire Kastler Brossel, Coll\`ege de France, CNRS, ENS-Universit\'e PSL, Sorbonne Universit\'e, 11 Place Marcelin Berthelot, 75005 Paris, France}
\affiliation{International Solvay Institutes, 1050 Brussels, Belgium}
\affiliation{Center for Nonlinear Phenomena and Complex Systems, Universit\'e Libre de Bruxelles, CP 231, Campus Plaine, B-1050 Brussels, Belgium}

\begin{abstract}
 
The St\v{r}eda formula establishes a fundamental physical connection between the topological invariants characterizing the bulk of topological matter and the presence of gapless edge modes. In this work, we extend the St\v{r}eda formula to periodically driven systems, providing a rigorous framework to elucidate the unconventional bulk-boundary correspondence of Floquet systems, while offering a physically grounded link between Floquet winding numbers and tractable response functions. Using the Sambe representation of periodically driven systems, we analyze the response of the unbounded Floquet density of states to a magnetic perturbation. This Floquet-St\v{r}eda response is regularized through Ces\`aro summation, yielding a well-defined, quantized result within spectral gaps. The response features two physically distinct contributions: a quantized charge flow between edge and bulk, and an anomalous energy flow between the system and the drive, offering new insight into the nature of anomalous edge states. This fundamental result rigorously connects Floquet winding numbers to the orbital magnetization density of Floquet states and holds broadly, from clean to disordered and inhomogeneous systems. This is further supported by providing a real-space formulation of the Floquet-St\v{r}eda response, which introduces a local topological marker suited for periodically driven settings. In translationally-invariant systems, the framework yields a remarkably simple expression for Floquet winding numbers involving geometric properties of Floquet-Bloch bands. A concrete experimental protocol is proposed to extract the Floquet-St\v{r}eda response via particle-density measurements in systems coupled to engineered baths. Finally, by expressing the topological invariants through the magnetic response of the Floquet density of states, this approach opens a promising route toward the topological characterization of interacting driven phases.
\end{abstract}

\date{May 19, 2025}
\maketitle
\section{Introduction}
Topology has led to an exceptionally fruitful cross-fertilization between different areas of research, establishing a bridge between concepts of  mathematics, quantum field theory and condensed matter physics. Recent developments in topological quantum matter have not only led to a deeper understanding of topological phenomena in thermodynamic equilibrium~\cite{Hasan2010,Qi_RMP,Haldane2017,Bradlyn2017,Sato2017,Yan2017}, but also initiated the ambitious task of generalizing these concepts to the realm of out-of-equilibrium settings~\cite{Ozawa_RMP,Cooper_RMP}.

Subjecting a system to a time-periodic modulation, also known as Floquet engineering, has emerged as one of the most promising research fronts to address this challenge~\cite{Kitagawa2010,Asboth2012,Asboth2013,Basov2017,Eckardt2017,Oka2019,Harper2020,Rudner2020,Weitenberg2021}. Formally, periodically-driven systems can be described by a time-independent Hamiltonian, at the expense of working with an infinite-dimensional space~\cite{Shirley1965,Sambe1973}. The spectrum of this unbounded Hamiltonian operator defines the \textit{quasienergies} of the system, an infinite set of eigenvalues that are repeated periodically according to the driving frequency, and can thus be organized within Floquet-Brillouin zones. In a lattice setting, the quasienergies are classified into Floquet-Bloch bands, whose geometric and topological properties can be precisely tailored by adjusting the driving protocol, potentially enabling the on-demand realization of topological phases~\cite{Sorensen2005,Oka2009,Lindner2011,Kitagawa2010,Asboth2012,Asboth2013,Rudner2013}. Signatures of both Floquet physics and driven-induced topology have been successfully detected in a variety of experimental platforms, ranging from solid-state~\cite{Wang2013a,Mahmood2016,McIver2020,Park2022,Zhou2023,Choi2025,Merboldt2025}, cold-atoms in optical lattices~\cite{Lignier2007,Struck2011,Aidelsburger2014,Jotzu2014,Flaschner2016,Tai2017,Asteria2019,Tarnowski2019,Wintersperger2020,Lu2022,Leonard2023,Braun2024}, and acoustic settings~\cite{Fleury2016,Peng2016,Cheng2022} to photonics~\cite{Kitagawa2012,Rechtsman2013,Hu2015,Gao2016,Cardano2017,Mukherjee2017,Maczewsky2017,Cheng2019,Roushan2017,Adiyatullin2023,ElSokhen2024}. 

The topological invariants that are commonly used to characterize bulk bands (e.g.~Chern numbers~\cite{Thouless1982}) have proven to be insufficient to capture the topological properties of Floquet systems~\cite{Kitagawa2010,Rudner2013}. In particular, the bulk-boundary correspondence, which establishes the edge-state structure in systems with boundaries~\cite{Hasan2010,Qi_RMP}, has to be treated with care:~In general, the net number of edge modes that are located within a quasienergy gap is dictated by a topological invariant that takes the time-dependent nature of the Hamiltonian into account~\cite{Asboth2012,Asboth2013,Rudner2013,Asboth2014,Nathan2015,Roy2017,Harper2020,Rudner2020,Vu2022}. The periodicity of the Floquet spectrum with respect to the driving frequency is central to this peculiarity, as it can result in drive-induced boundary modes that are located at the edges of the Floquet-Brillouin zone; see Fig.~\ref{conceptual_SF}(c). These exotic edge modes, which have no analogue in static systems, have been coined \textit{anomalous edge states}~\cite{Rudner2013}. The bulk-boundary correspondence of two-dimensional Floquet systems, which accurately dictates the total number of edge modes (both regular and anomalous), has been established through the introduction of a spatiotemporal winding number built from the full time-evolution operator~\cite{Rudner2013,Nathan2015,Roy2017,Harper2020}.  
The mathematical complexity of this topological index, combined with the difficulty of preparing a steady-state thermal-like occupation of the Floquet bands~\cite{DAlessio2014,Lazarides2014,Seetharam2015,Iadecola2015,Esin2018}, have hindered the identification of practical methods to probe this invariant and its physical consequences in generic settings. A tomography scheme, requiring the tracking of band-touching points and their corresponding topological charges in momentum space~\cite{Unal2019}, was recently implemented in an optical-lattice setup~\cite{Wintersperger2020} to deduce these intriguing topological numbers. Nonetheless, the measurement of a robustly quantized observable that is directly dictated by the spatiotemporal winding number of Ref.~\cite{Rudner2013} has remained elusive. 

Various strategies have been theoretically proposed in a relentless effort to solve this puzzle. Among these, transport probes have emerged as the most intuitive approach. Current proposals in this direction focus on detecting topological signatures that either arise from the driven-induced Floquet edge channels~\cite{Gu2011,Kundu2013,Piskunow2014,FoaTorres2014,Usaj2014,Farrell2015,Farrell2016,Kundu2020,Zhang2024_b} or from the bulk response of these out-of-equilibrium systems~\cite{Oka2009,Dehghani2015,Dauphin2017,PeraltaGavensky2018}. Approaches beyond transport are currently restricted to a very specific class of Floquet systems, where all the bulk states are strongly localized by disorder~\cite{Titum2016,Nathan2019}. These phases are predicted to exhibit a quantized bulk orbital magnetization density reflecting the number of anomalous edge channels~\cite{Nathan2017,Nathan2021,Glorioso2021}.
\subsection{Theoretical approach and main results}
This work introduces a physical framework that sheds light on the bulk-boundary correspondence of two-dimensional periodically-driven systems. As a key outcome, it connects the abstract winding numbers used to classify Floquet topological phases to physically tractable response functions. Our construction is entirely based on the elementary notion of spectral flow~\cite{Delplace_flow}, which lies at the heart of the St\v{r}eda formula introduced in the context of equilibrium systems~\cite{Streda1982,Widom1982,Streda1983,Prodan2016}.

\begin{figure}[t]
    \centering
    \includegraphics[width=\columnwidth]{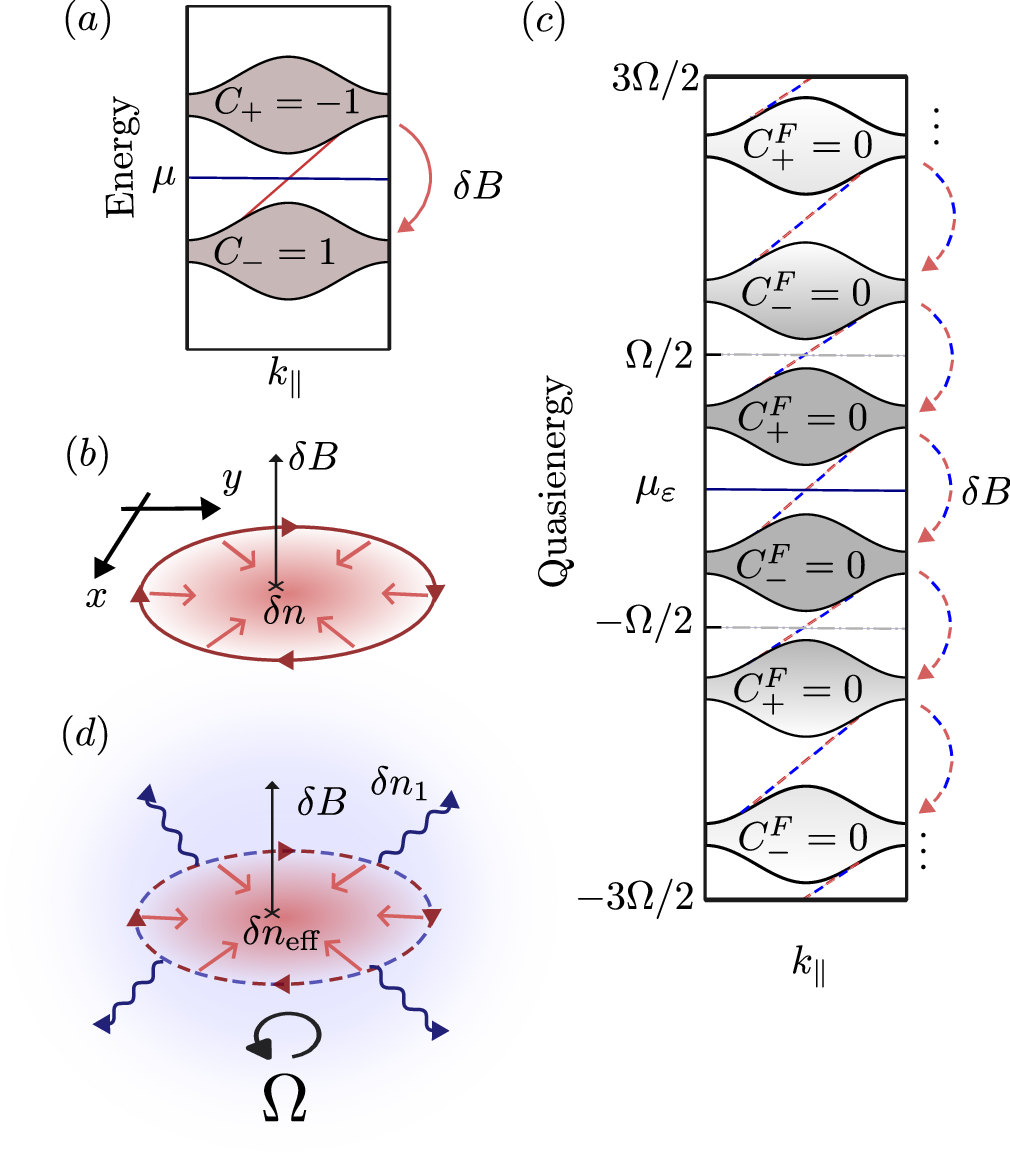}
    \caption{$(a)$ Sketch of the energy spectrum of a Chern insulator, defined on a semi-infinite ribbon geometry, where $k_{\parallel}$ denotes the quasimomentum along the direction of the ribbon. A chiral edge mode bridges the gap between the two bands, whose Chern numbers are $C_{\pm}\!=\!\mp 1$. Perturbing the system with a magnetic field $\delta B$ generates a spectral flow between the upper and lower band, physically carried by the chiral channel. $(b)$ Redistribution of the particle density (color gradient) of an open-boundary system, filled up to the chemical potential $\mu$ [blue line in $(a)$], when subjected to the magnetic perturbation. The arrows depict the flow of states from the edge to the bulk, where the density increases according to $\delta n\!=\!C_{-}\delta B/\Phi_0$. $(c)$ Quasienergy spectrum of an unbounded Floquet Hamiltonian, describing a  periodically-driven lattice at frequency $\Omega$. Even though all the Chern indices are equal to zero $C^{F}_{\pm}\!=\!0$, the system is topologically non-trivial, with chiral edge channels bridging all the gaps. The arrows depict the spectral flow between Floquet bands under the $\delta B$ field insertion. $(d)$ Sketch of the two contributions to the Floquet spectral flow:~the normal flow $\delta n_{\mathrm{eff}}$, dictated by the Chern number of the Floquet bands, and describing the flow of states between the edge and the bulk; and the `anomalous' flow, $\delta n_1$, encoding the quantized energy-flow between the system and the driving field. In the two-band model in panel $(c)$, the normal flow $\delta n_{\mathrm{eff}}\!=\!0$ because of the trivial Chern numbers.}
    \label{conceptual_SF}
\end{figure}
The St\v{r}eda formula is a remarkable thermodynamic relation that links the Hall conductivity of insulating states of matter to the magnetic response of their particle number $N(\mu)$ at a fixed chemical potential $\mu$, 
\begin{equation}
    \frac{\sigma_{H}}{\sigma_0} = \Phi_0 \frac{\partial N(\mu)}{\partial \Phi},
    \label{streda}
\end{equation}
where $\sigma_0\!=\!e^2/h$ is the conductance quantum, $\Phi_0\!=\!hc/e$ the normal flux quantum and $\Phi\!=\!B A$ the total flux through a system of area $A$ threaded by a constant magnetic field $B$.
Derived from non-perturbative thermodynamic arguments~\cite{Widom1982,MacDonald1989}, the St\v{r}eda relation has proven to be an extremely powerful tool to identify quantized bulk responses in a variety of different contexts~\cite{Yacoby1999,Umucalilar2008,Repellin2020,Leonard2023,Jamotte2023,Xie2021,Cai2023,PeraltaGavensky2023}. The guiding principle behind its validity is sketched in Fig.~\ref{conceptual_SF}$(a)$-$(b)$: Piercing a two-dimensional Chern insulator with a perturbing magnetic field induces a chiral spectral flow of states between Bloch bands of different Chern number. In an open-boundary sample, this flow is physically carried by the one-way edge channels bridging the gap between these bands, as shown in the schematic picture of Fig.~\ref{conceptual_SF}$(a)$. If such a lattice model is filled with particles up to the chemical potential $\mu$, a net flow of states will occur between the edge and the bulk upon activating the magnetic perturbation, leading to a redistribution of the particle density in the sample, as illustrated in Fig.~\ref{conceptual_SF}$(b)$. In the context of particle physics, this phenomenon is known as \textit{anomaly inflow} or Callan-Harvey mechanism~\cite{Callan1985,Fradkin1986,Stone1991}.
The particle flow into the insulating bulk, as dictated by Eq.~\eqref{streda}, is quantized according to the net number of chiral edge modes crossing the chemical potential. The quantized Hall conductivity in Eq.~\eqref{streda} being dictated by the Chern number of the filled Bloch bands~\cite{Thouless1982}, one naturally recovers the bulk-boundary correspondence through the St\v{r}eda relation. 

These arguments can be readily extended to two-dimensional Floquet systems, by making use of the aforementioned mapping between periodically-driven systems and their corresponding (unbounded) time-independent Hamiltonian~\cite{Shirley1965,Sambe1973}:~the so-called Sambe representation. At this level, the only crucial difference with the spectral flow that occurs at equilibrium in response to a magnetic perturbation is that it now takes place in an \emph{unbounded spectrum}, consisting of an infinite number of Floquet-Bloch bands. As illustrated in Fig.~\ref{conceptual_SF}$(c)$, the net number of driven-induced chiral edge channels that are located at quasienergy $\mu_\varepsilon$ can be determined by evaluating the number of states that have crossed this quasienergy via a ``bucket brigade mechanism". In this work, we demonstrate that this spectral flow, or Floquet-St\v{r}eda response, can be formally decomposed as a sum of two contributions:~the \textit{normal} and \textit{anomalous} flows. The normal flow is equivalent to the St\v{r}eda response exhibited by undriven systems; as previously mentioned, it  represents the quantized flow of charge between the edge and the bulk of the system. This flow 
can be entirely deduced from the Chern numbers of the Floquet-Bloch bands. In the special case represented in Fig.~\ref{conceptual_SF}($c$), all these Chern numbers are zero, such that the normal contribution vanishes. In contrast, the anomalous flow encodes the infinite summation that is required to accurately evaluate the net spectral flow at a given quasienergy. As illustrated in Fig.~\ref{conceptual_SF}$(d)$, we associate this striking phenomenon, inherent to out-of-equilibrium Floquet systems, to the emergence of a quantized energy flow between the system and the driving field, exclusively induced by the magnetic perturbation.  Interestingly, this picture connects to recent developments~\cite{Mondragon2018,Nakagawa2020},  which suggest that anomalous topological phases could find their origin in such energy flows. We also note that quantized energy-pumps were recently explored in zero- and one-dimensional Floquet systems~\cite{Martin2017,Kolodubretz2018,Crowley2019,Nathan2021_b,Sridhar2024}.

The core of our approach relies on the regularization of the Floquet-St\v{r}eda response defined in Sambe space, which turns out to be a mathematically ill-defined quantity. Notably, the anomalous flow can be identified as an integral version of a non-convergent Grandi-type series, which we regularize by means of a Ces\`aro summation method~\cite{Hardy2000,Titchmarsh1986}. This procedure leads to a series of key results, which we now list and summarize below.

\paragraph{\titles{\textbf{Classification of Floquet topological phases from their Floquet-St\v{r}eda response.}}}
{We show that the regularized Floquet spectral flow can be entirely obtained from the magnetic response of the Floquet density of states within a given Floquet zone [see Eq.~\eqref{flow_part_energy}]. This fundamental result not only pinpoints this quantity as a relevant physical observable to extract topological quantized responses, it also  highlights that the classification of Floquet topological phases does not require the knowledge of the full time-dynamics within a driving cycle, also known as micromotion, as long as the stroboscopic time-evolution is known in the presence of a perturbing magnetic field.}

\paragraph{\titles{\textbf{Bulk-boundary correspondence from the bulk properties of the Floquet-Bloch states.}}}{Our theory expresses the topological winding number of non-interacting periodically-driven lattice systems in terms of simple band properties, all deriving from the Floquet-Bloch Hamiltonian expressed in the Sambe representation. Indeed, we find a closed analytical and gauge-invariant formula for these topological indices, which is entirely expressed in terms of the quasienergies, Berry curvatures and intrinsic orbital magnetic moments of the Floquet-Bloch states defined in a given Floquet zone [see Eq.~\eqref{WF_Cesaro}].}

\paragraph{\titles{\textbf{Winding numbers in terms of the orbital magnetization of Floquet states.}}}{We derive an explicit relation between the Floquet-St\v{r}eda response and the orbital magnetization density associated to the bulk Floquet states [see Eqs.~\eqref{maxwell_rel_floquet} and~\eqref{Mztot}]. Remarkably, we find that the quantization of the anomalous spectral flow reflects the quantization of the total orbital magnetization density of the Floquet states within a Floquet zone. This key finding extends the result of Ref.~\cite{Nathan2017} beyond Floquet systems with fully localized bulk states, demonstrating its general applicability as long as the Floquet bulk gap at the zone edge remains open. In the case of translationally-invariant lattices, we demonstrate that the orbital magnetization density is strictly quantized upon fully occupying the Bloch-Floquet bands, in sharp contradiction with the results reported in Refs.~\cite{Topp2022,Dag2022}. 
}

\paragraph{\titles{\textbf{Generalized Floquet-St\v{r}eda formula:~Quantized charge and energy pump.}}}{The anomalous flow is explicitly identified as the  derivative of a first-order winding number $N_1[R]$ with respect to a magnetic field [see Eqs.~\eqref{N1_R} and~\eqref{WA_dN1}]. This mathematical observation allows us to draw a concrete parallelism between the Floquet-St\v{r}eda response stemming from our approach and the original St\v{r}eda formula in Eq.~\eqref{streda}. The generalized Floquet-St\v{r}eda formula  [see Eq.~\eqref{generalized_streda_floquet}] contains two physically distinguishable contributions [see Fig.~\ref{conceptual_SF}$(d)$]:~
the normal-flow contribution, which is captured by the Chern numbers of all the Floquet-Bloch bands that are located between the bottom edge of the Floquet-Brillouin zone and the gap at quasienergy  $\mu_\varepsilon$. This term is physically interpreted as the flow of particles (dressed by the driving field) between the edge and the bulk of the system. In the undriven limit, this contribution reduces to Eq.~\eqref{streda}. The second contribution corresponds to the anomalous flow, $\Phi_0 \partial N_1[R]/\partial \Phi$, which diagnoses the emergence of a quantized energy pump between the system and the driving field, an effect which is inextricably linked to the existence of resonant processes.} 

\paragraph{\titles{\textbf{A Floquet-St\v{r}eda sum-rule.}}}{We elucidate a protocol to extract Floquet topological invariants from particle density-response measurements, in an experimentally realistic setting that involves an engineered heat-bath. This proposal is based on a sum-rule scheme~[see Eq.~\eqref{sum_rule}], which can be seen as the counterpart of the Floquet sum-rule that was previously introduced in the context of edge transport~\cite{Kundu2013,Farrell2015,Farrell2016}.} 

\paragraph{\titles{\textbf{Real-space markers for Floquet winding numbers.}}}{Our formalism naturally leads to the definition of local Floquet winding numbers, which can be used to identify topological phases in different regions of space, e.g.~in inhomogeneous or disordered systems. Specifically, this is achieved by introducing real-space-resolved expressions for the normal and anomalous contributions to the Floquet spectral flow, which are constructed from the magnetic response of the local Floquet density of states [see Eqs.~\eqref{WN_local} and~\eqref{WA_local}]. }
\subsection{Outline}
The manuscript is structured as follows. Section~\ref{Sec_eq} is devoted to a study of the  St\v{r}eda response at thermodynamic equilibrium. Considering systems of non-interacting fermions on a lattice, we  derive a general expression that relates the magnetic response of the density of states to bulk properties. An emphasis is set on the quantum geometry of Bloch bands, which is exquisitely probed by this energy-resolved response function. 
We also obtain insightful expressions for the net spectral flow of states at a given chemical potential, upon threading an external flux through the lattice, and show how this physical quantity connects to well-defined topological invariants. Section~\ref{Sec_Floquet} generalizes this approach to the case of periodically-driven Floquet systems. We introduce the Sambe formalism in view of mapping these out-of-equilibrium problems to static (time-independent) Hamiltonian systems defined in an infinite-dimensional space. We define the net spectral flow of the corresponding unbounded spectrum, within a given Floquet spectral gap, identify the normal and anomalous flows, and introduce the Ces\`aro summation method to regularize the mathematically ill-defined anomalous contribution. In Sec.~\ref{Sec_PhysInt}, we discuss how our approach relates to other formalisms, previously introduced in the literature. We also provide a physical interpretation of our results, and elucidate the explicit connection between the regularized Floquet-St\v{r}eda response and the orbital magnetization density associated to the Floquet states. Then, in Sec.~\ref{Sec_Kitagawa}, we exhaustively analyze how the proposed formalism applies to a particular model that supports trivial, conventional and anomalous Floquet topological phases as a function of its parameters. In Sec.~\ref{Sec_sum-rule}, we introduce a protocol to measure winding numbers through particle-density measurements, in a setting that involves an engineered heat-bath, and show how it applies to a concrete model. In Sec.~\ref{Sec_local-marker}, we provide real-space-resolved markers for Floquet winding numbers, and we validate these definitions by performing numerical simulations for clean inhomogeneous systems (junctions) as well as for disordered samples with open-boundary conditions. Finally, Sec.~\ref{Conclusions} outlines possible directions for future research. The Appendices provide technical aspects of our derivations and alternative derivations of some of our results.
\section{Equilibrium St\v{r}eda response}\label{Sec_eq}
As a warm-up, we first investigate the St\v{r}eda response of a generic clean static system, consisting of non-interacting fermions on a lattice and described by a Hamiltonian $\hat{H}$. We note that a local St\v{r}eda response can be evaluated in the presence of disorder, as we discuss in Sec.~\ref{Sec_local-marker}. The particle density at finite temperature is given by
\begin{equation}
    n = \int_{-\infty}^{\infty}d\omega\, f(\omega)\rho(\omega)\,,
\end{equation}
where $f(\omega)$ is the Fermi-Dirac distribution function and $\rho(\omega)$ stands for the density of states (DOS) of the system, 
\begin{equation}
    \rho(\omega) = -\frac{1}{\pi A}\mathrm{Im}\mathrm{Tr}\left[\hat{G}(\omega+i 0^{+})\right]\, .
    \label{DOS}
\end{equation}
Here $A$ is the area of the sample, $\hat{G}(\omega)$ is the single-particle Green's function and $\mathrm{Tr}[\hdots]$ traces over spatial and internal degrees of freedom. Hence, evaluating the St\v{r}eda response in Eq.~\eqref{streda} boils down to determining how $\rho(\omega)$ gets modified upon applying a magnetic perturbation. In the presence of a small magnetic field, and considering the real-space representation, the matrix elements of the Hamiltonian $\hat H$ are modified as
\begin{equation}
    H_{\bm{R}_{\nu}\bm{R'}_{\nu'}} = e^{i\frac{2\pi}{\Phi_0}\varphi_{\bm{R}_{\nu}\bm{R'}_{\nu'}}}H^0_{\bm{R}_{\nu}\bm{R'}_{\nu'}},
\end{equation}
where
\begin{equation}
\label{Peierls_phases}
    \varphi_{\bm{R}_{\nu}\bm{R'}_{\nu'}} = \int_{\bm{R}_{\nu}}^{\bm{R'}_{\nu'}}\bm{A}(\bm{r})\cdot d\bm{r}\,,
\end{equation}
are the Peierls phase factors \cite{Peierls1933,Luttinger_p_phases}. Here the coordinates $\bm{R}_{\nu}$ denote the position of the site $\nu$ within the unit cell at $\bm{R}$, $\bm{A}(\bm{r})$ is the vector potential and $\hat{H}^0$ is the Hamiltonian in the absence of the perturbing field. 

Relevant physical quantities can be identified in the magnetic response of the DOS by imposing periodic boundary conditions in the lattice Hamiltonian. In this case, it is convenient to adopt the formalism used in Refs.~\cite{Kita2005,Onoda2006,Chen2011,PeraltaGavensky2023} by means of which a gauge-invariant propagator is introduced by factoring out the Peierls phase factors from the matrix elements of the Green's function
\begin{equation}
   G^{(B)}_{\bm{R}_{\nu}\bm{R'}_{\nu'}}(\omega) = e^{-i\frac{2\pi}{\Phi_0} \varphi_{\bm{R}_{\nu}\bm{R'}_{\nu'}}}G^{}_{\bm{R}^{}_{\nu}\bm{R'}_{\nu'}}(\omega)\,.
\end{equation}
In the non-interacting limit, the propagator reduces to $\hat{G}(\omega) = (\omega \hat{I}-\hat{H})^{-1}$, and the corresponding $\hat{G}^{(B)}(\omega)$ satisfies the modified Dyson's equation of motion
\begin{eqnarray}
\label{GinvDyson}
    \sum_{\bm{R'}\nu'}& &\left(\omega\, \delta_{\bm{R}_{\nu}\bm{R'}_{\nu'}}-H^{0}_{\bm{R}_{\nu}\bm{R'}_{\nu'}}\right)G^{(B)}_{\bm{R'}_{\nu'}\bm{R''}_{\nu''}}(\omega)\\
    \notag
    & &\times e^{i\frac{\pi}{\Phi_0}\bm{B}\cdot (\bm{R'}_{\nu'}-\bm{R}_{\nu})\times(\bm{R''}_{\nu''}-\bm{R'}_{\nu'})} = \delta_{\bm{R}_{\nu}\bm{R''}_{\nu''}}\,,
\end{eqnarray}
where $\bm{B}=\bm{\nabla}\times \bm{A}(\bm{r})$ is the external magnetic field. In\-te\-res\-tin\-gly, Eq.~\eqref{GinvDyson} indicates that $\hat{G}^{(B)}(\omega)$ is manifestly gauge and translationally invariant. In a translationally-invariant lattice, the propagator $\hat{G}^{(B)}(\omega)$ can then be Fourier transformed to quasimomentum space as

\begin{equation}
  G^{(B)}_{\bm{R}_{\nu}\bm{R'}_{\nu'}}(\omega) = \frac{1}{N_c}\sum_{\bm{k}}e^{i\bm{k}\cdot (\bm{R}_{\nu}-\bm{R'}_{\nu'})} [\hat{G}^{(B)}_{\bm{k}}(\omega)]_{\nu\nu'}\,,
\end{equation}
with $N_c$ the number of unit cells in the lattice. We note that the DOS in Eq.~\eqref{DOS} can be expressed in terms of this gauge-invariant propagator as
\begin{eqnarray}
\notag
    \rho(\omega) &=& -\frac{1}{\pi A}\mathrm{Im}\mathrm{Tr}\left[\hat{G}^{(B)}(\omega+i 0^{+})\right]\\
    &=&-\frac{1}{\pi}\int_{\mathrm{BZ}} \frac{d^2 k}{(2\pi)^2}\mathrm{Im}\,\mathrm{tr}\left[\hat{G}^{(B)}_{\bm{k}}(\omega+i 0^{+})\right]\,,
    \label{rhoPBC}
\end{eqnarray}
where the trace $\mathrm{tr}[\hdots]$ is restricted to summing over the internal degrees of freedom. Expanding Eq.~\eqref{GinvDyson} up to first order in the magnetic field, we find 
\begin{equation}
\hat{G}^{(B)}_{\bm{k}}(\omega) = \hat{G}^{}_{\bm{k}}(\omega) + \frac{i\pi}{\Phi_0} B^{i} \epsilon^{ijl}\hat{G}^{}_{\bm{k}}(\omega)\frac{\partial \hat{G}^{-1}_{\bm{k}}(\omega)}{\partial k_j}\frac{\partial \hat{G}^{}_{\bm{k}}(\omega)}{\partial k_l},
\label{GkB}
\end{equation}
where $\hat{G}_{\bm{k}}(\omega)$ is the Bloch Green's function in the absence of the applied magnetic field, namely $ \hat{G}^{}_{\bm{k}}(\omega) = (\omega\hat{I}_{\bm{k}}-\hat{H}_{\bm{k}})^{-1}$, where $\hat{H}_{\bm{k}}$ is the Bloch Hamiltonian. Here $\epsilon^{ijl}$ is the Levi-Civita tensor and a summation over indices is implicit. 

Considering, without loss of generality, a two-dimensional sample in the $x-y$ plane with the magnetic field applied along the $z$ direction, and inserting Eq.~\eqref{GkB} into Eq.~\eqref{rhoPBC}, we find that the magnetic response of the DOS evaluated at zero magnetic field is given by
\begin{eqnarray}
 \label{drho_dB}
   \Phi_0 \frac{\partial \rho(\omega)}{\partial B} &=&
    \int_{\mathrm{BZ}}\frac{d^2k}{2\pi}\sum_{\alpha}\bigg[\mathcal{F}_{xy}^{\alpha}(\bm{k})\delta(\omega-\varepsilon_{\alpha\bm{k}})\\
    \notag
    &+&\frac{\Phi_0}{2\pi}m^{\alpha}_z(\bm{k})\frac{\partial}{\partial \omega}\delta(\omega-\varepsilon_{\alpha\bm{k}})\bigg]\,,
\end{eqnarray}
where we identified the Berry curvature
\begin{equation}
   \mathcal{F}_{xy}^{\alpha}(\bm{k}) = i \left(\left\langle \partial_{k_x}u_{\alpha\bm{k}}|\partial_{k_y} u_{\alpha\bm{k}} \right\rangle-\left\langle \partial_{k_y}u_{\alpha\bm{k}}|\partial_{k_x} u_{\alpha\bm{k}} \right\rangle\right)\,,
\end{equation}
and the intrinsic orbital magnetic moment
\begin{equation}
    m_z^{\alpha}(\bm{k}) = \frac{2\pi}{\Phi_0}\mathrm{Im}\left[\left\langle \partial_{k_x} u_{\alpha\bm{k}}|\hat{H}^{}_{\bm{k}}-\varepsilon_{\alpha\bm{k}}|\partial_{k_y}u_{\alpha\bm{k}}\right\rangle\right]\,,
    \label{IOMM}
\end{equation}
of the $\alpha$-th Bloch eigenstate $|u_{\alpha\bm{k}}\rangle$ with energy $\varepsilon_{\alpha\bm{k}}$. The intrinsic orbital magnetic moment defined in Eq.~\eqref{IOMM} is semiclassically associated to the self-rotation of a Bloch wavepacket in band $\alpha$ around its center-of-mass~\cite{Sundaram1999,Xiao2005}. 

From hereon, we will refer to Eq.~\eqref{drho_dB} as the \textit{energy-resolved St\v{r}eda response}. To the best of our knowledge, Eq.~\eqref{drho_dB} has not been reported in the existing literature. We note that it can be alternatively derived from the modified phase-space density introduced in Ref.~\cite{Xiao2005}; see Appendix~\ref{A1}. 

Interestingly, the energy-resolved St\v{r}eda response can be used to identify ``hot spots" of Berry curvature, as we will illustrate below based on a specific example [see Fig.~\ref{dos_dB_haldane}]. In this sense, the physical quantity in Eq.~\eqref{drho_dB} provides an appealing and elegant way to probe the geometric fine-structure of Bloch bands~\cite{Liu2023}.

Throughout this derivation, we have assumed that the lattice Hamiltonian is defined on a torus geometry, such that the total flux threading the system $\Phi\!=\!B A$ should be quantized according to the Dirac quantization condition:~$\Phi/\Phi_0 \in \mathbb{Z}$. As a corollary, the minimum magnetic-field perturbation that can be applied corresponds to $\Delta B\!=\!\Phi_0/A$. In this sense, the magnetic-field derivative in Eq.~\eqref{drho_dB} should only be interpreted as such in the thermodynamic limit (i.e.~for a macroscopic system area). 

The spectral flow induced by the magnetic perturbation, at fixed chemical potential $\mu$, is given by the integral
\begin{equation}
    \mathcal{W}(\mu) = \Phi_0 \int_{-\infty}^{\infty}d\omega\, f(\omega)\frac{\partial \rho(\omega)}{\partial B} = \Phi_0 \left.\frac{\partial N(\mu)}{\partial \Phi}\right|_{B=0}\,.
    \label{sp_flow_eq}
\end{equation}
The spectral flow thus corresponds to the \textit{St\v{r}eda response}:~it quantifies the variation of the total number of particles at fixed chemical potential, $N(\mu)$, upon applying an external magnetic flux. Using Eq.~\eqref{drho_dB}, we readily find the expression
\begin{eqnarray}
\notag
   \mathcal{W}(\mu)\!&=&\!\!\int_{\mathrm{BZ}}\!\!\frac{d^2k}{2\pi}\sum_{\alpha}\bigg[\!\mathcal{F}_{xy}^{\alpha}(\bm{k})f(\varepsilon_{\alpha\bm{k}})+ \frac{\Phi_0}{2\pi}m_z^{\alpha}(\bm{k})\frac{\partial f(\varepsilon_{\alpha\bm{k}})}{\partial \mu}\!\bigg]\,.\\
   \label{eq_streda}
\end{eqnarray}
The first term in Eq.~\eqref{eq_streda} encodes the topological information of the filled Fermi sea, while the second term, which only involves the states near the Fermi surface, has a non-topological nature and it is generically finite for a metal. We also note that Eq.~\eqref{eq_streda} is in agreement with the modern theory of orbital magnetization in crystalline solids~\cite{Sundaram1999,Xiao2005,Thonhauser2005,Ceresoli2006,Shi2007,Resta2010,Atencia2024}. Indeed, it can be alternatively obtained by making use of the thermodynamic Maxwell relation $\partial N(\mu)/\partial \Phi = \partial \mathcal{M}(\mu)/\partial \mu$, where $\mathcal{M}(\mu)$ denotes the orbital magnetization density. 

In the limit of zero temperature, and for an insulating state, the topological term is the only surviving contribution in Eq.~\eqref{eq_streda}, and we recover the well-known result
\begin{eqnarray}
    \notag
    \mathcal{W}(\mu) &=& \frac{1}{2\pi} \int_{\mathrm{BZ}} d^2k  \sum_{\alpha}\Theta(\mu-\varepsilon_{\alpha\bm{k}})\mathcal{F}_{xy}^{\alpha}(\bm{k})\,,\\
    &=& \sum_{\alpha \in \mathrm{occ}} C_{\alpha}\,,
    \label{str_ins}
\end{eqnarray}
where $C_{\alpha}$ denotes the Chern number of the $\alpha$-th Bloch band, and where the sum runs over all occupied bands. Equation~\eqref{str_ins} quantifies the quantized Hall conductivity of non-interacting insulating states of matter~\cite{Thouless1982}. Importantly, when considering a system with a boundary, the spectral flow is carried by the chiral modes that propagate around the edge of the system. In this realistic geometry, the quantized response in Eq.~\eqref{str_ins} reflects the number of edge channels at the chemical potential $\mu$, as dictated by the bulk-edge correspondence. 

In this non-interacting framework, the number of particles $N(\mu)$ can be expressed as a first-order winding number of the $\mu$-dependent propagator $\hat{G}(\omega) = [(i\omega+\mu)\hat{I}-\hat{H}]^{-1}$ defined in frequency space~\cite{Seki2017}
\begin{equation}
    N(\mu) = N_1[G] = -\frac{1}{2\pi}\int_{-\infty}^{\infty}d\omega e^{i\omega 0^{+}}\mathrm{Tr}\left[\hat{G}^{-1}(\omega)\frac{\partial \hat{G}(\omega)}{\partial \omega}\right]\,.
    \label{N1_G}
\end{equation}
In this formalism, the St\v{r}eda response in Eq.~\eqref{sp_flow_eq} can be written in the form of a higher-order winding number of the Bloch-Green's function in the absence of magnetic field, namely~\cite{Prodan2016,PeraltaGavensky2023}
\begin{eqnarray}
\notag
   \mathcal{W}(\mu) &=& N_3[G]\\
   \notag
   &=&\!\!\frac{\epsilon^{z j l}}{8\pi^2}\int_{\mathrm{BZ}} d^2 k \!\!\int_{-\infty}^{\infty} d \omega e^{i\omega 0^{+}}\mathrm{tr}\left[ \hat{G}^{-1}_{\bm{k}}(i\omega)\frac{\partial \hat{G}^{}_{\bm{k}}(i\omega)}{\partial \omega}\right.\\
    & &\left. \hat{G}^{-1}_{\bm{k}}(i\omega)\frac{\partial \hat{G}^{}_{\bm{k}}(i\omega)}{\partial k_{j}}\hat{G}^{-1}_{\bm{k}}(i\omega)\frac{\partial \hat{G}^{}_{\bm{k}}(i\omega)}{\partial k_{l}}\right]\,,
    \label{N3}
\end{eqnarray}
where $\hat{G}_{\bm{k}}(i\omega)\!=\![(i\omega + \mu)\hat{I}_{\bm{k}}-\hat{H}^{}_{\bm{k}}]^{-1}$. This winding number remains quantized as long as this propagator does not have poles at zero frequency, i.e.~as long as the Hamiltonian does not have eigenvalues at the chemical potential $\mu$. We remark that Eq.~\eqref{N3} reduces to Eq.~\eqref{eq_streda} upon performing the frequency integration. Altogether, the St\v{r}eda response in Eq.~\eqref{sp_flow_eq} can be seen as a fundamental relation between a higher-order winding number and the flux-derivative of a first-order winding number~\cite{Prodan2016,PeraltaGavensky2023},
\begin{equation}
   \mathcal{W}(\mu) =  \Phi_0 \left.\frac{\partial N_1[G]}{\partial \Phi}\right|_{B=0} = N_3[G]\,.
   \label{eq_N1_N3}
\end{equation}
Interestingly, similar relations will be established below in the context of non-equilibrium Floquet systems; see Sec.~\ref{GSF}.
\subsection{Numerical analysis of the St\v{r}eda responses}
To illustrate the information encoded in both the energy-resolved St\v{r}eda response and the (integrated) St\v{r}eda response,  we now numerically evaluate the analytical expressions in Eqs.~\eqref{drho_dB} and~\eqref{eq_streda} using the Haldane model~\cite{Haldane1988}: a two-dimensional honeycomb lattice with on-site energies $\pm \Delta$ on each sublattice, first nearest-neighbour hoppings $J$ and complex next-nearest-neighbour hoppings $J' e^{\pm i\phi}$, which enable the breaking of time-reversal symmetry. At half-filling, the model has a topological phase transition at
\begin{equation}
  J' = J_c = \left|\frac{\Delta}{3\sqrt{3}\sin(\phi)}\right|\,, 
\end{equation}
the critical hopping strength separating a trivial insulator phase ($J'<J_c$), with zero Chern number bands, from a topological Chern insulator phase ($J'>J_c$), with the valence band having $C\!=\!\pm 1$ depending on the sign of $\phi$. For $J'\!=\!0$ and $\Delta \neq 0$ the system's spectrum has two equal-gap valleys exhibiting a finite Berry curvature, which is the same in magnitude but opposite in sign. As $J'$ is increased from zero, the gap in one of the valleys decreases while the other one increases monotonously, up to the critical point $J_c$ where the band gap closes and the topological phase transition occurs. 

\begin{figure}[t]
    \centering
    \includegraphics[width=0.95\columnwidth]{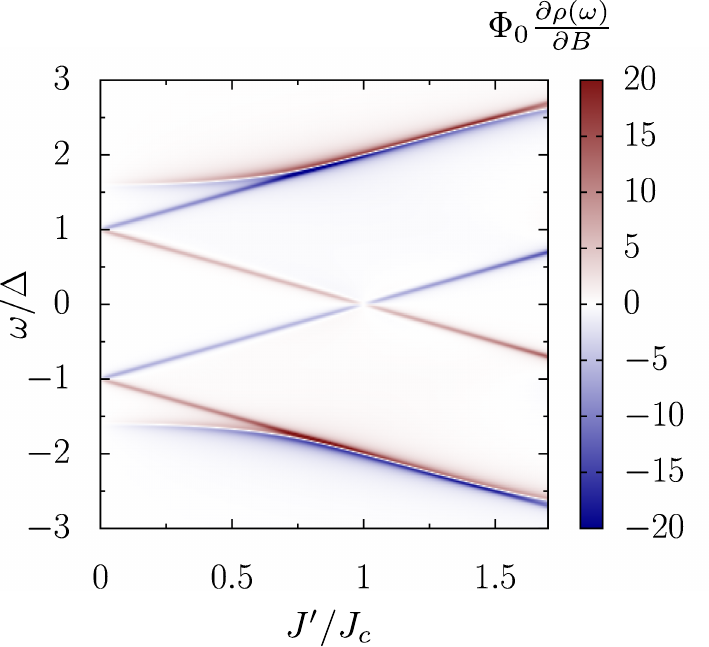}
    \caption{Energy-resolved St\v{r}eda response (in units of $1/J$) of the Haldane model with periodic boundary conditions [see Eq.~\eqref{drho_dB}] as a function of the next-nearest-neighbour hopping strength $J'/J_c$. The parameters are such that $\Delta=0.8\,J$ and $\phi = \pi/2$. }
    \label{dos_dB_haldane}
\end{figure}

This phenomenology is well captured by the energy-resolved St\v{r}eda response [Eq.~\eqref{drho_dB}], which is displayed in Fig.~\ref{dos_dB_haldane} as a function of $J'$, for $\phi\!=\!\pi/2$ and $\Delta\!=\!0.8\,J$.  The bifurcation of Berry curvature ``hot spots"  of opposite sign is clearly visible as the next-nearest-neighbour tunneling is increased. Indeed, when $J'<J_c$, the geometric fine structure of the topologically trivial bands is physically revealed by the flow of states from one valley to the other generated by the external magnetic field. The response is also highly sensitive to van-Hove singularities occurring at higher energies. For $J'>J_c$, the gap reopens but there is a relative change of sign of the low-energy response, a clear signature of the band-inversion phenomenology that occurs after the topological phase transition. 

\begin{figure}[t]
    \includegraphics[width=0.95\columnwidth]{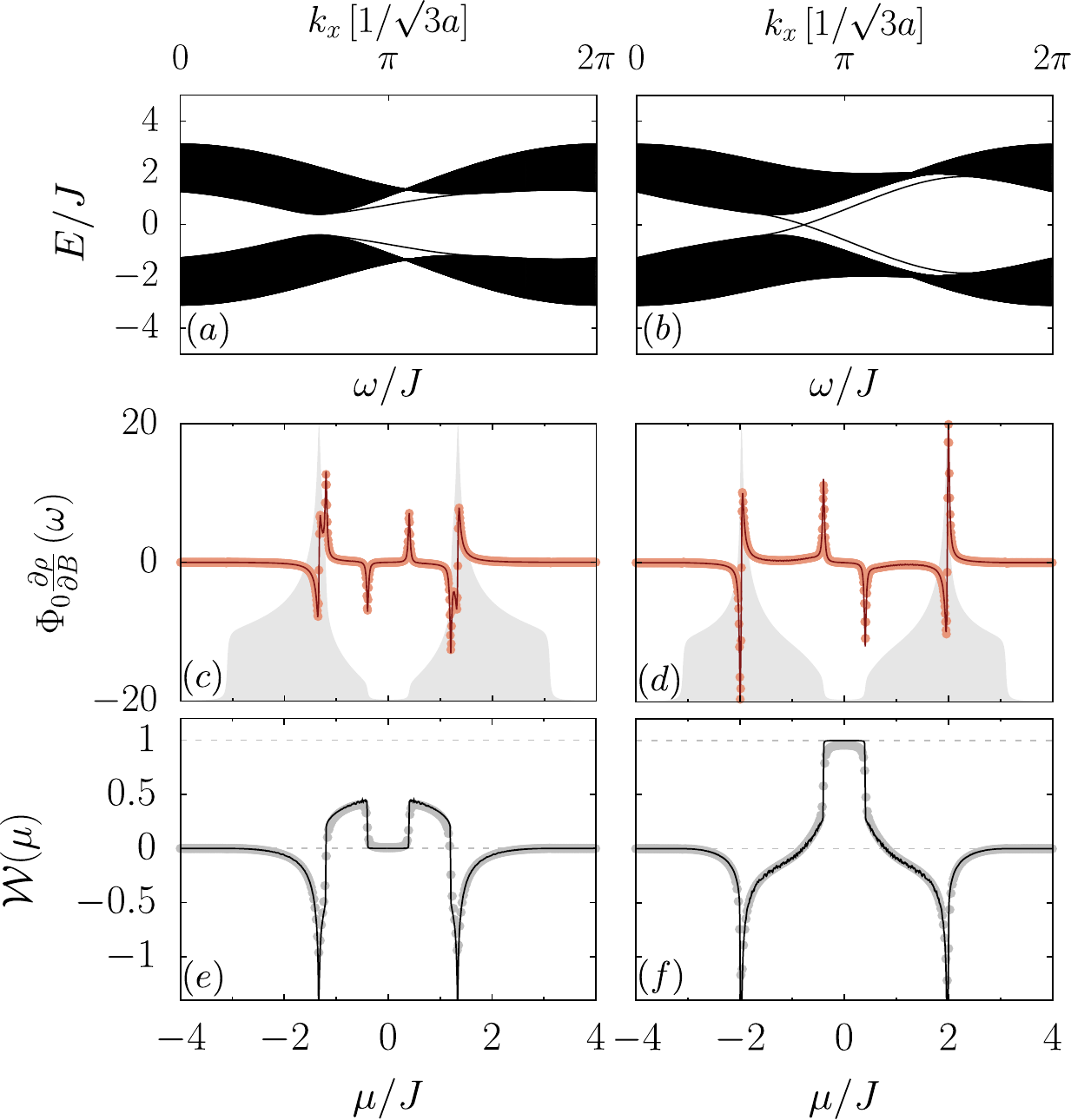}
    \caption{Energy spectrum of a zigzag ribbon of the Haldane model with $N_y = 150$ cells as a function of quasimomentum $k_x$ in $(a)$ the trivial phase ($J' = J_c/2$) and $(b)$ the topological phase ($J' = 1.5\,J_c$). Here $\phi=\pi/2$ and $\Delta=0.8\,J$. Panels $(c)$ and $(d)$ show, respectively, the energy-resolved St\v{r}eda response of the ribbon (filled points) and the same response computed with periodic boundary conditions (solid lines) [see Eq.~\eqref{drho_dB}], in units of $1/J$. To improve the visualization of the curves, we have chosen to broaden the poles of the propagator with an imaginary part of $0^{+}=0.02\,J$. The filled grey area indicates the DOS at $B=0$ in arbitrary units. Panels $(e)$ and $(f)$ show, respectively, the St\v{r}eda response as a function of chemical potential $\mu$ for the ribbon (filled points) and for the system with periodic boundary conditions (solid lines). The temperature has been here set to be zero, such that $f(\omega)=\Theta(\mu-\omega)$.}
    \label{dos_dB_finite}
\end{figure}

In the non-trivial topological regime, the magnetic field generates a net spectral flow from the upper band to the lower band, hence leading to a finite St\v{r}eda response [Eq.~\eqref{eq_streda}] when setting the chemical potential in the gap; this is shown in Fig.~\ref{dos_dB_finite}$(e)$-$(f)$. In an open-boundary sample, this flow of states is generated via the chiral edge states that bridge the gap [Fig.~\ref{dos_dB_finite}$(b)$], revealing the deep connection between St\v{r}eda's formula and the bulk-boundary correspondence. 

Although this analysis has been carried out by using the Bloch representation, the St\v{r}eda response of sufficiently large finite-size systems still preserves the features predicted by Eqs.~\eqref{drho_dB} and~\eqref{eq_streda}. In Figs.~\ref{dos_dB_finite}$(a)$ and \ref{dos_dB_finite}$(b)$, we show the spectrum of a  Haldane zigzag ribbon as a function of the quasimomentum in its longitudinal direction. In Fig.~\ref{dos_dB_finite}$(a)$ we have chosen $J'\!=\!J_c/2$ such that the system is in the trivial phase, while in Fig.~\ref{dos_dB_finite}$(b)$ the hopping strength is $J'\!=\!1.5\,J_c$, so that the system is in a non-trivial Chern insulating phase with chiral edge channels bridging the gap. The energy-resolved St\v{r}eda response of this open-boundary sample is shown for these two configurations with filled points in Figs.~\ref{dos_dB_finite}$(c)$ and \ref{dos_dB_finite}$(d)$, respectively, presenting an excellent agreement with the analytical response previously shown in Fig.~\ref{dos_dB_haldane} (solid lines). In Figs.~\ref{dos_dB_finite}$(e)$ and \ref{dos_dB_finite}$(f)$ we show its integrated version, the St\v{r}eda response $\mathcal{W}(\mu)$, as a function of the chemical potential in the limit of zero temperature. In the gapped regions, this response function approaches the Chern number of the occupied states ($C=0$ in the trivial phase and $C=1$ in the topological phase). The small difference between the response of the ribbon and that obtained from Eq.~\eqref{sp_flow_eq} is due to finite size effects and the broadening of the propagator poles.

The spectral flow under flux insertion, as defined by Eq.~\eqref{sp_flow_eq}, presents itself as a fundamental quantity for characterizing the topological properties of two-dimensional Chern insulators in thermodynamic equilibrium. It is the aim of the following section to generalize this powerful notion to out-of-equilibrium systems subjected to a time-periodic drive.
\section{Floquet-St\v{r}eda response in Sambe space}~\label{Sec_Floquet}
In this section, we generalize the formalism developed in Sec.~\ref{Sec_eq} with the aim of evaluating the St\v{r}eda response of a lattice system set out of equilibrium by a periodic drive of frequency $\Omega$. In Sec.~\ref{Sec_Floquet_Sambe}, we introduce the Sambe formalism that maps driven Floquet problems to static ones in an infinite dimensional extended (Sambe) space $\mathcal{S}$. In Sec.~\ref{Sec_unbounded_problem} we formulate the problem of evaluating the magnetic response of the Floquet density of states in Sambe space under flux insertion. Since this spectrum is unbounded from below, the introduction of regularization schemes is required to compute this quantity. A truncation approach akin to the one used in Ref.~\cite{Rudner2013} is discussed in Sec.~\ref{Sec_Truncation}. In Sec.~\ref{Sec_Floquet_cesaro}, we introduce a regularization procedure based on Ces\`aro summation methods, which better highlights the low-energy nature of the spectral flow. Finally, in Sec.~\ref{Sec_flow-Bloch}, we particularize the results of Sec.~\ref{Sec_Floquet_cesaro} for a translationally invariant lattice system.
\subsection{The Sambe space formulation}
\label{Sec_Floquet_Sambe}
The Schr\"odinger equation of Hamiltonians with discrete time-translational symmetry $\hat{H}(t)=\hat{H}(t+T)$, where $T=2\pi/\Omega$, admits stationary solutions called \textit{Floquet states}~\cite{Shirley1965,Sambe1973} $|\psi_{a}(t)\rangle = e^{-i\varepsilon_{a}t}|u_{a}(t)\rangle$; we take $\hbar\equiv1$ from hereon. Here $\varepsilon_{a}$ stands for the \textit{quasienergy} and $|u_{a}(t)\rangle$ is a $T$-periodic vector in Hilbert space, which satisfies
\begin{equation}
\hat{H}^F(t)|u_{a}(t)\rangle = \varepsilon_{a}|u_{a}(t)\rangle\,,
\label{Floquet_modes}
\end{equation}
where 
\begin{equation}
\hat{H}^{F}(t)=\hat{H}(t)-i\partial_t\,,    
\label{HF_Hilbert}
\end{equation}
is known as the Floquet Hamiltonian or quasienergy operator. The Floquet modes $|u_{a}(t)\rangle$ can be decomposed into Fourier harmonics as
\begin{equation}
    |u_{a}(t)\rangle = \sum_{n=-\infty}^\infty e^{-i n \Omega t}|u_{a}^{(n)}\rangle\,,
\end{equation}
where $n$ are integer numbers. They satisfy the time-independent equation
\begin{eqnarray}
 \sum_{n=-\infty}^\infty\left(\hat{H}_{m-n}-m\Omega\,\delta_{mn}\hat{I}\right)|u^{(n)}_{a}\rangle &=& \varepsilon_{a} |u_{a}^{(m)}\rangle\,,\quad \forall m,\,\,\,
\label{HFeq}   
\end{eqnarray}
where $\hat{H}_{m-n} = (1/T)\int_{0}^{T}dt e^{i(m-n)\Omega t}\hat{H}(t)$. Note that if the set $\{|u_{a}^{(n)}\rangle\}$ satisfies Eq.~\eqref{HFeq} with eigenvalue $\varepsilon_{a}$ then there is an infinite number of other solutions with eigenvalues $\varepsilon_{a}+s\Omega$ given by the set $\{\ket{u_{as}^{(n)}}\equiv|u_{a}^{(n+s)}\rangle\}$, with $s$ being an arbitrary integer number. Since $|\psi_{as}(t)\rangle=|\psi_{a}(t)\rangle$ all these solutions are physically equivalent, and hence it would be sufficient to consider only those that lie inside a given quasienergy window of width $\Omega$, also called a Floquet zone. Nevertheless, for convenience, one can also adopt an extended-zone picture and write all the solutions to Eq.~\eqref{HFeq} as the ones satisfying 
\begin{equation}
    \sum_{n=-\infty}^\infty \hat{\bm{H}}^{F}_{mn}|u^{(n+s)}_{a}\rangle = \varepsilon_{a s} |u_{a}^{(m+s)}\rangle\,, \quad \forall m,s ,
    \label{setEq}
\end{equation}
where $\hat{\bm{H}}^{F}_{mn} = \hat{H}_{m-n}-m\Omega\,\delta_{mn}\hat{I} $, $\varepsilon_{a s}= \varepsilon_{a}+s\Omega$ and $a$ labels the solutions within a given Floquet zone. 
Throughout this work, we will choose the convention $ \varepsilon_{a} \in (\varepsilon_{\pi}-\Omega,\varepsilon_{\pi}]$, which corresponds to the natural Floquet zone (NFZ) defined in Ref.~\cite{Nathan2015}. A special energy gap can be present at the edge of the NFZ, the so-called zone-edge gap~\cite{Nathan2015}; it is centered around $\varepsilon_{\pi}$ and becomes infinitely large in the limit  $\Omega\!\rightarrow\!\infty$. In systems with particle-hole or chiral symmetry, one naturally sets $\varepsilon_{\pi} = \Omega/2$~\cite{Roy2017}. 

These set of equations [Eq. \eqref{setEq}] can be written in the form of a time-independent Schr\"odinger equation by introducing the infinite-dimensional Sambe representation, namely 
\begin{equation}
    \hat{\bm{H}}^{F}|u^{}_{a s}\rangle\rangle = \varepsilon_{a s} |u_{a s}^{}\rangle\rangle\,.
    \label{boldHF}
\end{equation}
Here $|u_{a s}\rangle\rangle$ are eigenstates of the Floquet Hamiltonian $\hat{\bm{H}}^{F}$ in the extended Sambe space $\mathcal{S} = \mathcal{H}\otimes \mathcal{T}$, given by the product of the original Hilbert space $\mathcal{H}$ with the space of square-integrable $T$-periodic functions~\cite{Shirley1965,Sambe1973}. Namely, 
\be
\Ket{u_{as}}=\left(\dots,\ket{u_{as}^{(1)}},\ket{u_{as}^{(0)}},\ket{u_{as}^{(-1)}},\dots\right)^\mathrm{T}\,,
\ee
or, equivalently, the $n$-th block element of $|u_{as}\rangle\rangle$ is given by $\Ket{u_{as}}_n=|u_{as}^{(n)}\rangle=\ket{u_a^{(n+s)}}$. 
The orthogonality relation in this space is given by
\begin{equation}
\langle\langle u_{a s} | u_{b s'}\rangle\rangle = \sum_{n=-\infty}^\infty\langle u_{a s}^{(n)}|u_{b s'}^{(n)}\rangle = \delta_{ab}\delta_{ss'}.
\label{ort}
\end{equation}
In the following, we use the shorthanded notation $|u_{a }\rangle\rangle\!\equiv\!|u_{a 0}\rangle\rangle$ to denote states in the NFZ. 

In the presence of discrete translational symmetry, the Floquet Hamiltonian can be Fourier transformed to quasimomentum space, such that the quantum number $a = (\alpha,\bm{k})$, with $\bm{k}$ being the quasimomentum and $\alpha$ the Bloch band index. In this case, we have that,
\begin{equation}
 \hat{\bm{H}}^F_{\bm{k}}|u_{\alpha s \bm{k}}\rangle\rangle = \varepsilon_{\alpha s\bm{k}}|u_{\alpha s \bm{k}}\rangle\rangle\,,   
\end{equation}
where $\hat{\bm{H}}_{\bm{k}}^{F}$ is the corresponding Floquet-Bloch Hamiltonian in $\mathcal{S}$. Here $\varepsilon_{\alpha s\bm{k}} = \varepsilon_{\alpha\bm{k}} + s\Omega$ stand for the quasienergies of the Bloch system. In the following, we use the notation $|u_{\alpha \bm{k}}\rangle\rangle\!\equiv\! |u_{\alpha 0\bm{k}}\rangle\rangle$ to denote the $\alpha$-th Floquet-Bloch state in the NFZ, i.e.~ the $s\!=\!0$ state.
\subsection{Spectral flow of an unbounded operator}\label{Sec_unbounded_problem}
The core of this work relies on the evaluation of  the magnetic response of the $\Omega$-periodic Floquet density of states, which can be generically expressed in terms of the single-particle Floquet Green's function as
\begin{eqnarray}
\label{DOS_F_G}
    \rho^{F}(\omega) &=& -\frac{1}{\pi A}\mathrm{Im}\textbf{Tr}\left[\hat{\bm{G}}^{F}(\omega+i 0^{+})\right]\,,
\end{eqnarray}
where we have introduced the bold trace $\textbf{Tr}[\hdots]$ to define the trace in $\mathcal{S}$. In the absence of interactions, 
$\hat{\bm{G}}^{F}(\omega)=(\omega\bm{I} - \hat{\bm{H}}^{F})^{-1}$, such that Eq.~\eqref{DOS_F_G} takes the simple form
\begin{eqnarray}
\label{DOS_F}
    \rho^{F}(\omega) &=& \frac{1}{A}\sum_{a,s}\delta(\omega-\varepsilon_{a s})\,,
\end{eqnarray}
where we remind that the sum over $a$ is restricted such that $\varepsilon_a \in \mathrm{NFZ}$ and the sum over $s$ runs along all $\mathbb{Z}$.

We now introduce the central quantity of the present work: the net spectral flow of the unbounded Floquet spectrum under flux insertion. The \textit{Floquet spectral flow} at quasienergy $\mu_\varepsilon$ is formally defined as the following integral
\begin{equation}
    \mathcal{W}(\mu_\varepsilon) = \Phi_0\int_{-\infty}^{\mu_\varepsilon}d\omega\, \frac{\partial \rho^{F}(\omega)}{\partial B}\,.
    \label{WF}
\end{equation}
 We note that $\mathcal{W}(\mu_\varepsilon)\!=\!\mathcal{W}(\mu_\varepsilon+\Omega)$, such that one can restrict $\mu_\varepsilon$ to the NFZ, without loss of generality. We remark that Eq.~\eqref{WF} is well defined both on a torus and for samples with open boundary conditions (clean or disordered); see Sec. \ref{Sec_local-marker}.

For the particular case of a translationally invariant system, the Floquet propagator in Eq.~\eqref{DOS_F_G} can be expanded up to first order in the magnetic field by generalizing the formalism used in Sec.~\ref{Sec_eq} to Sambe space (see Appendix~\ref{Kita_Arai_Floquet}). As a result, 
one finds that the magnetic response of the Floquet DOS in Eq.~\eqref{DOS_F}, evaluated at zero magnetic field, is given by
\begin{eqnarray}
\notag
    \Phi_0 \frac{\partial \rho^{F}(\omega)}{\partial B} &=&
    \int_{\mathrm{BZ}}\frac{d^2k}{2\pi}\sum_{\alpha,s}\bigg[\mathcal{F}_{xy}^{\alpha}(\bm{k})\delta(\omega-\varepsilon_{\alpha s \bm{k}})\\
     \label{drhoF_dB}
    &+&\frac{\Phi_0}{2\pi}m^{\alpha}_z(\bm{k})\frac{\partial}{\partial \omega}\delta(\omega-\varepsilon_{\alpha s \bm{k}})\bigg]\,,
\end{eqnarray}
where we identified the Floquet Berry curvature
\begin{equation}
   \mathcal{F}_{xy}^{\alpha}(\bm{k}) = i \left(\left\langle\left\langle \partial_{k_x}u_{\alpha\bm{k}}|\partial_{k_y} u_{\alpha\bm{k}} \right\rangle\right\rangle-\left\langle\left\langle \partial_{k_y}u_{\alpha\bm{k}}|\partial_{k_x} u_{\alpha\bm{k}} \right\rangle\right\rangle\right)\,,
   \label{curvature_F}
\end{equation}
and the Floquet intrinsic orbital magnetic moment~\cite{Gao2022}
\begin{equation}
    m_z^{\alpha}(\bm{k}) = \frac{2\pi}{\Phi_0}\mathrm{Im}\left[\left\langle \left\langle\partial_{k_x} u_{\alpha\bm{k}}|\hat{\bm{H}}^{F}_{\bm{k}}-\varepsilon_{\alpha\bm{k}}|\partial_{k_y}u_{\alpha\bm{k}}\right\rangle\right\rangle\right]\,,
    \label{magnetization_F}
\end{equation}
of the $\alpha$-th Floquet-Bloch eigenstate in the NFZ. Here we have used that $\mathcal{F}^{\alpha s}_{xy}(\bm{k})\!=\!\mathcal{F}^{\alpha }_{xy}(\bm{k})$ and $m^{\alpha s}_{z}(\bm{k})\!=\!m^{\alpha }_{z}(\bm{k})$. From hereon, we will hence refer to Eq.~\eqref{drhoF_dB} as the \textit{energy-resolved Floquet-St\v{r}eda response}.

 In such a clean Bloch lattice setting, Eq.~\eqref{WF} can equally be expressed in terms of the winding number of the unbounded Floquet-Bloch Green's function, namely
\begin{equation}
    \mathcal{W}(\mu_\varepsilon)  =  N_3[\bm{G}^{F}]\,,
\end{equation}
with
\begin{eqnarray}
\notag
    N_3[\bm{G}^{F}] &=&\!\!\frac{\epsilon^{z j l}}{8\pi^2}\!\!\int_{\mathrm{BZ}} d^2 k \!\!\int_{-\infty}^{\infty}\!\!d \omega\, e^{i\omega 0^{+}}\textbf{tr}\left[ \hat{\bm{G}}^{F-1}_{\bm{k}}\!(i\omega)\frac{\partial \hat{\bm{G}}^{F}_{\bm{k}}\!(i\omega)}{\partial \omega}\right.\\
    \label{N3G_F}
    & &\!\!\left. \hat{\bm{G}}^{F-1}_{\bm{k}}\!(i\omega)\frac{\partial \hat{\bm{G}}^{F}_{\bm{k}}\!(i\omega)}{\partial k_{j}}\hat{\bm{G}}^{F-1}_{\bm{k}}\!(i\omega)\frac{\partial \hat{\bm{G}}^{F}_{\bm{k}}\!(i\omega)}{\partial k_{l}}\right]\,,
\end{eqnarray}
where $\textbf{tr}[\hdots]$ traces over both the Sambe indices and the internal degrees of freedom. In Eq.~\eqref{N3G_F}, we have implicitly defined a $\mu_{\varepsilon}$-dependent propagator $\hat{\bm{G}}^{F}_{\bm{k}}(i\omega) = [(i\omega + \mu_\varepsilon)\hat{\bm{I}}_{\bm{k}}-\hat{\bm{H}}^{F}_{\bm{k}}]^{-1}$. We note that Eqs.~\eqref{drhoF_dB} and~\eqref{N3G_F} have exactly the same form as Eqs.~\eqref{drho_dB} and~\eqref{N3}, with the sole difference being the use of the $\mathcal{S}$ space.

An important remark is in order: both Eqs.~\eqref{WF} and~\eqref{N3G_F} are ma\-the\-matically ill-defined, as they are formally expressed as non-convergent integrals in frequency space. Indeed, the integrand of Eq.~\eqref{WF} is an $\Omega$-periodic function, such that its integral down to $-\infty$ is not well-defined. This statement also holds true for the kernel in Eq.~\eqref{N3G_F}. This conundrum can be restated as a subtraction-of-infinities problem, since Eq.~\eqref{WF} can be recast in the form  
\begin{equation}
    \mathcal{W}(\mu_\varepsilon) = A \left(\int_{-\infty}^{\mu_\varepsilon}d\omega\, \rho^F(\omega,\Phi_0)-\int_{-\infty}^{\mu_\varepsilon}d\omega\, \rho^F(\omega,0)\right),
\end{equation}
where $\rho^{F}(\omega,\Phi_0)$ stands for the Floquet density of states of a system threaded by one flux quantum, and  $\rho^{F}(\omega,0)$ for the same quantity in the absence of flux. Here, we have considered the minimum difference $\Delta B\!=\! \Phi_0/A$ to evaluate the derivative in Eq.~\eqref{WF} as a discrete finite difference in a torus geometry, see Sec.~\ref{Sec_eq}. 

In the following, we will present different approaches to regularize these ill-defined quantities.
\subsection{The truncation approach}\label{Sec_Truncation}
One possible method to evaluate Eq.~\eqref{WF} would be to truncate the infinite dimensional Floquet Hamiltonian and Green's function to finite dimensional operators~\cite{Rudner2013}. These truncated operators only contain a finite number of multiplicities (or Floquet replicas) in Sambe space, such that their spectrum is bounded. In this scenario, the Floquet DOS would no longer be a strictly periodic function.

In particular, in a Bloch lattice, Eq.~\eqref{drhoF_dB} would get modified to
\begin{eqnarray}
\notag
    \Phi_0 \frac{\partial \widetilde{\rho}^{F}(\omega)}{\partial B} &=&
    \int_{\mathrm{BZ}}\frac{d^2k}{2\pi}\sum_{\alpha}\sum_{s=-S}^{S}\bigg[\widetilde{\mathcal{F}}_{xy}^{\alpha s}(\bm{k})\delta(\omega-\widetilde{\varepsilon}_{\alpha s \bm{k}})\\
     \label{drhoF_dB_truncated}
    &+&\frac{\Phi_0}{2\pi}\widetilde{m}^{\alpha s}_z(\bm{k})\frac{\partial}{\partial \omega}\delta(\omega-\widetilde{\varepsilon}_{\alpha s \bm{k}})\bigg]\,,
\end{eqnarray}
where $\widetilde{\mathcal{F}}^{\alpha s}_{xy}(\bm{k})$ and $\widetilde{m}^{\alpha s}_{z}(\bm{k})$ are, respectively, the Berry curvature and the intrinsic orbital magnetic moment of the modified Floquet-Bloch eigenstate $|\widetilde{u}_{\alpha s\bm{k}}\rangle\rangle$ of the truncated Hamiltonian, with quasienergy $\widetilde{\varepsilon}_{\alpha s \bm{k}}$. In Eq.~\eqref{drhoF_dB_truncated}, the index $s$ only runs over $2 S + 1$ terms, and Eq.~\eqref{drhoF_dB_truncated} reduces to Eq.~\eqref{drhoF_dB} in the limit $S \rightarrow \infty$. Since the components $|\widetilde{u}^{(n)}_{\alpha s \bm{k}}\rangle\rangle$ decay exponentially to zero when $|n-s| \gg 1$, the low-energy sector (i.e.~the Floquet zones around $s\!=\!0$) of the truncated Hamiltonian provides a reliable approximation for describing the low-energy physics of the exact Floquet Hamiltonian, and hence the presence of topological edge modes, as long as the truncation is performed using a large enough $S$~\cite{Rudner2013}. Specifically, $S$ should be such that $\langle u_{\alpha\bm{k}}^{(\pm S)}|u_{\alpha\bm{k}}^{(\pm S)}\rangle \ll 1$. 
In this case, one can safely state that, when $\mu_\varepsilon$ lies in a spectral gap, the Floquet spectral flow can be obtained as
\begin{eqnarray}
\notag
     \mathcal{W}(\mu_\varepsilon)&=& \Phi_0\int_{-\infty}^{\mu_\varepsilon}d\omega\frac{\partial \widetilde{\rho}^{F}(\omega)}{\partial B}\,,\\
     \notag
     &=& \int_{\mathrm{BZ}}\frac{d^2k}{2\pi}\sum_{\alpha}\sum_{s=-S}^{S}\widetilde{\mathcal{F}}_{xy}^{\alpha s}(\bm{k})\Theta(\mu_\varepsilon-\widetilde{\varepsilon}_{\alpha s \bm{k}})\,,\\
     &=&\sideset{}{'}\sum_{\alpha}\sideset{}{'}\sum_{s=-S}^{S}\widetilde{C}_{\alpha s}\,,
\end{eqnarray}
where the prime indicates that the summations only involve indices such that  $\widetilde{\varepsilon}_{\alpha s \bm{k}}\!<\!\mu_\varepsilon$, and where $\widetilde{C}_{\alpha s}$ denotes the Chern number associated with the modified Floquet-Bloch eigenstates $|\widetilde{u}_{\alpha s\bm{k}}\rangle\rangle$. 
While this procedure gives the correct answer for large enough $S$ \cite{Rudner2013,PerezPiskunow2015}, it relies on the computation of the modified Chern numbers $\widetilde{C}_{\alpha s}$ of \textit{all} the quasienergy bands of the truncated Floquet Hamiltonian that are located below $\mu_\varepsilon$. Quite generally, in the presence of anomalous Floquet modes, the Chern numbers associated with the lower part of the truncated spectrum (i.e.~bands with $s\!\sim\!-S$) turn out to be the more relevant ones to correctly quantify $\mathcal{W}(\mu_\varepsilon)$. 

In Sec.~\ref{Sec_Floquet_cesaro}, we develop a different method of regularization that does not rely on the truncation of the Floquet Hamiltonian and that genuinely reflects the low-energy properties of the bulk quasienergy bands of the system. This alternative approach will provide substantial physical intuition of the spectral flow of Floquet systems under flux insertion, as we discuss in Sec.~\ref{Sec_PhysInt}. 
\subsection{Ces\`aro regularization of the Floquet-St\v{r}eda response}\label{Sec_Floquet_cesaro}
It is useful to rewrite the Floquet spectral flow, given by Eq.~\eqref{WF}, as the sum of two contributions,
\begin{eqnarray}
\label{decomposition}
    \mathcal{W}(\mu_\varepsilon) = \mathcal{W}^{N}(\mu_\varepsilon) + \mathcal{W}^{A\,},
\end{eqnarray}
where
\begin{eqnarray}
\label{WFN}
   \mathcal{W}^{N}(\mu_\varepsilon)&=&\Phi_0 \int_{\varepsilon_{\pi}-\Omega}^{\mu_\varepsilon}d\omega\frac{\partial\rho^{F}(\omega)}{\partial B}\,,
\end{eqnarray}
and
\begin{equation}
     \mathcal{W}^A = \Phi_0\int_{-\infty}^{\varepsilon_{\pi}-\Omega}d\omega\frac{\partial\rho^{F}(\omega)}{\partial B}\, ,
    \label{WFA}
\end{equation}
and we remind that $\mu_\varepsilon$ is restricted to the interval $(\varepsilon_{\pi}-\Omega,\varepsilon_{\pi}]$ (NFZ). Throughout this derivation, we assume that there is a well defined  zone-edge gap located at $\varepsilon_\pi$ \footnote{If such a gap does not exist, that is, for the case of a driving protocol the does not open gaps at resonance, the Floquet zone must be chosen to locate another gap at its boundary~\cite{Roy2017}. Our approach can equally handle such a case, but for simplicity we do not consider it explicitly here}. Importantly, the NFZ is \textit{fixed} and remains unchanged as one activates the $B$ field, such that the derivative with respect to the magnetic field commutes with the integrals above.
From hereafter, we will refer to $\mathcal{W}^{N}(\mu_\varepsilon)$ as the \textit{normal spectral flow} and $\mathcal{W}^{A}$ as the \textit{anomalous spectral flow}.
To progress, it is important to note that the kernel of Eq.~\eqref{WFA} integrates to zero within any Floquet zone, 
\begin{equation}
 \Phi_0  \int_{\varepsilon_{\pi, s-1}}^{\varepsilon_{\pi,s}} d\omega  \frac{\partial \rho^{F}(\omega)}{\partial B} = 0\,, \quad \forall s\in \mathbb{Z} ,
 \label{prop}
\end{equation}
where we defined $\varepsilon_{\pi,s}=\varepsilon_{\pi}+ s\Omega$. This means that the kernel of the integral defining the anomalous spectral flow oscillates with a zero mean down to minus infinity, such that the integral in Eq.~\eqref{WFA} is formally not convergent. A similar pathological situation occurs for the sine function when considering the integral $\int_{-\infty}^{0}dy\sin(y)$. 
\begin{figure}[t]
    \centering
    \includegraphics[width=0.95\columnwidth]{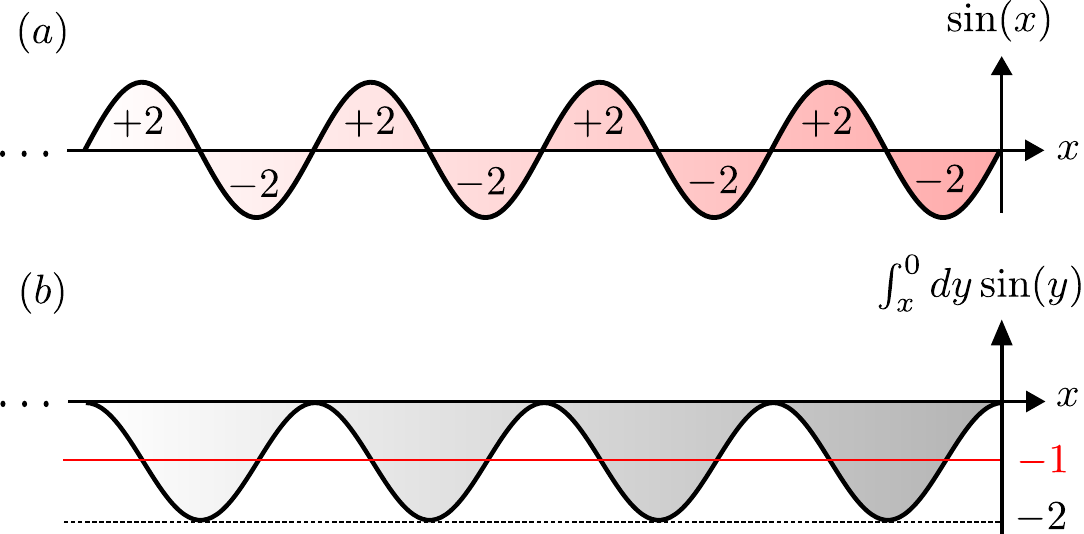}
    \caption{$(a)$ Illustration of $\sin(x)$, a periodic function with zero mean. The values $\pm 2$ indicate the area below each of the intervals $[(n-1)\pi/2,n\pi/2]$ for $n \in \mathbb{Z}$ and $n\leq 0$. The integral of this function between $-\infty$ and $0$ can then be thought of as the discrete Grandi series $2\sum_{n=1}^{\infty} (-1)^{n}$. $(b)$ Integral of the sine function in the interval $[x,0]$ as a function of $x$. The value $-1$, marked with a solid red line, corresponds to the Ces\`aro regularization of the non-convergent integral $\int_{-\infty}^{0}dy\sin(y)$.}
    \label{cesaro_scheme}
\end{figure}
As a matter of fact, this type of integrals correspond to integral versions of discrete Grandi type-series, such as $\sum_{n=0}^{\infty}(-1)^n$ (see Fig.~\ref{cesaro_scheme}). Mathematically, these non convergent series are treated by different summation methods, by means of which it is possible to define and assign values to them~\cite{Hardy2000}. 
In particular, a series $\sum_{n=1}^\infty a_n$ is called  ``Ces\`aro summable", with value $\mathcal{Q}$, if the arithmetic mean $\lim_{n\rightarrow\infty}\frac{1}{n}\sum_{k=1}^n s_k=\mathcal{Q}$,  with $s_k=a_1+\dots+a_k$ the $k$-th partial sum.
In the case represented in Fig.~\ref{cesaro_scheme}, the partial sums give $(-2,0,-2,0,\hdots)$ so that $\mathcal{Q}=-1$. More generally, a non-convergent integral is said to be $(C,1)$-summable if
\begin{equation}
    \int_{-\infty}^{0}dy\, g(y) \stackrel{(C,1)}{=} \lim_{\lambda\rightarrow\infty}\int_{-\lambda}^0 dy\,g(y)\left(1+\frac{y}{\lambda}\right)\,,
    \label{C1_reg}
\end{equation}
converges. For $g(y)=\sin(y)$, Eq.~\eqref{C1_reg} recovers the $\mathcal{Q}$-value $-1$.

In order to properly define the anomalous spectral flow in Eq.~\eqref{WFA}, we now perform a $(C,1)$ Ces\`aro regularization of the integral~\cite{Titchmarsh1986,Hardy2000}, namely
\begin{eqnarray}
\notag
\!\mathcal{W}^{A}\!\!&\stackrel{(C,1)}{=}&\!\!\lim_{\lambda \to \infty} \!\Phi_0\!\!\int_{-\lambda}^{\varepsilon_{\pi}-\Omega}\!\!d\omega\, \frac{\partial \rho^{F}(\omega)}{\partial B}\left(1+\frac{\omega}{\lambda}\right)\,,\\
    \notag
    \!\!&=&\!\!\!\!\lim_{S \to \infty}\!\Phi_0\!\sum_{s=1}^{S}\!\!\int_{\varepsilon_{\pi,-(s+1)}}^{\varepsilon_{\pi,-s}}\!\!\!\!d\omega\, \frac{\partial \rho^{F}(\omega)}{\partial B}\!\left(\!1+\frac{\omega}{(S+1)\Omega - \varepsilon_{\pi}}\!\right)\,,\\
    \notag
    &=& \lim_{S \to \infty} \Phi_0 \int_{\mathrm{NFZ}}d\omega\, \frac{\partial \rho^{F}(\omega)}{\partial B} \omega \frac{S}{(S+1)\Omega - \varepsilon_\pi}\,,
    \\
    \label{Cesaro}
    &=&\Phi_0 \int_{\mathrm{NFZ}}d\omega\, \frac{\partial \rho^{F}(\omega)}{\partial B} \frac{\omega }{\Omega}\,.
\end{eqnarray}
Note that we have taken the limit of $\lambda \rightarrow \infty$ by considering that an integer number of Floquet zones are contained within the interval $(-\lambda,\varepsilon_{\pi}-\Omega]$. We have also used the $\Omega$-periodicity of $\rho^F(\omega)$ and Eq.~\eqref{prop}. Even though the last integral in Eq.~\eqref{Cesaro} is performed within the NFZ, we remark that one can shift the interval of integration by any arbitrary multiple of the driving frequency and obtain the same result.

Importantly, the Ces\`aro regularization procedure performed in Eq.~\eqref{Cesaro} leads to an insightful expression for the net Floquet spectral flow in Eq.~\eqref{decomposition}, which highlights two physically distinguishable contributions
\begin{equation}
    \mathcal{W}(\mu_\varepsilon) = \Phi_0\int_{\varepsilon_{\pi}-\Omega}^{\mu_\varepsilon}d\omega \frac{\partial \rho^{F}(\omega)}{\partial B} + \frac{\Phi_0}{\Omega} \int_{\mathrm{NFZ}}^{}d\omega\, \left (\frac{\partial \rho^{F}(\omega)}{\partial B} \omega \right )\, .
    \label{flow_part_energy}
\end{equation}
The first term in Eq.~\eqref{flow_part_energy} represents a flow of \textit{dressed states} in response to the perturbing magnetic field, while the second term, which stems from the anomalous response, characterizes a flow of \textit{energy} within a given Floquet zone. Indeed, the regularized anomalous flow $\mathcal{W}^{A}$ in Eq.~\eqref{Cesaro} can be equally interpreted as the total orbital magnetization density of the Floquet states within a zone, measured in units of $\Omega/\Phi_0$. This key result, yet simple, has profound physical consequences, which will be further studied in Sec.~\ref{Sec_PhysInt}. 

We also note that the anomalous spectral flow can be expressed as
\begin{equation}
    \mathcal{W}^{A} = -\frac{1}{\Omega}\int_{\textrm{NFZ}}d\mu_\varepsilon\, \mathcal{W}^{N}(\mu_\varepsilon) \equiv  -\overline{\langle \mathcal{W}^{N}\rangle}_{\mathrm{NFZ}}\, ,
    \label{WN_average_WA}
\end{equation}
that is to say, it can be obtained as an average of the normal spectral flow within the natural Floquet zone. We note that Eq.~\eqref{WN_average_WA} implies a conservation law, $\int_\mathrm{NFZ}d\mu_\varepsilon\,\mathcal{W}(\mu_\varepsilon)=0$.

In an open-boundary sample with a bulk gap at quasienergy $\mu_\varepsilon$, the Floquet spectral flow is physically carried by the edge channels that bridge the Floquet spectral gaps [Fig.~\ref{conceptual_SF}]. As a corollary, Eq.~\eqref{flow_part_energy} fully quantifies the net number of chiral edge modes that are located in the gaps of periodically-driven settings. Importantly, this result does not rely on having a translationally invariant system, and hence remains valid even in the presence of disorder or inhomogeneities, as long as the bulk gaps remain open.

Evaluating the Floquet spectral flow precisely at the zone-edge gap ($\mu_\varepsilon = \varepsilon_{\pi}$) yields 
\begin{equation}
    \mathcal{W}(\varepsilon_{\pi}) = \mathcal{W}^{A}.
    \label{W_dyngap}
\end{equation}
Indeed, the normal flow vanishes $\mathcal{W}^{N}(\varepsilon_{\pi})\!=\!0$ due to Eq.~\eqref{prop}, and only the anomalous flow contributes to the Floquet-St\v{r}eda response. This general result indicates that the \textit{anomalous flow $\mathcal{W}^{A}$ in Eq.~\eqref{Cesaro} quantifies the number of anomalous edge states} that are located at the edge of the Floquet Brillouin zone.

\subsection{Evaluation of the Floquet spectral flow in a translationally invariant system}\label{Sec_flow-Bloch}

In a lattice system of Bloch particles, one can use Eq.~\eqref{drhoF_dB} to explicitly perform the integrals in Eqs.~\eqref{WFN} and~\eqref{Cesaro}. This yields
\begin{eqnarray}
\label{normal_flow}
    \mathcal{W}^{N}(\mu_\varepsilon) &=&\sum_{\alpha}\int_{\mathrm{BZ}}\frac{d^2k}{2\pi}\left[\mathcal{F}^{\alpha}_{xy}(\bm{k})\Theta(\mu_\varepsilon-\varepsilon_{\alpha\bm{k}})\right.\\
    \notag
    &+&\left.\frac{\Phi_0}{2\pi} m^{\alpha}_{z}(\bm{k})\delta(\mu_\varepsilon-\varepsilon_{\alpha\bm{k}})\!\right]\,,
\end{eqnarray}
and
\begin{equation}
   \mathcal{W}^A \stackrel{(C,1)}{=} \sum_{\alpha}\int_{\mathrm{BZ}}\frac{d^2 k}{2\pi \Omega}\left[\mathcal{F}_{xy}^{\alpha}(\bm{k})\varepsilon_{\alpha\bm{k}} - \frac{\Phi_0}{2\pi} m_z^{\alpha}(\bm{k})\right]\,,
   \label{anomaly}
\end{equation} 
where the Floquet Berry curvatures and intrinsic orbital magnetic moments are defined in Eqs.~\eqref{curvature_F} and~\eqref{magnetization_F}. We note that the form of Eq.~\eqref{anomaly} is independent of the choice of the Floquet Brillouin zone, since $\mathcal{F}^{\alpha s}_{xy}(\bm{k}) = \mathcal{F}^{\alpha}_{xy}(\bm{k})$, $m_z^{\alpha}(\bm{k})=m_z^{\alpha s}(\bm{k})$ and $\sum_{\alpha}\mathcal{F}_{xy}^{\alpha}(\bm{k})  = 0$. Even though we have chosen the $(C,1)$ regularization scheme to obtain this result, different regularizations also lead to the same conclusion, as discussed in Appendix~\ref{alt_regularizations}. 

Assuming that $\mu_\varepsilon$ lies within a gap of the Floquet spectrum, one can express the normal spectral flow $\mathcal{W}^{N}(\mu_\varepsilon)$ in Eq.~\eqref{normal_flow} as 
\begin{equation}
   \mathcal{W}^{N}(\mu_\varepsilon)=  \sum_{\alpha}\int_{\mathrm{BZ}}\frac{d^2k}{2\pi}\mathcal{F}^{\alpha}_{xy}(\bm{k})\Theta(\mu_\varepsilon-\varepsilon_{\alpha\bm{k}})= \sideset{}{'}\sum_{\alpha}C^{F}_{\alpha} ,
   \label{Cherns_FFZ}
\end{equation}
where we have taken $\varepsilon_{\pi}$ to be located in a well-defined zone-edge gap.
In the last equality of Eq.~\eqref{Cherns_FFZ}, we have introduced the Chern numbers $C^{F}_{\alpha}$ of the Floquet bands within the NFZ, and we indicated a prime in the summation on the right-hand side to stress that only the Chern numbers of the bands located below  $\mu_\varepsilon$ are considered. 
 
Combining Eqs.~\eqref{flow_part_energy},~\eqref{anomaly} and~\eqref{Cherns_FFZ} leads us to a general result: the Floquet spectral flow occurring in a spectral gap upon flux insertion is given by
\begin{eqnarray}
\notag
 \mathcal{W}(\mu_\varepsilon) &=& \sideset{}{'}\sum_{\alpha}C^{F}_{\alpha}+ \sum_{\alpha}\int_{\mathrm{BZ}}\frac{d^2 k}{2\pi \Omega}\left[\mathcal{F}_{xy}^{\alpha}(\bm{k})\varepsilon_{\alpha\bm{k}} - \frac{\Phi_0}{2\pi} m_z^{\alpha}(\bm{k})\right]\,.\\
\label{WF_Cesaro}
\end{eqnarray}
Equation~\eqref{WF_Cesaro} is a central result of the present work. It is worthy to emphasize that this formula only relies on the knowledge of low-energy bulk properties of the Floquet Hamiltonian in Sambe space, such as the quasienergies, Berry curvatures and intrinsic orbital magnetic moments of Floquet-Bloch bands within the a given Floquet zone. With these quantities at hand, it is possible to characterize the anomaly $\mathcal{W}^{A}$ that must be added to the Chern numbers of the Floquet-Bloch bands in order to correctly quantify the topology of these driven systems.\\

\section{Quantized anomaly as an orbital magnetization density and a generalized St\v{r}eda formula for Floquet systems}\label{Sec_PhysInt}
This section is primarily devoted to the physical interpretation of the formal results obtained in Sec.~\ref{Sec_Floquet_cesaro}. In Sec.~\ref{stroboscopic_micromotion}, we first discuss how our classification scheme of two-dimensional Floquet systems relates to the more conventional approach used in the literature, and which relies on the properties of the time-evolution operator~\cite{Kitagawa2010,Rudner2013}.  Then, in Sec.~\ref{Sec_OMD}, we present an explicit relation between Eq.~\eqref{flow_part_energy} and the orbital magnetization density of the Floquet states within a Floquet zone. We also show the correspondence between Eq.~\eqref{anomaly} and the recently derived expressions for the orbital magnetization density of Floquet-Bloch lattices. In Sec.~\ref{GSF}, we derive a way to reformulate Eq.~\eqref{flow_part_energy} as a generalized Floquet-St\v{r}eda formula, which better highlights the physical interpretation of the normal and anomalous contributions to the Floquet spectral flow. Finally, in Sec.~\ref{RLBL_Sec}, we illustrate how our theory applies in a simple paradigmatic model of an anomalous Floquet phase.
\subsection{From Sambe to Hilbert space: Stroboscopic dynamics and micromotion}\label{stroboscopic_micromotion}
The time-evolution operator of any periodically-driven system can always be decomposed as~\cite{Rahav2003,Goldman2014,Eckardt2015,Bukov2015}
\be
\hat{U}(t,t')=\hat{R}^{}(t) e^{-i \hat{H}_{\mathrm{eff}}(t-t')} \hat{R}^{\dagger}(t')\, ,
\label{evol_operator}
\ee
where $\hat{R}(t)$ is a unitary $T$-periodic operator, also known as \textit{micromotion operator}, and $\hat{H}_{\mathrm{eff}}$ is a time-independent Hamiltonian known as the \textit{effective Hamiltonian}. This Hamiltonian characterizes the stroboscopic time evolution of Floquet-driven systems, since $U(T,0)=e^{-i \hat{H}_{\mathrm{eff}}T}$ where we choose $t'\!=\!0$ such that $\hat{R}(0)\!=\!\hat{I}$, see Refs. \cite{Goldman2014,Eckardt2015} for more details. Conversely, $\hat{R}(t)$ characterizes the micromotion or short-time behavior within each period. Using Eq.~\eqref{evol_operator} and the equation of motion for the time-evolution operator, it is straightforward to show that
\begin{eqnarray}
\nonumber
    \hat{H}_{\mathrm{eff}} &=& \hat{R}^{\dagger}(t) \hat{H}(t) \hat{R}(t) - i \hat{R}^{\dagger}(t)\partial_t \hat{R}(t)\,,\\
    &=&\hat{R}^{\dagger}(t)\hat{H}^F(t)\hat{R}(t)\,.
    \label{Heff}
\end{eqnarray}
Namely, $\hat{R}(t)$ is the time-dependent unitary transformation that relates the effective Hamiltonian in Eq.~\eqref{evol_operator} to the original (time-dependent) Hamiltonian $\hat{H}(t)$~\cite{Rahav2003,Goldman2014,Eckardt2015,Bukov2015}. In Sambe space,  this transformation corresponds to the unitary transformation $\hat{\bm{R}}$ that block diagonalizes $\hat{\bm{H}}^F$ in Eq.~\eqref{boldHF}~\cite{Eckardt2015}. We note that $\hat{H}_{\mathrm{eff}}$ can always be chosen to have its eigenvalues within the $\mathrm{NFZ}$, which is the convention used throughout this work.
Furthermore,  it is easy to verify that if $\ket{u_a^\mathrm{eff}}$ is an eigenvector of $\hat{H}_{\mathrm{eff}}$ wih eigenvalue $\varepsilon_a \in \mathrm{NFZ}$, then $\ket{u_a(t)}=\hat{R}(t)\ket{u_a^\mathrm{eff}}$ is an eigenvector of $\hat{H}^F(t)$ with quasienergy $\varepsilon_a $ and viceversa.

The normal spectral flow in Eq.~\eqref{WFN} can be generically written as a St\v{r}eda-response of $\hat{H}_{\mathrm{eff}}$, namely,
\begin{eqnarray}
\notag
    \mathcal{W}^{N}(\mu_\varepsilon) &=& \Phi_0 \frac{\partial}{\partial \Phi} \int_{\varepsilon_{\pi}-\Omega}^{\mu_\varepsilon}d\omega\sum_{a,s}\delta(\omega-\varepsilon_{as})\,,\\
    \notag
    &=& \Phi_0 \frac{\partial}{\partial \Phi} \int_{\varepsilon_{\pi}-\Omega}^{\mu_\varepsilon}d\omega\mathrm{Tr}[\delta(\omega-\hat{H}_{\mathrm{eff}})]\,,\\
    \label{streda_eff}
    &=& \Phi_0 \frac{\partial N_{\mathrm{eff} }(\mu_\varepsilon)}{\partial \Phi}\,,
\end{eqnarray}
where $N_{\mathrm{eff}}(\mu_\varepsilon)$ simply counts the number of eigenstates that $\hat{H}_{\mathrm{eff}}$ exhibits between $\varepsilon_{\pi}-\Omega$ and $\mu_\varepsilon$. In complete analogy with the equilibrium context, one can re-express the number of states $N_{\mathrm{eff}}(\mu_\varepsilon)$ as the winding number of an effective propagator in frequency space [see Eq.~\eqref{N1_G}], namely,
\begin{eqnarray}
    \label{Neff_Geff}  
    N_{\mathrm{eff}}(\mu_\varepsilon) &=& N_1[G_{\mathrm{eff}}]\\
    \notag
    &=& -\frac{1}{2\pi}\int_{-\infty}^{\infty}d\omega e^{i\omega 0^{+}}\mathrm{Tr}\left[\hat{G}^{-1}_{\mathrm{eff}}(i\omega)\frac{\partial \hat{G}_{\mathrm{eff}}(i\omega)}{\partial \omega}\right],
\end{eqnarray}
where we have introduced the $\mu_\varepsilon$-dependent single-particle Green's function $\hat{G}_{\mathrm{eff}}(\omega)=[(i\omega+\mu_\varepsilon)\hat{I}-\hat{H}_{\mathrm{eff}}]^{-1}$. As long as $\hat{G}_{\mathrm{eff}}(i\omega)$ does not have poles at zero frequency (or, equivalently, $\hat{H}_{\mathrm{eff}}$ does not have eigenvalues at $\mu_\varepsilon$), Eq.~\eqref{Neff_Geff} remains quantized.

It is well-known that knowledge of the unperturbed stroboscopic time-evolution, as dictated by $\hat{H}_{\mathrm{eff}}$, is not enough to characterize anomalous topological Floquet phases
---see discussion at the end of this section.  Nevertheless, the formalism presented in Sec.~\ref{Sec_Floquet_cesaro} offers a method for determining the number of anomalous edge channels solely based on the knowledge of the magnetic response of 
$\hat{H}_{\mathrm{eff}}$. Indeed, by inserting Eq.~\eqref{DOS_F} into Eq.~\eqref{Cesaro}, we find that
\begin{eqnarray}
    \mathcal{W}^{A} = \frac{\Phi_0}{\Omega }\frac{\partial}{\partial \Phi} \sum_{a}\varepsilon_{a}= \frac{\Phi_0}{\Omega }\frac{\partial}{\partial \Phi} \mathrm{Tr}[\hat{H}_{\mathrm{eff}}]\,,
    \label{physical_insight}
\end{eqnarray}
where the sum is restricted to quasienergies $\{\varepsilon_{a}\}$ in a given Floquet zone. We remark that Eq.~\eqref{physical_insight} indeed signals an anomaly:~the expression on the right-hand side should vanish whenever $\hat{H}_{\mathrm{eff}}$ represents a conventional (local) Hamiltonian describing a closed physical system.

A couple of remarks are in order.  
The equality in Eq.~\eqref{physical_insight} is independent of the boundary conditions or the presence of inhomogeneities in the system, with the only caveat being that, in a torus geometry, the flux derivative must be interpreted as a finite difference due to the quantization of the magnetic monopole charge. In a realistic open-boundary sample, the flux can take continuous values. In this scenario, it is crucial to perform the summation over quasienergies in Eq.~\eqref{physical_insight} \textit{prior} to taking the flux derivative. Indeed, the restricted sum over a given Floquet zone and the magnetic-field derivative are not commuting operations, due to a non-trivial boundary term that appears under their exchange. This term leads to discontinuities or jumps (in multiples of $\Omega/\Phi_0$) generated by edge states that enter or leave the Floquet zone. This is essential to maintain the quantization  of $\mathcal{W}^A$ in open boundary samples. An example of this behavior is shown in Fig. \ref{N1vsN3fig} and discussed in more detail in Sec. \ref{RLBL_Sec}. 

We point out that expressions similar to Eq.~\eqref{physical_insight} were found in the context of Anderson localized Floquet phases~\cite{Nathan2017} and in the analysis of the spectral flow of the Floquet Hofstadter butterfly~\cite{Asboth2017}. In this sense, our approach offers a unifying framework, establishing the relation in Eq.~\eqref{physical_insight} for generic two-dimensional Floquet systems. For the particular case of a system of Bloch particles, it is worth recalling that Eq.~\eqref{anomaly} provides a closed analytical formula for $\mathcal{W}^{A}$,  which is entirely written in terms of bulk properties in the absence of the applied field.

Using Eqs.~\eqref{streda_eff} and ~\eqref{physical_insight}, we can therefore express the total Floquet spectral flow at quasienergy $\mu_\varepsilon$ as
\begin{equation}
    \mathcal{W}(\mu_\varepsilon) = \Phi_0 \frac{\partial N_{\mathrm{eff} }(\mu_\varepsilon)}{\partial \Phi} + \frac{\Phi_0}{\Omega }\frac{\partial}{\partial \Phi} \mathrm{Tr}[\hat{H}_{\mathrm{eff}}]\,.
    \label{W_strobo}
\end{equation}
Equation \eqref{W_strobo} clearly indicates that it is not necessary to have information on the micromotion operator $\hat{R}(t)$ to classify two-dimensional Floquet topological phases, provided the stroboscopic evolution is known in the presence of a small magnetic field. This is in full agreement with the conclusion of Ref.~\cite{Asboth2017}.

 At this stage, it is important to relate our results to the theoretical framework that initially guided the classification of Floquet topological systems \cite{Rudner2013,Roy2017,Harper2020}, in the absence of magnetic perturbations. In that context, the bulk-boundary correspondence was established by introducing a spatiotemporal winding number entirely built from the micromotion operator $\hat{R}(t)$, which we denote as $N_3[R]$. First, we note that the anomalous flow identified in Eq.~\eqref{anomaly} implicitly contains all the information about the micromotion, since it is written in terms of quantities that are defined in Sambe space. Importantly, as shown in detail in Appendix~\ref{proof_WR}, one can rigorously prove that the anomalous flow in Eq.~\eqref{anomaly} exactly corresponds to this spatiotemporal winding number,
\begin{eqnarray}
\notag
     \mathcal{W}^{A} &=& \frac{\epsilon^{zjl}}{8\pi^2}\!\!\int_{0}^{T}\!\!\!\!dt\!\!\int_{\mathrm{BZ}}\!\!d^2k \mathrm{tr}\left[\hat{R}^{\dagger}_{\bm{k}}(t)\frac{\partial \hat{R}_{\bm{k}}(t)}{\partial t}\right.\\
    & &\left.\hat{R}^{\dagger}_{\bm{k}}(t)\frac{\partial \hat{R}_{\bm{k}}(t)}{\partial k_l}\hat{R}^{\dagger}_{\bm{k}}(t)\frac{\partial \hat{R}_{\bm{k}}(t)}{\partial k_j}\right]\equiv N_3[R]\,. 
\label{WA_Rudner}
\end{eqnarray}
Here, we have introduced $\hat{R}_{\bm{k}}(t)$, the Fourier-transform of $\hat{R}(t)$ to quasimomentum space in the absence of the perturbing magnetic field. Equation~\eqref{WA_Rudner} can be equally expressed as the winding number of the two-point micromotion operator
$\hat{P}_{\bm{k}}(t,t') = \hat{R}_{\bm{k}}(t)\hat{R}_{\bm{k}}^{\dagger}(t')$~\cite{Eckardt2017}; see Appendix~\ref{proof_WR}. This result is in perfect agreement with Refs.~\cite{Rudner2013,Roy2017,Harper2020}, where Eq.~\eqref{WA_Rudner} was obtained entirely based on mathematical considerations. 

In Appendix~\ref{curvatureHeff_vs_curvatureSambe}, we also show that the Chern numbers associated to the Bloch bands of $\hat{H}_{\mathrm{eff}}$ are the same as the Chern numbers $C_{\alpha}^{F}$, which characterize the normal flow within a spectral gap [see Eq.~\eqref{Cherns_FFZ}]. 
\subsection{Floquet orbital magnetization density}\label{Sec_OMD}
In analogy with equilibrium systems, the orbital magnetization density associated to Floquet states with quasienergies $\varepsilon_{a} \in (\varepsilon_{\pi}-\Omega,\mu_{\varepsilon}]$ can be defined as 

\begin{equation}
    \mathcal{M}^{F}(\mu_\varepsilon) = -\int_{\textrm{NFZ}}d\omega\frac{\partial\rho^{F}(\omega)}{\partial B}(\omega-\mu_{\varepsilon})\Theta(\mu_{\varepsilon}-\omega).
\end{equation}
It is straightforward to verify that the normal and anomalous spectral flows, introduced in Eqs.~\eqref{WFN} and~\eqref{Cesaro}, can then be expressed in terms of $\mathcal{M}^{F}(\mu_\varepsilon)$ as
\begin{equation}
    \mathcal{W}^{N}(\mu_\varepsilon) = \Phi_0 \frac{\partial \mathcal{M}^{F}(\mu_\varepsilon)}{\partial \mu_\varepsilon}\,,
    \label{maxwell_rel_floquet}
\end{equation}
and
\begin{equation}
    \mathcal{W}^{A} = -\frac{\Phi_0}{\Omega}\int_{\mathrm{NFZ}}d\mu_\varepsilon\,\frac{\partial\mathcal{M}^{F}(\mu_\varepsilon)}{\partial \mu_\varepsilon}= -\frac{\Phi_0}{\Omega} \mathcal{M}^{F}(\varepsilon_{\pi})\,,
    \label{Mztot}
\end{equation}
where we used the fact that $\mathcal{M}^{F}(\varepsilon_{\pi} - \Omega)\!=\!0$, with our current conventions.
Altogether, the total Floquet spectral flow can be re-written in terms of the Floquet magnetization density as
\begin{equation}
    \mathcal{W}(\mu_\varepsilon) = \Phi_0 \frac{\partial \mathcal{M}^{F}(\mu_\varepsilon)}{\partial \mu_\varepsilon} -\frac{\Phi_0}{\Omega} \mathcal{M}^{F}(\varepsilon_{\pi}).
    \label{WF_rel_M}
\end{equation}
Equation~\eqref{Mztot} stands as a remarkable relation, which generically states that the quantized anomalous spectral flow $\mathcal{W}^{A}$, intrinsic to  Floquet systems, reflects the quantization of the total orbital magnetization density of the Floquet-Sambe states within a Floquet zone, in units of $\Omega/\Phi_0 = e/cT$. 
Through Eq. \eqref{physical_insight}, we can also interpret  $\mathcal{M}^{F}(\varepsilon_{\pi})$ in Eq. \eqref{Mztot} as being the orbital magnetization density of the \textit{fully occupied} spectrum of $\hat{H}_{\mathrm{eff}}$. In Sec.~\ref{GSF}, we will derive a series of relations which better highlight the physical interpretation of $\mathcal{W}^{A}$ as a measure of the exchange of quantized orbital angular momentum per cycle between the system of dressed particles, described by $\hat{H}_{\mathrm{eff}}$, and the driving field~\cite{Usaj2014,PerezPiskunow2015}. Such an effect can only take place in the resonant regime (i.e.~when $\Omega$ is of the order of the energy-scales of the time-averaged Hamiltonian~\cite{Kitagawa2010,Rudner2013,PerezPiskunow2015,GomezLeon2024}), since virtual processes cannot transfer angular momentum. This is consistent with the bulk-boundary correspondence principle:~edge modes at $\mu_\varepsilon=\varepsilon_{\pi}$ can only emerge through a gap-closing at the boundary of the Floquet zones.

 The quantization of orbital magnetization density in Floquet systems has previously been associated with anomalous Floquet Anderson insulator (AFAI) phases, as demonstrated in Ref.~\cite{Nathan2017}. These phases occur in strongly disordered systems, where topological chiral edge modes coexist with a bulk of fully localized Floquet states~\cite{Titum2016}.  In this sense, Eq.~\eqref{Mztot} extends the result of Ref.~\cite{Nathan2017} beyond Floquet systems with fully localized bulk states, demonstrating its general applicability provided there is a well-defined bulk gap at the Floquet zone edge. This statement is further supported by our real-space study of inhomogeneous systems in Sec.~\ref{Sec_local-marker}.

Interestingly, a general expression for the orbital magnetization density of Floquet systems with lattice translational symmetry, valid for any arbitrary stationary occupation of the Floquet-Bloch bands, was only recently derived in Refs.~\cite{Topp2022,Gao2022}. In order to connect these new developments to our spectral flow analysis, which entirely relies on  properties of the Floquet states, we will consider the stationary occupation of the $\alpha\bm{k}$-th Floquet mode to be given by a zero-temperature Fermi-Dirac distribution function with chemical potential $\mu\!=\!\mu_\varepsilon$, namely, $\Theta(\mu_\varepsilon-\varepsilon_{\alpha\bm{k}})$. In the original Hilbert space $\mathcal{H}$, the Floquet-Bloch orbital magnetization density can then be written as~\cite{Topp2022}
\begin{equation}
    \mathcal{M}^{F}(\mu_\varepsilon) = \mathcal{M}_1^{F}(\mu_\varepsilon) + \mathcal{M}_2^{F}(\mu_\varepsilon)\,,
\label{MF}
\end{equation}
where
\begin{eqnarray}
\notag
    \mathcal{M}_1^{F}(\mu_\varepsilon)&\!=\! & \frac{1}{\Phi_0}\sum_{\alpha}\int_{\mathrm{BZ}}\frac{d^2k}{2\pi}\,\Theta(\mu_\varepsilon-\varepsilon_{\alpha\bm{k}})\\
    \notag
    \times \frac{1}{T}\!\!\int_{0}^{T}\!\!dt\!\!\!& & \mathrm{Im}\left[\langle \partial_{k_x}u_{\alpha\bm{k}}(t)|\hat{H}_{\bm{k}}^{F}(t)-\varepsilon_{\alpha\bm{k}}|\partial_{k_y}u_{\alpha\bm{k}}(t)\rangle\right] , \\
    \label{M1F_tav}
\end{eqnarray}
and
\begin{eqnarray}
\notag
    \mathcal{M}_2^{F}(\mu_\varepsilon)& &= \frac{1}{\Phi_0}\sum_{\alpha}\int_{\mathrm{BZ}}\frac{d^2k}{\pi}\,\Theta(\mu_\varepsilon-\varepsilon_{\alpha\bm{k}})\\
    \notag
    & &\times \frac{1}{T}\int_{0}^{T}dt\,\mathrm{Im}\left[\langle \partial_{k_x}u_{\alpha\bm{k}}(t)|\varepsilon_{\alpha\bm{k}}-\mu_\varepsilon|\partial_{k_y}u_{\alpha\bm{k}}(t)\rangle\right]\,,\\
    \label{M2F_tav}
\end{eqnarray}
where $|u_{\alpha\bm{k}}(t)\rangle$ are the eigenmodes of the quasienergy operator with eigenenergy $\varepsilon_{\alpha\bm{k}}\!\in\!\mathrm{NFZ}$ and $\hat{H}_{\bm{k}}^{F}(t)\!=\!\hat{H}_{\bm{k}}(t)-i\partial_t$. 
In Eq.~\eqref{MF}, we introduced the standard decomposition of the orbital magnetization of Bloch systems, see Refs.~\cite{Xiao2005,Ceresoli2006,Resta2010,Atencia2024}. The $\mathcal{M}^{F}_1(\mu_\varepsilon)$ term is solely determined by the intrinsic orbital magnetic moment of the Floquet-Bloch bands, while the $\mathcal{M}^{F}_2(\mu_\varepsilon)$ contribution can be interpreted as the one stemming from the Berry curvature correction to the phase-space Floquet density of states~\cite{Xiao2005,Gao2022}.
We remind that $\mu_\varepsilon$ plays the role of the chemical potential $\mu$ in the present context.  

The two contributions given by Eqs.~\eqref{M1F_tav} and~\eqref{M2F_tav} can be alternatively written in terms of the Berry curvatures, quasienergies and intrinsic orbital magnetic moments of the Floquet-Bloch bands in Sambe space as
\be
\mathcal{M}_{1}^{F}(\mu_\varepsilon) =  \sum_{\alpha}\int_{\mathrm{BZ}}\frac{d^2 k}{(2\pi)^2}\Theta(\mu_\varepsilon-\varepsilon_{\alpha\bm{k}})\,m_z^\alpha(\bk)\,,
\label{M1F}
\ee
and
\be
\mathcal{M}_{2}^{F}(\mu_\varepsilon) =- \sum_{\alpha}\int_{\mathrm{BZ}}\frac{d^2 k}{2\pi\Phi_0}\Theta(\mu_\varepsilon-\varepsilon_{\alpha\bm{k}})(\varepsilon_{\alpha\bm{k}}-\mu_\varepsilon)\,\mathcal{F}_{xy}^{\alpha}(\bm{k})
\,.
\label{M2F}
\ee

We note that the $\bm{k}$-space Floquet orbital magnetization density computed on a torus geometry would physically represent the spatially averaged orbital magnetization density of an open boundary system, which generically has non-trivial contributions stemming both from the bulk and from the edge of the sample~\cite{Thonhauser2005}. Using Eqs.~\eqref{M1F} and ~\eqref{M2F}, it is easy to verify that the relations determined in Eqs.~\eqref{maxwell_rel_floquet} and~\eqref{Mztot} are in full agreement with the results found in Sec.~\ref{Sec_flow-Bloch}. In particular, the net number of chiral anomalous edge channels in a clean lattice, as given by Eq.~\eqref{anomaly}, is in one-to-one correspondence with the quantization of the total Floquet-Bloch orbital magnetization density, namely
\begin{eqnarray}
\notag
    \mathcal{W}^{A} &=& \sum_{\alpha}\int_{\textrm{BZ}}\frac{d^2k}{2\pi\Omega}\left[\mathcal{F}_{xy}^{\alpha}(\bm{k})\varepsilon_{\alpha\bm{k}} - \frac{\Phi_0}{2\pi} m_z^{\alpha}(\bm{k})\right]\\
    &=& -\frac{\Phi_0}{\Omega}\left[\mathcal{M}_1^{F}(\varepsilon_{\pi})+\mathcal{M}_2^{F}(\varepsilon_{\pi})\right].
\end{eqnarray}
We note that this fundamental relation stands in sharp contradiction with the findings of Refs.~\cite{Topp2022,Dag2022}, which report non-universal or trivial magnetic responses in the anomalous regime~\footnote{The origin of this disagreement can be traced back to an incorrect simplification of Eq.~\eqref{MF} in the two-band-model case, which was used in the numerical analysis of Refs.~\cite{Topp2022,Dag2022}.}.

\subsection{Generalized Floquet-St\v{r}eda formula \label{GSF}: From first order to higher-order winding numbers}
We now demonstrate that the anomalous flow in Eq.~\eqref{Cesaro} can be expressed as the flux-derivative of a first-order winding number built from the micromotion operator. This allows one to rewrite Eq.~\eqref{flow_part_energy} in the form of a generalized Floquet-St\v{r}eda response, which has the same mathematical form as Eq.~\eqref{eq_N1_N3}, and which reduces to that expression in the undriven limit.

Taking the trace of Eq.~\eqref{Heff}, and time-averaging the result over one driving cycle, one finds that
\begin{eqnarray}
\label{relHeff_Ht}
    \mathrm{Tr}[\hat{H}_{\mathrm{eff}}] &=& \frac{1}{T}\int_{0}^{T}dt\, \mathrm{Tr}[\hat{H}(t)] + N_1[R] \Omega \,,
\end{eqnarray}
where we have identified $N_1[R]$ as the first-order winding number of the micromotion operator $\hat{R}(t)$ over one period of the driving cycle,
\be
\label{N1_R}
    N_1[R] = - \frac{i}{2\pi}\int_{0}^{T}dt\, \mathrm{Tr}[\hat{R}^{\dagger}(t)\partial_t \hat{R}(t)]\,,
\ee
which guarantees that $N_1[R]\in \mathbb{Z}$. 
The first term on the right hand side of Eq. \eqref{relHeff_Ht} is nothing but the total mean energy of the driven system, which can be written as 
\begin{eqnarray}
\notag
\!\!\!\frac{1}{T}\int_{0}^{T}\!\! dt\, \mathrm{Tr}[\hat{H}(t)] &=&  \frac{1}{T}\int_{0}^{T}\!\! dt\, \mathrm{Tr}[\hat{H}^{F}(t) + i\partial_t]\\
  &=& \sum_a \left(\varepsilon_a+\Omega\sum_n n\,\bra{u_a^{(n)}}u_a^{(n)}\rangle\right)\, ,
\label{mean_energy}
\end{eqnarray}
where we remind that  $\hat{H}^{F}(t)$ is the Floquet Hamiltonian in $\mathcal{H}$ space, defined in Eq.~\eqref{HF_Hilbert}, and where the trace was taken in the eigenbasis of $\hat{H}^{F}(t)$ [see Eq.~\eqref{Floquet_modes}]. Comparing Eq.~\eqref{mean_energy} with Eq.~\eqref{relHeff_Ht}, one obtains the insightful identification
\be
    N_1[R] = -\sum_a \sum_{n=-\infty}^{\infty} n\,\bra{u_a^{(n)}}u_a^{(n)}\rangle\,.
    \label{N1_prob}
\ee
Moreover, using the eigenvectors $|u_{a}^{\mathrm{eff}}\rangle$ of $\hat{H}_{\mathrm{eff}}$ to take the trace in Eq.~\eqref{N1_R} and their relation with the eigenvectors of $\hat{H}^{F}(t)$, $|u_a(t)\rangle = \hat{R}(t)|u_{a}^{\mathrm{eff}}\rangle$, one arrives at the alternative rewriting of this winding number as
\begin{eqnarray}
\notag
    N_1[R] &=& -\frac{i}{2\pi}\sum_{a}\int_{0}^{T}dt \langle u_a(t)|\partial_t u_a(t)\rangle\,,\\
    &=& -\frac{1}{2\pi}\sum_a \gamma^{a}_{AA}\,,
    \label{N1_AA}
\end{eqnarray}
where $\gamma^{a}_{AA}$ denotes the Aharonov-Anandan phase of the time-periodic Floquet modes in the NFZ~\cite{Aharonov1987}, also known as non-adiabatic Berry phase~\cite{Moore1990,Moore1991}. Interestingly,  
Ref.~\cite{Mondragon2018} identified $\gamma^{a}_{AA}$ as the average ``position" of  the $a$-th Floquet state in frequency domain, and has therefore interpreted this phase as the polarization of these states in frequency space. Quite notably, our identification of Eqs.~\eqref{N1_prob} and~\eqref{N1_AA} with the first-order winding number of $\hat{R}(t)$ [see Eq.~\eqref{N1_R}], imposes that the sum of Aharonov-Anandan phases over a Floquet Brillouin zone should be quantized in integer numbers. These results are in complete analogy with the modern theory of electric polarization $\bm{P}$~\cite{Resta1992,KingSmith1993,Resta1994,Resta2010,Vanderbilt2018}. Here, $N_1[R]$ takes the role of $\bm{P}$ and the time domain that of quasimomentum space~\cite{PeraltaGavensky2025}. Similarly to $\bm{P}$, the value of $N_1[R]$ depends on the choice of the Floquet zone (see Appendix \ref{Appendix_gaugeN1}). In particular, our choice of the NFZ~\cite{Nathan2015} implies that $N_1[R]=0$ for $\Phi=0$.

The total magnetic response of the effective Hamiltonian can be easily related to the winding number $N_1[R]$ by taking the derivative of Eq.~\eqref{relHeff_Ht} with respect to flux. Importantly, the energy spectral flow of the time-averaged Hamiltonian is zero,
\begin{eqnarray}
    \frac{\Phi_0}{T}\frac{\partial}{\partial \Phi}\int_{0}^{T}dt\, \mathrm{Tr}[\hat{H}(t)] = \Phi_0 \frac{\partial}{\partial \Phi}\mathrm{Tr}[\hat{H}_0] = 0\,.
\end{eqnarray}
Indeed, the dc-component of the time-dependent Hamiltonian, $\hat{H}_0$, has essentially the same form as the undriven Hamiltonian, possibly with renormalized parameters. As  such,  its associated angular momentum operator, $\propto\partial \hat{H}_0/\partial\Phi$, is indeed traceless.
Therefore, using Eq.~\eqref{physical_insight}, we conclude that
\be
  \mathcal{W}^{A}= \frac{\Phi_0}{\Omega }\frac{\partial}{\partial \Phi} \mathrm{Tr}[\hat{H}_{\mathrm{eff}}] 
    = \Phi_0\frac{\partial  N_{1}[R]}{\partial \Phi}\,,
    \label{WA_dN1}
\ee
that is to say, the anomalous spectral flow  $\mathcal{W}^{A}$ is given by the derivative with respect to the magnetic flux of the winding number $N_1[R]$. 
We stress that, even though the value of $N_1[R]$ is defined modulo an integer number [see Appendix~\ref{Appendix_gaugeN1}], its variation with magnetic flux is gauge-independent and hence, physically meaningful. 

As discussed in Sec.~\ref{Sec_eq}, the flux-derivative of first-order winding numbers (formally understood as a finite-difference) is known to generate higher-order topological invariants~\cite{Prodan2016,PeraltaGavensky2023}. Indeed, Eq.~\eqref{WA_dN1} demonstrates that in a Bloch lattice
\begin{equation}
 \Phi_0 \frac{\partial N_1[R]}{\partial \Phi}= N_3[R]\,,   
 \label{WA=N3R}
\end{equation}
where $N_3[R]$  was introduced in Eq.~\eqref{WA_Rudner}. This observation, which naturally derives from our Str\v{e}da approach, provides yet another route to derive the winding numbers of Refs.~\cite{Rudner2013,Roy2017,Harper2020}.

Finally, we can use Eqs.~\eqref{streda_eff},~\eqref{Neff_Geff} and~\eqref{WA_dN1} to express the total Floquet spectral flow as
\begin{eqnarray}
\notag
    \mathcal{W}(\mu_\varepsilon) &=& \Phi_0 \left(\frac{\partial N_\mathrm{eff}(\mu_\varepsilon)}{\partial \Phi} + \frac{\partial N_1[R]}{\partial \Phi}\right)\,,\\
    &=& \Phi_0 \left(\frac{\partial N_1[G_{\mathrm{eff}}]}{\partial \Phi} + \frac{\partial N_1[R]}{\partial \Phi}\right)\, ,
    \label{generalized_streda_floquet}
\end{eqnarray}
which has a particularly appealing form as it is reminiscent of the equilibrium St\v{r}eda formula in Eq.~\eqref{eq_N1_N3}. As a matter of fact, the number of particles $N(\mu)$, or equivalently, the winding number $N_1[G]$, is replaced here by the sum of two contributions, $N_1[G_{\mathrm{eff}}]+N_1[R]$, arising from the normal and anomalous flow terms, respectively. We refer to Eq.~\eqref{generalized_streda_floquet} as the \textit{generalized Floquet-St\v{r}eda formula for periodically driven systems}, which trivially reduces to Eq.~\eqref{eq_N1_N3} in the absence of the driving field.  

The first term in Eq.~\eqref{generalized_streda_floquet} is analogous to the spectral flow exhibited by equilibrium systems, as it can be entirely obtained from the Chern numbers of the Floquet-Bloch bands; see Eq.~\eqref{Cherns_FFZ}. In a finite-size system with boundaries, we interpret this term as the one stemming from the flow of \textit{dressed states}, between the edge and the bulk of the sample. In contrast, the second term in Eq.~\eqref{generalized_streda_floquet} is specific to periodically-driven settings, and represents a quantized flow of \textit{energy} between the system and the driving field, see Eq.~\eqref{WA_dN1}.  The quantization of the orbital magnetization density found in Eq.~\eqref{Mztot} is nothing but a reflection of the quantization of this magnetic-field induced energy-flow; see also Eq.~\eqref{flow_part_energy}. 

In this framework, the driving field is treated classically, such that it can absorb or emit energy in a continuous fashion. Nonetheless, our results suggest that the magnetic-field induced energy exchange between the system and the driving field, averaged over one period, is quantized. In this sense, this effect can be physically understood as a quantized energy pump.

Inserting the relation between the anomalous flow and the Floquet orbital magnetization density [Eq.~\eqref{Mztot}] into Eq.~\eqref{WA_dN1}, and using Eq.~\eqref{WA=N3R}, one readily finds that
\begin{eqnarray}
\frac{\Phi_0}{\Omega}\mathcal{M}^{F}(\varepsilon_{\pi}) + \hbar N_3[R] = 0,
    \label{conservation_totalM}
\end{eqnarray}
where we momentarily restored $\hbar$. 
Equation~\eqref{conservation_totalM} is suggestive of a conservation law, reflecting the exchange of quantized angular momentum between the driving field and the system. 

Equation~\eqref{conservation_totalM} further suggests that the anomaly of these exotic Floquet phases could be cancelled if one properly incorporates the dynamics of the driving field into the global description of the driven system. This can be thought of as the generalization of the Callan-Harvey mechanism~\cite{Callan1985,Fradkin1986,Stone1991} for periodically-driven Floquet systems. This interpretation is consistent with the results found in Refs. \cite{Usaj2014,PerezPiskunow2015}, where the number of anomalous edge channels was found to be in one-to-one correspondence with the total number of photons involved in the resonant processes that are responsible for the opening of the Floquet gaps. A recent study, which fully incorporates the quantum mechanical description of the driving field, further supports this interpretation~\cite{Dag2024}.

\subsection{A simple illustrative example\label{RLBL_Sec}}
\begin{figure}[t]
    \centering
    \includegraphics[width=0.98\columnwidth]{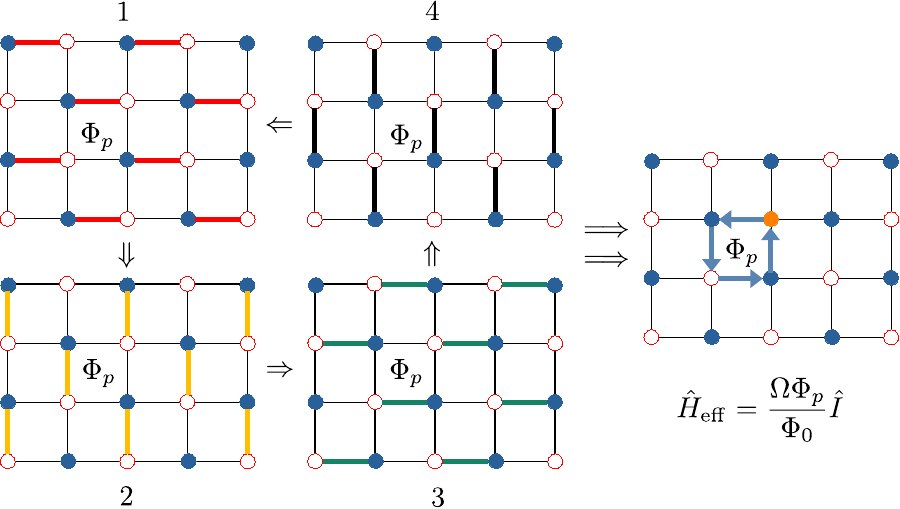}
    \caption{RLBL model of an anomalous Floquet insulator phase\cite{Rudner2013}. Nearest neighbour hoppings are turned on in a sequential manner ($1-2-3-4$) as indicated by the thick colored lines.  The parameters are such that a particle has probability one to jump between two connected sites on each step of the cycle. After a period, a particle initially on a given site (orange dot), makes a counter-clockwise loop around a lattice plaquette. If a magnetic flux $\Phi_p$ per plaquette is present, such a particle acquires a $-2\pi\Phi_p/\Phi_0$ phase after one period. }
    \label{RLBL}
\end{figure}

To grasp the physical content behind the results presented in this section, let us consider the Rudner-Lindner-Berg-Levin (RLBL) tight binding model, which was introduced in Ref.~\cite{Rudner2013} as a paradigmatic model for the anomalous Floquet phase. The system consists of a bipartite square lattice with nearest neighbour hoppings, which are turned on and off in time in a cyclic manner [see Fig.~\ref{RLBL}]. For an appropriate selection of  parameters, the bulk dynamics \textit{during} one cycle is such that a particle initially located at a given site will return with probability one to the initial site after performing a closed counter-clockwise loop throughout a lattice plaquette. Under these circumstances, the stroboscopic evolution operator reduces to the identity and the corresponding effective Hamiltonian $\hat{H}_\mathrm{eff}=0$. Clearly, this effective Hamiltonian is topologically trivial and cannot distinguish the non trivial dynamics described above from one where the hopping terms are always off and  particles do not move at all. Nevertheless, it is well-known \cite{Rudner2013} that this driving protocol generates chiral states, which are localized at the edges of open-boundary samples. This suggests that the knowledge of the  micromotion (i.e.~the time-evolution during a period) is required to properly characterize the topology of such a phase. However, this situation drastically changes when one applies a small perpendicular magnetic field to the lattice. In this scenario, a negatively charged particle that makes a counter-clockwise closed loop acquires an Aharonov–Bohm phase equal to $-2\pi \Phi_p/\Phi_0$, where $\Phi_p\!=\!\Phi/2N_c\!\ll\!\Phi_0$ is the  magnetic flux per plaquette and $N_c$ denotes the number of unit cells. In a torus geometry (i.e.~without edges), $\hat{H}_\mathrm{eff}\!=\!(\Omega/2N_c)(\Phi/\Phi_0) \hat{I}\neq0$, such that $\mathrm{Tr}[\hat{H}_\mathrm{eff}]=\Omega\Phi/\Phi_0$. Interestingly enough, this $\hat{H}_\mathrm{eff}$ with a nonzero trace cannot be obtained from a high-frequency expansion to any finite order.
Finally, from Eqs. \eqref{physical_insight} and \eqref{streda_eff} one obtains $\mathcal{W}^A= 1$ and $\mathcal{W}^{N}(\mu_\varepsilon)=0$ $\forall \mu_\varepsilon \neq 0$, respectively, such that $\mathcal{W}(\mu_\varepsilon)\!=\!1$ within all spectral gaps, in agreement with Ref.~\cite{Rudner2013}.

In this specific example, all bulk states are localized and the quasienergies are independent of $\bm{k}$; in particular they are equal to zero  for $\Phi=0$. Therefore, the quantization of $\mathcal{W}^A=1$ implies, by virtue of Eq. \eqref{anomaly}, or equivalently Eq. \eqref{Mztot}, that the total orbital magnetization density is also quantized (in units of $\Omega/\Phi_0$), 
\be
\mathcal{M}^{F}(\varepsilon_{\pi})=\mathcal{M}_1^{F}(\varepsilon_{\pi})=\sum_{\alpha}\int_{\mathrm{BZ}}\frac{d^2 k}{(2\pi)^2} m_z^\alpha(\bm{k})=- \frac{\Omega}{\Phi_0}\,,
\label{MF_RLBL}
\ee
since, in this fine-tuned case, the term $\mathcal{M}^{F}_2(\varepsilon_{\pi})=0$. This is consistent with the result found in Refs.~\cite{Nathan2017} for the  AFAI phase, where strong disorder was introduced to localize all bulk Floquet states. 

At this point, it is important to emphasize that the quantization derived in Eq.~\eqref{Mztot} does not require localized bulk states, which generically quenches $\mathcal{M}_2^{F}(\varepsilon_{\pi})$. In Sec.~\ref{Sec_Kitagawa}, we will present a deeper analysis of the behaviour of the two contributions, $\mathcal{M}_1(\varepsilon_{\pi})$ and $\mathcal{M}_2(\varepsilon_{\pi})$, in a more general situation [see Fig.~\ref{cherns_anomaly_kitagawa}]. In Sec.~\ref{Sec_local-marker}, we will additionally study the anomaly $\mathcal{W}^{A}$ in settings without translational invariance and in the presence of disorder.

\begin{figure}[t]
\centering
    \includegraphics[width=0.98\columnwidth]{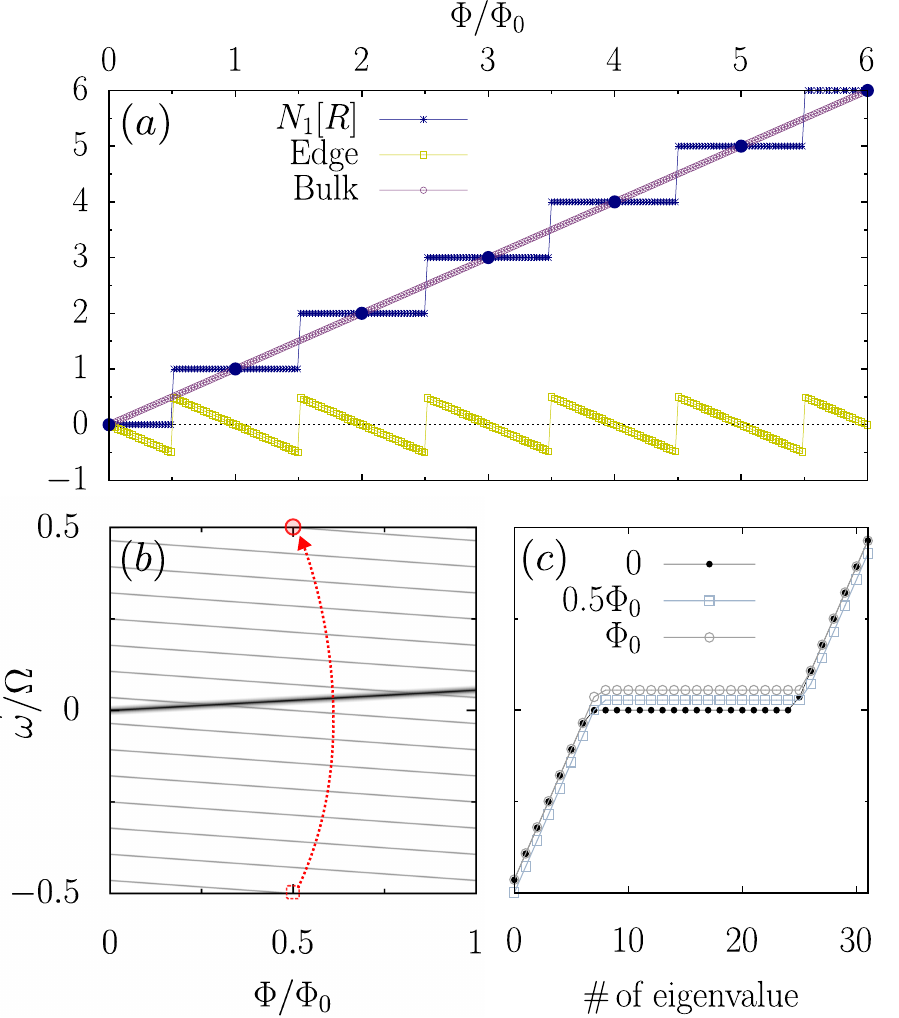}
    \caption{$(a)$ Winding number $N_1[R]$ for the RLBL model as a function of the total flux threading an open boundary sample ($20 \times 10$ unit cells) in units of $\Phi_0$. The parameters are the fined tuned ones illustrated in Fig.~\ref{RLBL}. The coarse grained slope of this step-like function is given by $N_3[R]=\mathcal{W}^{A}=+1$. We plot separately the contributions to $N_1[R]$ coming from the states localized around the edge of the system and that of the bulk states. (b) Flux dependence of the quasinenergies on the NFZ for a smaller sample ($4\times 4$ unit cells). At $\Phi/\Phi_0=0.5$ an edge mode `leaves' the NFZ from below and re-enters from above, as indicated by the arrow. (c) Quasienergy spectrum of the $4\times 4$ sample for three different values of flux $\Phi/\Phi_0 = 0, 0.5$ and $1$.}
    \label{N1vsN3fig}
\end{figure}    

In the fined-tuned RLBL model, considering a torus geometry and using Eq.~\eqref{WA_dN1}, one finds that $N_1[R]\!=\!\Phi/\Phi_0\!\in\!\mathbb{Z}$, the quantized magnetic monopole charge. Interestingly, the definition of the winding number in Eq.~\eqref{N1_R} does not rely on the use of periodic boundary conditions, such that its behavior can also be studied in an inhomogeneous system. In Fig.~\ref{N1vsN3fig}$(a)$, we plot $N_1[R]$ as a function of the flux (now a continuous variable) in an open-boundary sample of the RLBL model. The values of $N_1[R]$ are obtained by using Eq.~\eqref{WA_dN1}, with the effective Hamiltonian being numerically extracted from the stroboscopic time-evolution operator.
The parameters correspond to the driving protocol illustrated in Fig.~\ref{RLBL}, such that all bulk states are localized while chiral edge modes propagate along the system boundaries. The winding number $N_1[R]$ has a clear staircase behavior, with its coarse-grained slope being directly given by $N_3[R]=\mathcal{W}^{A}=+1$, as predicted by Eqs.~\eqref{WA_dN1} and~\eqref{WA=N3R}. The fine-tuned parameters of the model allow for an unambiguous separation of the contributions to $N_1[R]$ stemming from the bulk and from the edge of the sample, which we clearly discriminate in Fig.~\ref{N1vsN3fig}$(a)$. For this range of magnetic flux, the bulk states contribution increases monotonously in a smooth fashion, while the edge contribution is a discontinuous saw-tooth function with period $\Phi_0$. When both contributions are smooth, their slope is the same in magnitude but opposite in sign, reflecting the fact that the edge modes have the opposite direction of circulation with respect to the bulk states. 
This difference in the spectral flow of the two contributions is more clearly seen in Fig.~\ref{N1vsN3fig}$(b)$, where we plot the quasienergies of a smaller sample in the NFZ as a function of $\Phi/\Phi_0$. We remind that, within our analysis, the NFZ is fixed to be the one at zero field ($\Phi\!=\!0$). Here, the thick line corresponds to the bulk states while the rest of the lines correspond to the edge modes. When  $\Phi/\Phi_0$ reaches half an odd-integer value, the quasienergy corresponding to one of the edge modes `leaves' the NFZ boundary from below and `re-appears' from above leading to a jump of $+1$ in $N_1[R]$. This mechanism leads to a corresponding jump of $\Omega$ in $\mathrm{Tr}(\hat{H}_\mathrm{eff})$ [see Eq.~\eqref{WA_dN1}], reflecting the quantized energy pump discussed in Sec.~\ref{GSF}. 
In Fig.~\ref{N1vsN3fig}$(c)$, we plot the quasienergy spectrum for three different values of $\Phi/\Phi_0=0,0.5,1$. The degenerate quasienergies near the center of the NFZ correspond to bulk states. It is worth noticing the following:~(i) the total number of states inside the NFZ is independent of $\Phi/\Phi_0$ as a consequence of Eq. \eqref{prop}; (ii) the number of bulk states is also constant since the Chern number is zero and so there is no normal flow; (iii) the quasienergies of the edge modes for $\Phi/\Phi_0\!=\!1$ have the same value as for $\Phi\!=\!0$, while the quasienergies of the bulk states are shifted up.

This simple example illustrates several of our key results:~the possibility of extracting the winding number from the stroboscopic time-evolution of a lattice system; the correspondence between the quantization of the total orbital magnetization density of the Floquet bands on a torus geometry and the number of anomalous edge channels; and the magnetic-field-induced energy-pump mechanism leading to discontinuities in the behavior of $N_1[R]$, which are crucial to maintain the quantization of its coarse-grained slope as a function of flux in an open-boundary sample.

\section{Kitagawa model as a case of study}\label{Sec_Kitagawa}
\begin{figure}[t]
    \centering
    \includegraphics[width=\columnwidth]{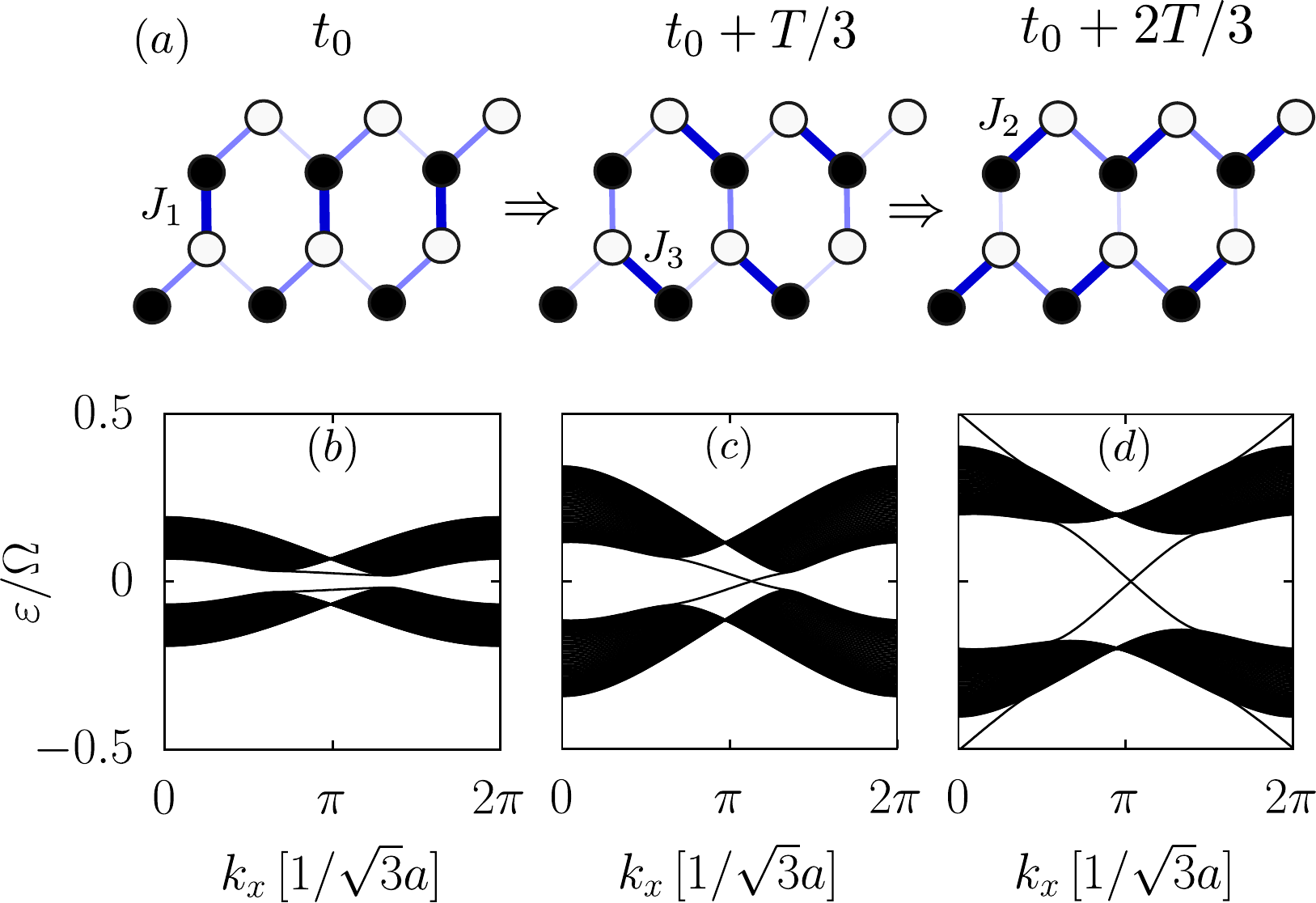}
    \caption{$(a)$ Driving protocol described by Eq.~\eqref{drv_prot}. The width and color of the lines connecting each lattice site represent the strength of the hopping amplitudes for different times $t_0$, $t_0 + T/3$ and $t_0 + 2T/3$. In the lower panels we show the quasienergy spectrum of a zigzag ribbon as a function of the quasimomentum along the translationally invariant direction. The parameters are such that $\Omega/J=20$, $\Delta/J = 0.5$ and $(b)$ $F=1.0$, $(c)$ $F=2.0$ and $(d)$ $F=2.75$.}
    \label{kitagawa_model}
\end{figure}

In order to illustrate how the general formulas derived in Secs.~\ref{Sec_unbounded_problem},~\ref{Sec_Floquet_cesaro},~\ref{Sec_flow-Bloch} and~\ref{Sec_OMD} apply in a concrete setting, we will consider a Kitagawa-type model~\cite{Kitagawa2010} as a case of study [Fig.~\ref{kitagawa_model}$(a)$]. The Hamiltonian is described within a tight-binding description of a honeycomb lattice
\begin{eqnarray}
\label{H_Kitagawa}
    \hat{H}(t) &=& \sum_{\bm{R}\in \mathcal{A}}\sum_{\nu=1}^{3}\left(J_{\nu}(t)\hat{c}^{\dagger}_{\bm{R}}\hat{c}^{}_{\bm{R}+\bm{\delta}_\nu}+h.c.\right)\\
    \notag
    &+& \Delta\left(\sum_{\bm{R}\in\mathcal{A}}\hat{c}^{\dagger}_{\bm{R}}\hat{c}^{}_{\bm{R}} -\sum_{\bm{R}\in\mathcal{B}}\hat{c}^{\dagger}_{\bm{R}}\hat{c}^{}_{\bm{R}}\right)\,,  
\end{eqnarray}
where $\bm{\delta}_1\!=\!(0,a_0)$, $\bm{\delta}_2\!=\!(-\sqrt{3}a_0/2,-a_0/2)$ and $\bm{\delta}_3\!=\!(\sqrt{3}a_0/2,-a_0/2)$. The distance between neighbouring sites is denoted as $a_0$. The second line in Eq.~\eqref{H_Kitagawa} corresponds to an inversion-symmetry breaking term,  an on-site energy mass term $+ \Delta$ in sublattice $\mathcal{A}$ and $-\Delta$ in sublattice $\mathcal{B}$. The hopping elements are modulated in a continuous fashion~\cite{Wintersperger2020} as
\begin{equation}
    J_{\nu}(t) = J\, e^{F \cos(\Omega t + \theta_{\nu})}\,,
    \label{drv_prot}
\end{equation}
where $F$ is a dimensionless parameter and $\theta_{\nu} = -\frac{4\pi}{3\sqrt{3}a_0}\bm{\delta}_\nu\cdot \bm{\check{x}}$. Different variants of this model have been experimentally realized in both photonics~\cite{Maczewsky2017,Mukherjee2017} and ultracold atom systems~\cite{Wintersperger2020,Braun2024}.

In Fig.~\ref{kitagawa_model}, we plot the dispersion relation of the Floquet quasienergies of this model in a ribbon geometry with zigzag terminations. Here we fix the driving frequency to $\Omega/J = 20$, $\Delta/J=0.5$, and modify the dimensionless parameter $F$. For $F=1.0$ [Fig.~\ref{kitagawa_model}$(b)$], the system is a trivial Floquet insulator, with both bands having a Chern number $C^{F}_{\pm}=0$. When $F\approx 1.56$, the gap at $\varepsilon=0$ closes and the system undergoes a transition to a conventional Floquet topological phase, with bands having $C^{F}_{\mp}=\pm 1$ and chiral edge channels bridging the gap [as seen in Fig.~\ref{kitagawa_model}$(c)$]. Indeed, the phase transition taking place between Fig.~\ref{kitagawa_model}$(b)$ and $(c)$ emulates the one of the Haldane model presented in Sec.~\ref{Sec_eq}. As $F$ is further increased, the zone-edge gap at $\varepsilon_{\pi}=\Omega/2$ vanishes and the transition to the anomalous Floquet phase occurs. As seen in Fig.~\ref{kitagawa_model}$(d)$, chiral edge channels emerge at both spectral gaps, while the bulk bands have trivial Chern numbers. 

\begin{figure}[t]
    \centering
    \includegraphics[width=\columnwidth]{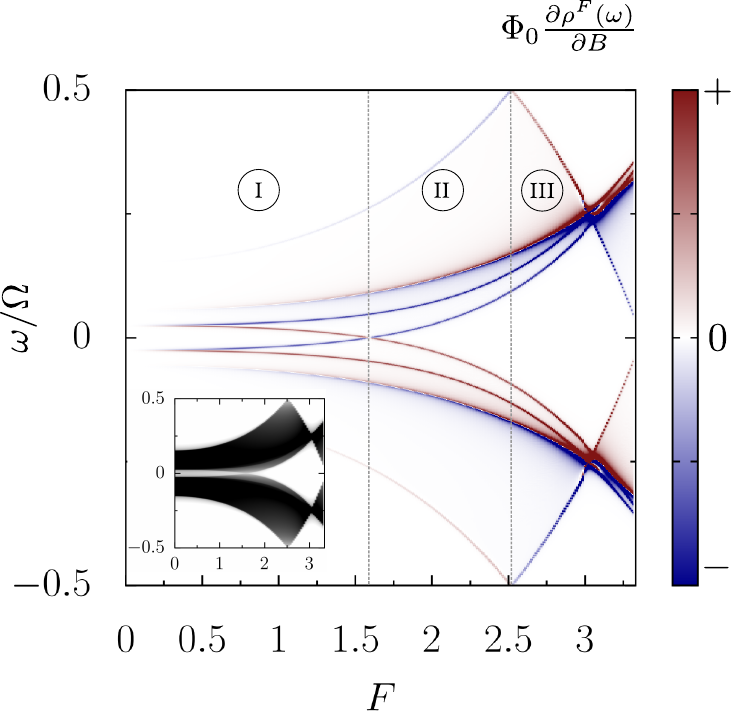}
    \caption{Energy-resolved Floquet-St\v{r}eda response [see Eq.~\eqref{drhoF_dB}] of the Kitagawa model given by Eq.~\eqref{H_Kitagawa} as a function of the dimensionless parameter $F$. The parameters are such that $\Delta/J=0.5$ and $\Omega/J=20$. The color scale is in arbitrary units, with the symbols $\pm$ indicating the sign of the response. In the inset we show the evolution of the DOS of this model (shaded area) at zero field. The dashed lines separate the different phases of the model: \textcircled{\tiny{I}} Trivial insulator phase, \textcircled{\tiny{II}} Floquet Chern insulator phase, \textcircled{\tiny{III}} Anomalous Floquet phase.}
    \label{drho_dB_kitagawa}
\end{figure}

In Fig.~\ref{drho_dB_kitagawa}, we plot the energy-resolved Floquet-St\v{r}eda response of this model [Eq.~\eqref{drhoF_dB}] as a function of the energy and the parameter $F$, for $\Omega/J\!=\!20$. Since this quantity is described by an $\Omega$-periodic function, one can safely restrict its analysis to the NFZ ($s=0$). Throughout the work, we numerically keep a maximum number of Floquet replicas such that $|s|<5$, which ensures numerical convergence of the results in the NFZ within the range of parameters used. The inset of Fig.~\ref{drho_dB_kitagawa} shows how the zero-field DOS evolves as a function of $F$, which essentially controls the bandwidth of the effective model. The physics taking place for $F\lesssim 2.5$ emulates the one discussed in Fig.~\ref{dos_dB_haldane} of Sec.~\ref{Sec_eq}. Indeed, one can clearly resolve the Berry curvature hot spots coming from the Dirac valleys and the band inversion taking place around $F\simeq 1.56$.
In region \textcircled{\tiny{I}} we have a trivial insulator phase, with no spectral flow between the upper and lower bands, while in region \textcircled{\tiny{II}} the spectral flow is the one of a topological Chern insulator. When $F \simeq 2.5$, the Floquet quasienergy bands touch the edge of the Floquet Brillouin zone and the system undergoes a transition to an anomalous Floquet phase \textcircled{\tiny{III}}, with chiral edge channels bridging all the gaps in an open boundary sample [see Fig.~\ref{kitagawa_model}$(d)$]. 

\begin{figure}[t]
    \centering
    \includegraphics[width=\columnwidth]{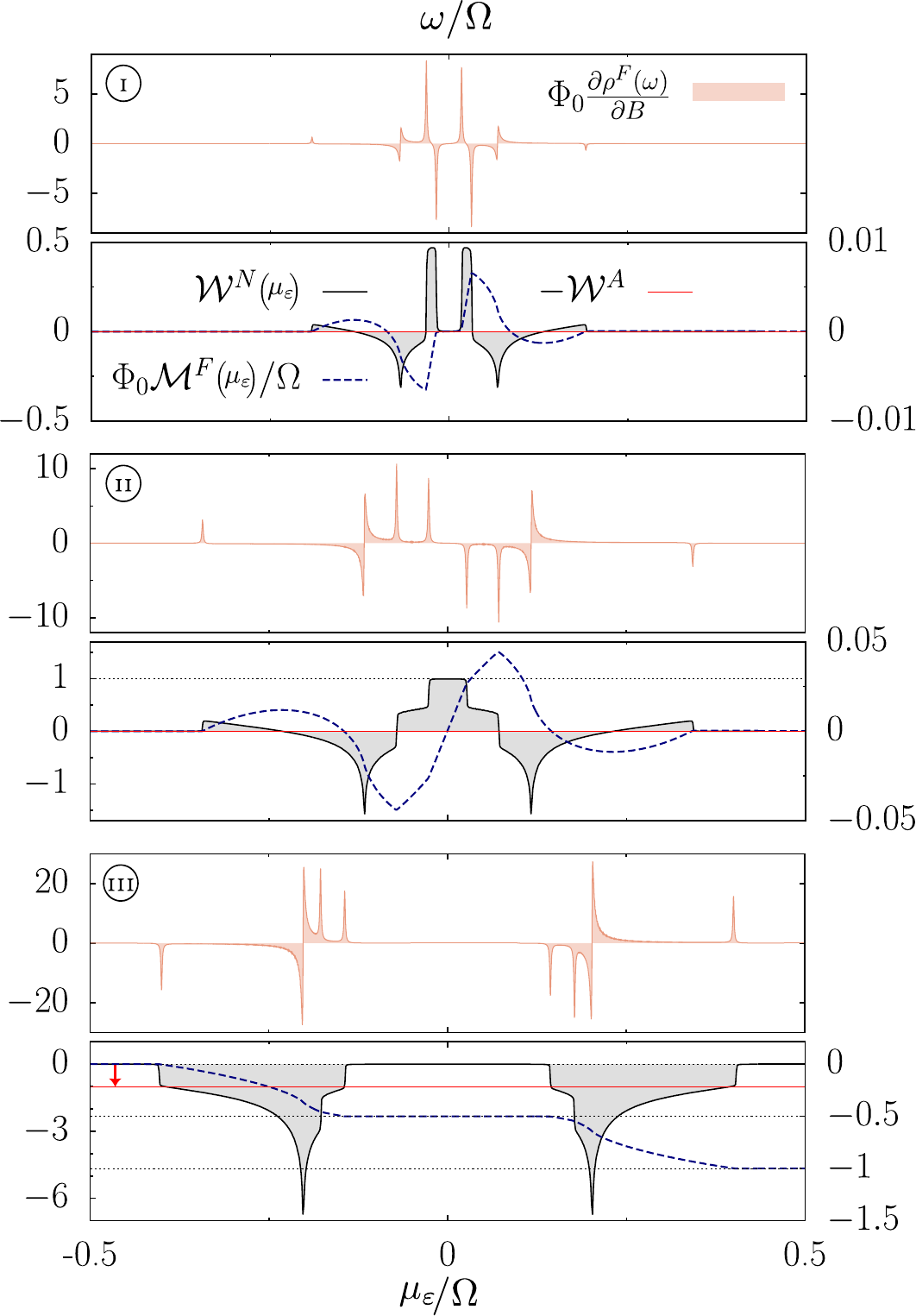}
    \caption{Energy-resolved Floquet-St\v{r}eda response [Eq.~\eqref{drhoF_dB}] (in units of $1/J$) and the corresponding normal spectral flow $\mathcal{W}^{N}(\mu_\varepsilon)$ (left y-axis), given by Eq.~\eqref{normal_flow}. We also plot the Floquet orbital magnetization density (right y-axis) with a dashed blue line in units of $\Omega/\Phi_0$ [see Eq.~\eqref{M2F_tav}]. We have chosen three representative parameters:
    \textcircled{\tiny{I}} $F=1.0$, trivial insulator phase, \textcircled{\tiny{II}} $F=2.0$, Floquet Chern insulator phase and \textcircled{\tiny{III}} $F=2.75$, anomalous Floquet phase. The rest of the parameters are the same as in Fig.~\ref{drho_dB_kitagawa}. The red solid line indicates the average of the normal spectral flow within the NFZ,  $\overline{\langle\mathcal{W}^{N}\rangle}_{\mathrm{NFZ}} = -\mathcal{W}^{A} = \Phi_0 \mathcal{M}^{F}(\Omega/2)/\Omega$ [see Eqs.~\eqref{WN_average_WA} and~\eqref{Mztot}]. The red arrow in \textcircled{\tiny{III}} highlights the anomaly.}
    \label{kitagawa_integrated}
\end{figure}
\begin{figure}[t]
    \centering
    \includegraphics[width=0.9\columnwidth]{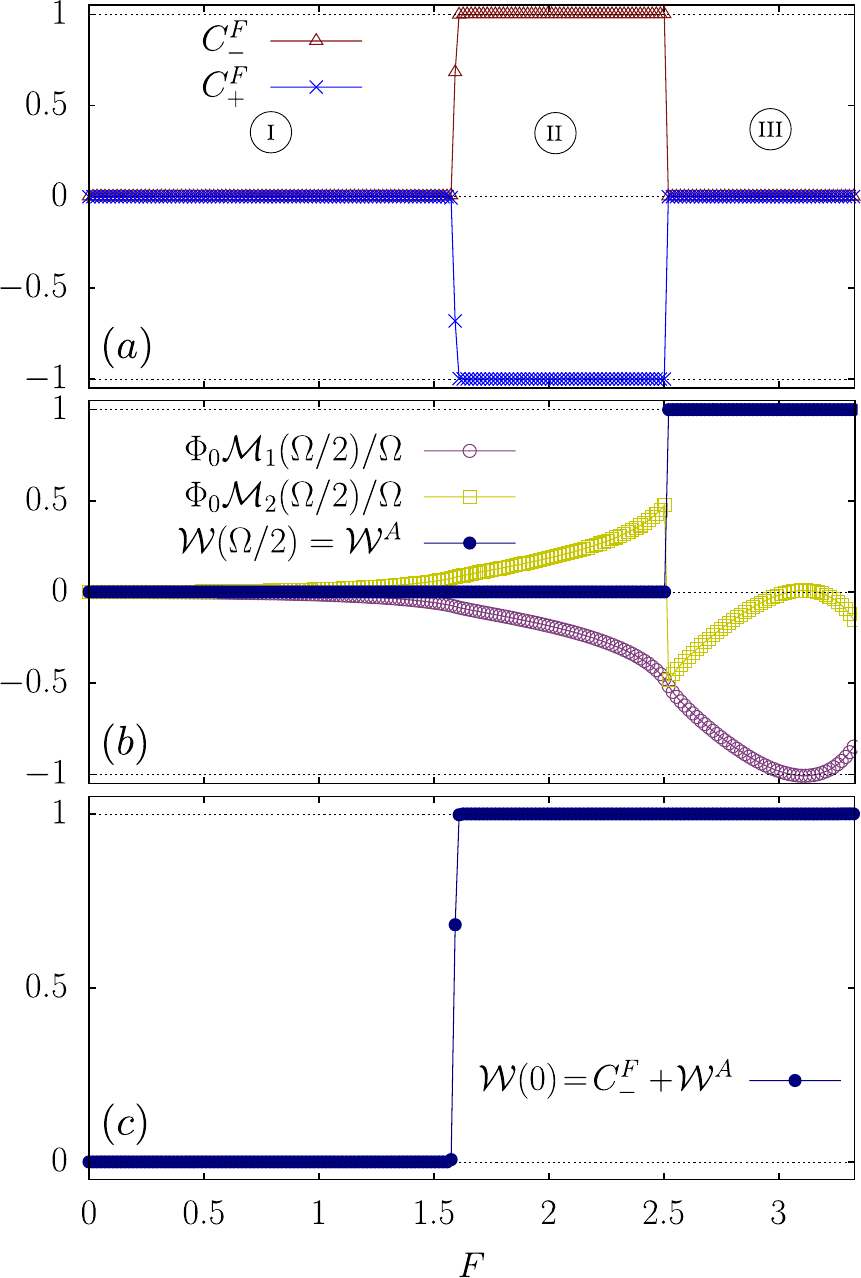}
    \caption{$(a)$ Chern numbers of the upper and lower quasienergy bands ($C^{F}_{\pm}$, respectively) of the Kitagawa model as a function of $F$. $(b)$ and $(c)$ Spectral flow  $\mathcal{W}(\mu_\varepsilon)$ as obtained from Eq.~\eqref{WF_Cesaro} for $\mu_\varepsilon=\Omega/2$ and $\mu_\varepsilon=0$, respectively, as a function of $F$. In panel $(b)$ we also plot the two contributions to the anomalous winding number $\mathcal{W}^{A}$, namely $\Phi_0\mathcal{M}^{F}_1(\Omega/2)/\Omega$ and $\Phi_0\mathcal{M}^{F}_2(\Omega/2)/\Omega$ [see Eqs.~\eqref{M1F} and~\eqref{M2F}]. In all panels $\Omega/J=20$ and $\Delta/J=0.5$.}
    \label{cherns_anomaly_kitagawa}
\end{figure}

In Fig.~\ref{kitagawa_integrated}, we show the energy-resolved Floquet-St\v{r}eda response for three representative values of $F$ that set the system deep within $\textcircled{\tiny{I}}$ the trivial insulator phase, $\textcircled{\tiny{II}}$ the Floquet Chern insulator phase, and  $\textcircled{\tiny{III}}$ the anomalous phase. These are vertical cuts of Fig.~\ref{drho_dB_kitagawa}. We also present its corresponding integrated version, i.e. the normal spectral flow $\mathcal{W}^{N}(\mu_\varepsilon)$ in Eq.~\eqref{normal_flow}, and the Floquet orbital magnetization density $\mathcal{M}^{F}(\mu_\varepsilon)$ in units of $\Omega/\Phi_0$ [Eq.~\eqref{MF}], for each of these values of $F$. In each of these plots, the solid red line indicates the average of $\mathcal{W}^{N}(\varepsilon)$ within the NFZ, $\overline{\langle\mathcal{W}^{N}\rangle}_{\mathrm{NFZ}}\!=\!-\mathcal{W}^{A}\!=\!\Phi_0 \mathcal{M}^{F}(\Omega/2)/\Omega$; see Eqs.~\eqref{WN_average_WA} and~\eqref{Mztot}. When $\mu_\varepsilon$ lies in gapped regions, the normal spectral flow takes quantized values, given  by the Chern numbers of the Floquet bands in the interval $(-\Omega/2,\mu_\varepsilon\,]$; see Eq.~\eqref{Cherns_FFZ}. These Chern numbers are identically equal to zero both in the trivial and the anomalous phases, as seen from Figs.~\ref{kitagawa_integrated} $\textcircled{\tiny{I}}$ and $\textcircled{\tiny{III}}$. In the Floquet Chern insulator phase, the lower and upper bands have Chern numbers $\pm 1$, as deduced from Fig.~\ref{kitagawa_integrated}$\textcircled{\tiny{II}}$. The most striking effect is reflected in the finite mean-value $\overline{\langle \mathcal{W}^{N}\rangle}_{\mathrm{NFZ}}=-1$ of the normal spectral flow in Fig.~\ref{kitagawa_integrated}$\textcircled{\tiny{III}}$, highlighted with the red arrow. This signature readily indicates the existence of a finite anomaly, as dictated by Eq.~\eqref{WN_average_WA}, and reflects the existence of a non-trivial topologically quantized Floquet spectral flow, even when the Chern numbers of the Floquet bands are identically equal to zero. The non-trivial quantization of the Floquet orbital magnetization density in units of $\Omega/\Phi_0$ is made evident in Fig.~\ref{kitagawa_integrated}$\textcircled{\tiny{III}}$. Indeed, the dashed-curve indicates that this quantity takes the half-quantized value of $\mathcal{M}^{F}(0)\!=\!-\Omega/2\Phi_0$ at the center of the Floquet zone, and the integer-quantized value $\mathcal{M}^{F}(\Omega/2)\!=\!-\Omega/\Phi_0$ at the zone-edge gap. The origin of the half-quantization occurring at the Floquet-zone center can be traced to the particle-hole symmetry of the problem.

The existence of a finite anomalous flow is further highlighted in Fig.~\ref{cherns_anomaly_kitagawa}. In Fig.~\ref{cherns_anomaly_kitagawa}$(a)$, we plot the Chern numbers of the upper and lower Floquet quasienergy bands ($C^{F}_{\pm}$, respectively) within the NFZ as a function of the parameter $F$. As already anticipated, the Chern numbers are only finite within region \textcircled{\tiny{II}}. In region \textcircled{\tiny{III}}, the Chern numbers are zero, but the system is still in a topologically non-trivial phase. This is clearly captured by the spectral flow calculated from Eq.~\eqref{WF_Cesaro}: In Figs.~\ref{cherns_anomaly_kitagawa}$(b)$ and $(c)$, we plot $\mathcal{W}(\mu_\varepsilon)$ for $\mu_\varepsilon=\Omega/2$ and $\mu_\varepsilon=0$, respectively, as a function of $F$. Sharp transitions from zero to finite values occur only when the corresponding gap closes and are consistent with the emergence of boundary edge modes in finite size samples [see Fig.~\ref{kitagawa_model}]. 

In Fig.~\ref{cherns_anomaly_kitagawa}$(b)$, we specifically plot the two contributions to the anomalous winding number $\mathcal{W}^{A}$, namely $\Phi_0\mathcal{M}_1^{F}(\Omega/2)/\Omega$ and $\Phi_0\mathcal{M}_2^{F}(\Omega/2)/\Omega$; see Eqs.~\eqref{M1F} and~\eqref{M2F}. Within phases \textcircled{\tiny{I}} and \textcircled{\tiny{II}}, these two terms are the same in magnitude but have opposite sign, exactly cancelling each other in such a way that $\mathcal{W}^{A}$ remains zero. Once the zone-edge gap vanishes, and the parameter $F$ is within region \textcircled{\tiny{III}}, these two contributions do not cancel anymore and exactly sum to an integer number (in this case $-1$), which correctly quantifies the quantized orbital magnetization density $\mathcal{M}^{F}(\Omega/2)$ in units of $\Omega/\Phi_0$ [see Eq.~\eqref{Mztot}]. Within the region where $\mathcal{M}_2(\Omega/2)$ is vanishingly small or zero, the Floquet bands become extremely narrow [see inset in Fig.~\ref{drho_dB_kitagawa}], hence reaching the limit of the fined-tuned RLBL model discussed in Eq.~\eqref{MF_RLBL} in Sec.~\ref{RLBL_Sec}. We remark that, when moving away from this finely tuned limit---i.e., for general parameter values---the bulk Floquet states are extended, while the total orbital magnetization density continues to be quantized. 
\section{Sum-rule procedure:~Winding numbers from the time-averaged Floquet DOS}\label{Sec_sum-rule}

In this section, we present an indirect way to access the quantized spectral flow $\mathcal{W}(\mu_\varepsilon)$, which is particularly appealing from the point of view of possible experimental explorations. In particular, this protocol which is based on an engineered-band scenario, provides a concrete scheme to measure the Floquet-St\v{r}eda response within a genuine many-body context.

The starting point is the alternative rewriting of the Floquet density of states defined in Eqs.~\eqref{DOS_F} and~\eqref{DOS_F_G} as the infinite sum
\begin{equation}
    \rho^{F}(\omega) = \sum_{m=-\infty}^\infty \rho^{F}_0(\omega+m\Omega)\,,
    \label{rho_F_series}
\end{equation}
where $m \in \mathbb{Z}$, and where we introduced
\begin{eqnarray} 
 \label{TADOS}
    \rho_{0}^{F}(\omega) &=& \frac{1}{A}\sum_{a,s} \langle u_{a}^{(s)}|u_{a}^{(s)}\rangle \delta(\omega-\varepsilon_{a s})\,.
\end{eqnarray}
Equation~\eqref{TADOS} defines the well-known \textit{time-averaged} Floquet density of states~\cite{Oka2009,Zhou2011,Usaj2014,Rudner2020,Rudner2020_b}, which projects the DOS of the extended Floquet Hamiltonian into its static component (the zeroth harmonic). We remark that $\rho_0^{F}(\omega)$ in Eq.~\eqref{TADOS} is localized in frequency space due to the localization of $|u_{a}^{(s)}\rangle$ in the index $s$. The Floquet spectral flow {at quasienergy $\mu_\varepsilon$}[see Eq.~\eqref{WF}] can hence be obtained as
\begin{eqnarray}
\label{illicit_switch}
    \mathcal{W}(\mu_\varepsilon) &=& \Phi_0 \int_{-\infty}^{\mu_\varepsilon}d\omega\,\sum_{m=-\infty}^\infty \frac{\partial \rho^{F}_0(\omega+m\Omega)}{\partial B}\,,\\
    \notag
    &\stackrel{(C,1)}{=}& \Phi_0 \sum_{m=-\infty}^\infty \int_{-\infty}^{\mu_\varepsilon+m\Omega} d\omega\, \frac{\partial \rho_0^{F}(\omega)}{\partial B}\,.
\end{eqnarray}
We note that in going from the first line to the second line of Eq.~\eqref{illicit_switch}, we have switched the unbounded integral and the summation over the infinite $m \in \mathbb{Z}$ via a Ces\`aro regularization procedure, see Appendix \ref{Appendix_demo}.
Importantly, even though the first line in Eq.~\eqref{illicit_switch} is an ill-defined (non-convergent) integral, the infinite sum in the second equality is completely convergent, in the sense that only a few values of $m$ around $m=0$ are needed to obtain $\mathcal{W}(\mu_\varepsilon)$ with $\mu_\varepsilon$ in the NFZ. Indeed, the terms with $m \ll -1$ are clearly small, due to the localization of $\rho_0^{F}(\omega)$ at low frequencies. On the other hand, using Eq.~\eqref{prop}, one can straightforwardly show that
\begin{equation}
    \int_{-\infty}^{\infty}d\omega\frac{\partial \rho^{F}_{0}(\omega)}{\partial B}=0\,,
    \label{prop_rho0}
\end{equation}
which means that the terms with $m \gg 1$ in Eq.~\eqref{illicit_switch} will also be vanishingly small. 


Equation~\eqref{illicit_switch} can be rewritten in a more appealing fashion, by introducing the time-averaged particle-density,
\begin{equation}
    \left.\overline{n}\right|_{\mu_\varepsilon} = \int_{-\infty}^{\mu_\varepsilon} d\omega \rho_0^{F}(\omega)\,.
    \label{n_average}
\end{equation}
In this way, the Floquet spectral flow at quasienergy $\mu_\varepsilon$, as obtained from Eq.~\eqref{illicit_switch}, can be written as the following summation
\begin{equation}
 \label{sum_rule}
   \mathcal{W}(\mu_\varepsilon) = \Phi_0 \sum_{m=-\infty}^\infty \left.\frac{\partial \overline{n}}{\partial B}\right|_{\mu_\varepsilon+m\Omega},
\end{equation}
from hereon referred to as the \textit{Floquet-St\v{r}eda sum rule procedure}.

Using the formalism developed in the Appendix~\ref{Kita_Arai_Floquet}, it is possible to obtain the magnetic response of the time-averaged Floquet density of states on a  translationally invariant system as
\begin{eqnarray}
\notag
    \Phi_0 \frac{\partial \rho_0^{F}(\omega)}{\partial B}&=& \mathrm{Im}\!\!\int_{\mathrm{BZ}}\!\!\frac{d^2k}{(2\pi)^2}\epsilon^{zjl}\textbf{tr}\left[\! i \hat{\bm{P}}_{0\bm{k}} \bm{G}^{F}_{\bm{k}}\!(\!\omega+i0^{+}\!)\!\frac{\partial \hat{\bm{H}}^{F}_{\bm{k}}}{\partial k_j}\right.\\
    &\times& \left. \bm{G}^{F}_{\bm{k}}\!(\omega+i0^{+})\!\frac{\partial \hat{\bm{H}}^{F}_{\bm{k}}}{\partial k_l}\bm{G}^{F}_{\bm{k}}\!(\omega+i0^{+})\right]\,,
    \label{drho0F_dB}
\end{eqnarray}
where $[\hat{\bm{P}}_{0\bm{k}}]_{nm} = \delta_{n 0}\delta_{m 0}\hat{I}_{\bm{k}}$, with $\hat{I}_{\bm{k}}$ the identity in $\bm{k}$-space. Equation~\eqref{drho0F_dB} is referred to as the \textit{energy-resolved Floquet-St\v{r}eda response of the time-averaged DOS}. 

 By means of Eqs.~\eqref{n_average} and ~\eqref{drho0F_dB}, we can numerically evaluate the different contributions appearing in Eq.~\eqref{sum_rule} entirely from the knowledge of the bulk Floquet-Bloch Green's functions $\hat{\bm{G}}^{F}_{\bm{k}}$ and Hamiltonian $\hat{\bm{H}}^{F}_{\bm{k}}$.

\begin{figure}[t]
    \centering
    \includegraphics[width=1.0\columnwidth]{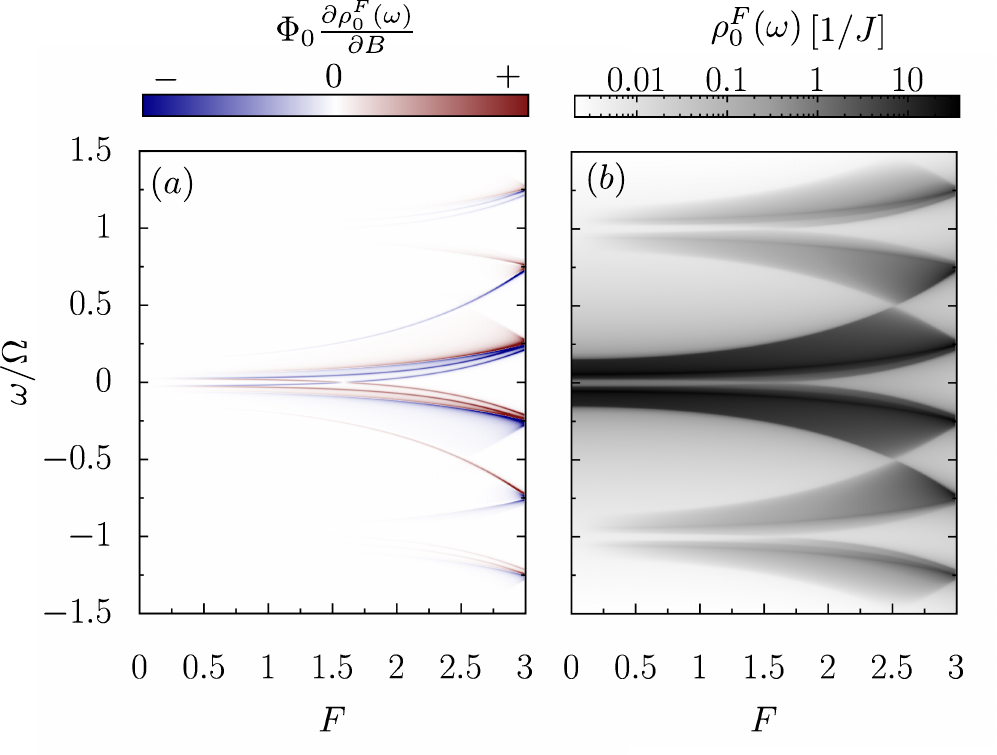}
    \caption{$(a)$ Energy-resolved Floquet-St\v{r}eda response of the time-averaged DOS, $\Phi_0 \frac{\partial \rho_0^{F}(\omega)}{\partial B}$ [see Eq.~\eqref{drho0F_dB}], as a function of the parameter $F$ in the Kitagawa model. The color scale is in arbitrary units, with $\pm$ indicating the sign of the response. $(b)$ Time-averaged Floquet DOS $\rho_0^{F}(\omega)$ as a function of frequency and the parameter $F$ for zero magnetic field. Note the logarithmic colorscale.}
    \label{streda_rho0}
\end{figure}

The restoration of quantized response functions via Floquet sum-rule schemes was originally discussed in Refs.~\cite{Kundu2013,Farrell2015,Farrell2016}. These theoretical works focused on the study of edge transport coefficients in open-boundary samples subjected to a time-periodic drive, and showed how quantized conductivities could be recovered by performing a series of measurements where the chemical potential of the leads is shifted in integer multiples of the driving frequency. As opposed to these transport settings, the Floquet-St\v{r}eda sum-rule procedure introduced in Eq.~\eqref{sum_rule} is entirely based on the measurement of a static response function that generically has non-trivial contributions from the edge and from the bulk of an open boundary sample.

As shown in Appendix~\ref{Appendix_bath}, the simplest model of a fermionic heat bath, described by featureless B\"uttiker-type reservoirs~\cite{Buttiker1985} connected to each site of the driven lattice, leads precisely to the steady-state particle-density given by Eq.~\eqref{n_average} when taking the limit of weak hybridization and zero temperature. In this case scenario, the value $\mu_\varepsilon$ in Eq.~\eqref{n_average} is physically determined by the chemical potential $\mu$ of the bath, such that the sum-rule in Eq.~\eqref{sum_rule} could be explicitly tested by changing this chemical potential in multiples of $\Omega$. Interestingly, this B\"uttiker bath does not lead to a thermal occupation of the Floquet bands in the NFZ [see Eq.~\eqref{D_occs} in Appendix~\ref{Appendix_bath}]. Notwithstanding, this highly non-thermal occupation is precisely the one that is needed to obtain quantization through the Floquet-St\v{r}eda sum rule procedure.  In this sense, this approach relaxes the typical requirement of preparing an insulator-like steady-state, which fully occupies the Floquet-Bloch bands. Engineering a bath as the one described in Appendix~\ref{Appendix_bath} could be possible with current experimental techniques in cold atom platforms, see Refs.~\cite{Schnell2023a,Schnell2024}. 

To further illustrate how the Floquet-St\v{r}eda sum-rule  procedure applies in practice, we now perform an explicit evaluation of these expressions for the Kitagawa model studied in Sec.~\ref{Sec_Kitagawa}. 
In Fig.~\ref{streda_rho0}$(a)$, we present the energy-resolved Floquet-St\v{r}eda response of the time-averaged Floquet DOS [Eq.~\eqref{drho0F_dB}] of this model as a function of the dimensionless parameter $F$. For the sake of clarity, we only plot its evolution up to one Floquet zone around the NFZ. In Fig.~\ref{streda_rho0}$(b)$, we present the corresponding time-averaged DOS,  $\rho_0^{F}(\omega)$, as a function of $F$ for zero magnetic field (note the logarithmic color scale). The parameters are the same as in Fig.~\ref{drho_dB_kitagawa}.
We can appreciate how both $\rho_0^{F}(\omega)$ and its magnetic field derivative are localized at low energies, with most of their weight in the NFZ for small values of $F$. As $F$ increases, the power spectrum of the time-periodic Floquet states is spread over more harmonics, leading to a corresponding spreading of the DOS and its magnetic response over the rest of the Floquet Brillouin zones. 

\begin{figure}[t]
    \centering
    \includegraphics[width=\columnwidth]{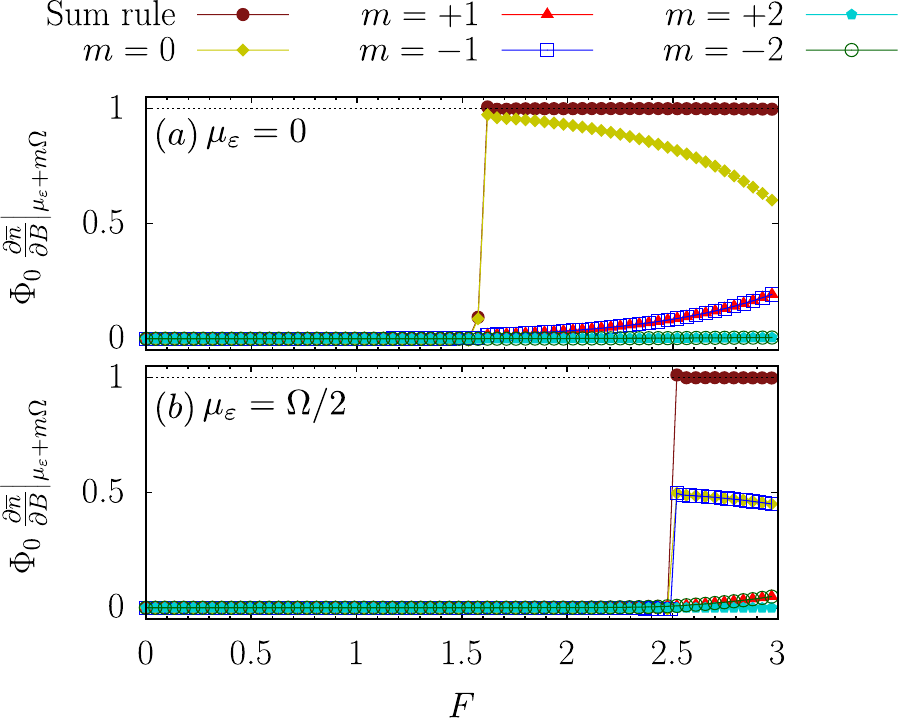}
    \caption{Floquet spectral flow obtained from the sum-rule procedure in Eq.~\eqref{sum_rule}.
    We present the contributions $m=0,\pm 1, \pm 2$ as a function of $F$ for $(a)$ $\mu_\varepsilon=0$ and $(b)$ $\mu_\varepsilon=\Omega/2$.}
    \label{sum_rule_kitagawa}
\end{figure}

The Floquet spectral flow $\mathcal{W}(\mu_\varepsilon)$ evaluated at the different spectral gaps of the NFZ, as obtained from the sum-rule in Eq.~\eqref{sum_rule}, is depicted in Fig.~\ref{sum_rule_kitagawa}. Here, we highlight the evolution of the $m=0,\pm 1, \pm 2$ contributions to this response function as a function of $F$. In Fig.~\ref{sum_rule_kitagawa}$(a)$, we fix $\mu_\varepsilon=0$ while  in Fig.~\ref{sum_rule_kitagawa}$(b)$, we set $\mu_\varepsilon=\Omega/2$. For this range of parameters, the response is remarkably well-converged and quantized by only summing these few values of $m$. The quantized values obtained for the different phases are in nice agreement with the ones in Fig.~\ref{cherns_anomaly_kitagawa}.
We note that the symmetry between the $\pm m$ contributions  in Fig.~\ref{sum_rule_kitagawa}$(a)$ is a consequence of particle-hole symmetry of the model at zero field.  This also explains the symmetry between the $m$ and $-m-1$ contributions in Fig.~\ref{sum_rule_kitagawa}$(b)$. The increase of the contributions with higher values of $m$ when increasing the parameter $F$ in Fig.~\ref{sum_rule_kitagawa} is consistent with the redistribution of spectral weight appreciated in Fig.~\ref{streda_rho0}.

These results demonstrate that the knowledge of the magnetic response of the time-averaged Floquet DOS is enough to fully characterize the topological properties of Floquet driven systems. This is another central result of the present work, as it provides an interesting route for experimentally accessing Floquet topological invariants in driven-dissipative systems with engineered heat-baths. In the next section, we will further develop this idea by defining a real-space version of the Floquet spectral flow. 

\section{Real-space approach to Floquet topological invariants}\label{Sec_local-marker}
In this section, we address the problem of promoting the bulk winding numbers that characterize the topological structure of two-dimensional Floquet driven systems to real-space resolved markers. While local topological markers have been extensively studied both in and out-of-equilibrium~\cite{Kitaev2006,Bianco2011,Tran2015,Privitera2016,Caio2019,Ulcakar2020}, their extension and generalization to Floquet topological phases still remains elusive. Indeed, the local Chern number identified in Ref.~\cite{Bianco2011}, has only been proven to be useful to characterize driven Floquet phases when $\Omega$ is larger than the bandwidth of the undriven model and when the initially occupied states evolve under a unitary time-evolution that adiabatically populates the Floquet bands~\cite{Privitera2016}.

We remark that, until this point, the Floquet spectral flow as obtained from Eq.~\eqref{WF_Cesaro}, and its two contributions, $\mathcal{W}^{N}(\mu_\varepsilon)$ and $\mathcal{W}^{A}$, as defined from Eqs.~\eqref{normal_flow} and~\eqref{anomaly}, have only been explicitly computed in lattices with discrete translational symmetry.  Nevertheless, the notion of spectral flow can be straightforwardly applied in a local fashion by defining the real-space resolved version of these quantities as
\begin{equation}
    \mathcal{W}(\mu_\varepsilon,\bm{R}) = \mathcal{W}^{N}(\mu_\varepsilon,\bm{R}) + \mathcal{W}^{A}(\bm{R})\,,
    \label{W_local}
\end{equation}
where
\begin{equation}
    \mathcal{W}^{N}(\mu_\varepsilon,\bm{R}) = \Phi_0 \int_{\varepsilon_{\pi}-\Omega}^{\mu_\varepsilon}d\omega \frac{\partial \rho^F(\omega,\bm{R})}{\partial B}\,,
    \label{WN_local}
\end{equation}
and
\begin{equation}
  \mathcal{W}^{A}(\bm{R}) = \Phi_0 \int_{\mathrm{NFZ}}^{}d\omega \frac{\partial \rho^F(\omega,\bm{R})}{\partial B}  \frac{\omega}{\Omega}\,.
  \label{WA_local}
\end{equation}
Here, we have defined the real-space versions of the normal and anomalous spectral flow by means of the local Floquet density of states at the unit cell $\bm{R}$, given by
\begin{equation}
    \rho^{F}(\omega,\bm{R}) = -\frac{1}{\pi}\sum_{n=-\infty}^\infty\sum_{\nu}\mathrm{Im}\left[\langle \bm{R}_{\nu}|\hat{\bm{G}}^{F}_{nn}(\omega+i 0^{+})|\bm{R}_{\nu}\rangle\right]\,,
\end{equation}
where the block elements of the Floquet Green's function are given in Eq.~\eqref{GRF} in Appendix~\ref{Kita_Arai_Floquet}. Interestingly, these markers do not deal with the problem of the occupation of the Floquet bands \cite{Privitera2016},
but only rely on its spectral properties. From hereon, we will refer to $\mathcal{W}(\mu_\varepsilon,\bm{R})$ as the local Floquet winding number.

\begin{figure}[t]
    \centering
    \includegraphics[width=\columnwidth]{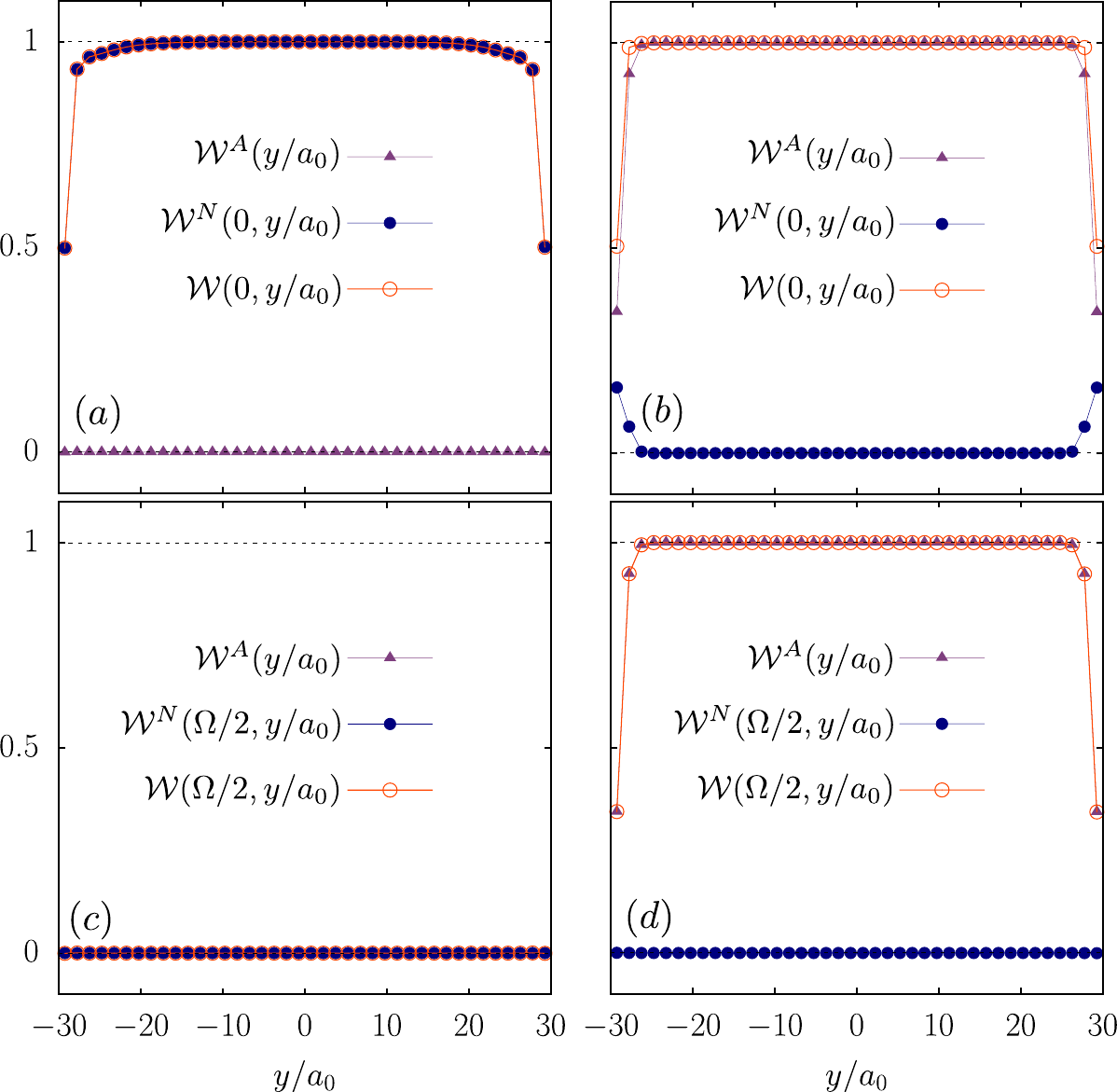}
    \caption{Local Floquet winding number $\mathcal{W}(\mu_\varepsilon,\bm{R})$ in an infinite zigzag ribbon of the Kitagawa model [Eq.~\eqref{H_Kitagawa}] as a function of the $y$-coordinate along its width ($N_y=40$ unit cells). We also show its two contributions $\mathcal{W}^{N}(\mu_{\varepsilon},\bm{R})$ and $\mathcal{W}^{A}(\bm{R})$. $(a)$  $\mu_\varepsilon=0$ and $F=2.0$, $(b)$ $\mu_\varepsilon=0$ and $F=2.75$, $(c)$ $\mu_\varepsilon=\Omega/2$ and $F=2.0$ and $(d)$  $\mu_\varepsilon=\Omega/2$ and $F=2.75$. In all the panels $\Omega/J=20$ and $\Delta/J = 0.5$.}
    \label{local_invariants_ribbon}
\end{figure}

\begin{figure}[t]
    \centering
    \includegraphics[width=0.75\columnwidth]{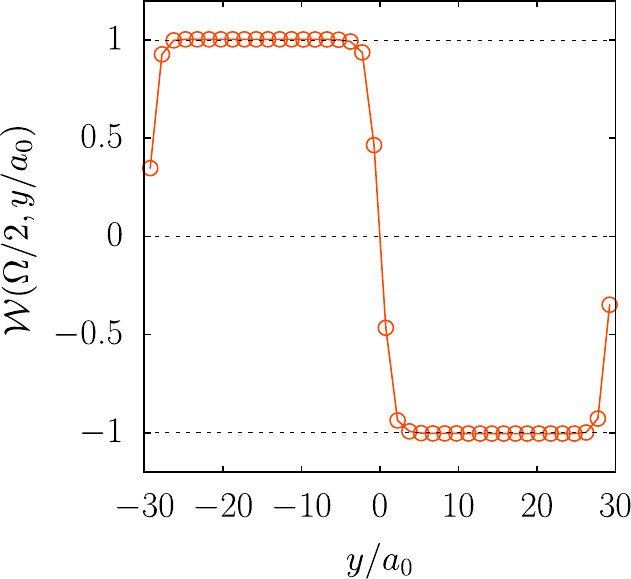}
    \caption{Local Floquet winding number $\mathcal{W}(\mu_\varepsilon,\bm{R})$ at $\mu_\varepsilon=\Omega/2$ in an infinite zigzag ribbon of the Kitagawa model [Eq.~\eqref{H_Kitagawa}] as a function of the $y$-coordinate along its width. The ribbon is divided into two infinite halves. While both are in the anomalous regime ($F=2.75$) and with parameters $\Delta/J=0.5$ and $\Omega/J=20$, the upper half ($y>0$) has hopping phases with a relative change of sign with respect to the lower half ($y<0$), i.e. $\theta_{\nu}(y<0)= -\theta_{\nu}(y>0)$, hence changing the chirality of the anomalous edge state.}
    \label{local_junction}
\end{figure}

Deep within the gapped bulk of an homogeneous sample, Eqs.~\eqref{W_local},~\eqref{WN_local} and~\eqref{WA_local} should coincide with the bulk invariants that we previously identified for lattice systems in a torus geometry. This is explicitly shown in Fig.~\ref{local_invariants_ribbon}, where we compute these quantities along the width of an infinite ribbon of the Kitagawa model studied in Sec.~\ref{Sec_Kitagawa}. In Figs.~\ref{local_invariants_ribbon}$(a)$ and $(b)$, we focus on the gap at the center of the NFZ ($\mu_\varepsilon=0$) and evaluate the local winding numbers for $F=2.0$ (Floquet Chern insulator phase \textcircled{\tiny{II}}) and $F=2.75$ (anomalous Floquet phase \textcircled{\tiny{III}}), respectively. The plots nicely confirm that the local winding number approaches the value $1$ in the bulk of the ribbon---an indication of the existence of non-trivial topological edge modes for both set of parameters (see Fig.~\ref{kitagawa_model})---while it deviates from the quantized value near the boundaries of the sample. In Fig.~\ref{local_invariants_ribbon}$(a)$, all the contribution comes from $\mathcal{W}^{N}(0,\bm{R})$, while in Fig.~\ref{local_invariants_ribbon}$(b)$ the main contribution comes from the local anomaly $\mathcal{W}^{A}(\bm{R})$, in agreement with the results in Fig.~\ref{cherns_anomaly_kitagawa}. We note that the local anomaly in Fig.~\ref{local_invariants_ribbon}$(a)$ is strictly zero across the entire ribbon, in agreement with the fact that the anomalous spectral flow should strictly vanish for $\Omega$ larger than the bandwidth of the model (as it is the case here). In Figs.~\ref{local_invariants_ribbon}$(c)$ and $(d)$ we focus on the evaluation of the local winding number at the zone-edge gap ( $\mu_\varepsilon=\varepsilon_\pi = \Omega/2$) and again consider the parameters $F=2.0$ and $F=2.75$, respectively. In Fig.~\ref{local_invariants_ribbon}$(c)$, we obtain a trivial local response, indicating the absence of edge states at $\Omega/2$ for these parameters. In contrast, in Fig.~\ref{local_invariants_ribbon}$(d)$ we recover the quantized value of $1$ deep in the bulk. Note that in these last two panels, the spectral flow given by $\mathcal{W}^{N}(\Omega/2,\bm{R})$ is strictly zero.

These local quantities can be used as well in the presence of disorder or to identify regions of different topological order in heterojunctions. This latter situation is shown in Fig.~\ref{local_junction}, were we plot the local Floquet winding number  $\mathcal{W}(\mu_\varepsilon,\bm{R})$ at $\mu_\varepsilon=\Omega/2$ in a Kitagawa ribbon that is divided at $y=0$ into two  halves. While both halves are in the anomalous regime ($F=2.75$) and with parameters $\Delta/J=0.5$ and $\Omega/J=20$, the upper half ($y>0$) has modulated nearest-neighbour hoppings [Eq.~\eqref{H_Kitagawa}] with phases that have a relative change of sign with respect to the lower half ($y<0$), i.e. $\theta_{\nu}(y<0)= -\theta_{\nu}(y>0)$ $\forall\, \nu$, hence changing the chirality of the anomalous edge states. This result nicely confirms that the local winding number in Eq.~\eqref{W_local} is able to distinguish phases with different topological order in different bulk-like regions. We have verified (not shown) that our approach correctly describes systems with arbitrary large winding numbers, as for example graphene in the presence of circularly polarized light with a frequency much smaller than the bandwidth~\cite{Usaj2014,PerezPiskunow2015}.

\begin{figure}[t]
    \centering
    \includegraphics[width=\columnwidth]{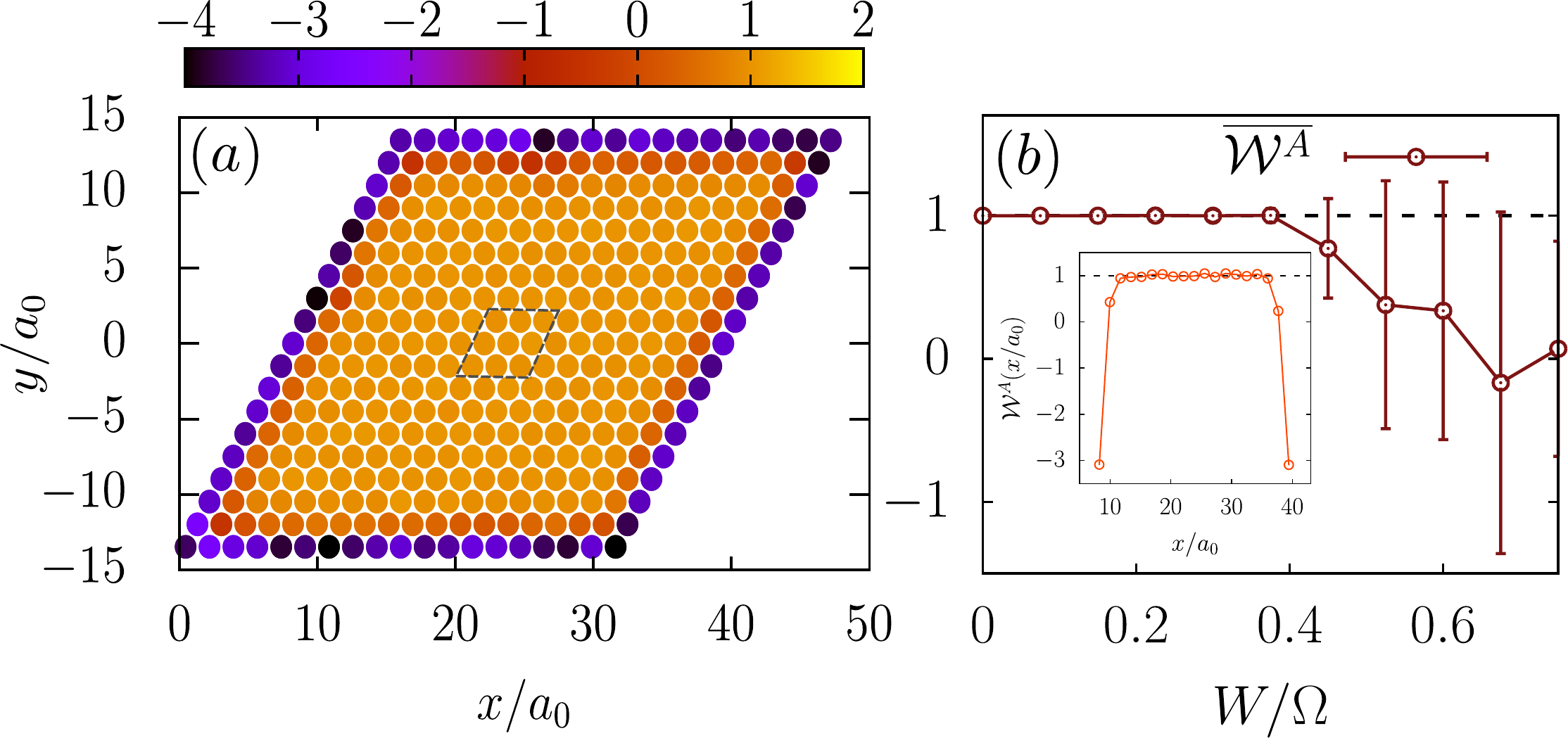}
    \caption{$(a)$ Local anomalous winding number $\mathcal{W}^{A}(\bm{R})$ in a finite flake ($19\times 19$ unit cells) of the Kitagawa model for $F=2.75$, $\Omega/J = 20$ and a given realization of random diagonal disorder of strength $W=0.22\,\Omega$. $(b)$ Average of the local anomalous winding number $\overline{\mathcal{W}}^{A}$ over the bulk region delimited with dashed grey-lines in panel $(a)$ as a function of disorder strength $W$. The error bars represent the standard deviation of the local marker within the bulk region. Each point has been obtained by averaging over $100$ disorder realizations. The inset shows a horizontal cut ($y=0$) of the local Floquet marker for a disorder strength of $W=0.22\,\Omega$ with the same disorder realization shown in panel $(a)$.}
    \label{disorder_figure}
\end{figure}

In Fig.~\ref{disorder_figure}, we study how disorder affects the local Floquet winding number in the anomalous regime ($F=2.75$). To that end, we add to the Hamiltonian in ~Eq.~\eqref{H_Kitagawa} a random time-independent potential of the form $\hat{V}_{\textrm{disorder}}=\sum_{\bm{R}\in \{\mathcal{A},\mathcal{B}\}}w_{\bm{R}}\hat{c}^{\dagger}_{\bm{R}}\hat{c}^{}_{\bm{R}}$, with the on-site energies $w_{\bm{R}}$ randomly drawn from a uniform distribution in the interval $[-W,W]$. The spatial dependence of the local anomalous winding number $\mathcal{W}^{A}(\bm{R})$ within a finite flake for $W=0.22\,\Omega$ and a given disorder realization is shown in Fig.~\ref{disorder_figure}$(a)$. It clearly appears that, within the bulk, this quantity fluctuates around a macroscopic value of $\sim 1$, the quantized value expected for the clean system ($W=0$). In Fig.~\ref{disorder_figure}$(b)$, we present the averaged local marker within a bulk region as a function of disorder strength $W$, showing how it remains remarkably well-quantized until a critical disorder strength of $W_c\sim 0.4\,\Omega$. We have verified that at this critical value, the bulk spectral gap of the flake closes at the Floquet zone-edge. 

These results confirm that the total bulk orbital magnetization density of Floquet states within a zone is quantized both for clean and disordered systems, provided the dynamical bulk gap at the Floquet zone-edge remains open.

\section{Discussion and Perspectives}\label{Conclusions}
This work introduced a general framework to classify topological Floquet systems based on the notion of spectral flow inherent to the equilibrium St\v{r}eda formula, hence offering a unified and coherent approach to the bulk-boundary correspondence of periodically-driven systems. 

Our theory provides a physical route to derive Floquet winding numbers, elucidating how these topological invariants can be entirely built from simple bulk properties of the Floquet states defined in Sambe space. Importantly, this theory identifies a fundamental mechanism at the origin of anomalous edge states:~the emergence of a magnetic-field-induced energy pump, between the system and the driving field, which is inextricably linked to the presence of resonant processes. 
Furthermore, our approach reveals that the classification of topological Floquet systems directly derives from the magnetic response of their density of states, such that all Floquet winding numbers can be fully extracted from the stroboscopic time-evolution of driven systems. 

Interestingly, periodically-driven systems belonging to certain symmetry classes~\cite{Roy2017} can exhibit anomalous edge modes with rich structures, which are not accounted for by the $N_3[R]$ invariant in Eq.~\eqref{WA_Rudner}. This is the case of the driven systems analyzed in Refs.~\cite{Umer2020,Shi2022,Zhang2023_d}, where particle-hole symmetry gives rise to helical edge modes that are reminiscent of valley Hall systems, i.e.~robust edge modes with opposite chirality that emerge at opposite valleys (gap closing points) in the Brillouin zone. As pointed out in Ref.~\cite{Shi2022}, the bulk-boundary correspondence of these symmetry-protected Floquet settings is correctly captured by a set of topological invariants that properly count the number of counter-propagating edge modes in each gap. In this context, a quantized St\v{r}eda response could be generated by a perturbation  that generates a helical spectral flow of edge states, i.e.~a momentum-dependent pseudo-magnetic field that acts on opposite valleys with an opposite sign. In static valley-Hall systems, Ref.~\cite{Jamotte2023} demonstrated that such an helical St\v{r}eda-type response can be generated by strain. Extending this approach to Floquet systems with anomalous helical edge states, and more generally, elucidating generalized St\v{r}eda responses for all topological symmetry classes~\cite{Roy2017}, constitute an exciting program.

Another promising direction concerns the generalization of our work to correlated systems.
Indeed, the classification of interacting Floquet topological systems and the stabilization of these dynamical phases is a subject of current research interest~\cite{Gross2012,Roy2016,Potter2016,Po2016,Else2016,Khemani2016,vonKeyserlingk2016,Harper2017,Moessner2017,Fidkowski2019,Nathan2019,Gong2020,Nathan2021,Zhang2023_c}. 
Our approach could be employed to characterize many-body Floquet systems in nonequilibrium steady states stabilized by coupling to thermal reservoirs~\cite{Tsuji2008,Aoki2014}. In such settings, a positive-definite Floquet density of states can be defined~\cite{Uhrig2019}, allowing for the extraction of the magnetic response from the single-particle correlators in the Floquet representation~\cite{Puviani2016}.
This could offer a new perspective on the physical interpretation of the topological indices that are relevant to classify correlated Floquet phases. Such a generalization of our approach would also be relevant for the experimental characterization of quantum-engineered Floquet settings, such as drive-induced fractional Chern insulators and chiral spin liquids~\cite{Cooper_RMP,Ozawa_RMP,Kalinowski2023,Sun2023,Nishad2023,Mambrini2024,Poliblanc2024}.

A protocol to dynamically detect the local Floquet density of states and to measure the quasienergy spectrum was recently introduced in Ref.~\cite{Zhang2024}. Having access to the magnetic response of these quantities represents a promising route to measure the energy-resolved Floquet-St\v{r}eda response and the local Floquet winding numbers introduced in this paper. Recent advances in the engineering of heat-baths in quantum simulator settings could also open the possibility of extracting the winding numbers from density measurements by means of the sum-rule scheme proposed in our work.

From a more fundamental point of view, it would be interesting to further explore the quantized energy-pump mechanism encoded in the anomalous flow, by theoretically considering the full quantum-mechanical description of the driving field~\cite{Wang2019,Sentef2020,Li2020,Dmytruk2022,Eckhardt2022,Appugliese2022,Nguyen2023,Masuki2023,Pasetti2023,PerezGonzalez2025,PerezGonzalez2023,Dag2024}. 
Within our semiclassical Floquet picture, the first-order winding number of the micromotion operator $N_1[R]$ is defined modulo an arbitrary integer number, but its absolute value could be meaningful in a fully quantized theory, where it should represent the average number of quanta of the driving field, and hence be bounded from below. Establishing a rigorous description of this topological index and its magnetic field derivative in the context of quantum Floquet-engineered systems could provide new insight on the physical origin of anomalous edge states in periodically-driven settings. In particular, the interpretation of $\hbar N_3[R]$ as being the angular momentum exchanged between the system and the driving field [Eq.~\eqref{conservation_totalM}]  should naturally emerge from a description in terms of driving-field quanta.

Let us conclude with a more formal remark, regarding the use of Ces\`aro summation in this physical context. The identification of a Grandi-type series at the very core of Floquet topological physics, but also the fact that its formal evaluation gives rise to a physical result are, at the very least, intriguing. In particular, the fact that the infinite summation of apparent zeros (i.e.~the trivial Chern numbers in Fig.~\ref{conceptual_SF}) leads to a \emph{finite} (and quantized) result, highlights the exquisite geometric fine-structure that is hidden within each Floquet-Bloch band. We are unaware of other applications of the Ces\`aro summation scheme as a method to regularize winding numbers of operators with an unbounded spectrum, as was the case for the Floquet Green's function index $N_3[\bm{G}^{F}]$ defined in Eq.~\eqref{N3G_F}. It would be interesting to explore whether this mathematical method could be further employed to identify new classes of topological indices and phenomena in physical settings. 

\begin{acknowledgments}
We thank I. Amelio, N. Dupont for helpful discussions. This research was financially supported by the FRS-FNRS (Belgium), the ERC Grant LATIS and the EOS project CHEQS. LPG  acknowledges support
provided by the L’Or\'eal-UNESCO for Women in Science Programme and the use of computational resources provided by the HPC cluster of the Physics Department at Centro At\'omico Bariloche (CNEA), Argentina. LPG would also like to thank the Institut Henri Poincaré (UAR 839 CNRS-Sorbonne Université) and the LabEx CARMIN (ANR-10-LABX-59-01) for their support. 
GU acknowledges ﬁnancial support from the ANPCyT-FONCyT (Argentina) under grants PICT 2018-1509 and PICT 2019-0371, SeCyT-UNCuyo grant 06/C053-T1 and the FRS-FNRS for a scientific research stay grant 2024/V 6/5/012.
\end{acknowledgments}
\vspace{0.5cm}
\textit{Authors' contributions.--} LPG conceived the project and developed the theoretical formalism with inputs from GU and NG. All authors contributed to the analysis and interpretation of the results, and to the writing of the manuscript.

\appendix
\section{Derivation of Eq.~\eqref{drho_dB} from the modified density of states in phase-space}\label{A1}
In Ref.~\cite{Xiao2005} the authors showed, using semiclassical arguments, that the phase-space density of states of Bloch electrons is modified in the presence of a both a finite Berry curvature and a weak external magnetic field. In a two-dimensional crystal, up to first order in the field, the phase-space DOS can be expressed as
\begin{equation}
    g_{\alpha}(\bm{k}) = \frac{1}{(2\pi)^2}\left(1 + \frac{2\pi}{\Phi_0} B \mathcal{F}^{\alpha}_{xy}(\bm{k})\right),
    \label{dos_Niu}
\end{equation}
with the index $\alpha$ indicating the Bloch band under consideration. 
In this appendix we show that the energy-resolved density of states response we report in Eq.~\eqref{drho_dB} is in full agreement with Eq.~\eqref{dos_Niu}. By definition, 
\begin{equation}
    \rho(\omega) = \int_{\mathrm{BZ}}d^2 k \sum_{\alpha} g_{\alpha}(\bm{k})\delta(\omega-\varepsilon_{\alpha\bm{k}}),
\end{equation}
such that
\begin{eqnarray}
    \label{A3}
    \Phi_0 \left.\frac{\partial \rho(\omega)}{\partial B}\right|_{B=0} &=& \int_{\mathrm{BZ}}\frac{d^2k}{2\pi}\sum_{\alpha} \mathcal{F}^{\alpha}_{xy}\delta(\omega-\varepsilon_{\alpha\bm{k}})\\
    \notag
    &-& \Phi_0\int_{\mathrm{BZ}} \frac{d^2k}{(2\pi)^2}\sum_{\alpha} \frac{\partial \delta(\omega-\varepsilon_{\alpha\bm{k}})}{\partial \omega}\frac{\partial \varepsilon_{\alpha\bm{k}}}{\partial B}. 
\end{eqnarray}
Using the semiclassical equation for the intrinsic orbital magnetic moment of a wavepacket centered at $\bm{k}$ in the $\alpha$-th Bloch band
\begin{equation}
    m^{\alpha}_z(\bm{k}) = -\frac{\partial \varepsilon_{\alpha\bm{k}}}{\partial B} = \frac{2\pi}{\Phi_0}\mathrm{Im}\left[\left\langle \partial_{k_x} u_{\alpha\bm{k}}|\hat{H}_{\bm{k}}-\varepsilon_{\alpha\bm{k}}|\partial_{k_y}u_{\alpha\bm{k}}\right\rangle\right],
\end{equation}
and replacing this expression in Eq.~\eqref{A3} we recover Eq.~\eqref{drho_dB} in the main text.
\section{Floquet Green's functions in the presence of a perturbing magnetic field}\label{Kita_Arai_Floquet}
In this Appendix, we generalize the formalism used in Sec.~\ref{Sec_eq} to obtain the first-order correction of the Floquet-Green's function in the presence of a perturbing magnetic field. The two-time propagators of systems subjected to a time-periodic drive are expressed as~\cite{Aoki2014}
\begin{equation}
    \hat{G}^{}(t,t') = \sum_{m,n} \int_{\mathrm{NFZ}} \frac{d\omega}{2\pi} e^{-i (\omega + m\Omega)t}e^{i (\omega+n\Omega)t'}\hat{\bm{G}}_{mn}^{F }(\omega)\,,
\end{equation}
where $\hat{\bm{G}}^{F}(\omega)$ stands for the time-independent propagator written in Sambe representation. In the absence of interactions,  
\begin{equation}
   \hat{\bm{G}}^{F}(\omega) = (\omega \hat{\bm{I}}-\hat{\bm{H}}^{F})^{-1}, 
\end{equation}
with $\hat{\bm{H}}^{F}$ the infinite-dimensional Floquet Hamiltonian. We note that
\begin{equation}
    \hat{\bm{G}}^{F}_{mn}(\omega) = \sum_{a, s}\frac{|u^{(m+s)}_{a}\rangle \langle u^{(n+s)}_{a}|}{\omega-\varepsilon_{a s}}\,,
    \label{GRF}
\end{equation}
where we remind that the index $a$ is restricted such that the quasienergies $\varepsilon_{a}$ lie in the NFZ and the sum over $s$ runs over all $\mathbb{Z}$. The Dyson equations of motion coupling these blocks are given by
\begin{equation}
    \sum_{m'}[\omega\, \delta_{m m'}\hat{I} - \hat{\bm{H}}^{F}_{m m'}]\hat{\bm{G}}^{F}_{m'n} = \delta_{m n}\hat{I}\,.
    \label{Dyson_blocks}
\end{equation}
In the presence of a small magnetic field, the matrix elements of the Floquet Hamiltonian in real-space are modified as
\begin{equation}
[\hat{\bm{H}}^{F}_{mn}]_{\bm{R}_{\nu}\bm{R'}_{\nu'}} = e^{i\frac{2\pi}{\Phi_0} \varphi_{\bm{R}_{\nu}\bm{R'}_{\nu'}}}[\hat{\bm{H}}^{F 0}_{mn}]_{\bm{R}_{\nu}\bm{R'}_{\nu'}}\,,
\label{Peierls_Floquet}
\end{equation}
with the Peierls phase factors defined as in Eq.~\eqref{Peierls_phases} and with $\hat{\bm{H}}^{F 0}$ being the Floquet Hamiltonian in the absence of the field. In the same spirit of the derivation in Sec.~\ref{Sec_eq}, we can introduce a gauge-invariant Floquet propagator with matrix elements defined as
\begin{equation}
    [\hat{\bm{G}}^{F (B)}_{mn}(\omega)]_{\bm{R}_{\nu}\bm{R'}_{\nu'}} = e^{-i\frac{2\pi}{\Phi_0} \varphi_{\bm{R}_{\nu}\bm{R'}_{\nu'}}}  [\hat{\bm{G}}^{F}_{mn}(\omega)]_{\bm{R}_{\nu}\bm{R'}_{\nu'}}\,.
    \label{GF_B_mn}
\end{equation} 
Introducing Eq.~\eqref{GF_B_mn} into Eq.~\eqref{Dyson_blocks}, we find that the gauge-invariant Floquet propagator satisfies the following modified Dyson's equation of motion, 
\begin{eqnarray}
\notag
    \sum_{\bm{R'}\nu'}\sum_{m'}& &\left(\omega \delta_{\bm{R}_{\nu}\bm{R'}_{\nu'}}\delta_{mm'}-[\hat{\bm{H}}^{F 0}_{mm'}]_{\bm{R}_{\nu}\bm{R'}_{\nu'}}\right)[\hat{\bm{G}}^{F(B)}_{m'n}(\omega)]_{\bm{R'}_{\nu'}\bm{R''}_{\nu''}}\\
\label{GinvDyson_F}
    & &\times e^{i\frac{\pi}{\Phi_0}\bm{B}\cdot (\bm{R'}_{\nu'}-\bm{R}_{\nu})\times(\bm{R''}_{\nu''}-\bm{R'}_{\nu'})} = \delta_{mn}\delta_{\bm{R}_{\nu}\bm{R''}_{\nu''}}\,.
\end{eqnarray}
Fourier transforming Eq.~\eqref{GinvDyson_F} to quasimomentum space and expanding up to first order in the magnetic perturbation, we find that
\begin{equation}
\label{GkB_F}
    \hat{\bm{G}}^{F (B)}_{\bm{k}}\!(\omega)\!=\!\hat{\bm{G}}^{F}_{\bm{k}}(\omega) + \frac{i\pi}{\Phi_0} B^{i} \epsilon^{ijl}\hat{\bm{G}}^{F}_{\bm{k}}\!(\omega)\frac{\partial \hat{\bm{G}}^{F -1}_{\bm{k}}\!(\omega)}{\partial k_j}\frac{\partial \hat{\bm{G}}^{F}_{\bm{k}}\!(\omega)}{\partial k_l}\,,
\end{equation}
where 
\begin{equation}
 \hat{\bm{G}}^{F}_{\bm{k}}(\omega) = (\omega\hat{\bm{I}}_{\bm{k}}-\hat{\bm{H}}_{\bm{k}}^{F})^{-1}   
\end{equation}
is the Floquet-Bloch Green's function in the absence of the field. We have here introduced the Floquet-Bloch Hamiltonian $\hat{\bm{H}}_{\bm{k}}^{F}$ in $\bm{k}$-space, related to $\hat{\bm{H}}^{F 0}$ in Eq.~\eqref{Peierls_Floquet} via
\begin{equation}
    [\hat{\bm{H}}^{F 0}_{mn}]_{\bm{R}_{\nu}\bm{R'}_{\nu'}} = \frac{1}{N_c}\sum_{\bm{k}}e^{i\bm{k}\cdot(\bm{R}_{\nu}-\bm{R'}_{\nu'})}[\hat{\bm{H}}_{\bm{k},mn}^{F}]_{\nu\nu'}\,.
\end{equation}
Considering that the external field $\bm{B}$ is applied along the $z$-direction, the magnetic response of the Floquet DOS is hence given by
\begin{eqnarray}
\label{drho_dB_F_appendix}
    \Phi_0 \frac{\partial \rho^{F}(\omega)}{\partial B} &=& -\frac{\Phi_0}{\pi}\frac{\partial}{\partial B}\mathrm{Im}\bm{\mathrm{Tr}}[\hat{\bm{G}}^{F}(\omega + i0^{+})]\\
    \notag
    &=& -\frac{\Phi_0}{\pi}\frac{\partial}{\partial B}\mathrm{Im}\bm{\mathrm{Tr}}[\hat{\bm{G}}^{F (B)}(\omega + i0^{+})]\\    
    \notag
    &=& -\frac{\Phi_0}{\pi}\int_{\mathrm{BZ}} \frac{d^2k}{(2\pi)^2}\mathrm{Im}\,\bm{\mathrm{tr}}\left[\frac{\partial \hat{G}^{F (B)}_{\bm{k}}(\omega+i 0^{+})}{\partial B}\right]\, ,
\end{eqnarray}
where we have used that $\bm{\mathrm{Tr}}[\hat{\bm{G}}^{F}] = \bm{\mathrm{Tr}}[\hat{\bm{G}}^{F (B)}]$, on account of the cancellation of the Peierls phase factors when performing the trace operation. Inserting Eq.~\eqref{GkB_F} into Eq.~\eqref{drho_dB_F_appendix} and using the spectral decomposition of the Floquet Hamiltonian and Green's function we finally arrive at Eq.~\eqref{drhoF_dB} in the main text.
\section{Alternative regularizations schemes}\label{alt_regularizations}

In Eq.~\eqref{Cesaro} in the main text, we have chosen to present a regularization of the divergent integral defining $\mathcal{W}^{A}$ [Eq.~\eqref{WFA}] via a Ces\`aro $(C,1)$ method. Notably, once an integral is $(C,1)$ summable, it is also $(C,\zeta)$ summable with $\zeta$ an integer larger than $1$. Even more, the result of this higher-order regularization procedure is exactly the same as the one obtained through the $(C,1)$ regularization~\cite{Titchmarsh1986}, such that
\begin{eqnarray}
\label{higher_cesaro}
\mathcal{W}^{A} &\stackrel{(C,\zeta)}{=}& \lim_{\lambda\to \infty}\Phi_0 \int_{-\infty}^{\varepsilon_{\pi}-\Omega}d\omega\frac{\partial \rho^{F}(\omega)}{\partial B}\left(1+\frac{\omega}{\lambda}\right)^{\zeta}\\
\notag
&=& \sum_{\alpha}\int_{\mathrm{BZ}}\frac{d^2 k}{2\pi \Omega}\left[\mathcal{F}_{xy}^{\alpha}(\bm{k})\varepsilon_{\alpha\bm{k}} - \frac{\Phi_0}{2\pi} m_z^{\alpha}(\bm{k})\right].
\end{eqnarray}
We could have also chosen to regularize Eq.~\eqref{WFA} by considering Abel's summation prescription, which consists on taking the following limit
\begin{eqnarray}
\label{Abel}
 \mathcal{W}^{A} &=&  \lim_{\kappa \to 0}\Phi_0\int_{-\infty}^{\varepsilon_{\pi}-\Omega} d\omega\frac{\partial \rho^{F}(\omega)}{\partial B} e^{\kappa \omega}\\
 \notag
 &=&  \lim_{\kappa \to 0}\sum_{\alpha}\sum_{s=-\infty}^{s=-1}\int_{\mathrm{BZ}}\frac{d^2 k}{2\pi}\left[\mathcal{F}_{xy}^{\alpha}(\bm{k}) - \kappa\frac{\Phi_0}{2\pi} m_z^{\alpha}(\bm{k})\right]e^{\kappa\varepsilon_{\alpha s \bm{k}}}.
\end{eqnarray}
Explicitly performing the infinite summation over Floquet Brillouin zones,
\begin{equation}
    \sum_{s=-\infty}^{s=-1}e^{\kappa s \Omega}= \frac{e^{-\kappa \Omega}}{e^{-\kappa \Omega}-1},
    \label{geometric_series}
\end{equation}
replacing Eq.~\eqref{geometric_series} in Eq.~\eqref{Abel}, and then taking the limit $\kappa \to 0$, we also re-obtain the result in Eq.~\eqref{higher_cesaro}. In this sense, the  result obtained in Eq.~\eqref{Cesaro} in the main text, expressed in terms of physical observables such as the Berry curvatures, quasienergies and intrinsic orbital magnetic moments of Floquet-Bloch bands, remains remarkably independent of the regularization scheme introduced to perform the calculation.

\section{Equivalence of $\mathcal{W}^{A}$ and the winding number of the micromotion operator\label{proof_WR}}
The anomalous winding number $\mathcal{W}^{A}$ in Eq.~\eqref{anomaly} can be expressed in the extended Sambe space as
\bea
\nonumber
\mathcal{W}^A&=&\frac{i\epsilon^{zjl}}{2\pi\Omega}\sum_{\alpha,\beta,s}\Biggl\{\\
\nonumber
&&\int_{\mathrm{BZ}}d^2k\, \varepsilon_{\alpha\bm{k}}\frac{\Bra{u_{\alpha\bk}}\frac{\partial \HFk}{\partial k_j}\Ket{u_{\beta s\bk}}\Bra{u_{\beta s\bk}}\frac{\partial\HFk}{\partial k_l}\Ket{u_{\alpha\bk}}}{(\varepsilon_{\alpha\bm{k}}-\varepsilon_{\beta s\bm{k}})^2}\\
\nonumber
&&-\frac{1}{2} \int_{\mathrm{BZ}}d^2k\,\frac{\Bra{u_{\alpha\bk}}\frac{\partial \HFk}{\partial k_j}\Ket{u_{\beta s \bk}}\Bra{u_{\beta s\bk}}\frac{\partial \HFk}{\partial k_l}\Ket{u_{\alpha\bk}}}{\varepsilon_{\alpha\bm{k}}-\varepsilon_{\beta s\bm{k}}}\Biggr\},\\
\label{WA}
\eea
where we have explicitly used the form of the Berry curvature and orbital magnetization given by Eqs.~\eqref{curvature_F} and~\eqref{magnetization_F}.
We recall that the notation $\Ket{\dots}$ indicates a vector in the extended Sambe space, $\varepsilon_{\beta s\bm{k}}=\varepsilon_{\beta\bm{k}}+s \Omega$, that $\Ket{u_{\alpha\bk}}\equiv\Ket{u_{\alpha0\bk}}$ and that the sum over $s\in \mathbb{Z}$ represents the infinite number of Floquet Brillouin zones. It is straightforward to verify that the two main terms in Eq.~\eqref{WA} cancel out for $s=0$ (NFZ)---this is easily seen by symmetrizing the first term with respect to the indexes $\alpha$ and $\beta$. Therefore, the sum over $s$ can be restricted to $s\neq0$.
In such a case, one can safely write
\be
\Bra{u_{\alpha\bk}}\frac{\partial \HFk}{\partial k_j}\Ket{u_{\beta s\bk}}=(\varepsilon_{\alpha\bm{k}}-\varepsilon_{\beta s\bm{k}})\langle\langle \frac{\partial u_{\alpha\bk}}{\partial k_j}\Ket{u_{\beta s\bk}}\,.
\label{element_F}
\ee
Let us introduce a new basis set $\{\Ket{v_{\xi s\bk}}\}$ that block diagonalizes $\HFk$.
This new basis can related to a ${\bk}$-independent basis $\{\Ket{pm}\}$ through an unitary transformation defined by the operator $\UFk$,  that is  $\{\Ket{pm}\}\xrightarrow{\UFk}\{\Ket{v_{\xi s\bk}}\}$,  so  that  
\be
\Bra{q m} \UFk^\dagger\HFk\UFk^{} \Ket{p m'}=\delta_{mm'}\left(\langle q|\hat{H}^{}_{\mathrm{eff}}(\bk)|p\rangle-\delta_{q p}\,m\Omega\right)\,.
\ee
This defines the effective Hamitonian $\hat{H}^{}_{\mathrm{eff}}(\bk)$ in the Hilbert space whose matrix elements in the $\{\ket{p}\}$ basis are $\langle q|\hat{H}^{}_{\mathrm{eff}}(\bk)|p\rangle$. 
The matrix elements in Eq.~\eqref{element_F} can be rewritten in terms of  $\hat{\bm{R}}_{\bm{k}}$. Indeed, we note that
\begin{eqnarray}
\label{C5}
 \langle\langle \frac{\partial u_{\alpha\bk}}{\partial k_j}\Ket{u_{\beta s\bk}} &=&\!\sum_{\xi,s'}\langle\langle \frac{\partial u_{\alpha\bk}}{\partial k_j}\Ket{v_{\xi s'\bk}}\langle\langle v_{\xi s'\bk}\Ket{u_{\beta s\bk}}\\
 \label{C6}
 &=&\!-\!\!\sum_{\xi,s'}\langle\langle u_{\alpha\bk}\Ket{\frac{\partial v_{\xi s'\bk}}{\partial k_j}}\langle\langle v_{\xi s'\bk}\Ket{u_{\beta s\bk}}\\
 \label{C7}
 &=& -\Bra{ u_{\alpha\bk}}\frac{\partial \UFk}{\partial k_j}\UFk^\dagger\Ket{u_{\beta s\bk}}.
\end{eqnarray}
In Eq.~\eqref{C5} we have simply introduced an identity operator and in Eq.~\eqref{C6} we  used that $s\neq 0$ so that $\langle\langle u_{\alpha\bk}\Ket{ v_{\xi s\bk}}=0$. Finally, in Eq.~\eqref{C7} we have used that, because $\{\Ket{p m }\}$ is independent of ${\bk}$,
\be
\frac{\partial \UFk}{\partial k_j}\UFk^\dagger=\sum_{\xi,s}  |\frac{\partial v_{\xi s\bk}}{\partial k_j}\rangle\rangle\Bra{v_{\xi s\bk}}\,.
\ee
Replacing Eq.~\eqref{C7} in Eq.~\eqref{element_F}, we then find that
\be
\Bra{u_{\alpha\bk}}\frac{\partial \HFk}{\partial k_j}\Ket{u_{\beta s\bk}}=(\varepsilon_{\beta s\bm{k}}-\varepsilon_{\alpha\bm{k}})\Bra{ u_{\alpha\bk}}\frac{\partial \UFk}{\partial k_j}\UFk^\dagger\Ket{u_{\beta s\bk}}\,.
\label{element_F_U}
\ee
Replacing Eq. \eqref{element_F_U} in Eq. \eqref{WA} and using that $\Bra{ u_{\alpha\bk}}\frac{\partial \UFk}{\partial k_j}\UFk^\dagger\Ket{u_{\beta s\bk}}$ is invariant under the traslation of the FBZ index, $\Bra{ u_{\alpha\bk}}\frac{\partial \UFk}{\partial k_j}\UFk^\dagger\Ket{u_{\beta s\bk}}=\Bra{ u_{\alpha(-s)\bk}}\frac{\partial \UFk}{\partial k_j}\UFk^\dagger\Ket{u_{\beta \bk}}$ one gets, after some simple but tedious algebra, that
\bea
\mathcal{W}^A&=&-\frac{i\epsilon^{zjl}}{4\pi\Omega}\int_{\mathrm{BZ}}d^2k \,\sum_{\alpha,\beta,s}s\Omega\times\\
\nonumber
&&\Bra{ u_{\alpha\bk}}\frac{\partial \UFk}{\partial k_j}\UFk^\dagger\Ket{u_{\beta s\bk}}\Bra{ u_{\beta s\bk}}\frac{\partial \UFk}{\partial k_l}\UFk^\dagger\Ket{u_{\alpha \bk}}\,.    
\eea
We now define a new operator $\tUFk$ such that $\tUFk^\dagger\Ket{u_{\beta s\bk}}=-is\Omega\,\UFk^\dagger\Ket{u_{\beta s\bk}}$. Then
\bea
\nonumber
\mathcal{W}^A
&=&\frac{\epsilon^{zjl}}{4\pi\Omega}\int_{\mathrm{BZ}}d^2k\sum_{\alpha} \Bra{ u_{\alpha\bk}}\frac{\partial \UFk}{\partial k_j}\tUFk^\dagger\frac{\partial \UFk}{\partial k_l}\UFk^\dagger\Ket{u_{\alpha \bk}}\,.\\
\label{WA_F}
\eea
In order to proceed, we now use the fact that any operator $\hat{\bm{O}}_{\bk}$ in $\mathcal{S}$ that is translationally invariant with respect to the Floquet index $m$ can be related to a periodic operator $\hat{O}_{\bk}(t)$ acting on $\mathcal{H}$ by the relation \cite{Eckardt2015} 
\be
\hat{O}_{\bk}(t)=\sum_{p,q}\sum_m \ket{q_{\bk}} \Bra{q_{\bk} m} \hat{\bm{O}}_{\bk}\Ket{p_{\bk} 0}\bra{p_{\bk}} e^{-im\Omega t}\,.
\label{operator_H_u}
\ee
Hence, we can make the following identification $\UFk\rightarrow \Uk(t)$ and $\tUFk\rightarrow -\partial \Uk(t)/\partial t$ where $\Uk(t)$ is nothing but a unitary operator that relates the time dependent Hamiltonian $\hat{H}_{\bm{k}}(t)$ with the time independent effective Hamiltonian \cite{Eckardt2015}
\be
\hat{H}^{}_{\mathrm{eff}}(\bm{k})=\Uk^\dagger(t) \hat{H}_{\bm{k}}(t)\Uk(t)-i \Uk^\dagger(t)\frac{d\Uk(t)}{dt}\,.
\ee
In addition, from Eq.~\eqref{operator_H_u} we can write 
\be
\frac{1}{T}\int_0^Tdt\,\mathrm{tr}\left(\hat{O}_{\bk}(t)\right)=\sum_{\alpha}  \Bra{u_{\alpha \bk}} \hat{\bm{O}}_{\bk}\Ket{u_{\alpha\bk}}\,,
\ee
while it can be easily shown that a product of operators in $\mathcal{S}$ translates on the same product of the corresponding operators in $\mathcal{H}$, $\hat{\bm{O}}_1\hat{\bm{O}}_2\rightarrow \hat{O}_1(t)\hat{O}_2(t)$.
Hence, after some rearranging, the final expression for $\mathcal{W}^A$ in terms of $\Uk(t)$ reads 
\begin{eqnarray}
\notag
\mathcal{W}^{A} &=& \frac{\epsilon^{zjl}}{8\pi^2}\!\!\int_{0}^{T}\!\!\!\!dt\!\!\int_{\mathrm{BZ}}\!\!d^2k\, \mathrm{tr}\left[\hat{R}^{\dagger}_{\bm{k}}(t)\frac{\partial \hat{R}_{\bm{k}}(t)}{\partial t}\right.\\
\nonumber
    & &\left.\hat{R}^{\dagger}_{\bm{k}}(t)\frac{\partial \hat{R}_{\bm{k}}(t)}{\partial k_l}\hat{R}^{\dagger}_{\bm{k}}(t)\frac{\partial \hat{R}_{\bm{k}}(t)}{\partial k_j}\right]\,,\\
    &=& N_3[R]\,,
\end{eqnarray}
which is the higher-order winding number $N_3[R]$ associated to the micromotion operator defined by $\hat{R}_{\bm{k}}(t)$ in time-quasimomenta space.
This invariant can also be equally written in terms of the two-point micromotion operator $\Pk(t,t')=\Uk(t)\Uk^\dagger(t')$ \cite{Eckardt2015,Goldman2014},
\begin{eqnarray}
\notag
\mathcal{W}^{A} &=& \frac{\epsilon^{zjl}}{8\pi^2}\!\!\int_{0}^{T}\!\!\!\!dt\!\!\int_{\mathrm{BZ}}\!\!d^2k \mathrm{tr}\left[\hat{P}^{\dagger}_{\bm{k}}(t,t')\frac{\partial \hat{P}_{\bm{k}}(t,t')}{\partial t}\right.\\
\nonumber
    & &\left.\hat{P}^{\dagger}_{\bm{k}}(t,t')\frac{\partial \hat{P}_{\bm{k}}(t,t')}{\partial k_l}\hat{P}^{\dagger}_{\bm{k}}(t,t')\frac{\partial \hat{P}_{\bm{k}}(t,t')}{\partial k_j}\right]\,,\\
    &=&N_3[P]\,,
    \label{R}
\end{eqnarray}
because the winding number of the product of two operators is given by the sum of the winding numbers of each operator \cite{Wagner2023}, $N_3[O_1O_2]=N_3[O_1]+N_3[O_2]$ and $\partial \Uk^\dagger(t')/\partial t=0$.
Equation \eqref{R} is the topological invariant introduced in Refs. \cite{Rudner2013,Roy2017,Harper2020} and hence proofs the equivalence between both approaches.

\section{Relation between the Berry curvature defined in Sambe space and that associated with $\hat{H}_{\mathrm{eff}}$}\label{curvatureHeff_vs_curvatureSambe}
Here we provide an explicit formula relating the Berry curvature  calculated in $\mathcal{S}$ space, given by Eq.~\eqref{curvature_F} in the main text, and the corresponding expression obtained form the eigenvectors of the effective Hamiltonian $\hat{H}_{\mathrm{eff}}$. Any matrix element of an operator $\hat{\bm{O}}$ in $\mathcal{S}$ can be written as 
\be
\Bra{u_a} \hat{\bm{O}}\Ket{u_b}=\frac{1}{T}\int_0^Tdt\,\bra{u_a(t)} \hat{O}(t)\ket{u_b(t)} 
\ee
where $\hat{O}(t)$ is the associated time dependent operator in the Hilbert space and $\ket{u_a(t)}$ are the Floquet modes, see Eq.~\eqref{Floquet_modes}. In the particular case of 
 the Floquet Berry curvature in Eq.~\eqref{curvature_F} we have
\bea
\label{Fxy_time_averaged}
   \mathcal{F}_{xy}^{\alpha}(\bm{k}) &=& i\epsilon^{zjl}\left\langle\left\langle \partial_{k_j}u_{\alpha\bm{k}}|\partial_{k_l} u_{\alpha\bm{k}} \right\rangle\right\rangle\,,\\
\nonumber
&=&i\epsilon^{zjl}\frac{1}{T}\int_0^T dt\,\bra{\partial_{k_j}u_{\alpha\bm{k}}(t)}\partial_{k_l} u_{\alpha\bm{k}}(t)
 \rangle\,.
\eea
Using Eq.~\eqref{Heff} to relate the eigenvectors of the Floquet Hamiltonian in $\bk$ space, $\hat{H}_{\bk}^F(t)$, with those of $\hat{H}_{\mathrm{eff}}(\bk)$ via the relation $\ket{u_{\alpha\bm{k}}(t)}=\Uk(t)\ket{u_{\alpha\bm{k}}^\mathrm{eff}}$, with $\Uk(t)$ satisfying $
\hat{H}^{}_{\mathrm{eff}}(\bm{k})=\Uk^\dagger(t) \hat{H}_{\bk}^F(t)\Uk(t)
$, we readily find that
\begin{equation}
    \mathcal{F}_{xy}^{\alpha}(\bm{k}) = \mathcal{F}_{xy}^{\alpha,\mathrm{eff}}(\bm{k}) + \mathcal{D}_{xy}^{\alpha,\mathrm{eff}}(\bm{k})\,.
   \label{curvature_F_Ap}
\end{equation}
In Eq.~\eqref{curvature_F_Ap} we have defined the Berry curvature of the $\alpha$-th band of the effective Hamiltonian,
\begin{equation}
    \mathcal{F}_{xy}^{\alpha,\mathrm{eff}}(\bm{k})=i\epsilon^{zjl}\bra{\partial_{k_j}u_{\alpha\bm{k}}^\mathrm{eff}}\partial_{k_l} u_{\alpha\bm{k}}^\mathrm{eff}\rangle\, ,
\end{equation}
and
\begin{equation}
    \mathcal{D}_{xy}^{\alpha,\mathrm{eff}}(\bm{k})=i\epsilon^{zjl}\partial_{k_j}\left(\!\bra{u_{\alpha\bm{k}}^\mathrm{eff}}\frac{1}{T}\!\int_0^T\!\!\!dt\,\Uk^\dagger(t)\partial_{k_l}\Uk(t) \ket{u_{\alpha\bm{k}}^\mathrm{eff}}\!\right)\,.
    \label{total_derivative}
\end{equation}
Notice that the difference between $\mathcal{F}_{xy}^{\alpha}(\bm{k})$ and $\mathcal{F}_{xy}^{\alpha,\mathrm{eff}}(\bm{k})$, given by Eq.~\eqref{total_derivative}, is written in terms of total derivatives with respect to different quasimomementa of a $\bk$-periodic function and hence vanishes when integrated over the Bloch Brillouin zone.  Namely,
\bea
\nonumber
C^F_\alpha&=&\frac{1}{2\pi}\int_{\mathrm{BZ}}d^2k\,  \mathcal{F}_{xy}^{\alpha}(\bm{k})\,\\
\nonumber
&=&\frac{1}{2\pi}\int_{\mathrm{BZ}}d^2k\, \left(\mathcal{F}_{xy}^{\alpha,\mathrm{eff}}(\bm{k}) + \mathcal{D}_{xy}^{\alpha,\mathrm{eff}}(\bm{k})\right)\,,\\
&=&\frac{1}{2\pi}\int_{\mathrm{BZ}}d^2k\, \mathcal{F}_{xy}^{\alpha,\mathrm{eff}}(\bm{k})=C^F_{\alpha,\mathrm{eff}}\,.
\label{CF:Ceff}
\eea
This guarantees that the Chern number of a given Floquet band, $C^F_\alpha$, can be equally calculated with $\mathcal{F}_{xy}^{\alpha}(\bm{k})$ or $\mathcal{F}_{xy}^{\alpha,\mathrm{eff}}(\bm{k})$.
Importantly, this is not the case for the calculation of the anomalous spectral flow $\mathcal{W}^A$ as it involves terms as the product $\varepsilon_{\alpha\bk}\mathcal{F}_{xy}^{\alpha}(\bm{k})$. 
A similar analysis can be done for the case of the intrinsic orbital magnetic moment [see Eq.~\eqref{magnetization_F}], where differences between $m_z^{\alpha}(\bm{k})$ and $m_z^{\alpha,\mathrm{eff}}(\bm{k})$ are also found.

Finally, we call attention on the fact that 
\begin{equation}
    \mathcal{W}^A_\mathrm{eff}= \sum_{\alpha}\int_{\mathrm{BZ}}\frac{d^2k}{2\pi\Omega}\left[\mathcal{F}_{xy}^{\alpha,\mathrm{eff}}\varepsilon_{\alpha\bm{k}}-\frac{\Phi_0}{2\pi}m_z^{\alpha,\mathrm{eff}}(\bm{k})\right]=0
\end{equation}
That means that, in general, $\mathcal{W}^{A}_{\mathrm{eff}}\neq \mathcal{W}^{A}$.

\section{Gauge dependence of $N_1[R]$}\label{Appendix_gaugeN1}
As shown in Eq.~\eqref{N1_R}, the first-order winding number of the micromotion operator $\hat{R}(t)$ over one period of the driving cycle is defined as
\be
\label{N1_R_Ap}
    N_1[R] = - \frac{i}{2\pi}\int_{0}^{T}dt\, \mathrm{Tr}[\hat{R}^{\dagger}(t)\partial_t \hat{R}(t)]\,,
\ee
which imposes  that $N_1[R]\in \mathbb{Z}$. This quantity is related to the effective Hamiltonian by the equation
\begin{eqnarray}
\label{relHeff_Ht_Ap}
    \mathrm{Tr}[\hat{H}_{\mathrm{eff}}] &=& \frac{1}{T}\int_{0}^{T}dt\, \mathrm{Tr}[\hat{H}(t)] + N_1[R] \Omega \,.
\end{eqnarray}
It is important to notice that the value of $N_1[R]$  has no physical meaning by itself as it includes trivial contributions that make it gauge-dependent. To show this, let us first denote the total time averaged energy as $\bar{E}=\frac{1}{T}\int_{0}^{T}dt\, \mathrm{Tr}[\hat{H}(t)]$. Then, given $\hat{R}(t)$, we can introduce a new, equally valid, time-periodic operator
\be
\hat{R}'(t)=e^{-i[\bar{\varepsilon}/\Omega]\Omega t\hat{I}}\hat{R}(t)\,,
\label{Rp}
\ee
where $[x]$ denotes the closest integer number to $x$, $\bar{\varepsilon}=\bar{E}/d_\mathcal{H}$ and $d_\mathcal{H}=\mathrm{Tr}(\hat{I})$ is the dimension of the Hilbert space. The corresponding effective Hamiltonian is then given by
\begin{eqnarray}
\nonumber
    \hat{H}_{\mathrm{eff}}' &=& \hat{R}'^{\dagger}(t)\hat{H}(t)\hat{R}'^{}(t)-i\hat{R}'^{\dagger}(t)\partial_t \hat{R}'^{}(t)\\
    &=& \hat{H}_{\mathrm{eff}}-[\bar{\varepsilon}/\Omega]\Omega \hat{I}\,.
\end{eqnarray}
It is then clear that
\be
N_1[R']=N_1[R]-[\bar{\varepsilon}/\Omega]d_\mathcal{H}\,.
\label{N1_Rp}
\ee
Equation~\eqref{N1_Rp} highlights the fact that $N_1[R']$ can present discrete jumps when a global energy shift is introduced in the problem (which has no physical relevance). The change of $N_1[R']$ is a consequence 
of the intrinsic indeterminate nature of $\hat{H}_{\mathrm{eff}}'$ up to integer multiples of $\Omega$. 
Note, however, that $\partial_\Phi N_1[R']=\partial_\Phi N_1[R]$ since 
$\partial_\Phi \bar{\varepsilon}=0$ as discussed in the main text.

This gauge-dependence can also be traced back to the properties of the Floquet states. In fact, Eq.~\eqref{N1_prob} 
shows that the value of $N_1[R]$ depends on the Floquet zone of reference. If we had chosen the effective Hamiltonian as the one having eigenvalues in the $s$-th Floquet zone, namely $\varepsilon_{as} \in (-\Omega/2+s\Omega,\Omega/2 + s\Omega]$, then its corresponding winding would be given by
\bea
\nonumber
    N_1^s[R] &=& -\sum_a \sum_{n=-\infty}^{\infty} n\,\bra{u_{as}^{(n)}}u_{as}^{(n)}\rangle\,,\\
\nonumber
    &=&-\sum_a \sum_{n=-\infty}^{\infty} n\,\bra{u_{a}^{(n+s)}}u_{a}^{(n+s)}\rangle\,,\\
    &=&N_1^0[R]+s\,d_\mathcal{H}=N_1[R]+s\,d_\mathcal{H}\,,
\eea
for all values of $s\in\mathbb{Z}$. 

The unitary transformation, Eq.~\eqref{Heff}, from where Eq.~\eqref{relHeff_Ht_Ap} is derived, does not necessarily implies that all the eigenvalues of $\hat{H}_{\mathrm{eff}}$ belong to the same Floquet zone for an arbitrary $\hat{R}(t)$. In order to guarantee that, an additional `folding' procedure might be required. As shown below, this folding can always be done using a T-periodic unitary operator, then assuring that there is  a $\hat{R}_\mathrm{fold}(t)$ that transforms $H(t)$ into $\hat{H}'_{\mathrm{eff}}$ with its eigenvalues inside a single Floquet zone.  
To prove that, let us denote $\{\varepsilon_a\}$ the set of eigenvalues of $\hat{H}_{\mathrm{eff}}$, obtained after a rotation with a given $\hat{R}(t)$. We can then define the unitary operator $\hat{R}'(t)$, which is diagonal in the eigenbasis of $\hat{H}_{\mathrm{eff}}$ and has the following diagonal matrix elements
\be
\left[\hat{R}'(t)\right]_{aa}=\exp\left(-i\left[\frac{\varepsilon_a-\epsilon}{\Omega}\right]\Omega t\right)\,,
\ee
where $\epsilon$ denotes the center of the chosen Floquet zone. Then, $\hat{R}_\mathrm{fold}(t)=\hat{R}'(t)\hat{R}(t)$ is the desired unitary transformation that takes $H(t)$ into  $\hat{H}'_{\mathrm{eff}}$ with eigenvalues $\varepsilon'_a=\varepsilon_a-[(\varepsilon_a-\epsilon)/\Omega]\Omega$ such that $|\varepsilon'_a-\epsilon|\le\Omega/2\,\forall a$. Correspondingly
\be
N_1[R_\mathrm{fold}]=N_1[R]-\sum_a \left[\frac{\varepsilon_a-\epsilon}{\Omega}\right]\,.
\ee
The freedom to chose the Floquet zone around any particular quasienergy can be exploited to set $N_1[R]$ equal to zero for a fixed value of the system's parameters. This is done by adopting the `natural Floquet zone' (NFZ) as defined in Ref.~\cite{Nathan2015}. Namely, a zone such that
\be
 \mathrm{Tr}[\hat{H}_{\mathrm{eff}}] = \frac{1}{T}\int_{0}^{T}dt\, \mathrm{Tr}[\hat{H}(t)]\,,
\ee
with $\hat{H}_{\mathrm{eff}}$ having all its eigenvalues inside such a zone.
Making such a choice for $\Phi=0$, and using Eq.~\eqref{WA=N3R}, we have that
\be
N_1[R,\Phi/\Phi_0]=N_3[R]\frac{\Phi}{\Phi_0}\, .
\ee
As a final remark, we notice that the expression for the time-averaged energy, Eq.~\eqref{mean_energy}, can also be written as~\cite{Fainshtein1978,Reynoso2013} 
\be
\frac{1}{T}\int_{0}^{T} dt\, \mathrm{Tr}[\hat{H}(t)] =  \sum_a \left(\varepsilon_a-\Omega\frac{\partial\varepsilon_a}{\partial\Omega}\right)\, ,
\label{mean_energy_Ap}
\ee
where we assumed that $H(t)$ depends on time though the combination $\Omega t$. Then, using Eq.~\eqref{relHeff_Ht}, we have yet another expression for $N_1[R]$, namely
\be
N_1[R]=\sum_a \frac{\partial\varepsilon_a}{\partial\Omega}\,.
\ee
\section{Demonstration of Eq.~\eqref{illicit_switch}}\label{Appendix_demo}
The spectral flow $\mathcal{W}(\mu_\varepsilon)$ is defined as
\begin{eqnarray}
\nonumber
   \mathcal{W}(\mu_\varepsilon) &=& \Phi_0 \int_{-\infty}^{\mu_\varepsilon} d\omega\,\frac{\partial \rho^{F}(\omega)}{\partial B}\,,\\
    &=& \Phi_0 \int_{-\infty}^{\mu_\varepsilon}d\omega\,\sum_{m=-\infty}^\infty \frac{\partial \rho^{F}_0(\omega+m\Omega)}{\partial B}\,,
\end{eqnarray}
where Eq.~\eqref{rho_F_series} was used in the second equality. Introducing the Ces\`aro renormalization we get 
\begin{equation}
    \mathcal{W}(\mu_\varepsilon) \stackrel{(C,1)}{=} \Phi_0\lim_{\lambda\rightarrow\infty}
     \int_{-\lambda}^{\mu_\varepsilon}d\omega\sum_{m=-\infty}^\infty \frac{\partial \rho^{F}_0(\omega+m\Omega)}{\partial B}\left(1+\frac{\omega}{\lambda}\right)\,.\\
\end{equation}
For a given (finite) value of $\lambda$  this expression has the form $\int_a^b d\omega\sum_m f_n(\omega)$, which can be rewritten as $\sum_m\int_a^b d\omega f_m(\omega)$ provided $\sum_m\int_a^b d\omega |f_m(\omega)|<\infty$ (Fubini-Tonelli's theorem \cite{Rudin1986,*Royden2010}). In our case we have, 
\begin{equation}
    f_m(\omega)= \Phi_0
 \frac{\partial \rho^{F}_0(\omega+m\Omega)}{\partial B}\left(1+\frac{\omega}{\lambda}\right)\,.
\end{equation}
Due to the exponential decay of $\rho_0(\omega)$ away from the NFZ, the integral of $|f_m(\omega)|$ in the interval $[-\lambda,\mu_\varepsilon]$ is finite $\forall m$. Furthermore, for the same reason, such an integral goes to zero exponentially fast for large $m$. Hence, the condition of the Fubini-Tonelli's theorem is fulfilled so that
\begin{equation}
    \mathcal{W}(\mu_\varepsilon) \stackrel{(C,1)}{=} \Phi_0\lim_{\lambda\rightarrow\infty}\sum_{m=-\infty}^\infty
     \int_{-\lambda}^{\mu_\varepsilon}d\omega \frac{\partial \rho^{F}_0(\omega+m\Omega)}{\partial B}\left(1+\frac{\omega}{\lambda}\right)\,.
     \label{G4}
\end{equation}
We now define
\begin{equation}
    A_m(\lambda) =\Phi_0
     \int_{-\lambda}^{\mu_\varepsilon}d\omega \frac{\partial \rho^{F}_0(\omega+m\Omega)}{\partial B}\left(1+\frac{\omega}{\lambda}\right)\,,
\end{equation}
and
\begin{equation}
    A_m^\infty =\Phi_0
     \int_{-\infty}^{\mu_\varepsilon}d\omega \frac{\partial \rho^{F}_0(\omega+m\Omega)}{\partial B}\,.
     \label{Am}
\end{equation}
Since the integral in Eq.~\eqref{Am} is well defined and finite in the standard sense,  the Ces\`aro renormalization must agree with it so that $\lim_{\lambda\rightarrow\infty} A_m(\lambda)=A_m^\infty$. It is straightforward to verify that, due to Eq.~\eqref{prop_rho0} and the fast (exponential) decay of $\rho_0(\omega)$, the series $\sum_m |A_m^\infty|<\infty$. Hence, the sum and the limit in Eq.~\eqref{G4} can be exchanged~\cite{Rudin1986,*Royden2010} 
\begin{eqnarray}
\nonumber
    \mathcal{W}(\mu_\varepsilon) &\stackrel{(C,1)}{=}& \sum_{m=-\infty}^\infty\lim_{\lambda\rightarrow\infty}A_m(\lambda)\,,\\
    &\stackrel{(C,1)}{=}&\sum_{m=-\infty}^\infty A_m^\infty\,,\\
    \nonumber
     &\stackrel{(C,1)}{=}&\Phi_0\sum_{m=-\infty}^\infty
     \int_{-\infty}^{\mu_\varepsilon}d\omega\, \frac{\partial \rho^{F}_0(\omega+m\Omega)}{\partial B}\,,
\end{eqnarray}
which reduces to Eq.~\eqref{illicit_switch} after a change of variable.  

\section{Time-averaged Floquet DOS with B\"uttiker-type reservoirs}\label{Appendix_bath}
\begin{figure}[t]
    \centering
    \includegraphics[width=0.85\columnwidth]{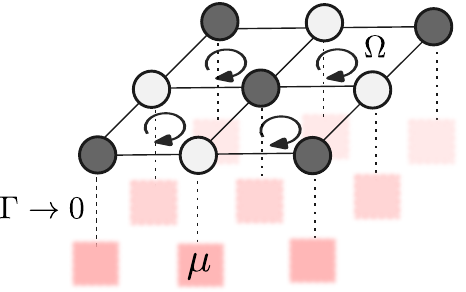}
    \caption{Scheme of the lattice system, driven with frequency $\Omega$, coupled to a B\"uttiker-type fermionic heat bath at chemical potential $\mu$. Each lattice site is connected to its own reservoir with a vanishing strength $\Gamma\rightarrow 0$.}
    \label{bath_scheme}
\end{figure}

In this Appendix, we will introduce the simplest model of a heat bath that leads to a steady-state occupation which is determined by the time-averaged Floquet density of states introduced in Eq.~\eqref{TADOS}. In particular, we will consider a B\"uttiker-type bath~\cite{Buttiker1985}: featureless fermionic reservoirs which are attached to every site of the driven lattice but uncoupled from each other [see Fig.~\ref{bath_scheme}].

The steady-state time-dependent particle-density of such a driven dissipative system can be obtained by making use of standard Floquet-Keldysh Green's functions techniques~\cite{Tsuji2008,Aoki2014}. The particle-density can be expressed in terms of its Fourier harmonics as
\begin{equation}
    n(t) = \sum_{p}e^{-i p\Omega t}n_{p},
\end{equation}
where
\begin{eqnarray}
    n_p &=& -\frac{i}{A} \int_{\mathrm{NFZ}}\frac{d\omega}{2\pi}\sum_{n}\mathrm{Tr}\left[\hat{\bm{G}}^{F <}_{p+n,n}(\omega)\right].
    \label{np}
\end{eqnarray}
We have here introduced the lesser Green's function in Floquet representation, given by~\cite{Aoki2014}
\begin{equation}
 \hat{\bm{G}}^{F<}(\omega) = \hat{\bm{G}}^{F}(\omega+i\Gamma)\hat{\bm{\Sigma}}^{F<}(\omega)\hat{\bm{G}}^{F}(\omega-i\Gamma).
\label{LGF}
\end{equation} 
The parameter $\Gamma$ characterizes the hybridization with the reservoirs. For simplicity, we have omitted the frequency dependence of the latter, equivalent to considering a bath with a constant density of states in the wide-band limit.
The lesser self-energy due to these featureless reservoirs is diagonal in Sambe space, namely
\begin{equation}
    \hat{\bm{\Sigma}}^{F,<}_{mn}(\omega) = \delta_{mn}\hat{I} f(\omega + n\Omega) 2i \Gamma.
    \label{LSF}
\end{equation}
Replacing Eqs.~\eqref{LGF} and~\eqref{LSF} in Eq.~\eqref{np}, making use of Eq.~\eqref{GRF} and the orthogonality relation in Eq.~\eqref{ort}, we find that
\begin{eqnarray}
    n_p &=& \frac{\Gamma}{\pi A}\int_{\mathrm{NFZ}}^{}d\omega \sum_{a}\sum_{l,s} f(\omega+l\Omega)\times \\
    \notag
    & &\frac{\langle u_{a}^{(l+s)}|u_{a}^{(l+p+s)}\rangle}{\left[\omega-(\varepsilon_{a}+s\Omega)+i\Gamma\right]\left[\omega-\left(\varepsilon_{a}+\left(s+p\right)\Omega\right)-i\Gamma\right]}.
\end{eqnarray}
When considering the weak hybridization limit ($\Gamma \to 0$) all the higher order harmonics ($p\neq 0$) vanish, meaning that in the steady-state the system reaches a time-independent particle-density, given by
\begin{eqnarray}
 \overline{n}|_{\mu} =  \lim_{\Gamma\to 0}n_0 &=& \sum_{a} D(\varepsilon_{a}), 
  \label{averaged_nF}
\end{eqnarray}
with
\begin{equation}
    D(\varepsilon_{a})=\sum_l f(\varepsilon_{a}+l\Omega)\langle u_{a}^{(l)}|u_{a}^{(l)}\rangle.
    \label{D_occs}
\end{equation}
We recall that in the equations above the sum over $a$ runs over states in the NFZ. In Eq.~\eqref{averaged_nF} we emphasize that $\overline{n}$ is computed at the chemical potential $\mu$, which is fixed by the heat bath.
The occupation function $D(\varepsilon)$ is self-consistent, in the sense that it depends itself on the Floquet eigenstates and thus changes when modifying the driving parameters. We also note that Eqs.~\eqref{averaged_nF} and~\eqref{D_occs} have been reported in Refs.~\cite{Matsyshyn2023,Shi2024,Kumari2024}.  In the undriven case, the Floquet states in the natural Floquet Brillouin zone are such that $|u_{a}^{(l)}\rangle = \delta_{l0} |u_{a}^{(0)}\rangle$ which makes $D(\varepsilon_{a}) = f(\varepsilon_{a})$. 

Interestingly, Eq.~\eqref{averaged_nF} can be written as an integral over frequencies of the equilibrium Fermi-Dirac distribution function times an effective density of states as
\begin{equation}
   \overline{n}|_{\mu} = \int_{-\infty}^{\infty}d\omega f(\omega)\rho_0^{F}(\omega),
   \label{n_bath}
\end{equation}
where
\begin{equation}
    \rho_0^{F}(\omega) = \sum_{a,l} \langle u_a^{(l)}|u_a^{(l)}\rangle \delta(\omega-(\varepsilon_a+l\Omega)).
\end{equation}
This is precisely the time-averaged Floquet density of states introduced in Eq.~\eqref{TADOS} of the main text. Eq.~\eqref{n_bath} suggests that we can describe the effect of this featureless bath by considering that the system has a thermal occupation given by the Fermi-Dirac function but with an effective density of states described by $\rho^{F}_0(\omega)$. In the limit of zero temperature, Eq.~\eqref{n_bath} reduces to
\begin{equation}
    \overline{n}|_{\mu} = \int_{-\infty}^{\mu}d\omega\rho_0^{F}(\omega),
\end{equation}
which is the time-averaged particle density introduced in Eq.~\eqref{n_average} in the main text for $\mu_\varepsilon=\mu$.

%


\begin{thebibliography}{200}%
\makeatletter
\providecommand \@ifxundefined [1]{%
 \@ifx{#1\undefined}
}%
\providecommand \@ifnum [1]{%
 \ifnum #1\expandafter \@firstoftwo
 \else \expandafter \@secondoftwo
 \fi
}%
\providecommand \@ifx [1]{%
 \ifx #1\expandafter \@firstoftwo
 \else \expandafter \@secondoftwo
 \fi
}%
\providecommand \natexlab [1]{#1}%
\providecommand \enquote  [1]{``#1''}%
\providecommand \bibnamefont  [1]{#1}%
\providecommand \bibfnamefont [1]{#1}%
\providecommand \citenamefont [1]{#1}%
\providecommand \href@noop [0]{\@secondoftwo}%
\providecommand \href [0]{\begingroup \@sanitize@url \@href}%
\providecommand \@href[1]{\@@startlink{#1}\@@href}%
\providecommand \@@href[1]{\endgroup#1\@@endlink}%
\providecommand \@sanitize@url [0]{\catcode `\\12\catcode `\$12\catcode `\&12\catcode `\#12\catcode `\^12\catcode `\_12\catcode `\%12\relax}%
\providecommand \@@startlink[1]{}%
\providecommand \@@endlink[0]{}%
\providecommand \url  [0]{\begingroup\@sanitize@url \@url }%
\providecommand \@url [1]{\endgroup\@href {#1}{\urlprefix }}%
\providecommand \urlprefix  [0]{URL }%
\providecommand \Eprint [0]{\href }%
\providecommand \doibase [0]{https://doi.org/}%
\providecommand \selectlanguage [0]{\@gobble}%
\providecommand \bibinfo  [0]{\@secondoftwo}%
\providecommand \bibfield  [0]{\@secondoftwo}%
\providecommand \translation [1]{[#1]}%
\providecommand \BibitemOpen [0]{}%
\providecommand \bibitemStop [0]{}%
\providecommand \bibitemNoStop [0]{.\EOS\space}%
\providecommand \EOS [0]{\spacefactor3000\relax}%
\providecommand \BibitemShut  [1]{\csname bibitem#1\endcsname}%
\let\auto@bib@innerbib\@empty
\bibitem [{\citenamefont {Hasan}\ and\ \citenamefont {Kane}(2010)}]{Hasan2010}%
  \BibitemOpen
  \bibfield  {author} {\bibinfo {author} {\bibfnamefont {M.~Z.}\ \bibnamefont {Hasan}}\ and\ \bibinfo {author} {\bibfnamefont {C.~L.}\ \bibnamefont {Kane}},\ }\bibfield  {title} {\bibinfo {title} {{Colloquium: Topological insulators}},\ }\href {https://doi.org/10.1103/RevModPhys.82.3045} {\bibfield  {journal} {\bibinfo  {journal} {Rev. Mod. Phys.}\ }\textbf {\bibinfo {volume} {82}},\ \bibinfo {pages} {3045} (\bibinfo {year} {2010})}\BibitemShut {NoStop}%
\bibitem [{\citenamefont {Qi}\ and\ \citenamefont {Zhang}(2011)}]{Qi_RMP}%
  \BibitemOpen
  \bibfield  {author} {\bibinfo {author} {\bibfnamefont {X.-L.}\ \bibnamefont {Qi}}\ and\ \bibinfo {author} {\bibfnamefont {S.-C.}\ \bibnamefont {Zhang}},\ }\bibfield  {title} {\bibinfo {title} {Topological insulators and superconductors},\ }\href {https://doi.org/10.1103/RevModPhys.83.1057} {\bibfield  {journal} {\bibinfo  {journal} {Rev. Mod. Phys.}\ }\textbf {\bibinfo {volume} {83}},\ \bibinfo {pages} {1057} (\bibinfo {year} {2011})}\BibitemShut {NoStop}%
\bibitem [{\citenamefont {Haldane}(2017)}]{Haldane2017}%
  \BibitemOpen
  \bibfield  {author} {\bibinfo {author} {\bibfnamefont {F.~D.~M.}\ \bibnamefont {Haldane}},\ }\bibfield  {title} {\bibinfo {title} {{Nobel Lecture: Topological quantum matter}},\ }\href {https://doi.org/10.1103/RevModPhys.89.040502} {\bibfield  {journal} {\bibinfo  {journal} {Rev. Mod. Phys.}\ }\textbf {\bibinfo {volume} {89}},\ \bibinfo {pages} {040502} (\bibinfo {year} {2017})}\BibitemShut {NoStop}%
\bibitem [{\citenamefont {Bradlyn}\ \emph {et~al.}(2017)\citenamefont {Bradlyn}, \citenamefont {Elcoro}, \citenamefont {Cano}, \citenamefont {Vergniory}, \citenamefont {Wang}, \citenamefont {Felser}, \citenamefont {Aroyo},\ and\ \citenamefont {Bernevig}}]{Bradlyn2017}%
  \BibitemOpen
  \bibfield  {author} {\bibinfo {author} {\bibfnamefont {B.}~\bibnamefont {Bradlyn}}, \bibinfo {author} {\bibfnamefont {L.}~\bibnamefont {Elcoro}}, \bibinfo {author} {\bibfnamefont {J.}~\bibnamefont {Cano}}, \bibinfo {author} {\bibfnamefont {M.~G.}\ \bibnamefont {Vergniory}}, \bibinfo {author} {\bibfnamefont {Z.}~\bibnamefont {Wang}}, \bibinfo {author} {\bibfnamefont {C.}~\bibnamefont {Felser}}, \bibinfo {author} {\bibfnamefont {M.~I.}\ \bibnamefont {Aroyo}},\ and\ \bibinfo {author} {\bibfnamefont {B.~A.}\ \bibnamefont {Bernevig}},\ }\bibfield  {title} {\bibinfo {title} {Topological quantum chemistry},\ }\href {https://doi.org/10.1038/nature23268} {\bibfield  {journal} {\bibinfo  {journal} {Nature}\ }\textbf {\bibinfo {volume} {547}},\ \bibinfo {pages} {298–305} (\bibinfo {year} {2017})}\BibitemShut {NoStop}%
\bibitem [{\citenamefont {Sato}\ and\ \citenamefont {Ando}(2017)}]{Sato2017}%
  \BibitemOpen
  \bibfield  {author} {\bibinfo {author} {\bibfnamefont {M.}~\bibnamefont {Sato}}\ and\ \bibinfo {author} {\bibfnamefont {Y.}~\bibnamefont {Ando}},\ }\bibfield  {title} {\bibinfo {title} {Topological superconductors: a review},\ }\href {https://doi.org/10.1088/1361-6633/aa6ac7} {\bibfield  {journal} {\bibinfo  {journal} {Reports on Progress in Physics}\ }\textbf {\bibinfo {volume} {80}},\ \bibinfo {pages} {076501} (\bibinfo {year} {2017})}\BibitemShut {NoStop}%
\bibitem [{\citenamefont {Yan}\ and\ \citenamefont {Felser}(2017)}]{Yan2017}%
  \BibitemOpen
  \bibfield  {author} {\bibinfo {author} {\bibfnamefont {B.}~\bibnamefont {Yan}}\ and\ \bibinfo {author} {\bibfnamefont {C.}~\bibnamefont {Felser}},\ }\bibfield  {title} {\bibinfo {title} {{Topological Materials: Weyl Semimetals}},\ }\href {https://doi.org/10.1146/annurev-conmatphys-031016-025458} {\bibfield  {journal} {\bibinfo  {journal} {Annual Review of Condensed Matter Physics}\ }\textbf {\bibinfo {volume} {8}},\ \bibinfo {pages} {337–354} (\bibinfo {year} {2017})}\BibitemShut {NoStop}%
\bibitem [{\citenamefont {Ozawa}\ \emph {et~al.}(2019)\citenamefont {Ozawa}, \citenamefont {Price}, \citenamefont {Amo}, \citenamefont {Goldman}, \citenamefont {Hafezi}, \citenamefont {Lu}, \citenamefont {Rechtsman}, \citenamefont {Schuster}, \citenamefont {Simon}, \citenamefont {Zilberberg},\ and\ \citenamefont {Carusotto}}]{Ozawa_RMP}%
  \BibitemOpen
  \bibfield  {author} {\bibinfo {author} {\bibfnamefont {T.}~\bibnamefont {Ozawa}}, \bibinfo {author} {\bibfnamefont {H.~M.}\ \bibnamefont {Price}}, \bibinfo {author} {\bibfnamefont {A.}~\bibnamefont {Amo}}, \bibinfo {author} {\bibfnamefont {N.}~\bibnamefont {Goldman}}, \bibinfo {author} {\bibfnamefont {M.}~\bibnamefont {Hafezi}}, \bibinfo {author} {\bibfnamefont {L.}~\bibnamefont {Lu}}, \bibinfo {author} {\bibfnamefont {M.~C.}\ \bibnamefont {Rechtsman}}, \bibinfo {author} {\bibfnamefont {D.}~\bibnamefont {Schuster}}, \bibinfo {author} {\bibfnamefont {J.}~\bibnamefont {Simon}}, \bibinfo {author} {\bibfnamefont {O.}~\bibnamefont {Zilberberg}},\ and\ \bibinfo {author} {\bibfnamefont {I.}~\bibnamefont {Carusotto}},\ }\bibfield  {title} {\bibinfo {title} {Topological photonics},\ }\href {https://doi.org/10.1103/RevModPhys.91.015006} {\bibfield  {journal} {\bibinfo  {journal} {Rev. Mod. Phys.}\ }\textbf {\bibinfo {volume} {91}},\ \bibinfo {pages} {015006} (\bibinfo {year} {2019})}\BibitemShut {NoStop}%
\bibitem [{\citenamefont {Cooper}\ \emph {et~al.}(2019)\citenamefont {Cooper}, \citenamefont {Dalibard},\ and\ \citenamefont {Spielman}}]{Cooper_RMP}%
  \BibitemOpen
  \bibfield  {author} {\bibinfo {author} {\bibfnamefont {N.~R.}\ \bibnamefont {Cooper}}, \bibinfo {author} {\bibfnamefont {J.}~\bibnamefont {Dalibard}},\ and\ \bibinfo {author} {\bibfnamefont {I.~B.}\ \bibnamefont {Spielman}},\ }\bibfield  {title} {\bibinfo {title} {Topological bands for ultracold atoms},\ }\href {https://doi.org/10.1103/RevModPhys.91.015005} {\bibfield  {journal} {\bibinfo  {journal} {Rev. Mod. Phys.}\ }\textbf {\bibinfo {volume} {91}},\ \bibinfo {pages} {015005} (\bibinfo {year} {2019})}\BibitemShut {NoStop}%
\bibitem [{\citenamefont {Kitagawa}\ \emph {et~al.}(2010)\citenamefont {Kitagawa}, \citenamefont {Berg}, \citenamefont {Rudner},\ and\ \citenamefont {Demler}}]{Kitagawa2010}%
  \BibitemOpen
  \bibfield  {author} {\bibinfo {author} {\bibfnamefont {T.}~\bibnamefont {Kitagawa}}, \bibinfo {author} {\bibfnamefont {E.}~\bibnamefont {Berg}}, \bibinfo {author} {\bibfnamefont {M.}~\bibnamefont {Rudner}},\ and\ \bibinfo {author} {\bibfnamefont {E.}~\bibnamefont {Demler}},\ }\bibfield  {title} {\bibinfo {title} {Topological characterization of periodically driven quantum systems},\ }\href {https://doi.org/10.1103/PhysRevB.82.235114} {\bibfield  {journal} {\bibinfo  {journal} {Phys. Rev. B}\ }\textbf {\bibinfo {volume} {82}},\ \bibinfo {pages} {235114} (\bibinfo {year} {2010})}\BibitemShut {NoStop}%
\bibitem [{\citenamefont {Asb\'oth}(2012)}]{Asboth2012}%
  \BibitemOpen
  \bibfield  {author} {\bibinfo {author} {\bibfnamefont {J.~K.}\ \bibnamefont {Asb\'oth}},\ }\bibfield  {title} {\bibinfo {title} {Symmetries, topological phases, and bound states in the one-dimensional quantum walk},\ }\href {https://doi.org/10.1103/PhysRevB.86.195414} {\bibfield  {journal} {\bibinfo  {journal} {Phys. Rev. B}\ }\textbf {\bibinfo {volume} {86}},\ \bibinfo {pages} {195414} (\bibinfo {year} {2012})}\BibitemShut {NoStop}%
\bibitem [{\citenamefont {Asb\'oth}\ and\ \citenamefont {Obuse}(2013)}]{Asboth2013}%
  \BibitemOpen
  \bibfield  {author} {\bibinfo {author} {\bibfnamefont {J.~K.}\ \bibnamefont {Asb\'oth}}\ and\ \bibinfo {author} {\bibfnamefont {H.}~\bibnamefont {Obuse}},\ }\bibfield  {title} {\bibinfo {title} {Bulk-boundary correspondence for chiral symmetric quantum walks},\ }\href {https://doi.org/10.1103/PhysRevB.88.121406} {\bibfield  {journal} {\bibinfo  {journal} {Phys. Rev. B}\ }\textbf {\bibinfo {volume} {88}},\ \bibinfo {pages} {121406} (\bibinfo {year} {2013})}\BibitemShut {NoStop}%
\bibitem [{\citenamefont {Basov}\ \emph {et~al.}(2017)\citenamefont {Basov}, \citenamefont {Averitt},\ and\ \citenamefont {Hsieh}}]{Basov2017}%
  \BibitemOpen
  \bibfield  {author} {\bibinfo {author} {\bibfnamefont {D.~N.}\ \bibnamefont {Basov}}, \bibinfo {author} {\bibfnamefont {R.~D.}\ \bibnamefont {Averitt}},\ and\ \bibinfo {author} {\bibfnamefont {D.}~\bibnamefont {Hsieh}},\ }\bibfield  {title} {\bibinfo {title} {Towards properties on demand in quantum materials},\ }\href {https://doi.org/10.1038/nmat5017} {\bibfield  {journal} {\bibinfo  {journal} {Nature Materials}\ }\textbf {\bibinfo {volume} {16}},\ \bibinfo {pages} {1077} (\bibinfo {year} {2017})}\BibitemShut {NoStop}%
\bibitem [{\citenamefont {Eckardt}(2017)}]{Eckardt2017}%
  \BibitemOpen
  \bibfield  {author} {\bibinfo {author} {\bibfnamefont {A.}~\bibnamefont {Eckardt}},\ }\bibfield  {title} {\bibinfo {title} {{Colloquium: Atomic quantum gases in periodically driven optical lattices}},\ }\href {https://doi.org/10.1103/RevModPhys.89.011004} {\bibfield  {journal} {\bibinfo  {journal} {Rev. Mod. Phys.}\ }\textbf {\bibinfo {volume} {89}},\ \bibinfo {pages} {011004} (\bibinfo {year} {2017})}\BibitemShut {NoStop}%
\bibitem [{\citenamefont {Oka}\ and\ \citenamefont {Kitamura}(2019)}]{Oka2019}%
  \BibitemOpen
  \bibfield  {author} {\bibinfo {author} {\bibfnamefont {T.}~\bibnamefont {Oka}}\ and\ \bibinfo {author} {\bibfnamefont {S.}~\bibnamefont {Kitamura}},\ }\bibfield  {title} {\bibinfo {title} {{Floquet Engineering of Quantum Materials}},\ }\href {https://doi.org/10.1146/annurev-conmatphys-031218-013423} {\bibfield  {journal} {\bibinfo  {journal} {Annual Review of Condensed Matter Physics}\ }\textbf {\bibinfo {volume} {10}},\ \bibinfo {pages} {387–408} (\bibinfo {year} {2019})}\BibitemShut {NoStop}%
\bibitem [{\citenamefont {Harper}\ \emph {et~al.}(2020)\citenamefont {Harper}, \citenamefont {Roy}, \citenamefont {Rudner},\ and\ \citenamefont {Sondhi}}]{Harper2020}%
  \BibitemOpen
  \bibfield  {author} {\bibinfo {author} {\bibfnamefont {F.}~\bibnamefont {Harper}}, \bibinfo {author} {\bibfnamefont {R.}~\bibnamefont {Roy}}, \bibinfo {author} {\bibfnamefont {M.~S.}\ \bibnamefont {Rudner}},\ and\ \bibinfo {author} {\bibfnamefont {S.}~\bibnamefont {Sondhi}},\ }\bibfield  {title} {\bibinfo {title} {{Topology and Broken Symmetry in Floquet Systems}},\ }\href {https://doi.org/https://doi.org/10.1146/annurev-conmatphys-031218-013721} {\bibfield  {journal} {\bibinfo  {journal} {Annual Review of Condensed Matter Physics}\ }\textbf {\bibinfo {volume} {11}},\ \bibinfo {pages} {345} (\bibinfo {year} {2020})}\BibitemShut {NoStop}%
\bibitem [{\citenamefont {Rudner}\ and\ \citenamefont {Lindner}(2020{\natexlab{a}})}]{Rudner2020}%
  \BibitemOpen
  \bibfield  {author} {\bibinfo {author} {\bibfnamefont {M.~S.}\ \bibnamefont {Rudner}}\ and\ \bibinfo {author} {\bibfnamefont {N.~H.}\ \bibnamefont {Lindner}},\ }\bibfield  {title} {\bibinfo {title} {{Band structure engineering and non-equilibrium dynamics in Floquet topological insulators}},\ }\href {https://doi.org/10.1038/s42254-020-0170-z} {\bibfield  {journal} {\bibinfo  {journal} {Nature Reviews Physics}\ }\textbf {\bibinfo {volume} {2}},\ \bibinfo {pages} {229} (\bibinfo {year} {2020}{\natexlab{a}})}\BibitemShut {NoStop}%
\bibitem [{\citenamefont {Weitenberg}\ and\ \citenamefont {Simonet}(2021)}]{Weitenberg2021}%
  \BibitemOpen
  \bibfield  {author} {\bibinfo {author} {\bibfnamefont {C.}~\bibnamefont {Weitenberg}}\ and\ \bibinfo {author} {\bibfnamefont {J.}~\bibnamefont {Simonet}},\ }\bibfield  {title} {\bibinfo {title} {{Tailoring quantum gases by Floquet engineering}},\ }\href {https://doi.org/10.1038/s41567-021-01316-x} {\bibfield  {journal} {\bibinfo  {journal} {Nature Physics}\ }\textbf {\bibinfo {volume} {17}},\ \bibinfo {pages} {1342} (\bibinfo {year} {2021})}\BibitemShut {NoStop}%
\bibitem [{\citenamefont {Shirley}(1965)}]{Shirley1965}%
  \BibitemOpen
  \bibfield  {author} {\bibinfo {author} {\bibfnamefont {J.~H.}\ \bibnamefont {Shirley}},\ }\bibfield  {title} {\bibinfo {title} {{Solution of the Schr\"odinger Equation with a Hamiltonian Periodic in Time}},\ }\href {https://doi.org/10.1103/PhysRev.138.B979} {\bibfield  {journal} {\bibinfo  {journal} {Phys. Rev.}\ }\textbf {\bibinfo {volume} {138}},\ \bibinfo {pages} {B979} (\bibinfo {year} {1965})}\BibitemShut {NoStop}%
\bibitem [{\citenamefont {Sambe}(1973)}]{Sambe1973}%
  \BibitemOpen
  \bibfield  {author} {\bibinfo {author} {\bibfnamefont {H.}~\bibnamefont {Sambe}},\ }\bibfield  {title} {\bibinfo {title} {{Steady States and Quasienergies of a Quantum-Mechanical System in an Oscillating Field}},\ }\href {https://doi.org/10.1103/PhysRevA.7.2203} {\bibfield  {journal} {\bibinfo  {journal} {Phys. Rev. A}\ }\textbf {\bibinfo {volume} {7}},\ \bibinfo {pages} {2203} (\bibinfo {year} {1973})}\BibitemShut {NoStop}%
\bibitem [{\citenamefont {{S\o{}rensen, Anders S. and Demler, Eugene and Lukin, Mikhail D.}}(2005)}]{Sorensen2005}%
  \BibitemOpen
  \bibfield  {author} {\bibinfo {author} {\bibnamefont {{S\o{}rensen, Anders S. and Demler, Eugene and Lukin, Mikhail D.}}},\ }\bibfield  {title} {\bibinfo {title} {{Fractional Quantum Hall States of Atoms in Optical Lattices}},\ }\href {https://doi.org/10.1103/PhysRevLett.94.086803} {\bibfield  {journal} {\bibinfo  {journal} {Phys. Rev. Lett.}\ }\textbf {\bibinfo {volume} {94}},\ \bibinfo {pages} {086803} (\bibinfo {year} {2005})}\BibitemShut {NoStop}%
\bibitem [{\citenamefont {Oka}\ and\ \citenamefont {Aoki}(2009)}]{Oka2009}%
  \BibitemOpen
  \bibfield  {author} {\bibinfo {author} {\bibfnamefont {T.}~\bibnamefont {Oka}}\ and\ \bibinfo {author} {\bibfnamefont {H.}~\bibnamefont {Aoki}},\ }\bibfield  {title} {\bibinfo {title} {{Photovoltaic Hall effect in graphene}},\ }\href {https://doi.org/10.1103/PhysRevB.79.081406} {\bibfield  {journal} {\bibinfo  {journal} {Phys. Rev. B}\ }\textbf {\bibinfo {volume} {79}},\ \bibinfo {pages} {081406} (\bibinfo {year} {2009})}\BibitemShut {NoStop}%
\bibitem [{\citenamefont {Lindner}\ \emph {et~al.}(2011)\citenamefont {Lindner}, \citenamefont {Refael},\ and\ \citenamefont {Galitski}}]{Lindner2011}%
  \BibitemOpen
  \bibfield  {author} {\bibinfo {author} {\bibfnamefont {N.~H.}\ \bibnamefont {Lindner}}, \bibinfo {author} {\bibfnamefont {G.}~\bibnamefont {Refael}},\ and\ \bibinfo {author} {\bibfnamefont {V.}~\bibnamefont {Galitski}},\ }\bibfield  {title} {\bibinfo {title} {Floquet topological insulator in semiconductor quantum wells},\ }\href@noop {} {\bibfield  {journal} {\bibinfo  {journal} {Nature Physics}\ }\textbf {\bibinfo {volume} {7}},\ \bibinfo {pages} {490} (\bibinfo {year} {2011})}\BibitemShut {NoStop}%
\bibitem [{\citenamefont {Rudner}\ \emph {et~al.}(2013)\citenamefont {Rudner}, \citenamefont {Lindner}, \citenamefont {Berg},\ and\ \citenamefont {Levin}}]{Rudner2013}%
  \BibitemOpen
  \bibfield  {author} {\bibinfo {author} {\bibfnamefont {M.~S.}\ \bibnamefont {Rudner}}, \bibinfo {author} {\bibfnamefont {N.~H.}\ \bibnamefont {Lindner}}, \bibinfo {author} {\bibfnamefont {E.}~\bibnamefont {Berg}},\ and\ \bibinfo {author} {\bibfnamefont {M.}~\bibnamefont {Levin}},\ }\bibfield  {title} {\bibinfo {title} {{Anomalous Edge States and the Bulk-Edge Correspondence for Periodically Driven Two-Dimensional Systems}},\ }\href {https://doi.org/10.1103/PhysRevX.3.031005} {\bibfield  {journal} {\bibinfo  {journal} {Phys. Rev. X}\ }\textbf {\bibinfo {volume} {3}},\ \bibinfo {pages} {031005} (\bibinfo {year} {2013})}\BibitemShut {NoStop}%
\bibitem [{\citenamefont {Wang}\ \emph {et~al.}(2013)\citenamefont {Wang}, \citenamefont {Steinberg}, \citenamefont {Jarillo-Herrero},\ and\ \citenamefont {Gedik}}]{Wang2013a}%
  \BibitemOpen
  \bibfield  {author} {\bibinfo {author} {\bibfnamefont {Y.~H.}\ \bibnamefont {Wang}}, \bibinfo {author} {\bibfnamefont {H.}~\bibnamefont {Steinberg}}, \bibinfo {author} {\bibfnamefont {P.}~\bibnamefont {Jarillo-Herrero}},\ and\ \bibinfo {author} {\bibfnamefont {N.}~\bibnamefont {Gedik}},\ }\bibfield  {title} {\bibinfo {title} {Observation of {Floquet-Bloch} states on the surface of a topological insulator},\ }\href {https://doi.org/10.1126/science.1239834} {\bibfield  {journal} {\bibinfo  {journal} {Science}\ }\textbf {\bibinfo {volume} {342}},\ \bibinfo {pages} {453} (\bibinfo {year} {2013})}\BibitemShut {NoStop}%
\bibitem [{\citenamefont {Mahmood}\ \emph {et~al.}(2016)\citenamefont {Mahmood}, \citenamefont {Chan}, \citenamefont {Alpichshev}, \citenamefont {Gardner}, \citenamefont {Lee}, \citenamefont {Lee},\ and\ \citenamefont {Gedik}}]{Mahmood2016}%
  \BibitemOpen
  \bibfield  {author} {\bibinfo {author} {\bibfnamefont {F.}~\bibnamefont {Mahmood}}, \bibinfo {author} {\bibfnamefont {C.-K.}\ \bibnamefont {Chan}}, \bibinfo {author} {\bibfnamefont {Z.}~\bibnamefont {Alpichshev}}, \bibinfo {author} {\bibfnamefont {D.}~\bibnamefont {Gardner}}, \bibinfo {author} {\bibfnamefont {Y.}~\bibnamefont {Lee}}, \bibinfo {author} {\bibfnamefont {P.~A.}\ \bibnamefont {Lee}},\ and\ \bibinfo {author} {\bibfnamefont {N.}~\bibnamefont {Gedik}},\ }\bibfield  {title} {\bibinfo {title} {{Selective scattering between Floquet--Bloch and Volkov states in a topological insulator}},\ }\href {https://doi.org/10.1038/nphys3609} {\bibfield  {journal} {\bibinfo  {journal} {Nature Physics}\ }\textbf {\bibinfo {volume} {12}},\ \bibinfo {pages} {306} (\bibinfo {year} {2016})}\BibitemShut {NoStop}%
\bibitem [{\citenamefont {McIver}\ \emph {et~al.}(2020)\citenamefont {McIver}, \citenamefont {Schulte}, \citenamefont {Stein}, \citenamefont {Matsuyama}, \citenamefont {Jotzu}, \citenamefont {Meier},\ and\ \citenamefont {Cavalleri}}]{McIver2020}%
  \BibitemOpen
  \bibfield  {author} {\bibinfo {author} {\bibfnamefont {J.~W.}\ \bibnamefont {McIver}}, \bibinfo {author} {\bibfnamefont {B.}~\bibnamefont {Schulte}}, \bibinfo {author} {\bibfnamefont {F.~U.}\ \bibnamefont {Stein}}, \bibinfo {author} {\bibfnamefont {T.}~\bibnamefont {Matsuyama}}, \bibinfo {author} {\bibfnamefont {G.}~\bibnamefont {Jotzu}}, \bibinfo {author} {\bibfnamefont {G.}~\bibnamefont {Meier}},\ and\ \bibinfo {author} {\bibfnamefont {A.}~\bibnamefont {Cavalleri}},\ }\bibfield  {title} {\bibinfo {title} {{Light-induced anomalous Hall effect in graphene}},\ }\href {https://doi.org/10.1038/s41567-019-0698-y} {\bibfield  {journal} {\bibinfo  {journal} {Nature Physics}\ }\textbf {\bibinfo {volume} {16}},\ \bibinfo {pages} {38} (\bibinfo {year} {2020})}\BibitemShut {NoStop}%
\bibitem [{\citenamefont {Park}\ \emph {et~al.}(2022)\citenamefont {Park}, \citenamefont {Lee}, \citenamefont {Jang}, \citenamefont {Choi}, \citenamefont {Park}, \citenamefont {Jung}, \citenamefont {Watanabe}, \citenamefont {Taniguchi}, \citenamefont {Cho},\ and\ \citenamefont {Lee}}]{Park2022}%
  \BibitemOpen
  \bibfield  {author} {\bibinfo {author} {\bibfnamefont {S.}~\bibnamefont {Park}}, \bibinfo {author} {\bibfnamefont {W.}~\bibnamefont {Lee}}, \bibinfo {author} {\bibfnamefont {S.}~\bibnamefont {Jang}}, \bibinfo {author} {\bibfnamefont {Y.-B.}\ \bibnamefont {Choi}}, \bibinfo {author} {\bibfnamefont {J.}~\bibnamefont {Park}}, \bibinfo {author} {\bibfnamefont {W.}~\bibnamefont {Jung}}, \bibinfo {author} {\bibfnamefont {K.}~\bibnamefont {Watanabe}}, \bibinfo {author} {\bibfnamefont {T.}~\bibnamefont {Taniguchi}}, \bibinfo {author} {\bibfnamefont {G.~Y.}\ \bibnamefont {Cho}},\ and\ \bibinfo {author} {\bibfnamefont {G.-H.}\ \bibnamefont {Lee}},\ }\bibfield  {title} {\bibinfo {title} {{Steady Floquet--Andreev states in graphene Josephson junctions}},\ }\href {https://doi.org/10.1038/s41586-021-04364-8} {\bibfield  {journal} {\bibinfo  {journal} {Nature}\ }\textbf {\bibinfo {volume} {603}},\ \bibinfo {pages} {421} (\bibinfo {year} {2022})}\BibitemShut {NoStop}%
\bibitem [{\citenamefont {Zhou}\ \emph {et~al.}(2023)\citenamefont {Zhou}, \citenamefont {Bao}, \citenamefont {Fan}, \citenamefont {Zhou}, \citenamefont {Gao}, \citenamefont {Zhong}, \citenamefont {Lin}, \citenamefont {Liu}, \citenamefont {Yu}, \citenamefont {Tang}, \citenamefont {Meng}, \citenamefont {Duan},\ and\ \citenamefont {Zhou}}]{Zhou2023}%
  \BibitemOpen
  \bibfield  {author} {\bibinfo {author} {\bibfnamefont {S.}~\bibnamefont {Zhou}}, \bibinfo {author} {\bibfnamefont {C.}~\bibnamefont {Bao}}, \bibinfo {author} {\bibfnamefont {B.}~\bibnamefont {Fan}}, \bibinfo {author} {\bibfnamefont {H.}~\bibnamefont {Zhou}}, \bibinfo {author} {\bibfnamefont {Q.}~\bibnamefont {Gao}}, \bibinfo {author} {\bibfnamefont {H.}~\bibnamefont {Zhong}}, \bibinfo {author} {\bibfnamefont {T.}~\bibnamefont {Lin}}, \bibinfo {author} {\bibfnamefont {H.}~\bibnamefont {Liu}}, \bibinfo {author} {\bibfnamefont {P.}~\bibnamefont {Yu}}, \bibinfo {author} {\bibfnamefont {P.}~\bibnamefont {Tang}}, \bibinfo {author} {\bibfnamefont {S.}~\bibnamefont {Meng}}, \bibinfo {author} {\bibfnamefont {W.}~\bibnamefont {Duan}},\ and\ \bibinfo {author} {\bibfnamefont {S.}~\bibnamefont {Zhou}},\ }\bibfield  {title} {\bibinfo {title} {{Pseudospin-selective Floquet band engineering in black phosphorus}},\ }\href {https://doi.org/10.1038/s41586-022-05610-3} {\bibfield  {journal} {\bibinfo  {journal} {Nature}\
  }\textbf {\bibinfo {volume} {614}},\ \bibinfo {pages} {75} (\bibinfo {year} {2023})}\BibitemShut {NoStop}%
\bibitem [{\citenamefont {Choi}\ \emph {et~al.}(2025)\citenamefont {Choi}, \citenamefont {Mogi}, \citenamefont {De~Giovannini}, \citenamefont {Azoury}, \citenamefont {Lv}, \citenamefont {Su}, \citenamefont {H\"{u}bener}, \citenamefont {Rubio},\ and\ \citenamefont {Gedik}}]{Choi2025}%
  \BibitemOpen
  \bibfield  {author} {\bibinfo {author} {\bibfnamefont {D.}~\bibnamefont {Choi}}, \bibinfo {author} {\bibfnamefont {M.}~\bibnamefont {Mogi}}, \bibinfo {author} {\bibfnamefont {U.}~\bibnamefont {De~Giovannini}}, \bibinfo {author} {\bibfnamefont {D.}~\bibnamefont {Azoury}}, \bibinfo {author} {\bibfnamefont {B.}~\bibnamefont {Lv}}, \bibinfo {author} {\bibfnamefont {Y.}~\bibnamefont {Su}}, \bibinfo {author} {\bibfnamefont {H.}~\bibnamefont {H\"{u}bener}}, \bibinfo {author} {\bibfnamefont {A.}~\bibnamefont {Rubio}},\ and\ \bibinfo {author} {\bibfnamefont {N.}~\bibnamefont {Gedik}},\ }\bibfield  {title} {\bibinfo {title} {{Observation of Floquet–Bloch states in monolayer graphene}},\ }\bibfield  {journal} {\bibinfo  {journal} {Nature Physics}\ }\href {https://doi.org/10.1038/s41567-025-02888-8} {10.1038/s41567-025-02888-8} (\bibinfo {year} {2025})\BibitemShut {NoStop}%
\bibitem [{\citenamefont {Merboldt}\ \emph {et~al.}(2025)\citenamefont {Merboldt}, \citenamefont {Sch\"{u}ler}, \citenamefont {Schmitt}, \citenamefont {Bange}, \citenamefont {Bennecke}, \citenamefont {Gadge}, \citenamefont {Pierz}, \citenamefont {Schumacher}, \citenamefont {Momeni}, \citenamefont {Steil}, \citenamefont {Manmana}, \citenamefont {Sentef}, \citenamefont {Reutzel},\ and\ \citenamefont {Mathias}}]{Merboldt2025}%
  \BibitemOpen
  \bibfield  {author} {\bibinfo {author} {\bibfnamefont {M.}~\bibnamefont {Merboldt}}, \bibinfo {author} {\bibfnamefont {M.}~\bibnamefont {Sch\"{u}ler}}, \bibinfo {author} {\bibfnamefont {D.}~\bibnamefont {Schmitt}}, \bibinfo {author} {\bibfnamefont {J.~P.}\ \bibnamefont {Bange}}, \bibinfo {author} {\bibfnamefont {W.}~\bibnamefont {Bennecke}}, \bibinfo {author} {\bibfnamefont {K.}~\bibnamefont {Gadge}}, \bibinfo {author} {\bibfnamefont {K.}~\bibnamefont {Pierz}}, \bibinfo {author} {\bibfnamefont {H.~W.}\ \bibnamefont {Schumacher}}, \bibinfo {author} {\bibfnamefont {D.}~\bibnamefont {Momeni}}, \bibinfo {author} {\bibfnamefont {D.}~\bibnamefont {Steil}}, \bibinfo {author} {\bibfnamefont {S.~R.}\ \bibnamefont {Manmana}}, \bibinfo {author} {\bibfnamefont {M.~A.}\ \bibnamefont {Sentef}}, \bibinfo {author} {\bibfnamefont {M.}~\bibnamefont {Reutzel}},\ and\ \bibinfo {author} {\bibfnamefont {S.}~\bibnamefont {Mathias}},\ }\bibfield  {title} {\bibinfo {title} {{Observation of Floquet states in graphene}},\ }\bibfield
  {journal} {\bibinfo  {journal} {Nature Physics}\ }\href {https://doi.org/10.1038/s41567-025-02889-7} {10.1038/s41567-025-02889-7} (\bibinfo {year} {2025})\BibitemShut {NoStop}%
\bibitem [{\citenamefont {Lignier}\ \emph {et~al.}(2007)\citenamefont {Lignier}, \citenamefont {Sias}, \citenamefont {Ciampini}, \citenamefont {Singh}, \citenamefont {Zenesini}, \citenamefont {Morsch},\ and\ \citenamefont {Arimondo}}]{Lignier2007}%
  \BibitemOpen
  \bibfield  {author} {\bibinfo {author} {\bibfnamefont {H.}~\bibnamefont {Lignier}}, \bibinfo {author} {\bibfnamefont {C.}~\bibnamefont {Sias}}, \bibinfo {author} {\bibfnamefont {D.}~\bibnamefont {Ciampini}}, \bibinfo {author} {\bibfnamefont {Y.}~\bibnamefont {Singh}}, \bibinfo {author} {\bibfnamefont {A.}~\bibnamefont {Zenesini}}, \bibinfo {author} {\bibfnamefont {O.}~\bibnamefont {Morsch}},\ and\ \bibinfo {author} {\bibfnamefont {E.}~\bibnamefont {Arimondo}},\ }\bibfield  {title} {\bibinfo {title} {{Dynamical Control of Matter-Wave Tunneling in Periodic Potentials}},\ }\href {https://doi.org/10.1103/PhysRevLett.99.220403} {\bibfield  {journal} {\bibinfo  {journal} {Phys. Rev. Lett.}\ }\textbf {\bibinfo {volume} {99}},\ \bibinfo {pages} {220403} (\bibinfo {year} {2007})}\BibitemShut {NoStop}%
\bibitem [{\citenamefont {Struck}\ \emph {et~al.}(2011)\citenamefont {Struck}, \citenamefont {\"{O}lschl\"{a}ger}, \citenamefont {Le~Targat}, \citenamefont {Soltan-Panahi}, \citenamefont {Eckardt}, \citenamefont {Lewenstein}, \citenamefont {Windpassinger},\ and\ \citenamefont {Sengstock}}]{Struck2011}%
  \BibitemOpen
  \bibfield  {author} {\bibinfo {author} {\bibfnamefont {J.}~\bibnamefont {Struck}}, \bibinfo {author} {\bibfnamefont {C.}~\bibnamefont {\"{O}lschl\"{a}ger}}, \bibinfo {author} {\bibfnamefont {R.}~\bibnamefont {Le~Targat}}, \bibinfo {author} {\bibfnamefont {P.}~\bibnamefont {Soltan-Panahi}}, \bibinfo {author} {\bibfnamefont {A.}~\bibnamefont {Eckardt}}, \bibinfo {author} {\bibfnamefont {M.}~\bibnamefont {Lewenstein}}, \bibinfo {author} {\bibfnamefont {P.}~\bibnamefont {Windpassinger}},\ and\ \bibinfo {author} {\bibfnamefont {K.}~\bibnamefont {Sengstock}},\ }\bibfield  {title} {\bibinfo {title} {{Quantum Simulation of Frustrated Classical Magnetism in Triangular Optical Lattices}},\ }\href {https://doi.org/10.1126/science.1207239} {\bibfield  {journal} {\bibinfo  {journal} {Science}\ }\textbf {\bibinfo {volume} {333}},\ \bibinfo {pages} {996–999} (\bibinfo {year} {2011})}\BibitemShut {NoStop}%
\bibitem [{\citenamefont {Aidelsburger}\ \emph {et~al.}(2014)\citenamefont {Aidelsburger}, \citenamefont {Lohse}, \citenamefont {Schweizer}, \citenamefont {Atala}, \citenamefont {Barreiro}, \citenamefont {Nascimbène}, \citenamefont {Cooper}, \citenamefont {Bloch},\ and\ \citenamefont {Goldman}}]{Aidelsburger2014}%
  \BibitemOpen
  \bibfield  {author} {\bibinfo {author} {\bibfnamefont {M.}~\bibnamefont {Aidelsburger}}, \bibinfo {author} {\bibfnamefont {M.}~\bibnamefont {Lohse}}, \bibinfo {author} {\bibfnamefont {C.}~\bibnamefont {Schweizer}}, \bibinfo {author} {\bibfnamefont {M.}~\bibnamefont {Atala}}, \bibinfo {author} {\bibfnamefont {J.~T.}\ \bibnamefont {Barreiro}}, \bibinfo {author} {\bibfnamefont {S.}~\bibnamefont {Nascimbène}}, \bibinfo {author} {\bibfnamefont {N.~R.}\ \bibnamefont {Cooper}}, \bibinfo {author} {\bibfnamefont {I.}~\bibnamefont {Bloch}},\ and\ \bibinfo {author} {\bibfnamefont {N.}~\bibnamefont {Goldman}},\ }\bibfield  {title} {\bibinfo {title} {{Measuring the Chern number of Hofstadter bands with ultracold bosonic atoms}},\ }\href {https://doi.org/10.1038/nphys3171} {\bibfield  {journal} {\bibinfo  {journal} {Nature Physics}\ }\textbf {\bibinfo {volume} {11}},\ \bibinfo {pages} {162–166} (\bibinfo {year} {2014})}\BibitemShut {NoStop}%
\bibitem [{\citenamefont {Jotzu}\ \emph {et~al.}(2014)\citenamefont {Jotzu}, \citenamefont {Messer}, \citenamefont {Desbuquois}, \citenamefont {Lebrat}, \citenamefont {Uehlinger}, \citenamefont {Greif},\ and\ \citenamefont {Esslinger}}]{Jotzu2014}%
  \BibitemOpen
  \bibfield  {author} {\bibinfo {author} {\bibfnamefont {G.}~\bibnamefont {Jotzu}}, \bibinfo {author} {\bibfnamefont {M.}~\bibnamefont {Messer}}, \bibinfo {author} {\bibfnamefont {R.}~\bibnamefont {Desbuquois}}, \bibinfo {author} {\bibfnamefont {M.}~\bibnamefont {Lebrat}}, \bibinfo {author} {\bibfnamefont {T.}~\bibnamefont {Uehlinger}}, \bibinfo {author} {\bibfnamefont {D.}~\bibnamefont {Greif}},\ and\ \bibinfo {author} {\bibfnamefont {T.}~\bibnamefont {Esslinger}},\ }\bibfield  {title} {\bibinfo {title} {{Experimental realization of the topological Haldane model with ultracold fermions}},\ }\href {https://doi.org/10.1038/nature13915} {\bibfield  {journal} {\bibinfo  {journal} {Nature}\ }\textbf {\bibinfo {volume} {515}},\ \bibinfo {pages} {237} (\bibinfo {year} {2014})}\BibitemShut {NoStop}%
\bibitem [{\citenamefont {{N. Fläschner and B. S. Rem and M. Tarnowski and D. Vogel and D.-S. Lühmann and K. Sengstock and C. Weitenberg}}(2016)}]{Flaschner2016}%
  \BibitemOpen
  \bibfield  {author} {\bibinfo {author} {\bibnamefont {{N. Fläschner and B. S. Rem and M. Tarnowski and D. Vogel and D.-S. Lühmann and K. Sengstock and C. Weitenberg}}},\ }\bibfield  {title} {\bibinfo {title} {{Experimental reconstruction of the Berry curvature in a Floquet Bloch band}},\ }\href {https://doi.org/10.1126/science.aad4568} {\bibfield  {journal} {\bibinfo  {journal} {Science}\ }\textbf {\bibinfo {volume} {352}},\ \bibinfo {pages} {1091} (\bibinfo {year} {2016})},\ \Eprint {https://arxiv.org/abs/https://www.science.org/doi/pdf/10.1126/science.aad4568} {https://www.science.org/doi/pdf/10.1126/science.aad4568} \BibitemShut {NoStop}%
\bibitem [{\citenamefont {Tai}\ \emph {et~al.}(2017)\citenamefont {Tai}, \citenamefont {Lukin}, \citenamefont {Rispoli}, \citenamefont {Schittko}, \citenamefont {Menke}, \citenamefont {Borgnia}, \citenamefont {Preiss}, \citenamefont {Grusdt}, \citenamefont {Kaufman},\ and\ \citenamefont {Greiner}}]{Tai2017}%
  \BibitemOpen
  \bibfield  {author} {\bibinfo {author} {\bibfnamefont {M.~E.}\ \bibnamefont {Tai}}, \bibinfo {author} {\bibfnamefont {A.}~\bibnamefont {Lukin}}, \bibinfo {author} {\bibfnamefont {M.}~\bibnamefont {Rispoli}}, \bibinfo {author} {\bibfnamefont {R.}~\bibnamefont {Schittko}}, \bibinfo {author} {\bibfnamefont {T.}~\bibnamefont {Menke}}, \bibinfo {author} {\bibfnamefont {D.}~\bibnamefont {Borgnia}}, \bibinfo {author} {\bibfnamefont {P.~M.}\ \bibnamefont {Preiss}}, \bibinfo {author} {\bibfnamefont {F.}~\bibnamefont {Grusdt}}, \bibinfo {author} {\bibfnamefont {A.~M.}\ \bibnamefont {Kaufman}},\ and\ \bibinfo {author} {\bibfnamefont {M.}~\bibnamefont {Greiner}},\ }\bibfield  {title} {\bibinfo {title} {{Microscopy of the interacting Harper–Hofstadter model in the two-body limit}},\ }\href {https://doi.org/10.1038/nature22811} {\bibfield  {journal} {\bibinfo  {journal} {Nature}\ }\textbf {\bibinfo {volume} {546}},\ \bibinfo {pages} {519–523} (\bibinfo {year} {2017})}\BibitemShut {NoStop}%
\bibitem [{\citenamefont {Asteria}\ \emph {et~al.}(2019)\citenamefont {Asteria}, \citenamefont {Tran}, \citenamefont {Ozawa}, \citenamefont {Tarnowski}, \citenamefont {Rem}, \citenamefont {Fl\"{a}schner}, \citenamefont {Sengstock}, \citenamefont {Goldman},\ and\ \citenamefont {Weitenberg}}]{Asteria2019}%
  \BibitemOpen
  \bibfield  {author} {\bibinfo {author} {\bibfnamefont {L.}~\bibnamefont {Asteria}}, \bibinfo {author} {\bibfnamefont {D.~T.}\ \bibnamefont {Tran}}, \bibinfo {author} {\bibfnamefont {T.}~\bibnamefont {Ozawa}}, \bibinfo {author} {\bibfnamefont {M.}~\bibnamefont {Tarnowski}}, \bibinfo {author} {\bibfnamefont {B.~S.}\ \bibnamefont {Rem}}, \bibinfo {author} {\bibfnamefont {N.}~\bibnamefont {Fl\"{a}schner}}, \bibinfo {author} {\bibfnamefont {K.}~\bibnamefont {Sengstock}}, \bibinfo {author} {\bibfnamefont {N.}~\bibnamefont {Goldman}},\ and\ \bibinfo {author} {\bibfnamefont {C.}~\bibnamefont {Weitenberg}},\ }\bibfield  {title} {\bibinfo {title} {{Measuring quantized circular dichroism in ultracold topological matter}},\ }\href {https://doi.org/10.1038/s41567-019-0417-8} {\bibfield  {journal} {\bibinfo  {journal} {Nature Physics}\ }\textbf {\bibinfo {volume} {15}},\ \bibinfo {pages} {449–454} (\bibinfo {year} {2019})}\BibitemShut {NoStop}%
\bibitem [{\citenamefont {Tarnowski}\ \emph {et~al.}(2019)\citenamefont {Tarnowski}, \citenamefont {{\"U}nal}, \citenamefont {Fl{\"a}schner}, \citenamefont {Rem}, \citenamefont {Eckardt}, \citenamefont {Sengstock},\ and\ \citenamefont {Weitenberg}}]{Tarnowski2019}%
  \BibitemOpen
  \bibfield  {author} {\bibinfo {author} {\bibfnamefont {M.}~\bibnamefont {Tarnowski}}, \bibinfo {author} {\bibfnamefont {F.~N.}\ \bibnamefont {{\"U}nal}}, \bibinfo {author} {\bibfnamefont {N.}~\bibnamefont {Fl{\"a}schner}}, \bibinfo {author} {\bibfnamefont {B.~S.}\ \bibnamefont {Rem}}, \bibinfo {author} {\bibfnamefont {A.}~\bibnamefont {Eckardt}}, \bibinfo {author} {\bibfnamefont {K.}~\bibnamefont {Sengstock}},\ and\ \bibinfo {author} {\bibfnamefont {C.}~\bibnamefont {Weitenberg}},\ }\bibfield  {title} {\bibinfo {title} {{Measuring topology from dynamics by obtaining the Chern number from a linking number}},\ }\href {https://doi.org/10.1038/s41467-019-09668-y} {\bibfield  {journal} {\bibinfo  {journal} {Nature Communications}\ }\textbf {\bibinfo {volume} {10}},\ \bibinfo {pages} {1728} (\bibinfo {year} {2019})}\BibitemShut {NoStop}%
\bibitem [{\citenamefont {Wintersperger}\ \emph {et~al.}(2020)\citenamefont {Wintersperger}, \citenamefont {Braun}, \citenamefont {{\"U}nal}, \citenamefont {Eckardt}, \citenamefont {Liberto}, \citenamefont {Goldman}, \citenamefont {Bloch},\ and\ \citenamefont {Aidelsburger}}]{Wintersperger2020}%
  \BibitemOpen
  \bibfield  {author} {\bibinfo {author} {\bibfnamefont {K.}~\bibnamefont {Wintersperger}}, \bibinfo {author} {\bibfnamefont {C.}~\bibnamefont {Braun}}, \bibinfo {author} {\bibfnamefont {F.~N.}\ \bibnamefont {{\"U}nal}}, \bibinfo {author} {\bibfnamefont {A.}~\bibnamefont {Eckardt}}, \bibinfo {author} {\bibfnamefont {M.~D.}\ \bibnamefont {Liberto}}, \bibinfo {author} {\bibfnamefont {N.}~\bibnamefont {Goldman}}, \bibinfo {author} {\bibfnamefont {I.}~\bibnamefont {Bloch}},\ and\ \bibinfo {author} {\bibfnamefont {M.}~\bibnamefont {Aidelsburger}},\ }\bibfield  {title} {\bibinfo {title} {{Realization of an anomalous Floquet topological system with ultracold atoms}},\ }\href {https://doi.org/10.1038/s41567-020-0949-y} {\bibfield  {journal} {\bibinfo  {journal} {Nature Physics}\ }\textbf {\bibinfo {volume} {16}},\ \bibinfo {pages} {1058} (\bibinfo {year} {2020})}\BibitemShut {NoStop}%
\bibitem [{\citenamefont {Lu}\ \emph {et~al.}(2022)\citenamefont {Lu}, \citenamefont {Reid}, \citenamefont {Fritsch}, \citenamefont {Pi\~neiro},\ and\ \citenamefont {Spielman}}]{Lu2022}%
  \BibitemOpen
  \bibfield  {author} {\bibinfo {author} {\bibfnamefont {M.}~\bibnamefont {Lu}}, \bibinfo {author} {\bibfnamefont {G.~H.}\ \bibnamefont {Reid}}, \bibinfo {author} {\bibfnamefont {A.~R.}\ \bibnamefont {Fritsch}}, \bibinfo {author} {\bibfnamefont {A.~M.}\ \bibnamefont {Pi\~neiro}},\ and\ \bibinfo {author} {\bibfnamefont {I.~B.}\ \bibnamefont {Spielman}},\ }\bibfield  {title} {\bibinfo {title} {{Floquet Engineering Topological Dirac Bands}},\ }\href {https://doi.org/10.1103/PhysRevLett.129.040402} {\bibfield  {journal} {\bibinfo  {journal} {Phys. Rev. Lett.}\ }\textbf {\bibinfo {volume} {129}},\ \bibinfo {pages} {040402} (\bibinfo {year} {2022})}\BibitemShut {NoStop}%
\bibitem [{\citenamefont {L{\'e}onard}\ \emph {et~al.}(2023)\citenamefont {L{\'e}onard}, \citenamefont {Kim}, \citenamefont {Kwan}, \citenamefont {Segura}, \citenamefont {Grusdt}, \citenamefont {Repellin}, \citenamefont {Goldman},\ and\ \citenamefont {Greiner}}]{Leonard2023}%
  \BibitemOpen
  \bibfield  {author} {\bibinfo {author} {\bibfnamefont {J.}~\bibnamefont {L{\'e}onard}}, \bibinfo {author} {\bibfnamefont {S.}~\bibnamefont {Kim}}, \bibinfo {author} {\bibfnamefont {J.}~\bibnamefont {Kwan}}, \bibinfo {author} {\bibfnamefont {P.}~\bibnamefont {Segura}}, \bibinfo {author} {\bibfnamefont {F.}~\bibnamefont {Grusdt}}, \bibinfo {author} {\bibfnamefont {C.}~\bibnamefont {Repellin}}, \bibinfo {author} {\bibfnamefont {N.}~\bibnamefont {Goldman}},\ and\ \bibinfo {author} {\bibfnamefont {M.}~\bibnamefont {Greiner}},\ }\bibfield  {title} {\bibinfo {title} {{Realization of a fractional quantum Hall state with ultracold atoms}},\ }\href {https://doi.org/10.1038/s41586-023-06122-4} {\bibfield  {journal} {\bibinfo  {journal} {Nature}\ }\textbf {\bibinfo {volume} {619}},\ \bibinfo {pages} {495} (\bibinfo {year} {2023})}\BibitemShut {NoStop}%
\bibitem [{\citenamefont {Braun}\ \emph {et~al.}(2024)\citenamefont {Braun}, \citenamefont {Saint-Jalm}, \citenamefont {Hesse}, \citenamefont {Arceri}, \citenamefont {Bloch},\ and\ \citenamefont {Aidelsburger}}]{Braun2024}%
  \BibitemOpen
  \bibfield  {author} {\bibinfo {author} {\bibfnamefont {C.}~\bibnamefont {Braun}}, \bibinfo {author} {\bibfnamefont {R.}~\bibnamefont {Saint-Jalm}}, \bibinfo {author} {\bibfnamefont {A.}~\bibnamefont {Hesse}}, \bibinfo {author} {\bibfnamefont {J.}~\bibnamefont {Arceri}}, \bibinfo {author} {\bibfnamefont {I.}~\bibnamefont {Bloch}},\ and\ \bibinfo {author} {\bibfnamefont {M.}~\bibnamefont {Aidelsburger}},\ }\bibfield  {title} {\bibinfo {title} {{Real-space detection and manipulation of topological edge modes with ultracold atoms}},\ }\bibfield  {journal} {\bibinfo  {journal} {Nature Physics}\ }\href {https://doi.org/10.1038/s41567-024-02506-z} {10.1038/s41567-024-02506-z} (\bibinfo {year} {2024})\BibitemShut {NoStop}%
\bibitem [{\citenamefont {Fleury}\ \emph {et~al.}(2016)\citenamefont {Fleury}, \citenamefont {Khanikaev},\ and\ \citenamefont {Al{\`u}}}]{Fleury2016}%
  \BibitemOpen
  \bibfield  {author} {\bibinfo {author} {\bibfnamefont {R.}~\bibnamefont {Fleury}}, \bibinfo {author} {\bibfnamefont {A.~B.}\ \bibnamefont {Khanikaev}},\ and\ \bibinfo {author} {\bibfnamefont {A.}~\bibnamefont {Al{\`u}}},\ }\bibfield  {title} {\bibinfo {title} {Floquet topological insulators for sound},\ }\href {https://doi.org/10.1038/ncomms11744} {\bibfield  {journal} {\bibinfo  {journal} {Nature Communications}\ }\textbf {\bibinfo {volume} {7}},\ \bibinfo {pages} {11744} (\bibinfo {year} {2016})}\BibitemShut {NoStop}%
\bibitem [{\citenamefont {Peng}\ \emph {et~al.}(2016)\citenamefont {Peng}, \citenamefont {Qin}, \citenamefont {Zhao}, \citenamefont {Shen}, \citenamefont {Xu}, \citenamefont {Bao}, \citenamefont {Jia},\ and\ \citenamefont {Zhu}}]{Peng2016}%
  \BibitemOpen
  \bibfield  {author} {\bibinfo {author} {\bibfnamefont {Y.-G.}\ \bibnamefont {Peng}}, \bibinfo {author} {\bibfnamefont {C.-Z.}\ \bibnamefont {Qin}}, \bibinfo {author} {\bibfnamefont {D.-G.}\ \bibnamefont {Zhao}}, \bibinfo {author} {\bibfnamefont {Y.-X.}\ \bibnamefont {Shen}}, \bibinfo {author} {\bibfnamefont {X.-Y.}\ \bibnamefont {Xu}}, \bibinfo {author} {\bibfnamefont {M.}~\bibnamefont {Bao}}, \bibinfo {author} {\bibfnamefont {H.}~\bibnamefont {Jia}},\ and\ \bibinfo {author} {\bibfnamefont {X.-F.}\ \bibnamefont {Zhu}},\ }\bibfield  {title} {\bibinfo {title} {{Experimental demonstration of anomalous Floquet topological insulator for sound}},\ }\href {https://doi.org/10.1038/ncomms13368} {\bibfield  {journal} {\bibinfo  {journal} {Nature Communications}\ }\textbf {\bibinfo {volume} {7}},\ \bibinfo {pages} {13368} (\bibinfo {year} {2016})}\BibitemShut {NoStop}%
\bibitem [{\citenamefont {Cheng}\ \emph {et~al.}(2022)\citenamefont {Cheng}, \citenamefont {Bomantara}, \citenamefont {Xue}, \citenamefont {Zhu}, \citenamefont {Gong},\ and\ \citenamefont {Zhang}}]{Cheng2022}%
  \BibitemOpen
  \bibfield  {author} {\bibinfo {author} {\bibfnamefont {Z.}~\bibnamefont {Cheng}}, \bibinfo {author} {\bibfnamefont {R.~W.}\ \bibnamefont {Bomantara}}, \bibinfo {author} {\bibfnamefont {H.}~\bibnamefont {Xue}}, \bibinfo {author} {\bibfnamefont {W.}~\bibnamefont {Zhu}}, \bibinfo {author} {\bibfnamefont {J.}~\bibnamefont {Gong}},\ and\ \bibinfo {author} {\bibfnamefont {B.}~\bibnamefont {Zhang}},\ }\bibfield  {title} {\bibinfo {title} {{Observation of $\ensuremath{\pi}/2$ Modes in an Acoustic Floquet System}},\ }\href {https://doi.org/10.1103/PhysRevLett.129.254301} {\bibfield  {journal} {\bibinfo  {journal} {Phys. Rev. Lett.}\ }\textbf {\bibinfo {volume} {129}},\ \bibinfo {pages} {254301} (\bibinfo {year} {2022})}\BibitemShut {NoStop}%
\bibitem [{\citenamefont {Kitagawa}\ \emph {et~al.}(2012)\citenamefont {Kitagawa}, \citenamefont {Broome}, \citenamefont {Fedrizzi}, \citenamefont {Rudner}, \citenamefont {Berg}, \citenamefont {Kassal}, \citenamefont {Aspuru-Guzik}, \citenamefont {Demler},\ and\ \citenamefont {White}}]{Kitagawa2012}%
  \BibitemOpen
  \bibfield  {author} {\bibinfo {author} {\bibfnamefont {T.}~\bibnamefont {Kitagawa}}, \bibinfo {author} {\bibfnamefont {M.~A.}\ \bibnamefont {Broome}}, \bibinfo {author} {\bibfnamefont {A.}~\bibnamefont {Fedrizzi}}, \bibinfo {author} {\bibfnamefont {M.~S.}\ \bibnamefont {Rudner}}, \bibinfo {author} {\bibfnamefont {E.}~\bibnamefont {Berg}}, \bibinfo {author} {\bibfnamefont {I.}~\bibnamefont {Kassal}}, \bibinfo {author} {\bibfnamefont {A.}~\bibnamefont {Aspuru-Guzik}}, \bibinfo {author} {\bibfnamefont {E.}~\bibnamefont {Demler}},\ and\ \bibinfo {author} {\bibfnamefont {A.~G.}\ \bibnamefont {White}},\ }\bibfield  {title} {\bibinfo {title} {Observation of topologically protected bound states in photonic quantum walks},\ }\href {https://doi.org/10.1038/ncomms1872} {\bibfield  {journal} {\bibinfo  {journal} {Nature Communications}\ }\textbf {\bibinfo {volume} {3}},\ \bibinfo {pages} {882} (\bibinfo {year} {2012})}\BibitemShut {NoStop}%
\bibitem [{\citenamefont {Rechtsman}\ \emph {et~al.}(2013)\citenamefont {Rechtsman}, \citenamefont {Zeuner}, \citenamefont {Plotnik}, \citenamefont {Lumer}, \citenamefont {Podolsky}, \citenamefont {Dreisow}, \citenamefont {Nolte}, \citenamefont {Segev},\ and\ \citenamefont {Szameit}}]{Rechtsman2013}%
  \BibitemOpen
  \bibfield  {author} {\bibinfo {author} {\bibfnamefont {M.~C.}\ \bibnamefont {Rechtsman}}, \bibinfo {author} {\bibfnamefont {J.~M.}\ \bibnamefont {Zeuner}}, \bibinfo {author} {\bibfnamefont {Y.}~\bibnamefont {Plotnik}}, \bibinfo {author} {\bibfnamefont {Y.}~\bibnamefont {Lumer}}, \bibinfo {author} {\bibfnamefont {D.}~\bibnamefont {Podolsky}}, \bibinfo {author} {\bibfnamefont {F.}~\bibnamefont {Dreisow}}, \bibinfo {author} {\bibfnamefont {S.}~\bibnamefont {Nolte}}, \bibinfo {author} {\bibfnamefont {M.}~\bibnamefont {Segev}},\ and\ \bibinfo {author} {\bibfnamefont {A.}~\bibnamefont {Szameit}},\ }\bibfield  {title} {\bibinfo {title} {{Photonic Floquet topological insulators}},\ }\href {https://doi.org/10.1038/nature12066} {\bibfield  {journal} {\bibinfo  {journal} {Nature}\ }\textbf {\bibinfo {volume} {496}},\ \bibinfo {pages} {196–200} (\bibinfo {year} {2013})}\BibitemShut {NoStop}%
\bibitem [{\citenamefont {Hu}\ \emph {et~al.}(2015)\citenamefont {Hu}, \citenamefont {Pillay}, \citenamefont {Wu}, \citenamefont {Pasek}, \citenamefont {Shum},\ and\ \citenamefont {Chong}}]{Hu2015}%
  \BibitemOpen
  \bibfield  {author} {\bibinfo {author} {\bibfnamefont {W.}~\bibnamefont {Hu}}, \bibinfo {author} {\bibfnamefont {J.~C.}\ \bibnamefont {Pillay}}, \bibinfo {author} {\bibfnamefont {K.}~\bibnamefont {Wu}}, \bibinfo {author} {\bibfnamefont {M.}~\bibnamefont {Pasek}}, \bibinfo {author} {\bibfnamefont {P.~P.}\ \bibnamefont {Shum}},\ and\ \bibinfo {author} {\bibfnamefont {Y.~D.}\ \bibnamefont {Chong}},\ }\bibfield  {title} {\bibinfo {title} {{Measurement of a Topological Edge Invariant in a Microwave Network}},\ }\href {https://doi.org/10.1103/PhysRevX.5.011012} {\bibfield  {journal} {\bibinfo  {journal} {Phys. Rev. X}\ }\textbf {\bibinfo {volume} {5}},\ \bibinfo {pages} {011012} (\bibinfo {year} {2015})}\BibitemShut {NoStop}%
\bibitem [{\citenamefont {Gao}\ \emph {et~al.}(2016)\citenamefont {Gao}, \citenamefont {Gao}, \citenamefont {Shi}, \citenamefont {Yang}, \citenamefont {Lin}, \citenamefont {Xu}, \citenamefont {Joannopoulos}, \citenamefont {Solja{\v c}i{\'c}}, \citenamefont {Chen}, \citenamefont {Lu}, \citenamefont {Chong},\ and\ \citenamefont {Zhang}}]{Gao2016}%
  \BibitemOpen
  \bibfield  {author} {\bibinfo {author} {\bibfnamefont {F.}~\bibnamefont {Gao}}, \bibinfo {author} {\bibfnamefont {Z.}~\bibnamefont {Gao}}, \bibinfo {author} {\bibfnamefont {X.}~\bibnamefont {Shi}}, \bibinfo {author} {\bibfnamefont {Z.}~\bibnamefont {Yang}}, \bibinfo {author} {\bibfnamefont {X.}~\bibnamefont {Lin}}, \bibinfo {author} {\bibfnamefont {H.}~\bibnamefont {Xu}}, \bibinfo {author} {\bibfnamefont {J.~D.}\ \bibnamefont {Joannopoulos}}, \bibinfo {author} {\bibfnamefont {M.}~\bibnamefont {Solja{\v c}i{\'c}}}, \bibinfo {author} {\bibfnamefont {H.}~\bibnamefont {Chen}}, \bibinfo {author} {\bibfnamefont {L.}~\bibnamefont {Lu}}, \bibinfo {author} {\bibfnamefont {Y.}~\bibnamefont {Chong}},\ and\ \bibinfo {author} {\bibfnamefont {B.}~\bibnamefont {Zhang}},\ }\bibfield  {title} {\bibinfo {title} {Probing topological protection using a designer surface plasmon structure},\ }\href {https://doi.org/10.1038/ncomms11619} {\bibfield  {journal} {\bibinfo  {journal} {Nature Communications}\ }\textbf {\bibinfo
  {volume} {7}},\ \bibinfo {pages} {11619} (\bibinfo {year} {2016})}\BibitemShut {NoStop}%
\bibitem [{\citenamefont {Cardano}\ \emph {et~al.}(2017)\citenamefont {Cardano}, \citenamefont {D'Errico}, \citenamefont {Dauphin}, \citenamefont {Maffei}, \citenamefont {Piccirillo}, \citenamefont {de~Lisio}, \citenamefont {De~Filippis}, \citenamefont {Cataudella}, \citenamefont {Santamato}, \citenamefont {Marrucci}, \citenamefont {Lewenstein},\ and\ \citenamefont {Massignan}}]{Cardano2017}%
  \BibitemOpen
  \bibfield  {author} {\bibinfo {author} {\bibfnamefont {F.}~\bibnamefont {Cardano}}, \bibinfo {author} {\bibfnamefont {A.}~\bibnamefont {D'Errico}}, \bibinfo {author} {\bibfnamefont {A.}~\bibnamefont {Dauphin}}, \bibinfo {author} {\bibfnamefont {M.}~\bibnamefont {Maffei}}, \bibinfo {author} {\bibfnamefont {B.}~\bibnamefont {Piccirillo}}, \bibinfo {author} {\bibfnamefont {C.}~\bibnamefont {de~Lisio}}, \bibinfo {author} {\bibfnamefont {G.}~\bibnamefont {De~Filippis}}, \bibinfo {author} {\bibfnamefont {V.}~\bibnamefont {Cataudella}}, \bibinfo {author} {\bibfnamefont {E.}~\bibnamefont {Santamato}}, \bibinfo {author} {\bibfnamefont {L.}~\bibnamefont {Marrucci}}, \bibinfo {author} {\bibfnamefont {M.}~\bibnamefont {Lewenstein}},\ and\ \bibinfo {author} {\bibfnamefont {P.}~\bibnamefont {Massignan}},\ }\bibfield  {title} {\bibinfo {title} {{Detection of Zak phases and topological invariants in a chiral quantum walk of twisted photons}},\ }\href {https://doi.org/10.1038/ncomms15516} {\bibfield  {journal} {\bibinfo
  {journal} {Nature Communications}\ }\textbf {\bibinfo {volume} {8}},\ \bibinfo {pages} {15516} (\bibinfo {year} {2017})}\BibitemShut {NoStop}%
\bibitem [{\citenamefont {Mukherjee}\ \emph {et~al.}(2017)\citenamefont {Mukherjee}, \citenamefont {Spracklen}, \citenamefont {Valiente}, \citenamefont {Andersson}, \citenamefont {{\"O}hberg}, \citenamefont {Goldman},\ and\ \citenamefont {Thomson}}]{Mukherjee2017}%
  \BibitemOpen
  \bibfield  {author} {\bibinfo {author} {\bibfnamefont {S.}~\bibnamefont {Mukherjee}}, \bibinfo {author} {\bibfnamefont {A.}~\bibnamefont {Spracklen}}, \bibinfo {author} {\bibfnamefont {M.}~\bibnamefont {Valiente}}, \bibinfo {author} {\bibfnamefont {E.}~\bibnamefont {Andersson}}, \bibinfo {author} {\bibfnamefont {P.}~\bibnamefont {{\"O}hberg}}, \bibinfo {author} {\bibfnamefont {N.}~\bibnamefont {Goldman}},\ and\ \bibinfo {author} {\bibfnamefont {R.~R.}\ \bibnamefont {Thomson}},\ }\bibfield  {title} {\bibinfo {title} {Experimental observation of anomalous topological edge modes in a slowly driven photonic lattice},\ }\href {https://doi.org/10.1038/ncomms13918} {\bibfield  {journal} {\bibinfo  {journal} {Nature Communications}\ }\textbf {\bibinfo {volume} {8}},\ \bibinfo {pages} {13918} (\bibinfo {year} {2017})}\BibitemShut {NoStop}%
\bibitem [{\citenamefont {Maczewsky}\ \emph {et~al.}(2017)\citenamefont {Maczewsky}, \citenamefont {Zeuner}, \citenamefont {Nolte},\ and\ \citenamefont {Szameit}}]{Maczewsky2017}%
  \BibitemOpen
  \bibfield  {author} {\bibinfo {author} {\bibfnamefont {L.~J.}\ \bibnamefont {Maczewsky}}, \bibinfo {author} {\bibfnamefont {J.~M.}\ \bibnamefont {Zeuner}}, \bibinfo {author} {\bibfnamefont {S.}~\bibnamefont {Nolte}},\ and\ \bibinfo {author} {\bibfnamefont {A.}~\bibnamefont {Szameit}},\ }\bibfield  {title} {\bibinfo {title} {{Observation of photonic anomalous Floquet topological insulators}},\ }\href {https://doi.org/10.1038/ncomms13756} {\bibfield  {journal} {\bibinfo  {journal} {Nature Communications}\ }\textbf {\bibinfo {volume} {8}},\ \bibinfo {pages} {13756} (\bibinfo {year} {2017})}\BibitemShut {NoStop}%
\bibitem [{\citenamefont {Cheng}\ \emph {et~al.}(2019)\citenamefont {Cheng}, \citenamefont {Pan}, \citenamefont {Wang}, \citenamefont {Zhang}, \citenamefont {Yu}, \citenamefont {Gover}, \citenamefont {Zhang}, \citenamefont {Li}, \citenamefont {Zhou},\ and\ \citenamefont {Zhu}}]{Cheng2019}%
  \BibitemOpen
  \bibfield  {author} {\bibinfo {author} {\bibfnamefont {Q.}~\bibnamefont {Cheng}}, \bibinfo {author} {\bibfnamefont {Y.}~\bibnamefont {Pan}}, \bibinfo {author} {\bibfnamefont {H.}~\bibnamefont {Wang}}, \bibinfo {author} {\bibfnamefont {C.}~\bibnamefont {Zhang}}, \bibinfo {author} {\bibfnamefont {D.}~\bibnamefont {Yu}}, \bibinfo {author} {\bibfnamefont {A.}~\bibnamefont {Gover}}, \bibinfo {author} {\bibfnamefont {H.}~\bibnamefont {Zhang}}, \bibinfo {author} {\bibfnamefont {T.}~\bibnamefont {Li}}, \bibinfo {author} {\bibfnamefont {L.}~\bibnamefont {Zhou}},\ and\ \bibinfo {author} {\bibfnamefont {S.}~\bibnamefont {Zhu}},\ }\bibfield  {title} {\bibinfo {title} {{Observation of Anomalous $\ensuremath{\pi}$ Modes in Photonic Floquet Engineering}},\ }\href {https://doi.org/10.1103/PhysRevLett.122.173901} {\bibfield  {journal} {\bibinfo  {journal} {Phys. Rev. Lett.}\ }\textbf {\bibinfo {volume} {122}},\ \bibinfo {pages} {173901} (\bibinfo {year} {2019})}\BibitemShut {NoStop}%
\bibitem [{\citenamefont {Roushan}\ \emph {et~al.}(2017)\citenamefont {Roushan}, \citenamefont {Neill}, \citenamefont {Megrant}, \citenamefont {Chen}, \citenamefont {Babbush}, \citenamefont {Barends}, \citenamefont {Campbell}, \citenamefont {Chen}, \citenamefont {Chiaro}, \citenamefont {Dunsworth}, \citenamefont {Fowler}, \citenamefont {Jeffrey}, \citenamefont {Kelly}, \citenamefont {Lucero}, \citenamefont {Mutus}, \citenamefont {O'Malley}, \citenamefont {Neeley}, \citenamefont {Quintana}, \citenamefont {Sank}, \citenamefont {Vainsencher}, \citenamefont {Wenner}, \citenamefont {White}, \citenamefont {Kapit}, \citenamefont {Neven},\ and\ \citenamefont {Martinis}}]{Roushan2017}%
  \BibitemOpen
  \bibfield  {author} {\bibinfo {author} {\bibfnamefont {P.}~\bibnamefont {Roushan}}, \bibinfo {author} {\bibfnamefont {C.}~\bibnamefont {Neill}}, \bibinfo {author} {\bibfnamefont {A.}~\bibnamefont {Megrant}}, \bibinfo {author} {\bibfnamefont {Y.}~\bibnamefont {Chen}}, \bibinfo {author} {\bibfnamefont {R.}~\bibnamefont {Babbush}}, \bibinfo {author} {\bibfnamefont {R.}~\bibnamefont {Barends}}, \bibinfo {author} {\bibfnamefont {B.}~\bibnamefont {Campbell}}, \bibinfo {author} {\bibfnamefont {Z.}~\bibnamefont {Chen}}, \bibinfo {author} {\bibfnamefont {B.}~\bibnamefont {Chiaro}}, \bibinfo {author} {\bibfnamefont {A.}~\bibnamefont {Dunsworth}}, \bibinfo {author} {\bibfnamefont {A.}~\bibnamefont {Fowler}}, \bibinfo {author} {\bibfnamefont {E.}~\bibnamefont {Jeffrey}}, \bibinfo {author} {\bibfnamefont {J.}~\bibnamefont {Kelly}}, \bibinfo {author} {\bibfnamefont {E.}~\bibnamefont {Lucero}}, \bibinfo {author} {\bibfnamefont {J.}~\bibnamefont {Mutus}}, \bibinfo {author} {\bibfnamefont {P.~J.~J.}\ \bibnamefont
  {O'Malley}}, \bibinfo {author} {\bibfnamefont {M.}~\bibnamefont {Neeley}}, \bibinfo {author} {\bibfnamefont {C.}~\bibnamefont {Quintana}}, \bibinfo {author} {\bibfnamefont {D.}~\bibnamefont {Sank}}, \bibinfo {author} {\bibfnamefont {A.}~\bibnamefont {Vainsencher}}, \bibinfo {author} {\bibfnamefont {J.}~\bibnamefont {Wenner}}, \bibinfo {author} {\bibfnamefont {T.}~\bibnamefont {White}}, \bibinfo {author} {\bibfnamefont {E.}~\bibnamefont {Kapit}}, \bibinfo {author} {\bibfnamefont {H.}~\bibnamefont {Neven}},\ and\ \bibinfo {author} {\bibfnamefont {J.}~\bibnamefont {Martinis}},\ }\bibfield  {title} {\bibinfo {title} {{Chiral ground-state currents of interacting photons in a synthetic magnetic field}},\ }\href {https://doi.org/10.1038/nphys3930} {\bibfield  {journal} {\bibinfo  {journal} {Nature Physics}\ }\textbf {\bibinfo {volume} {13}},\ \bibinfo {pages} {146} (\bibinfo {year} {2017})}\BibitemShut {NoStop}%
\bibitem [{\citenamefont {Adiyatullin}\ \emph {et~al.}(2023)\citenamefont {Adiyatullin}, \citenamefont {Upreti}, \citenamefont {Lechevalier}, \citenamefont {Evain}, \citenamefont {Copie}, \citenamefont {Suret}, \citenamefont {Randoux}, \citenamefont {Delplace},\ and\ \citenamefont {Amo}}]{Adiyatullin2023}%
  \BibitemOpen
  \bibfield  {author} {\bibinfo {author} {\bibfnamefont {A.~F.}\ \bibnamefont {Adiyatullin}}, \bibinfo {author} {\bibfnamefont {L.~K.}\ \bibnamefont {Upreti}}, \bibinfo {author} {\bibfnamefont {C.}~\bibnamefont {Lechevalier}}, \bibinfo {author} {\bibfnamefont {C.}~\bibnamefont {Evain}}, \bibinfo {author} {\bibfnamefont {F.}~\bibnamefont {Copie}}, \bibinfo {author} {\bibfnamefont {P.}~\bibnamefont {Suret}}, \bibinfo {author} {\bibfnamefont {S.}~\bibnamefont {Randoux}}, \bibinfo {author} {\bibfnamefont {P.}~\bibnamefont {Delplace}},\ and\ \bibinfo {author} {\bibfnamefont {A.}~\bibnamefont {Amo}},\ }\bibfield  {title} {\bibinfo {title} {{Topological Properties of Floquet Winding Bands in a Photonic Lattice}},\ }\href {https://doi.org/10.1103/PhysRevLett.130.056901} {\bibfield  {journal} {\bibinfo  {journal} {Phys. Rev. Lett.}\ }\textbf {\bibinfo {volume} {130}},\ \bibinfo {pages} {056901} (\bibinfo {year} {2023})}\BibitemShut {NoStop}%
\bibitem [{\citenamefont {El~Sokhen}\ \emph {et~al.}(2024)\citenamefont {El~Sokhen}, \citenamefont {G\'omez-Le\'on}, \citenamefont {Adiyatullin}, \citenamefont {Randoux}, \citenamefont {Delplace},\ and\ \citenamefont {Amo}}]{ElSokhen2024}%
  \BibitemOpen
  \bibfield  {author} {\bibinfo {author} {\bibfnamefont {R.}~\bibnamefont {El~Sokhen}}, \bibinfo {author} {\bibfnamefont {A.}~\bibnamefont {G\'omez-Le\'on}}, \bibinfo {author} {\bibfnamefont {A.~F.}\ \bibnamefont {Adiyatullin}}, \bibinfo {author} {\bibfnamefont {S.}~\bibnamefont {Randoux}}, \bibinfo {author} {\bibfnamefont {P.}~\bibnamefont {Delplace}},\ and\ \bibinfo {author} {\bibfnamefont {A.}~\bibnamefont {Amo}},\ }\bibfield  {title} {\bibinfo {title} {{Edge-dependent anomalous topology in synthetic photonic lattices subject to discrete step walks}},\ }\href {https://doi.org/10.1103/PhysRevResearch.6.023282} {\bibfield  {journal} {\bibinfo  {journal} {Phys. Rev. Res.}\ }\textbf {\bibinfo {volume} {6}},\ \bibinfo {pages} {023282} (\bibinfo {year} {2024})}\BibitemShut {NoStop}%
\bibitem [{\citenamefont {Thouless}\ \emph {et~al.}(1982)\citenamefont {Thouless}, \citenamefont {Kohmoto}, \citenamefont {Nightingale},\ and\ \citenamefont {den Nijs}}]{Thouless1982}%
  \BibitemOpen
  \bibfield  {author} {\bibinfo {author} {\bibfnamefont {D.~J.}\ \bibnamefont {Thouless}}, \bibinfo {author} {\bibfnamefont {M.}~\bibnamefont {Kohmoto}}, \bibinfo {author} {\bibfnamefont {M.~P.}\ \bibnamefont {Nightingale}},\ and\ \bibinfo {author} {\bibfnamefont {M.}~\bibnamefont {den Nijs}},\ }\bibfield  {title} {\bibinfo {title} {{Quantized Hall Conductance in a Two-Dimensional Periodic Potential}},\ }\href {https://doi.org/10.1103/PhysRevLett.49.405} {\bibfield  {journal} {\bibinfo  {journal} {Phys. Rev. Lett.}\ }\textbf {\bibinfo {volume} {49}},\ \bibinfo {pages} {405} (\bibinfo {year} {1982})}\BibitemShut {NoStop}%
\bibitem [{\citenamefont {Asb\'oth}\ \emph {et~al.}(2014)\citenamefont {Asb\'oth}, \citenamefont {Tarasinski},\ and\ \citenamefont {Delplace}}]{Asboth2014}%
  \BibitemOpen
  \bibfield  {author} {\bibinfo {author} {\bibfnamefont {J.~K.}\ \bibnamefont {Asb\'oth}}, \bibinfo {author} {\bibfnamefont {B.}~\bibnamefont {Tarasinski}},\ and\ \bibinfo {author} {\bibfnamefont {P.}~\bibnamefont {Delplace}},\ }\bibfield  {title} {\bibinfo {title} {Chiral symmetry and bulk-boundary correspondence in periodically driven one-dimensional systems},\ }\href {https://doi.org/10.1103/PhysRevB.90.125143} {\bibfield  {journal} {\bibinfo  {journal} {Phys. Rev. B}\ }\textbf {\bibinfo {volume} {90}},\ \bibinfo {pages} {125143} (\bibinfo {year} {2014})}\BibitemShut {NoStop}%
\bibitem [{\citenamefont {Nathan}\ and\ \citenamefont {Rudner}(2015)}]{Nathan2015}%
  \BibitemOpen
  \bibfield  {author} {\bibinfo {author} {\bibfnamefont {F.}~\bibnamefont {Nathan}}\ and\ \bibinfo {author} {\bibfnamefont {M.~S.}\ \bibnamefont {Rudner}},\ }\bibfield  {title} {\bibinfo {title} {{Topological singularities and the general classification of Floquet–Bloch systems}},\ }\href {https://doi.org/10.1088/1367-2630/17/12/125014} {\bibfield  {journal} {\bibinfo  {journal} {New Journal of Physics}\ }\textbf {\bibinfo {volume} {17}},\ \bibinfo {pages} {125014} (\bibinfo {year} {2015})}\BibitemShut {NoStop}%
\bibitem [{\citenamefont {Roy}\ and\ \citenamefont {Harper}(2017)}]{Roy2017}%
  \BibitemOpen
  \bibfield  {author} {\bibinfo {author} {\bibfnamefont {R.}~\bibnamefont {Roy}}\ and\ \bibinfo {author} {\bibfnamefont {F.}~\bibnamefont {Harper}},\ }\bibfield  {title} {\bibinfo {title} {{Periodic table for Floquet topological insulators}},\ }\href {https://doi.org/10.1103/PhysRevB.96.155118} {\bibfield  {journal} {\bibinfo  {journal} {Phys. Rev. B}\ }\textbf {\bibinfo {volume} {96}},\ \bibinfo {pages} {155118} (\bibinfo {year} {2017})}\BibitemShut {NoStop}%
\bibitem [{\citenamefont {Vu}(2022)}]{Vu2022}%
  \BibitemOpen
  \bibfield  {author} {\bibinfo {author} {\bibfnamefont {D.}~\bibnamefont {Vu}},\ }\bibfield  {title} {\bibinfo {title} {{Dynamic bulk-boundary correspondence for anomalous Floquet topology}},\ }\href {https://doi.org/10.1103/PhysRevB.105.064304} {\bibfield  {journal} {\bibinfo  {journal} {Phys. Rev. B}\ }\textbf {\bibinfo {volume} {105}},\ \bibinfo {pages} {064304} (\bibinfo {year} {2022})}\BibitemShut {NoStop}%
\bibitem [{\citenamefont {D'Alessio}\ and\ \citenamefont {Rigol}(2014)}]{DAlessio2014}%
  \BibitemOpen
  \bibfield  {author} {\bibinfo {author} {\bibfnamefont {L.}~\bibnamefont {D'Alessio}}\ and\ \bibinfo {author} {\bibfnamefont {M.}~\bibnamefont {Rigol}},\ }\bibfield  {title} {\bibinfo {title} {{Long-time Behavior of Isolated Periodically Driven Interacting Lattice Systems}},\ }\href {https://doi.org/10.1103/PhysRevX.4.041048} {\bibfield  {journal} {\bibinfo  {journal} {Phys. Rev. X}\ }\textbf {\bibinfo {volume} {4}},\ \bibinfo {pages} {041048} (\bibinfo {year} {2014})}\BibitemShut {NoStop}%
\bibitem [{\citenamefont {Lazarides}\ \emph {et~al.}(2014)\citenamefont {Lazarides}, \citenamefont {Das},\ and\ \citenamefont {Moessner}}]{Lazarides2014}%
  \BibitemOpen
  \bibfield  {author} {\bibinfo {author} {\bibfnamefont {A.}~\bibnamefont {Lazarides}}, \bibinfo {author} {\bibfnamefont {A.}~\bibnamefont {Das}},\ and\ \bibinfo {author} {\bibfnamefont {R.}~\bibnamefont {Moessner}},\ }\bibfield  {title} {\bibinfo {title} {{Equilibrium states of generic quantum systems subject to periodic driving}},\ }\href {https://doi.org/10.1103/PhysRevE.90.012110} {\bibfield  {journal} {\bibinfo  {journal} {Phys. Rev. E}\ }\textbf {\bibinfo {volume} {90}},\ \bibinfo {pages} {012110} (\bibinfo {year} {2014})}\BibitemShut {NoStop}%
\bibitem [{\citenamefont {Seetharam}\ \emph {et~al.}(2015)\citenamefont {Seetharam}, \citenamefont {Bardyn}, \citenamefont {Lindner}, \citenamefont {Rudner},\ and\ \citenamefont {Refael}}]{Seetharam2015}%
  \BibitemOpen
  \bibfield  {author} {\bibinfo {author} {\bibfnamefont {K.~I.}\ \bibnamefont {Seetharam}}, \bibinfo {author} {\bibfnamefont {C.-E.}\ \bibnamefont {Bardyn}}, \bibinfo {author} {\bibfnamefont {N.~H.}\ \bibnamefont {Lindner}}, \bibinfo {author} {\bibfnamefont {M.~S.}\ \bibnamefont {Rudner}},\ and\ \bibinfo {author} {\bibfnamefont {G.}~\bibnamefont {Refael}},\ }\bibfield  {title} {\bibinfo {title} {{Controlled Population of Floquet-Bloch States via Coupling to Bose and Fermi Baths}},\ }\href {https://doi.org/10.1103/PhysRevX.5.041050} {\bibfield  {journal} {\bibinfo  {journal} {Phys. Rev. X}\ }\textbf {\bibinfo {volume} {5}},\ \bibinfo {pages} {041050} (\bibinfo {year} {2015})}\BibitemShut {NoStop}%
\bibitem [{\citenamefont {Iadecola}\ \emph {et~al.}(2015)\citenamefont {Iadecola}, \citenamefont {Neupert},\ and\ \citenamefont {Chamon}}]{Iadecola2015}%
  \BibitemOpen
  \bibfield  {author} {\bibinfo {author} {\bibfnamefont {T.}~\bibnamefont {Iadecola}}, \bibinfo {author} {\bibfnamefont {T.}~\bibnamefont {Neupert}},\ and\ \bibinfo {author} {\bibfnamefont {C.}~\bibnamefont {Chamon}},\ }\bibfield  {title} {\bibinfo {title} {{Occupation of topological Floquet bands in open systems}},\ }\href {https://doi.org/10.1103/PhysRevB.91.235133} {\bibfield  {journal} {\bibinfo  {journal} {Phys. Rev. B}\ }\textbf {\bibinfo {volume} {91}},\ \bibinfo {pages} {235133} (\bibinfo {year} {2015})}\BibitemShut {NoStop}%
\bibitem [{\citenamefont {Esin}\ \emph {et~al.}(2018)\citenamefont {Esin}, \citenamefont {Rudner}, \citenamefont {Refael},\ and\ \citenamefont {Lindner}}]{Esin2018}%
  \BibitemOpen
  \bibfield  {author} {\bibinfo {author} {\bibfnamefont {I.}~\bibnamefont {Esin}}, \bibinfo {author} {\bibfnamefont {M.~S.}\ \bibnamefont {Rudner}}, \bibinfo {author} {\bibfnamefont {G.}~\bibnamefont {Refael}},\ and\ \bibinfo {author} {\bibfnamefont {N.~H.}\ \bibnamefont {Lindner}},\ }\bibfield  {title} {\bibinfo {title} {{Quantized transport and steady states of Floquet topological insulators}},\ }\href {https://doi.org/10.1103/PhysRevB.97.245401} {\bibfield  {journal} {\bibinfo  {journal} {Phys. Rev. B}\ }\textbf {\bibinfo {volume} {97}},\ \bibinfo {pages} {245401} (\bibinfo {year} {2018})}\BibitemShut {NoStop}%
\bibitem [{\citenamefont {\"Unal}\ \emph {et~al.}(2019)\citenamefont {\"Unal}, \citenamefont {Seradjeh},\ and\ \citenamefont {Eckardt}}]{Unal2019}%
  \BibitemOpen
  \bibfield  {author} {\bibinfo {author} {\bibfnamefont {F.~N.}\ \bibnamefont {\"Unal}}, \bibinfo {author} {\bibfnamefont {B.}~\bibnamefont {Seradjeh}},\ and\ \bibinfo {author} {\bibfnamefont {A.}~\bibnamefont {Eckardt}},\ }\bibfield  {title} {\bibinfo {title} {{How to Directly Measure Floquet Topological Invariants in Optical Lattices}},\ }\href {https://doi.org/10.1103/PhysRevLett.122.253601} {\bibfield  {journal} {\bibinfo  {journal} {Phys. Rev. Lett.}\ }\textbf {\bibinfo {volume} {122}},\ \bibinfo {pages} {253601} (\bibinfo {year} {2019})}\BibitemShut {NoStop}%
\bibitem [{\citenamefont {Gu}\ \emph {et~al.}(2011)\citenamefont {Gu}, \citenamefont {Fertig}, \citenamefont {Arovas},\ and\ \citenamefont {Auerbach}}]{Gu2011}%
  \BibitemOpen
  \bibfield  {author} {\bibinfo {author} {\bibfnamefont {Z.}~\bibnamefont {Gu}}, \bibinfo {author} {\bibfnamefont {H.~A.}\ \bibnamefont {Fertig}}, \bibinfo {author} {\bibfnamefont {D.~P.}\ \bibnamefont {Arovas}},\ and\ \bibinfo {author} {\bibfnamefont {A.}~\bibnamefont {Auerbach}},\ }\bibfield  {title} {\bibinfo {title} {Floquet spectrum and transport through an irradiated graphene ribbon},\ }\href {https://doi.org/10.1103/PhysRevLett.107.216601} {\bibfield  {journal} {\bibinfo  {journal} {Phys. Rev. Lett.}\ }\textbf {\bibinfo {volume} {107}},\ \bibinfo {pages} {216601} (\bibinfo {year} {2011})}\BibitemShut {NoStop}%
\bibitem [{\citenamefont {Kundu}\ and\ \citenamefont {Seradjeh}(2013)}]{Kundu2013}%
  \BibitemOpen
  \bibfield  {author} {\bibinfo {author} {\bibfnamefont {A.}~\bibnamefont {Kundu}}\ and\ \bibinfo {author} {\bibfnamefont {B.}~\bibnamefont {Seradjeh}},\ }\bibfield  {title} {\bibinfo {title} {{Transport Signatures of Floquet Majorana Fermions in Driven Topological Superconductors}},\ }\href {https://doi.org/10.1103/PhysRevLett.111.136402} {\bibfield  {journal} {\bibinfo  {journal} {Phys. Rev. Lett.}\ }\textbf {\bibinfo {volume} {111}},\ \bibinfo {pages} {136402} (\bibinfo {year} {2013})}\BibitemShut {NoStop}%
\bibitem [{\citenamefont {Perez-Piskunow}\ \emph {et~al.}(2014)\citenamefont {Perez-Piskunow}, \citenamefont {Usaj}, \citenamefont {Balseiro},\ and\ \citenamefont {Torres}}]{Piskunow2014}%
  \BibitemOpen
  \bibfield  {author} {\bibinfo {author} {\bibfnamefont {P.~M.}\ \bibnamefont {Perez-Piskunow}}, \bibinfo {author} {\bibfnamefont {G.}~\bibnamefont {Usaj}}, \bibinfo {author} {\bibfnamefont {C.~A.}\ \bibnamefont {Balseiro}},\ and\ \bibinfo {author} {\bibfnamefont {L.~E. F.~F.}\ \bibnamefont {Torres}},\ }\bibfield  {title} {\bibinfo {title} {Floquet chiral edge states in graphene},\ }\href {https://doi.org/10.1103/PhysRevB.89.121401} {\bibfield  {journal} {\bibinfo  {journal} {Phys. Rev. B}\ }\textbf {\bibinfo {volume} {89}},\ \bibinfo {pages} {121401} (\bibinfo {year} {2014})}\BibitemShut {NoStop}%
\bibitem [{\citenamefont {Foa~Torres}\ \emph {et~al.}(2014)\citenamefont {Foa~Torres}, \citenamefont {Perez-Piskunow}, \citenamefont {Balseiro},\ and\ \citenamefont {Usaj}}]{FoaTorres2014}%
  \BibitemOpen
  \bibfield  {author} {\bibinfo {author} {\bibfnamefont {L.~E.~F.}\ \bibnamefont {Foa~Torres}}, \bibinfo {author} {\bibfnamefont {P.~M.}\ \bibnamefont {Perez-Piskunow}}, \bibinfo {author} {\bibfnamefont {C.~A.}\ \bibnamefont {Balseiro}},\ and\ \bibinfo {author} {\bibfnamefont {G.}~\bibnamefont {Usaj}},\ }\bibfield  {title} {\bibinfo {title} {Multiterminal conductance of a floquet topological insulator},\ }\href {https://doi.org/10.1103/PhysRevLett.113.266801} {\bibfield  {journal} {\bibinfo  {journal} {Phys. Rev. Lett.}\ }\textbf {\bibinfo {volume} {113}},\ \bibinfo {pages} {266801} (\bibinfo {year} {2014})}\BibitemShut {NoStop}%
\bibitem [{\citenamefont {Usaj}\ \emph {et~al.}(2014)\citenamefont {Usaj}, \citenamefont {Perez-Piskunow}, \citenamefont {Foa~Torres},\ and\ \citenamefont {Balseiro}}]{Usaj2014}%
  \BibitemOpen
  \bibfield  {author} {\bibinfo {author} {\bibfnamefont {G.}~\bibnamefont {Usaj}}, \bibinfo {author} {\bibfnamefont {P.~M.}\ \bibnamefont {Perez-Piskunow}}, \bibinfo {author} {\bibfnamefont {L.~E.~F.}\ \bibnamefont {Foa~Torres}},\ and\ \bibinfo {author} {\bibfnamefont {C.~A.}\ \bibnamefont {Balseiro}},\ }\bibfield  {title} {\bibinfo {title} {{Irradiated graphene as a tunable Floquet topological insulator}},\ }\href {https://doi.org/10.1103/PhysRevB.90.115423} {\bibfield  {journal} {\bibinfo  {journal} {Phys. Rev. B}\ }\textbf {\bibinfo {volume} {90}},\ \bibinfo {pages} {115423} (\bibinfo {year} {2014})}\BibitemShut {NoStop}%
\bibitem [{\citenamefont {Farrell}\ and\ \citenamefont {Pereg-Barnea}(2015)}]{Farrell2015}%
  \BibitemOpen
  \bibfield  {author} {\bibinfo {author} {\bibfnamefont {A.}~\bibnamefont {Farrell}}\ and\ \bibinfo {author} {\bibfnamefont {T.}~\bibnamefont {Pereg-Barnea}},\ }\bibfield  {title} {\bibinfo {title} {{Photon-Inhibited Topological Transport in Quantum Well Heterostructures}},\ }\href {https://doi.org/10.1103/PhysRevLett.115.106403} {\bibfield  {journal} {\bibinfo  {journal} {Phys. Rev. Lett.}\ }\textbf {\bibinfo {volume} {115}},\ \bibinfo {pages} {106403} (\bibinfo {year} {2015})}\BibitemShut {NoStop}%
\bibitem [{\citenamefont {Farrell}\ and\ \citenamefont {Pereg-Barnea}(2016)}]{Farrell2016}%
  \BibitemOpen
  \bibfield  {author} {\bibinfo {author} {\bibfnamefont {A.}~\bibnamefont {Farrell}}\ and\ \bibinfo {author} {\bibfnamefont {T.}~\bibnamefont {Pereg-Barnea}},\ }\bibfield  {title} {\bibinfo {title} {{Edge-state transport in Floquet topological insulators}},\ }\href {https://doi.org/10.1103/PhysRevB.93.045121} {\bibfield  {journal} {\bibinfo  {journal} {Phys. Rev. B}\ }\textbf {\bibinfo {volume} {93}},\ \bibinfo {pages} {045121} (\bibinfo {year} {2016})}\BibitemShut {NoStop}%
\bibitem [{\citenamefont {Kundu}\ \emph {et~al.}(2020)\citenamefont {Kundu}, \citenamefont {Rudner}, \citenamefont {Berg},\ and\ \citenamefont {Lindner}}]{Kundu2020}%
  \BibitemOpen
  \bibfield  {author} {\bibinfo {author} {\bibfnamefont {A.}~\bibnamefont {Kundu}}, \bibinfo {author} {\bibfnamefont {M.}~\bibnamefont {Rudner}}, \bibinfo {author} {\bibfnamefont {E.}~\bibnamefont {Berg}},\ and\ \bibinfo {author} {\bibfnamefont {N.~H.}\ \bibnamefont {Lindner}},\ }\bibfield  {title} {\bibinfo {title} {{Quantized large-bias current in the anomalous Floquet-Anderson insulator}},\ }\href {https://doi.org/10.1103/PhysRevB.101.041403} {\bibfield  {journal} {\bibinfo  {journal} {Phys. Rev. B}\ }\textbf {\bibinfo {volume} {101}},\ \bibinfo {pages} {041403} (\bibinfo {year} {2020})}\BibitemShut {NoStop}%
\bibitem [{\citenamefont {Zhang}\ \emph {et~al.}(2024{\natexlab{a}})\citenamefont {Zhang}, \citenamefont {Nathan}, \citenamefont {Lindner},\ and\ \citenamefont {Rudner}}]{Zhang2024_b}%
  \BibitemOpen
  \bibfield  {author} {\bibinfo {author} {\bibfnamefont {R.}~\bibnamefont {Zhang}}, \bibinfo {author} {\bibfnamefont {F.}~\bibnamefont {Nathan}}, \bibinfo {author} {\bibfnamefont {N.~H.}\ \bibnamefont {Lindner}},\ and\ \bibinfo {author} {\bibfnamefont {M.~S.}\ \bibnamefont {Rudner}},\ }\bibfield  {title} {\bibinfo {title} {{Achieving quantized transport in Floquet topological insulators via energy filters}},\ }\href {https://doi.org/10.1103/PhysRevB.110.075428} {\bibfield  {journal} {\bibinfo  {journal} {Phys. Rev. B}\ }\textbf {\bibinfo {volume} {110}},\ \bibinfo {pages} {075428} (\bibinfo {year} {2024}{\natexlab{a}})}\BibitemShut {NoStop}%
\bibitem [{\citenamefont {Dehghani}\ \emph {et~al.}(2015)\citenamefont {Dehghani}, \citenamefont {Oka},\ and\ \citenamefont {Mitra}}]{Dehghani2015}%
  \BibitemOpen
  \bibfield  {author} {\bibinfo {author} {\bibfnamefont {H.}~\bibnamefont {Dehghani}}, \bibinfo {author} {\bibfnamefont {T.}~\bibnamefont {Oka}},\ and\ \bibinfo {author} {\bibfnamefont {A.}~\bibnamefont {Mitra}},\ }\bibfield  {title} {\bibinfo {title} {{Out-of-equilibrium electrons and the Hall conductance of a Floquet topological insulator}},\ }\href {https://doi.org/10.1103/PhysRevB.91.155422} {\bibfield  {journal} {\bibinfo  {journal} {Phys. Rev. B}\ }\textbf {\bibinfo {volume} {91}},\ \bibinfo {pages} {155422} (\bibinfo {year} {2015})}\BibitemShut {NoStop}%
\bibitem [{\citenamefont {Dauphin}\ \emph {et~al.}(2017)\citenamefont {Dauphin}, \citenamefont {Tran}, \citenamefont {Lewenstein},\ and\ \citenamefont {Goldman}}]{Dauphin2017}%
  \BibitemOpen
  \bibfield  {author} {\bibinfo {author} {\bibfnamefont {A.}~\bibnamefont {Dauphin}}, \bibinfo {author} {\bibfnamefont {D.-T.}\ \bibnamefont {Tran}}, \bibinfo {author} {\bibfnamefont {M.}~\bibnamefont {Lewenstein}},\ and\ \bibinfo {author} {\bibfnamefont {N.}~\bibnamefont {Goldman}},\ }\bibfield  {title} {\bibinfo {title} {{Loading ultracold gases in topological Floquet bands: the fate of current and center-of-mass responses}},\ }\href {https://doi.org/10.1088/2053-1583/aa6a3b} {\bibfield  {journal} {\bibinfo  {journal} {2D Materials}\ }\textbf {\bibinfo {volume} {4}},\ \bibinfo {pages} {024010} (\bibinfo {year} {2017})}\BibitemShut {NoStop}%
\bibitem [{\citenamefont {Peralta~Gavensky}\ \emph {et~al.}(2018)\citenamefont {Peralta~Gavensky}, \citenamefont {Usaj},\ and\ \citenamefont {Balseiro}}]{PeraltaGavensky2018}%
  \BibitemOpen
  \bibfield  {author} {\bibinfo {author} {\bibfnamefont {L.}~\bibnamefont {Peralta~Gavensky}}, \bibinfo {author} {\bibfnamefont {G.}~\bibnamefont {Usaj}},\ and\ \bibinfo {author} {\bibfnamefont {C.~A.}\ \bibnamefont {Balseiro}},\ }\bibfield  {title} {\bibinfo {title} {{Time-resolved Hall conductivity of pulse-driven topological quantum systems}},\ }\href {https://doi.org/10.1103/PhysRevB.98.165414} {\bibfield  {journal} {\bibinfo  {journal} {Phys. Rev. B}\ }\textbf {\bibinfo {volume} {98}},\ \bibinfo {pages} {165414} (\bibinfo {year} {2018})}\BibitemShut {NoStop}%
\bibitem [{\citenamefont {Titum}\ \emph {et~al.}(2016)\citenamefont {Titum}, \citenamefont {Berg}, \citenamefont {Rudner}, \citenamefont {Refael},\ and\ \citenamefont {Lindner}}]{Titum2016}%
  \BibitemOpen
  \bibfield  {author} {\bibinfo {author} {\bibfnamefont {P.}~\bibnamefont {Titum}}, \bibinfo {author} {\bibfnamefont {E.}~\bibnamefont {Berg}}, \bibinfo {author} {\bibfnamefont {M.~S.}\ \bibnamefont {Rudner}}, \bibinfo {author} {\bibfnamefont {G.}~\bibnamefont {Refael}},\ and\ \bibinfo {author} {\bibfnamefont {N.~H.}\ \bibnamefont {Lindner}},\ }\bibfield  {title} {\bibinfo {title} {{Anomalous Floquet-Anderson Insulator as a Nonadiabatic Quantized Charge Pump}},\ }\href {https://doi.org/10.1103/PhysRevX.6.021013} {\bibfield  {journal} {\bibinfo  {journal} {Phys. Rev. X}\ }\textbf {\bibinfo {volume} {6}},\ \bibinfo {pages} {021013} (\bibinfo {year} {2016})}\BibitemShut {NoStop}%
\bibitem [{\citenamefont {Nathan}\ \emph {et~al.}(2019)\citenamefont {Nathan}, \citenamefont {Abanin}, \citenamefont {Berg}, \citenamefont {Lindner},\ and\ \citenamefont {Rudner}}]{Nathan2019}%
  \BibitemOpen
  \bibfield  {author} {\bibinfo {author} {\bibfnamefont {F.}~\bibnamefont {Nathan}}, \bibinfo {author} {\bibfnamefont {D.}~\bibnamefont {Abanin}}, \bibinfo {author} {\bibfnamefont {E.}~\bibnamefont {Berg}}, \bibinfo {author} {\bibfnamefont {N.~H.}\ \bibnamefont {Lindner}},\ and\ \bibinfo {author} {\bibfnamefont {M.~S.}\ \bibnamefont {Rudner}},\ }\bibfield  {title} {\bibinfo {title} {{Anomalous Floquet insulators}},\ }\href {https://doi.org/10.1103/PhysRevB.99.195133} {\bibfield  {journal} {\bibinfo  {journal} {Phys. Rev. B}\ }\textbf {\bibinfo {volume} {99}},\ \bibinfo {pages} {195133} (\bibinfo {year} {2019})}\BibitemShut {NoStop}%
\bibitem [{\citenamefont {Nathan}\ \emph {et~al.}(2017)\citenamefont {Nathan}, \citenamefont {Rudner}, \citenamefont {Lindner}, \citenamefont {Berg},\ and\ \citenamefont {Refael}}]{Nathan2017}%
  \BibitemOpen
  \bibfield  {author} {\bibinfo {author} {\bibfnamefont {F.}~\bibnamefont {Nathan}}, \bibinfo {author} {\bibfnamefont {M.~S.}\ \bibnamefont {Rudner}}, \bibinfo {author} {\bibfnamefont {N.~H.}\ \bibnamefont {Lindner}}, \bibinfo {author} {\bibfnamefont {E.}~\bibnamefont {Berg}},\ and\ \bibinfo {author} {\bibfnamefont {G.}~\bibnamefont {Refael}},\ }\bibfield  {title} {\bibinfo {title} {{Quantized Magnetization Density in Periodically Driven Systems}},\ }\href {https://doi.org/10.1103/PhysRevLett.119.186801} {\bibfield  {journal} {\bibinfo  {journal} {Phys. Rev. Lett.}\ }\textbf {\bibinfo {volume} {119}},\ \bibinfo {pages} {186801} (\bibinfo {year} {2017})}\BibitemShut {NoStop}%
\bibitem [{\citenamefont {Nathan}\ \emph {et~al.}(2021{\natexlab{a}})\citenamefont {Nathan}, \citenamefont {Abanin}, \citenamefont {Lindner}, \citenamefont {Berg},\ and\ \citenamefont {Rudner}}]{Nathan2021}%
  \BibitemOpen
  \bibfield  {author} {\bibinfo {author} {\bibfnamefont {F.}~\bibnamefont {Nathan}}, \bibinfo {author} {\bibfnamefont {D.~A.}\ \bibnamefont {Abanin}}, \bibinfo {author} {\bibfnamefont {N.~H.}\ \bibnamefont {Lindner}}, \bibinfo {author} {\bibfnamefont {E.}~\bibnamefont {Berg}},\ and\ \bibinfo {author} {\bibfnamefont {M.~S.}\ \bibnamefont {Rudner}},\ }\bibfield  {title} {\bibinfo {title} {{Hierarchy of many-body invariants and quantized magnetization in anomalous Floquet insulators}},\ }\href {https://doi.org/10.21468/SciPostPhys.10.6.128} {\bibfield  {journal} {\bibinfo  {journal} {SciPost Phys.}\ }\textbf {\bibinfo {volume} {10}},\ \bibinfo {pages} {128} (\bibinfo {year} {2021}{\natexlab{a}})}\BibitemShut {NoStop}%
\bibitem [{\citenamefont {Glorioso}\ \emph {et~al.}(2021)\citenamefont {Glorioso}, \citenamefont {Gromov},\ and\ \citenamefont {Ryu}}]{Glorioso2021}%
  \BibitemOpen
  \bibfield  {author} {\bibinfo {author} {\bibfnamefont {P.}~\bibnamefont {Glorioso}}, \bibinfo {author} {\bibfnamefont {A.}~\bibnamefont {Gromov}},\ and\ \bibinfo {author} {\bibfnamefont {S.}~\bibnamefont {Ryu}},\ }\bibfield  {title} {\bibinfo {title} {{Effective response theory for Floquet topological systems}},\ }\href {https://doi.org/10.1103/PhysRevResearch.3.013117} {\bibfield  {journal} {\bibinfo  {journal} {Phys. Rev. Res.}\ }\textbf {\bibinfo {volume} {3}},\ \bibinfo {pages} {013117} (\bibinfo {year} {2021})}\BibitemShut {NoStop}%
\bibitem [{\citenamefont {Delplace}(2022)}]{Delplace_flow}%
  \BibitemOpen
  \bibfield  {author} {\bibinfo {author} {\bibfnamefont {P.}~\bibnamefont {Delplace}},\ }\bibfield  {title} {\bibinfo {title} {{Berry-Chern monopoles and spectral flows}},\ }\href {https://doi.org/10.21468/SciPostPhysLectNotes.39} {\bibfield  {journal} {\bibinfo  {journal} {SciPost Phys. Lect. Notes}\ ,\ \bibinfo {pages} {39}} (\bibinfo {year} {2022})}\BibitemShut {NoStop}%
\bibitem [{\citenamefont {Streda}(1982)}]{Streda1982}%
  \BibitemOpen
  \bibfield  {author} {\bibinfo {author} {\bibfnamefont {P.}~\bibnamefont {Streda}},\ }\bibfield  {title} {\bibinfo {title} {Theory of quantised {Hall} conductivity in two dimensions},\ }\href {https://doi.org/10.1088/0022-3719/15/22/005} {\bibfield  {journal} {\bibinfo  {journal} {Journal of Physics C: Solid State Physics}\ }\textbf {\bibinfo {volume} {15}},\ \bibinfo {pages} {L717} (\bibinfo {year} {1982})}\BibitemShut {NoStop}%
\bibitem [{\citenamefont {Widom}(1982)}]{Widom1982}%
  \BibitemOpen
  \bibfield  {author} {\bibinfo {author} {\bibfnamefont {A.}~\bibnamefont {Widom}},\ }\bibfield  {title} {\bibinfo {title} {Thermodynamic derivation of the {Hall} effect current},\ }\href {https://doi.org/https://doi.org/10.1016/0375-9601(82)90401-7} {\bibfield  {journal} {\bibinfo  {journal} {Physics Letters A}\ }\textbf {\bibinfo {volume} {90}},\ \bibinfo {pages} {474} (\bibinfo {year} {1982})}\BibitemShut {NoStop}%
\bibitem [{\citenamefont {Streda}\ and\ \citenamefont {Smrcka}(1983)}]{Streda1983}%
  \BibitemOpen
  \bibfield  {author} {\bibinfo {author} {\bibfnamefont {P.}~\bibnamefont {Streda}}\ and\ \bibinfo {author} {\bibfnamefont {L.}~\bibnamefont {Smrcka}},\ }\bibfield  {title} {\bibinfo {title} {{Thermodynamic derivation of the Hall current and the thermopower in quantising magnetic field}},\ }\href {https://doi.org/10.1088/0022-3719/16/24/005} {\bibfield  {journal} {\bibinfo  {journal} {Journal of Physics C: Solid State Physics}\ }\textbf {\bibinfo {volume} {16}},\ \bibinfo {pages} {L895} (\bibinfo {year} {1983})}\BibitemShut {NoStop}%
\bibitem [{\citenamefont {Prodan}\ and\ \citenamefont {Schulz-Baldes}(2016)}]{Prodan2016}%
  \BibitemOpen
  \bibfield  {author} {\bibinfo {author} {\bibfnamefont {E.}~\bibnamefont {Prodan}}\ and\ \bibinfo {author} {\bibfnamefont {H.}~\bibnamefont {Schulz-Baldes}},\ }\href {https://doi.org/10.1007/978-3-319-29351-6} {\emph {\bibinfo {title} {Bulk and Boundary Invariants for Complex Topological Insulators: From K-Theory to Physics}}}\ (\bibinfo  {publisher} {Springer International Publishing},\ \bibinfo {year} {2016})\BibitemShut {NoStop}%
\bibitem [{\citenamefont {MacDonald}(1989)}]{MacDonald1989}%
  \BibitemOpen
  \bibinfo {editor} {\bibfnamefont {A.~H.}\ \bibnamefont {MacDonald}},\ ed.,\ \href@noop {} {\emph {\bibinfo {title} {Quantum Hall Effect: A Perspective}}},\ Perspectives in Condensed Matter Physics\ (\bibinfo  {publisher} {Springer},\ \bibinfo {address} {Dordrecht, Netherlands},\ \bibinfo {year} {1989})\BibitemShut {NoStop}%
\bibitem [{\citenamefont {Yacoby}\ \emph {et~al.}(1999)\citenamefont {Yacoby}, \citenamefont {Hess}, \citenamefont {Fulton}, \citenamefont {Pfeiffer},\ and\ \citenamefont {West}}]{Yacoby1999}%
  \BibitemOpen
  \bibfield  {author} {\bibinfo {author} {\bibfnamefont {A.}~\bibnamefont {Yacoby}}, \bibinfo {author} {\bibfnamefont {H.}~\bibnamefont {Hess}}, \bibinfo {author} {\bibfnamefont {T.}~\bibnamefont {Fulton}}, \bibinfo {author} {\bibfnamefont {L.}~\bibnamefont {Pfeiffer}},\ and\ \bibinfo {author} {\bibfnamefont {K.}~\bibnamefont {West}},\ }\bibfield  {title} {\bibinfo {title} {{Electrical imaging of the quantum Hall state}},\ }\href {https://doi.org/10.1016/s0038-1098(99)00139-8} {\bibfield  {journal} {\bibinfo  {journal} {Solid State Communications}\ }\textbf {\bibinfo {volume} {111}},\ \bibinfo {pages} {1–13} (\bibinfo {year} {1999})}\BibitemShut {NoStop}%
\bibitem [{\citenamefont {{Umucal\ifmmode \imath \else \i \fi{}lar, R. O. and Zhai, Hui and Oktel, M. \"O.}}(2008)}]{Umucalilar2008}%
  \BibitemOpen
  \bibfield  {author} {\bibinfo {author} {\bibnamefont {{Umucal\ifmmode \imath \else \i \fi{}lar, R. O. and Zhai, Hui and Oktel, M. \"O.}}},\ }\bibfield  {title} {\bibinfo {title} {{Trapped Fermi Gases in Rotating Optical Lattices: Realization and Detection of the Topological Hofstadter Insulator}},\ }\href {https://doi.org/10.1103/PhysRevLett.100.070402} {\bibfield  {journal} {\bibinfo  {journal} {Phys. Rev. Lett.}\ }\textbf {\bibinfo {volume} {100}},\ \bibinfo {pages} {070402} (\bibinfo {year} {2008})}\BibitemShut {NoStop}%
\bibitem [{\citenamefont {Repellin}\ \emph {et~al.}(2020)\citenamefont {Repellin}, \citenamefont {L\'eonard},\ and\ \citenamefont {Goldman}}]{Repellin2020}%
  \BibitemOpen
  \bibfield  {author} {\bibinfo {author} {\bibfnamefont {C.}~\bibnamefont {Repellin}}, \bibinfo {author} {\bibfnamefont {J.}~\bibnamefont {L\'eonard}},\ and\ \bibinfo {author} {\bibfnamefont {N.}~\bibnamefont {Goldman}},\ }\bibfield  {title} {\bibinfo {title} {{Fractional Chern insulators of few bosons in a box: Hall plateaus from center-of-mass drifts and density profiles}},\ }\href {https://doi.org/10.1103/PhysRevA.102.063316} {\bibfield  {journal} {\bibinfo  {journal} {Phys. Rev. A}\ }\textbf {\bibinfo {volume} {102}},\ \bibinfo {pages} {063316} (\bibinfo {year} {2020})}\BibitemShut {NoStop}%
\bibitem [{\citenamefont {Jamotte}\ \emph {et~al.}(2023)\citenamefont {Jamotte}, \citenamefont {Peralta~Gavensky}, \citenamefont {Morais~Smith}, \citenamefont {Di~Liberto},\ and\ \citenamefont {Goldman}}]{Jamotte2023}%
  \BibitemOpen
  \bibfield  {author} {\bibinfo {author} {\bibfnamefont {M.}~\bibnamefont {Jamotte}}, \bibinfo {author} {\bibfnamefont {L.}~\bibnamefont {Peralta~Gavensky}}, \bibinfo {author} {\bibfnamefont {C.}~\bibnamefont {Morais~Smith}}, \bibinfo {author} {\bibfnamefont {M.}~\bibnamefont {Di~Liberto}},\ and\ \bibinfo {author} {\bibfnamefont {N.}~\bibnamefont {Goldman}},\ }\bibfield  {title} {\bibinfo {title} {{Quantized valley Hall response from local bulk density variations}},\ }\href {https://doi.org/10.1038/s42005-023-01377-9} {\bibfield  {journal} {\bibinfo  {journal} {Communications Physics}\ }\textbf {\bibinfo {volume} {6}},\ \bibinfo {pages} {264} (\bibinfo {year} {2023})}\BibitemShut {NoStop}%
\bibitem [{\citenamefont {Xie}\ \emph {et~al.}(2021)\citenamefont {Xie}, \citenamefont {Pierce}, \citenamefont {Park}, \citenamefont {Parker}, \citenamefont {Khalaf}, \citenamefont {Ledwith}, \citenamefont {Cao}, \citenamefont {Lee}, \citenamefont {Chen}, \citenamefont {Forrester}, \citenamefont {Watanabe}, \citenamefont {Taniguchi}, \citenamefont {Vishwanath}, \citenamefont {Jarillo-Herrero},\ and\ \citenamefont {Yacoby}}]{Xie2021}%
  \BibitemOpen
  \bibfield  {author} {\bibinfo {author} {\bibfnamefont {Y.}~\bibnamefont {Xie}}, \bibinfo {author} {\bibfnamefont {A.~T.}\ \bibnamefont {Pierce}}, \bibinfo {author} {\bibfnamefont {J.~M.}\ \bibnamefont {Park}}, \bibinfo {author} {\bibfnamefont {D.~E.}\ \bibnamefont {Parker}}, \bibinfo {author} {\bibfnamefont {E.}~\bibnamefont {Khalaf}}, \bibinfo {author} {\bibfnamefont {P.}~\bibnamefont {Ledwith}}, \bibinfo {author} {\bibfnamefont {Y.}~\bibnamefont {Cao}}, \bibinfo {author} {\bibfnamefont {S.~H.}\ \bibnamefont {Lee}}, \bibinfo {author} {\bibfnamefont {S.}~\bibnamefont {Chen}}, \bibinfo {author} {\bibfnamefont {P.~R.}\ \bibnamefont {Forrester}}, \bibinfo {author} {\bibfnamefont {K.}~\bibnamefont {Watanabe}}, \bibinfo {author} {\bibfnamefont {T.}~\bibnamefont {Taniguchi}}, \bibinfo {author} {\bibfnamefont {A.}~\bibnamefont {Vishwanath}}, \bibinfo {author} {\bibfnamefont {P.}~\bibnamefont {Jarillo-Herrero}},\ and\ \bibinfo {author} {\bibfnamefont {A.}~\bibnamefont {Yacoby}},\ }\bibfield  {title} {\bibinfo
  {title} {Fractional chern insulators in magic-angle twisted bilayer graphene},\ }\href {https://doi.org/10.1038/s41586-021-04002-3} {\bibfield  {journal} {\bibinfo  {journal} {Nature}\ }\textbf {\bibinfo {volume} {600}},\ \bibinfo {pages} {439} (\bibinfo {year} {2021})}\BibitemShut {NoStop}%
\bibitem [{\citenamefont {Cai}\ \emph {et~al.}(2023)\citenamefont {Cai}, \citenamefont {Anderson}, \citenamefont {Wang}, \citenamefont {Zhang}, \citenamefont {Liu}, \citenamefont {Holtzmann}, \citenamefont {Zhang}, \citenamefont {Fan}, \citenamefont {Taniguchi}, \citenamefont {Watanabe}, \citenamefont {Ran}, \citenamefont {Cao}, \citenamefont {Fu}, \citenamefont {Xiao}, \citenamefont {Yao},\ and\ \citenamefont {Xu}}]{Cai2023}%
  \BibitemOpen
  \bibfield  {author} {\bibinfo {author} {\bibfnamefont {J.}~\bibnamefont {Cai}}, \bibinfo {author} {\bibfnamefont {E.}~\bibnamefont {Anderson}}, \bibinfo {author} {\bibfnamefont {C.}~\bibnamefont {Wang}}, \bibinfo {author} {\bibfnamefont {X.}~\bibnamefont {Zhang}}, \bibinfo {author} {\bibfnamefont {X.}~\bibnamefont {Liu}}, \bibinfo {author} {\bibfnamefont {W.}~\bibnamefont {Holtzmann}}, \bibinfo {author} {\bibfnamefont {Y.}~\bibnamefont {Zhang}}, \bibinfo {author} {\bibfnamefont {F.}~\bibnamefont {Fan}}, \bibinfo {author} {\bibfnamefont {T.}~\bibnamefont {Taniguchi}}, \bibinfo {author} {\bibfnamefont {K.}~\bibnamefont {Watanabe}}, \bibinfo {author} {\bibfnamefont {Y.}~\bibnamefont {Ran}}, \bibinfo {author} {\bibfnamefont {T.}~\bibnamefont {Cao}}, \bibinfo {author} {\bibfnamefont {L.}~\bibnamefont {Fu}}, \bibinfo {author} {\bibfnamefont {D.}~\bibnamefont {Xiao}}, \bibinfo {author} {\bibfnamefont {W.}~\bibnamefont {Yao}},\ and\ \bibinfo {author} {\bibfnamefont {X.}~\bibnamefont {Xu}},\ }\bibfield  {title}
  {\bibinfo {title} {{Signatures of fractional quantum anomalous Hall states in twisted MoTe2}},\ }\href {https://doi.org/10.1038/s41586-023-06289-w} {\bibfield  {journal} {\bibinfo  {journal} {Nature}\ }\textbf {\bibinfo {volume} {622}},\ \bibinfo {pages} {63} (\bibinfo {year} {2023})}\BibitemShut {NoStop}%
\bibitem [{\citenamefont {Peralta~Gavensky}\ \emph {et~al.}(2023)\citenamefont {Peralta~Gavensky}, \citenamefont {Sachdev},\ and\ \citenamefont {Goldman}}]{PeraltaGavensky2023}%
  \BibitemOpen
  \bibfield  {author} {\bibinfo {author} {\bibfnamefont {L.}~\bibnamefont {Peralta~Gavensky}}, \bibinfo {author} {\bibfnamefont {S.}~\bibnamefont {Sachdev}},\ and\ \bibinfo {author} {\bibfnamefont {N.}~\bibnamefont {Goldman}},\ }\bibfield  {title} {\bibinfo {title} {{Connecting the Many-Body Chern Number to Luttinger's Theorem through St\v{r}eda's Formula}},\ }\href {https://doi.org/10.1103/PhysRevLett.131.236601} {\bibfield  {journal} {\bibinfo  {journal} {Phys. Rev. Lett.}\ }\textbf {\bibinfo {volume} {131}},\ \bibinfo {pages} {236601} (\bibinfo {year} {2023})}\BibitemShut {NoStop}%
\bibitem [{\citenamefont {Callan}\ and\ \citenamefont {Harvey}(1985)}]{Callan1985}%
  \BibitemOpen
  \bibfield  {author} {\bibinfo {author} {\bibfnamefont {C.}~\bibnamefont {Callan}}\ and\ \bibinfo {author} {\bibfnamefont {J.}~\bibnamefont {Harvey}},\ }\bibfield  {title} {\bibinfo {title} {Anomalies and fermion zero modes on strings and domain walls},\ }\href {https://doi.org/https://doi.org/10.1016/0550-3213(85)90489-4} {\bibfield  {journal} {\bibinfo  {journal} {Nuclear Physics B}\ }\textbf {\bibinfo {volume} {250}},\ \bibinfo {pages} {427} (\bibinfo {year} {1985})}\BibitemShut {NoStop}%
\bibitem [{\citenamefont {Fradkin}\ \emph {et~al.}(1986)\citenamefont {Fradkin}, \citenamefont {Dagotto},\ and\ \citenamefont {Boyanovsky}}]{Fradkin1986}%
  \BibitemOpen
  \bibfield  {author} {\bibinfo {author} {\bibfnamefont {E.}~\bibnamefont {Fradkin}}, \bibinfo {author} {\bibfnamefont {E.}~\bibnamefont {Dagotto}},\ and\ \bibinfo {author} {\bibfnamefont {D.}~\bibnamefont {Boyanovsky}},\ }\bibfield  {title} {\bibinfo {title} {{Physical Realization of the Parity Anomaly in Condensed Matter Physics}},\ }\href {https://doi.org/10.1103/PhysRevLett.57.2967} {\bibfield  {journal} {\bibinfo  {journal} {Phys. Rev. Lett.}\ }\textbf {\bibinfo {volume} {57}},\ \bibinfo {pages} {2967} (\bibinfo {year} {1986})}\BibitemShut {NoStop}%
\bibitem [{\citenamefont {Stone}(1991)}]{Stone1991}%
  \BibitemOpen
  \bibfield  {author} {\bibinfo {author} {\bibfnamefont {M.}~\bibnamefont {Stone}},\ }\bibfield  {title} {\bibinfo {title} {{Edge waves in the quantum Hall effect}},\ }\href {https://doi.org/https://doi.org/10.1016/0003-4916(91)90177-A} {\bibfield  {journal} {\bibinfo  {journal} {Annals of Physics}\ }\textbf {\bibinfo {volume} {207}},\ \bibinfo {pages} {38} (\bibinfo {year} {1991})}\BibitemShut {NoStop}%
\bibitem [{\citenamefont {Mondragon-Shem}\ \emph {et~al.}(2019)\citenamefont {Mondragon-Shem}, \citenamefont {Martin}, \citenamefont {Alexandradinata},\ and\ \citenamefont {Cheng}}]{Mondragon2018}%
  \BibitemOpen
  \bibfield  {author} {\bibinfo {author} {\bibfnamefont {I.}~\bibnamefont {Mondragon-Shem}}, \bibinfo {author} {\bibfnamefont {I.}~\bibnamefont {Martin}}, \bibinfo {author} {\bibfnamefont {A.}~\bibnamefont {Alexandradinata}},\ and\ \bibinfo {author} {\bibfnamefont {M.}~\bibnamefont {Cheng}},\ }\href {https://arxiv.org/abs/1811.10632} {\bibinfo {title} {Quantized frequency-domain polarization of driven phases of matter}} (\bibinfo {year} {2019}),\ \Eprint {https://arxiv.org/abs/1811.10632} {arXiv:1811.10632 [cond-mat.mes-hall]} \BibitemShut {NoStop}%
\bibitem [{\citenamefont {Nakagawa}\ \emph {et~al.}(2020)\citenamefont {Nakagawa}, \citenamefont {Slager}, \citenamefont {Higashikawa},\ and\ \citenamefont {Oka}}]{Nakagawa2020}%
  \BibitemOpen
  \bibfield  {author} {\bibinfo {author} {\bibfnamefont {M.}~\bibnamefont {Nakagawa}}, \bibinfo {author} {\bibfnamefont {R.-J.}\ \bibnamefont {Slager}}, \bibinfo {author} {\bibfnamefont {S.}~\bibnamefont {Higashikawa}},\ and\ \bibinfo {author} {\bibfnamefont {T.}~\bibnamefont {Oka}},\ }\bibfield  {title} {\bibinfo {title} {{Wannier representation of Floquet topological states}},\ }\href {https://doi.org/10.1103/PhysRevB.101.075108} {\bibfield  {journal} {\bibinfo  {journal} {Phys. Rev. B}\ }\textbf {\bibinfo {volume} {101}},\ \bibinfo {pages} {075108} (\bibinfo {year} {2020})}\BibitemShut {NoStop}%
\bibitem [{\citenamefont {Martin}\ \emph {et~al.}(2017)\citenamefont {Martin}, \citenamefont {Refael},\ and\ \citenamefont {Halperin}}]{Martin2017}%
  \BibitemOpen
  \bibfield  {author} {\bibinfo {author} {\bibfnamefont {I.}~\bibnamefont {Martin}}, \bibinfo {author} {\bibfnamefont {G.}~\bibnamefont {Refael}},\ and\ \bibinfo {author} {\bibfnamefont {B.}~\bibnamefont {Halperin}},\ }\bibfield  {title} {\bibinfo {title} {{Topological Frequency Conversion in Strongly Driven Quantum Systems}},\ }\href {https://doi.org/10.1103/PhysRevX.7.041008} {\bibfield  {journal} {\bibinfo  {journal} {Phys. Rev. X}\ }\textbf {\bibinfo {volume} {7}},\ \bibinfo {pages} {041008} (\bibinfo {year} {2017})}\BibitemShut {NoStop}%
\bibitem [{\citenamefont {Kolodrubetz}\ \emph {et~al.}(2018)\citenamefont {Kolodrubetz}, \citenamefont {Nathan}, \citenamefont {Gazit}, \citenamefont {Morimoto},\ and\ \citenamefont {Moore}}]{Kolodubretz2018}%
  \BibitemOpen
  \bibfield  {author} {\bibinfo {author} {\bibfnamefont {M.~H.}\ \bibnamefont {Kolodrubetz}}, \bibinfo {author} {\bibfnamefont {F.}~\bibnamefont {Nathan}}, \bibinfo {author} {\bibfnamefont {S.}~\bibnamefont {Gazit}}, \bibinfo {author} {\bibfnamefont {T.}~\bibnamefont {Morimoto}},\ and\ \bibinfo {author} {\bibfnamefont {J.~E.}\ \bibnamefont {Moore}},\ }\bibfield  {title} {\bibinfo {title} {{Topological Floquet-Thouless Energy Pump}},\ }\href {https://doi.org/10.1103/PhysRevLett.120.150601} {\bibfield  {journal} {\bibinfo  {journal} {Phys. Rev. Lett.}\ }\textbf {\bibinfo {volume} {120}},\ \bibinfo {pages} {150601} (\bibinfo {year} {2018})}\BibitemShut {NoStop}%
\bibitem [{\citenamefont {Crowley}\ \emph {et~al.}(2019)\citenamefont {Crowley}, \citenamefont {Martin},\ and\ \citenamefont {Chandran}}]{Crowley2019}%
  \BibitemOpen
  \bibfield  {author} {\bibinfo {author} {\bibfnamefont {P.~J.~D.}\ \bibnamefont {Crowley}}, \bibinfo {author} {\bibfnamefont {I.}~\bibnamefont {Martin}},\ and\ \bibinfo {author} {\bibfnamefont {A.}~\bibnamefont {Chandran}},\ }\bibfield  {title} {\bibinfo {title} {{Topological classification of quasiperiodically driven quantum systems}},\ }\href {https://doi.org/10.1103/PhysRevB.99.064306} {\bibfield  {journal} {\bibinfo  {journal} {Phys. Rev. B}\ }\textbf {\bibinfo {volume} {99}},\ \bibinfo {pages} {064306} (\bibinfo {year} {2019})}\BibitemShut {NoStop}%
\bibitem [{\citenamefont {Nathan}\ \emph {et~al.}(2021{\natexlab{b}})\citenamefont {Nathan}, \citenamefont {Ge}, \citenamefont {Gazit}, \citenamefont {Rudner},\ and\ \citenamefont {Kolodrubetz}}]{Nathan2021_b}%
  \BibitemOpen
  \bibfield  {author} {\bibinfo {author} {\bibfnamefont {F.}~\bibnamefont {Nathan}}, \bibinfo {author} {\bibfnamefont {R.}~\bibnamefont {Ge}}, \bibinfo {author} {\bibfnamefont {S.}~\bibnamefont {Gazit}}, \bibinfo {author} {\bibfnamefont {M.}~\bibnamefont {Rudner}},\ and\ \bibinfo {author} {\bibfnamefont {M.}~\bibnamefont {Kolodrubetz}},\ }\bibfield  {title} {\bibinfo {title} {{Quasiperiodic Floquet-Thouless Energy Pump}},\ }\href {https://doi.org/10.1103/PhysRevLett.127.166804} {\bibfield  {journal} {\bibinfo  {journal} {Phys. Rev. Lett.}\ }\textbf {\bibinfo {volume} {127}},\ \bibinfo {pages} {166804} (\bibinfo {year} {2021}{\natexlab{b}})}\BibitemShut {NoStop}%
\bibitem [{\citenamefont {Sridhar}\ \emph {et~al.}(2024)\citenamefont {Sridhar}, \citenamefont {Ghosh}, \citenamefont {Srinivasan}, \citenamefont {Miller},\ and\ \citenamefont {Dutt}}]{Sridhar2024}%
  \BibitemOpen
  \bibfield  {author} {\bibinfo {author} {\bibfnamefont {S.~K.}\ \bibnamefont {Sridhar}}, \bibinfo {author} {\bibfnamefont {S.}~\bibnamefont {Ghosh}}, \bibinfo {author} {\bibfnamefont {D.}~\bibnamefont {Srinivasan}}, \bibinfo {author} {\bibfnamefont {A.~R.}\ \bibnamefont {Miller}},\ and\ \bibinfo {author} {\bibfnamefont {A.}~\bibnamefont {Dutt}},\ }\bibfield  {title} {\bibinfo {title} {{Quantized topological pumping in Floquet synthetic dimensions with a driven dissipative photonic molecule}},\ }\href {https://doi.org/10.1038/s41567-024-02413-3} {\bibfield  {journal} {\bibinfo  {journal} {Nature Physics}\ }\textbf {\bibinfo {volume} {20}},\ \bibinfo {pages} {843} (\bibinfo {year} {2024})}\BibitemShut {NoStop}%
\bibitem [{\citenamefont {{Hardy, Godfrey H}}(2000)}]{Hardy2000}%
  \BibitemOpen
  \bibfield  {author} {\bibinfo {author} {\bibnamefont {{Hardy, Godfrey H}}},\ }\href@noop {} {\emph {\bibinfo {title} {{Divergent Series}}}},\ \bibinfo {edition} {2nd}\ ed.,\ {AMS Chelsea Publishing}\ (\bibinfo  {publisher} {{American Mathematical Society}},\ \bibinfo {address} {{Providence, RI}},\ \bibinfo {year} {2000})\BibitemShut {NoStop}%
\bibitem [{\citenamefont {{Titchmarsh, E}}(1986)}]{Titchmarsh1986}%
  \BibitemOpen
  \bibfield  {author} {\bibinfo {author} {\bibnamefont {{Titchmarsh, E}}},\ }\href@noop {} {\emph {\bibinfo {title} {{Introduction to the theory of Fourier integrals}}}},\ \bibinfo {edition} {3rd}\ ed.\ (\bibinfo  {publisher} {{Chelsea House}},\ \bibinfo {year} {1986})\BibitemShut {NoStop}%
\bibitem [{\citenamefont {Topp}\ \emph {et~al.}(2022)\citenamefont {Topp}, \citenamefont {T\"orm\"a}, \citenamefont {Kennes},\ and\ \citenamefont {Mitra}}]{Topp2022}%
  \BibitemOpen
  \bibfield  {author} {\bibinfo {author} {\bibfnamefont {G.~E.}\ \bibnamefont {Topp}}, \bibinfo {author} {\bibfnamefont {P.}~\bibnamefont {T\"orm\"a}}, \bibinfo {author} {\bibfnamefont {D.~M.}\ \bibnamefont {Kennes}},\ and\ \bibinfo {author} {\bibfnamefont {A.}~\bibnamefont {Mitra}},\ }\bibfield  {title} {\bibinfo {title} {{Orbital magnetization of Floquet topological systems}},\ }\href {https://doi.org/10.1103/PhysRevB.105.195426} {\bibfield  {journal} {\bibinfo  {journal} {Phys. Rev. B}\ }\textbf {\bibinfo {volume} {105}},\ \bibinfo {pages} {195426} (\bibinfo {year} {2022})}\BibitemShut {NoStop}%
\bibitem [{\citenamefont {Dag}\ and\ \citenamefont {Mitra}(2022)}]{Dag2022}%
  \BibitemOpen
  \bibfield  {author} {\bibinfo {author} {\bibfnamefont {C.~B.}\ \bibnamefont {Dag}}\ and\ \bibinfo {author} {\bibfnamefont {A.}~\bibnamefont {Mitra}},\ }\bibfield  {title} {\bibinfo {title} {{Floquet topological systems with flat bands: Edge modes, Berry curvature, and orbital magnetization}},\ }\href {https://doi.org/10.1103/PhysRevB.105.245136} {\bibfield  {journal} {\bibinfo  {journal} {Phys. Rev. B}\ }\textbf {\bibinfo {volume} {105}},\ \bibinfo {pages} {245136} (\bibinfo {year} {2022})}\BibitemShut {NoStop}%
\bibitem [{\citenamefont {Peierls}(1933)}]{Peierls1933}%
  \BibitemOpen
  \bibfield  {author} {\bibinfo {author} {\bibfnamefont {R.}~\bibnamefont {Peierls}},\ }\bibfield  {title} {\bibinfo {title} {{Zur Theorie des Diamagnetismus von Leitungselektronen}},\ }\href {https://doi.org/10.1007/BF01342591} {\bibfield  {journal} {\bibinfo  {journal} {Zeitschrift f{\"u}r Physik}\ }\textbf {\bibinfo {volume} {80}},\ \bibinfo {pages} {763} (\bibinfo {year} {1933})}\BibitemShut {NoStop}%
\bibitem [{\citenamefont {Luttinger}(1951)}]{Luttinger_p_phases}%
  \BibitemOpen
  \bibfield  {author} {\bibinfo {author} {\bibfnamefont {J.~M.}\ \bibnamefont {Luttinger}},\ }\bibfield  {title} {\bibinfo {title} {The effect of a magnetic field on electrons in a periodic potential},\ }\href {https://doi.org/10.1103/PhysRev.84.814} {\bibfield  {journal} {\bibinfo  {journal} {Phys. Rev.}\ }\textbf {\bibinfo {volume} {84}},\ \bibinfo {pages} {814} (\bibinfo {year} {1951})}\BibitemShut {NoStop}%
\bibitem [{\citenamefont {Kita}\ and\ \citenamefont {Arai}(2005)}]{Kita2005}%
  \BibitemOpen
  \bibfield  {author} {\bibinfo {author} {\bibfnamefont {T.}~\bibnamefont {Kita}}\ and\ \bibinfo {author} {\bibfnamefont {M.}~\bibnamefont {Arai}},\ }\bibfield  {title} {\bibinfo {title} {Theory of {Interacting} {Bloch} {Electrons} in a {Magnetic} {Field}},\ }\href {https://doi.org/10.1143/JPSJ.74.2813} {\bibfield  {journal} {\bibinfo  {journal} {Journal of the Physical Society of Japan}\ }\textbf {\bibinfo {volume} {74}},\ \bibinfo {pages} {2813} (\bibinfo {year} {2005})}\BibitemShut {NoStop}%
\bibitem [{\citenamefont {Onoda}\ \emph {et~al.}(2006)\citenamefont {Onoda}, \citenamefont {Sugimoto},\ and\ \citenamefont {Nagaosa}}]{Onoda2006}%
  \BibitemOpen
  \bibfield  {author} {\bibinfo {author} {\bibfnamefont {S.}~\bibnamefont {Onoda}}, \bibinfo {author} {\bibfnamefont {N.}~\bibnamefont {Sugimoto}},\ and\ \bibinfo {author} {\bibfnamefont {N.}~\bibnamefont {Nagaosa}},\ }\bibfield  {title} {\bibinfo {title} {{Theory of Non-Equilibirum States Driven by Constant Electromagnetic Fields: — Non-Commutative Quantum Mechanics in the Keldysh Formalism —}},\ }\href {https://doi.org/10.1143/PTP.116.61} {\bibfield  {journal} {\bibinfo  {journal} {Progress of Theoretical Physics}\ }\textbf {\bibinfo {volume} {116}},\ \bibinfo {pages} {61} (\bibinfo {year} {2006})}\BibitemShut {NoStop}%
\bibitem [{\citenamefont {Chen}\ and\ \citenamefont {Lee}(2011)}]{Chen2011}%
  \BibitemOpen
  \bibfield  {author} {\bibinfo {author} {\bibfnamefont {K.-T.}\ \bibnamefont {Chen}}\ and\ \bibinfo {author} {\bibfnamefont {P.~A.}\ \bibnamefont {Lee}},\ }\bibfield  {title} {\bibinfo {title} {Unified formalism for calculating polarization, magnetization, and more in a periodic insulator},\ }\href {https://doi.org/10.1103/PhysRevB.84.205137} {\bibfield  {journal} {\bibinfo  {journal} {Phys. Rev. B}\ }\textbf {\bibinfo {volume} {84}},\ \bibinfo {pages} {205137} (\bibinfo {year} {2011})}\BibitemShut {NoStop}%
\bibitem [{\citenamefont {Sundaram}\ and\ \citenamefont {Niu}(1999)}]{Sundaram1999}%
  \BibitemOpen
  \bibfield  {author} {\bibinfo {author} {\bibfnamefont {G.}~\bibnamefont {Sundaram}}\ and\ \bibinfo {author} {\bibfnamefont {Q.}~\bibnamefont {Niu}},\ }\bibfield  {title} {\bibinfo {title} {{Wave-packet dynamics in slowly perturbed crystals: Gradient corrections and Berry-phase effects}},\ }\href {https://doi.org/10.1103/PhysRevB.59.14915} {\bibfield  {journal} {\bibinfo  {journal} {Phys. Rev. B}\ }\textbf {\bibinfo {volume} {59}},\ \bibinfo {pages} {14915} (\bibinfo {year} {1999})}\BibitemShut {NoStop}%
\bibitem [{\citenamefont {Xiao}\ \emph {et~al.}(2005)\citenamefont {Xiao}, \citenamefont {Shi},\ and\ \citenamefont {Niu}}]{Xiao2005}%
  \BibitemOpen
  \bibfield  {author} {\bibinfo {author} {\bibfnamefont {D.}~\bibnamefont {Xiao}}, \bibinfo {author} {\bibfnamefont {J.}~\bibnamefont {Shi}},\ and\ \bibinfo {author} {\bibfnamefont {Q.}~\bibnamefont {Niu}},\ }\bibfield  {title} {\bibinfo {title} {{Berry Phase Correction to Electron Density of States in Solids}},\ }\href {https://doi.org/10.1103/PhysRevLett.95.137204} {\bibfield  {journal} {\bibinfo  {journal} {Phys. Rev. Lett.}\ }\textbf {\bibinfo {volume} {95}},\ \bibinfo {pages} {137204} (\bibinfo {year} {2005})}\BibitemShut {NoStop}%
\bibitem [{\citenamefont {Liu}\ \emph {et~al.}(2023)\citenamefont {Liu}, \citenamefont {Fulga}, \citenamefont {Bergholtz},\ and\ \citenamefont {Asboth}}]{Liu2023}%
  \BibitemOpen
  \bibfield  {author} {\bibinfo {author} {\bibfnamefont {H.}~\bibnamefont {Liu}}, \bibinfo {author} {\bibfnamefont {C.}~\bibnamefont {Fulga}}, \bibinfo {author} {\bibfnamefont {E.~J.}\ \bibnamefont {Bergholtz}},\ and\ \bibinfo {author} {\bibfnamefont {J.}~\bibnamefont {Asboth}},\ }\href {https://arxiv.org/abs/2312.08436} {\bibinfo {title} {Topological fine structure of an energy band}} (\bibinfo {year} {2023}),\ \Eprint {https://arxiv.org/abs/2312.08436} {arXiv:2312.08436 [cond-mat.mes-hall]} \BibitemShut {NoStop}%
\bibitem [{\citenamefont {Thonhauser}\ \emph {et~al.}(2005)\citenamefont {Thonhauser}, \citenamefont {Ceresoli}, \citenamefont {Vanderbilt},\ and\ \citenamefont {Resta}}]{Thonhauser2005}%
  \BibitemOpen
  \bibfield  {author} {\bibinfo {author} {\bibfnamefont {T.}~\bibnamefont {Thonhauser}}, \bibinfo {author} {\bibfnamefont {D.}~\bibnamefont {Ceresoli}}, \bibinfo {author} {\bibfnamefont {D.}~\bibnamefont {Vanderbilt}},\ and\ \bibinfo {author} {\bibfnamefont {R.}~\bibnamefont {Resta}},\ }\bibfield  {title} {\bibinfo {title} {{Orbital Magnetization in Periodic Insulators}},\ }\href {https://doi.org/10.1103/PhysRevLett.95.137205} {\bibfield  {journal} {\bibinfo  {journal} {Phys. Rev. Lett.}\ }\textbf {\bibinfo {volume} {95}},\ \bibinfo {pages} {137205} (\bibinfo {year} {2005})}\BibitemShut {NoStop}%
\bibitem [{\citenamefont {Ceresoli}\ \emph {et~al.}(2006)\citenamefont {Ceresoli}, \citenamefont {Thonhauser}, \citenamefont {Vanderbilt},\ and\ \citenamefont {Resta}}]{Ceresoli2006}%
  \BibitemOpen
  \bibfield  {author} {\bibinfo {author} {\bibfnamefont {D.}~\bibnamefont {Ceresoli}}, \bibinfo {author} {\bibfnamefont {T.}~\bibnamefont {Thonhauser}}, \bibinfo {author} {\bibfnamefont {D.}~\bibnamefont {Vanderbilt}},\ and\ \bibinfo {author} {\bibfnamefont {R.}~\bibnamefont {Resta}},\ }\bibfield  {title} {\bibinfo {title} {{Orbital magnetization in crystalline solids: Multi-band insulators, Chern insulators, and metals}},\ }\href {https://doi.org/10.1103/PhysRevB.74.024408} {\bibfield  {journal} {\bibinfo  {journal} {Phys. Rev. B}\ }\textbf {\bibinfo {volume} {74}},\ \bibinfo {pages} {024408} (\bibinfo {year} {2006})}\BibitemShut {NoStop}%
\bibitem [{\citenamefont {Shi}\ \emph {et~al.}(2007)\citenamefont {Shi}, \citenamefont {Vignale}, \citenamefont {Xiao},\ and\ \citenamefont {Niu}}]{Shi2007}%
  \BibitemOpen
  \bibfield  {author} {\bibinfo {author} {\bibfnamefont {J.}~\bibnamefont {Shi}}, \bibinfo {author} {\bibfnamefont {G.}~\bibnamefont {Vignale}}, \bibinfo {author} {\bibfnamefont {D.}~\bibnamefont {Xiao}},\ and\ \bibinfo {author} {\bibfnamefont {Q.}~\bibnamefont {Niu}},\ }\bibfield  {title} {\bibinfo {title} {{Quantum Theory of Orbital Magnetization and Its Generalization to Interacting Systems}},\ }\href {https://doi.org/10.1103/PhysRevLett.99.197202} {\bibfield  {journal} {\bibinfo  {journal} {Phys. Rev. Lett.}\ }\textbf {\bibinfo {volume} {99}},\ \bibinfo {pages} {197202} (\bibinfo {year} {2007})}\BibitemShut {NoStop}%
\bibitem [{\citenamefont {Resta}(2010)}]{Resta2010}%
  \BibitemOpen
  \bibfield  {author} {\bibinfo {author} {\bibfnamefont {R.}~\bibnamefont {Resta}},\ }\bibfield  {title} {\bibinfo {title} {Electrical polarization and orbital magnetization: the modern theories},\ }\href {https://doi.org/10.1088/0953-8984/22/12/123201} {\bibfield  {journal} {\bibinfo  {journal} {Journal of Physics: Condensed Matter}\ }\textbf {\bibinfo {volume} {22}},\ \bibinfo {pages} {123201} (\bibinfo {year} {2010})}\BibitemShut {NoStop}%
\bibitem [{\citenamefont {Rhonald Burgos~Atencia}\ and\ \citenamefont {Culcer}(2024)}]{Atencia2024}%
  \BibitemOpen
  \bibfield  {author} {\bibinfo {author} {\bibfnamefont {A.~A.}\ \bibnamefont {Rhonald Burgos~Atencia}}\ and\ \bibinfo {author} {\bibfnamefont {D.}~\bibnamefont {Culcer}},\ }\bibfield  {title} {\bibinfo {title} {{Orbital angular momentum of Bloch electrons: equilibrium formulation, magneto-electric phenomena, and the orbital Hall effect}},\ }\href {https://doi.org/10.1080/23746149.2024.2371972} {\bibfield  {journal} {\bibinfo  {journal} {Advances in Physics: X}\ }\textbf {\bibinfo {volume} {9}},\ \bibinfo {pages} {2371972} (\bibinfo {year} {2024})},\ \Eprint {https://arxiv.org/abs/https://doi.org/10.1080/23746149.2024.2371972} {https://doi.org/10.1080/23746149.2024.2371972} \BibitemShut {NoStop}%
\bibitem [{\citenamefont {Seki}\ and\ \citenamefont {Yunoki}(2017)}]{Seki2017}%
  \BibitemOpen
  \bibfield  {author} {\bibinfo {author} {\bibfnamefont {K.}~\bibnamefont {Seki}}\ and\ \bibinfo {author} {\bibfnamefont {S.}~\bibnamefont {Yunoki}},\ }\bibfield  {title} {\bibinfo {title} {{Topological interpretation of the Luttinger theorem}},\ }\href {https://doi.org/10.1103/PhysRevB.96.085124} {\bibfield  {journal} {\bibinfo  {journal} {Phys. Rev. B}\ }\textbf {\bibinfo {volume} {96}},\ \bibinfo {pages} {085124} (\bibinfo {year} {2017})}\BibitemShut {NoStop}%
\bibitem [{\citenamefont {Haldane}(1988)}]{Haldane1988}%
  \BibitemOpen
  \bibfield  {author} {\bibinfo {author} {\bibfnamefont {F.~D.~M.}\ \bibnamefont {Haldane}},\ }\bibfield  {title} {\bibinfo {title} {{Model for a Quantum Hall Effect without Landau Levels: Condensed-Matter Realization of the "Parity Anomaly"}},\ }\href {https://doi.org/10.1103/PhysRevLett.61.2015} {\bibfield  {journal} {\bibinfo  {journal} {Phys. Rev. Lett.}\ }\textbf {\bibinfo {volume} {61}},\ \bibinfo {pages} {2015} (\bibinfo {year} {1988})}\BibitemShut {NoStop}%
\bibitem [{\citenamefont {Gao}\ and\ \citenamefont {Niu}(2022)}]{Gao2022}%
  \BibitemOpen
  \bibfield  {author} {\bibinfo {author} {\bibfnamefont {Q.}~\bibnamefont {Gao}}\ and\ \bibinfo {author} {\bibfnamefont {Q.}~\bibnamefont {Niu}},\ }\bibfield  {title} {\bibinfo {title} {Semiclassical dynamics of electrons in a space-time crystal: Magnetization, polarization, and current response},\ }\href {https://doi.org/10.1103/PhysRevB.106.224311} {\bibfield  {journal} {\bibinfo  {journal} {Phys. Rev. B}\ }\textbf {\bibinfo {volume} {106}},\ \bibinfo {pages} {224311} (\bibinfo {year} {2022})}\BibitemShut {NoStop}%
\bibitem [{\citenamefont {Perez-Piskunow}\ \emph {et~al.}(2015)\citenamefont {Perez-Piskunow}, \citenamefont {Foa~Torres},\ and\ \citenamefont {Usaj}}]{PerezPiskunow2015}%
  \BibitemOpen
  \bibfield  {author} {\bibinfo {author} {\bibfnamefont {P.~M.}\ \bibnamefont {Perez-Piskunow}}, \bibinfo {author} {\bibfnamefont {L.~E.~F.}\ \bibnamefont {Foa~Torres}},\ and\ \bibinfo {author} {\bibfnamefont {G.}~\bibnamefont {Usaj}},\ }\bibfield  {title} {\bibinfo {title} {Hierarchy of floquet gaps and edge states for driven honeycomb lattices},\ }\href {https://doi.org/10.1103/PhysRevA.91.043625} {\bibfield  {journal} {\bibinfo  {journal} {Phys. Rev. A}\ }\textbf {\bibinfo {volume} {91}},\ \bibinfo {pages} {043625} (\bibinfo {year} {2015})}\BibitemShut {NoStop}%
\bibitem [{Note1()}]{Note1}%
  \BibitemOpen
  \bibinfo {note} {If such a gap does not exist, that is, for the case of a driving protocol the does not open gaps at resonance, the Floquet zone must be chosen to locate another gap at its boundary~\cite {Roy2017}. Our approach can equally handle such a case, but for simplicity we do not consider it explicitly here}\BibitemShut {NoStop}%
\bibitem [{\citenamefont {Rahav}\ \emph {et~al.}(2003)\citenamefont {Rahav}, \citenamefont {Gilary},\ and\ \citenamefont {Fishman}}]{Rahav2003}%
  \BibitemOpen
  \bibfield  {author} {\bibinfo {author} {\bibfnamefont {S.}~\bibnamefont {Rahav}}, \bibinfo {author} {\bibfnamefont {I.}~\bibnamefont {Gilary}},\ and\ \bibinfo {author} {\bibfnamefont {S.}~\bibnamefont {Fishman}},\ }\bibfield  {title} {\bibinfo {title} {{Effective Hamiltonians for periodically driven systems}},\ }\href {https://doi.org/10.1103/PhysRevA.68.013820} {\bibfield  {journal} {\bibinfo  {journal} {Phys. Rev. A}\ }\textbf {\bibinfo {volume} {68}},\ \bibinfo {pages} {013820} (\bibinfo {year} {2003})}\BibitemShut {NoStop}%
\bibitem [{\citenamefont {Goldman}\ and\ \citenamefont {Dalibard}(2014)}]{Goldman2014}%
  \BibitemOpen
  \bibfield  {author} {\bibinfo {author} {\bibfnamefont {N.}~\bibnamefont {Goldman}}\ and\ \bibinfo {author} {\bibfnamefont {J.}~\bibnamefont {Dalibard}},\ }\bibfield  {title} {\bibinfo {title} {{Periodically Driven Quantum Systems: Effective Hamiltonians and Engineered Gauge Fields}},\ }\href {https://doi.org/10.1103/PhysRevX.4.031027} {\bibfield  {journal} {\bibinfo  {journal} {Phys. Rev. X}\ }\textbf {\bibinfo {volume} {4}},\ \bibinfo {pages} {031027} (\bibinfo {year} {2014})}\BibitemShut {NoStop}%
\bibitem [{\citenamefont {Eckardt}\ and\ \citenamefont {Anisimovas}(2015)}]{Eckardt2015}%
  \BibitemOpen
  \bibfield  {author} {\bibinfo {author} {\bibfnamefont {A.}~\bibnamefont {Eckardt}}\ and\ \bibinfo {author} {\bibfnamefont {E.}~\bibnamefont {Anisimovas}},\ }\bibfield  {title} {\bibinfo {title} {{High-frequency approximation for periodically driven quantum systems from a Floquet-space perspective}},\ }\href {https://doi.org/10.1088/1367-2630/17/9/093039} {\bibfield  {journal} {\bibinfo  {journal} {New Journal of Physics}\ }\textbf {\bibinfo {volume} {17}},\ \bibinfo {pages} {093039} (\bibinfo {year} {2015})}\BibitemShut {NoStop}%
\bibitem [{\citenamefont {Marin~Bukov}\ and\ \citenamefont {Polkovnikov}(2015)}]{Bukov2015}%
  \BibitemOpen
  \bibfield  {author} {\bibinfo {author} {\bibfnamefont {L.~D.}\ \bibnamefont {Marin~Bukov}}\ and\ \bibinfo {author} {\bibfnamefont {A.}~\bibnamefont {Polkovnikov}},\ }\bibfield  {title} {\bibinfo {title} {{Universal high-frequency behavior of periodically driven systems: from dynamical stabilization to Floquet engineering}},\ }\href {https://doi.org/10.1080/00018732.2015.1055918} {\bibfield  {journal} {\bibinfo  {journal} {Advances in Physics}\ }\textbf {\bibinfo {volume} {64}},\ \bibinfo {pages} {139} (\bibinfo {year} {2015})}\BibitemShut {NoStop}%
\bibitem [{\citenamefont {Asb\'oth}\ and\ \citenamefont {Alberti}(2017)}]{Asboth2017}%
  \BibitemOpen
  \bibfield  {author} {\bibinfo {author} {\bibfnamefont {J.~K.}\ \bibnamefont {Asb\'oth}}\ and\ \bibinfo {author} {\bibfnamefont {A.}~\bibnamefont {Alberti}},\ }\bibfield  {title} {\bibinfo {title} {{Spectral Flow and Global Topology of the Hofstadter Butterfly}},\ }\href {https://doi.org/10.1103/PhysRevLett.118.216801} {\bibfield  {journal} {\bibinfo  {journal} {Phys. Rev. Lett.}\ }\textbf {\bibinfo {volume} {118}},\ \bibinfo {pages} {216801} (\bibinfo {year} {2017})}\BibitemShut {NoStop}%
\bibitem [{\citenamefont {G\'omez-Le\'on}(2024)}]{GomezLeon2024}%
  \BibitemOpen
  \bibfield  {author} {\bibinfo {author} {\bibfnamefont {A.}~\bibnamefont {G\'omez-Le\'on}},\ }\bibfield  {title} {\bibinfo {title} {{Anomalous Floquet Phases. A resonance phenomena}},\ }\href {https://doi.org/10.22331/q-2024-11-13-1522} {\bibfield  {journal} {\bibinfo  {journal} {Quantum}\ }\textbf {\bibinfo {volume} {8}},\ \bibinfo {pages} {1522} (\bibinfo {year} {2024})}\BibitemShut {NoStop}%
\bibitem [{Note2()}]{Note2}%
  \BibitemOpen
  \bibinfo {note} {The origin of this disagreement can be traced back to an incorrect simplification of Eq.~\protect \eqref {MF} in the two-band-model case, which was used in the numerical analysis of Refs.~\cite {Topp2022,Dag2022}.}\BibitemShut {Stop}%
\bibitem [{\citenamefont {Aharonov}\ and\ \citenamefont {Anandan}(1987)}]{Aharonov1987}%
  \BibitemOpen
  \bibfield  {author} {\bibinfo {author} {\bibfnamefont {Y.}~\bibnamefont {Aharonov}}\ and\ \bibinfo {author} {\bibfnamefont {J.}~\bibnamefont {Anandan}},\ }\bibfield  {title} {\bibinfo {title} {Phase change during a cyclic quantum evolution},\ }\href {https://doi.org/10.1103/PhysRevLett.58.1593} {\bibfield  {journal} {\bibinfo  {journal} {Phys. Rev. Lett.}\ }\textbf {\bibinfo {volume} {58}},\ \bibinfo {pages} {1593} (\bibinfo {year} {1987})}\BibitemShut {NoStop}%
\bibitem [{\citenamefont {Moore}(1990)}]{Moore1990}%
  \BibitemOpen
  \bibfield  {author} {\bibinfo {author} {\bibfnamefont {D.~J.}\ \bibnamefont {Moore}},\ }\bibfield  {title} {\bibinfo {title} {{Floquet theory and the non-adiabatic Berry phase}},\ }\href {https://doi.org/10.1088/0305-4470/23/13/006} {\bibfield  {journal} {\bibinfo  {journal} {Journal of Physics A: Mathematical and General}\ }\textbf {\bibinfo {volume} {23}},\ \bibinfo {pages} {L665} (\bibinfo {year} {1990})}\BibitemShut {NoStop}%
\bibitem [{\citenamefont {Moore}(1991)}]{Moore1991}%
  \BibitemOpen
  \bibfield  {author} {\bibinfo {author} {\bibfnamefont {D.}~\bibnamefont {Moore}},\ }\bibfield  {title} {\bibinfo {title} {The calculation of nonadiabatic berry phases},\ }\href {https://doi.org/https://doi.org/10.1016/0370-1573(91)90089-5} {\bibfield  {journal} {\bibinfo  {journal} {Physics Reports}\ }\textbf {\bibinfo {volume} {210}},\ \bibinfo {pages} {1} (\bibinfo {year} {1991})}\BibitemShut {NoStop}%
\bibitem [{\citenamefont {Resta}(1992)}]{Resta1992}%
  \BibitemOpen
  \bibfield  {author} {\bibinfo {author} {\bibfnamefont {R.}~\bibnamefont {Resta}},\ }\bibfield  {title} {\bibinfo {title} {{Theory of the electric polarization in crystals}},\ }\href {https://doi.org/10.1080/00150199208016065} {\bibfield  {journal} {\bibinfo  {journal} {Ferroelectrics}\ }\textbf {\bibinfo {volume} {136}},\ \bibinfo {pages} {51–55} (\bibinfo {year} {1992})}\BibitemShut {NoStop}%
\bibitem [{\citenamefont {King-Smith}\ and\ \citenamefont {Vanderbilt}(1993)}]{KingSmith1993}%
  \BibitemOpen
  \bibfield  {author} {\bibinfo {author} {\bibfnamefont {R.~D.}\ \bibnamefont {King-Smith}}\ and\ \bibinfo {author} {\bibfnamefont {D.}~\bibnamefont {Vanderbilt}},\ }\bibfield  {title} {\bibinfo {title} {Theory of polarization of crystalline solids},\ }\href {https://doi.org/10.1103/PhysRevB.47.1651} {\bibfield  {journal} {\bibinfo  {journal} {Phys. Rev. B}\ }\textbf {\bibinfo {volume} {47}},\ \bibinfo {pages} {1651} (\bibinfo {year} {1993})}\BibitemShut {NoStop}%
\bibitem [{\citenamefont {Resta}(1994)}]{Resta1994}%
  \BibitemOpen
  \bibfield  {author} {\bibinfo {author} {\bibfnamefont {R.}~\bibnamefont {Resta}},\ }\bibfield  {title} {\bibinfo {title} {{Macroscopic polarization in crystalline dielectrics: the geometric phase approach}},\ }\href {https://doi.org/10.1103/RevModPhys.66.899} {\bibfield  {journal} {\bibinfo  {journal} {Rev. Mod. Phys.}\ }\textbf {\bibinfo {volume} {66}},\ \bibinfo {pages} {899} (\bibinfo {year} {1994})}\BibitemShut {NoStop}%
\bibitem [{\citenamefont {Vanderbilt}(2018)}]{Vanderbilt2018}%
  \BibitemOpen
  \bibfield  {author} {\bibinfo {author} {\bibfnamefont {D.}~\bibnamefont {Vanderbilt}},\ }\href {https://doi.org/10.1017/9781316662205} {\emph {\bibinfo {title} {{Berry Phases in Electronic Structure Theory: Electric Polarization, Orbital Magnetization and Topological Insulators}}}}\ (\bibinfo  {publisher} {Cambridge University Press},\ \bibinfo {year} {2018})\BibitemShut {NoStop}%
\bibitem [{\citenamefont {{Peralta Gavensky}}\ \emph {et~al.}(2025)\citenamefont {{Peralta Gavensky}}, \citenamefont {Goldman},\ and\ \citenamefont {Usaj}}]{PeraltaGavensky2025}%
  \BibitemOpen
  \bibfield  {author} {\bibinfo {author} {\bibfnamefont {L.}~\bibnamefont {{Peralta Gavensky}}}, \bibinfo {author} {\bibfnamefont {N.}~\bibnamefont {Goldman}},\ and\ \bibinfo {author} {\bibfnamefont {G.}~\bibnamefont {Usaj}},\ }\href {https://arxiv.org/abs/2506.20719} {\bibinfo {title} {{Quantized Chern-Simons Axion Coupling in Anomalous Floquet Systems}}} (\bibinfo {year} {2025}),\ \Eprint {https://arxiv.org/abs/2506.20719} {arXiv:2506.20719 [cond-mat.mes-hall]} \BibitemShut {NoStop}%
\bibitem [{\citenamefont {Dag}\ and\ \citenamefont {Rokaj}(2024)}]{Dag2024}%
  \BibitemOpen
  \bibfield  {author} {\bibinfo {author} {\bibfnamefont {C.~B.}\ \bibnamefont {Dag}}\ and\ \bibinfo {author} {\bibfnamefont {V.}~\bibnamefont {Rokaj}},\ }\bibfield  {title} {\bibinfo {title} {{Engineering topology in graphene with chiral cavities}},\ }\href {https://doi.org/10.1103/PhysRevB.110.L121101} {\bibfield  {journal} {\bibinfo  {journal} {Phys. Rev. B}\ }\textbf {\bibinfo {volume} {110}},\ \bibinfo {pages} {L121101} (\bibinfo {year} {2024})}\BibitemShut {NoStop}%
\bibitem [{\citenamefont {Zhou}\ and\ \citenamefont {Wu}(2011)}]{Zhou2011}%
  \BibitemOpen
  \bibfield  {author} {\bibinfo {author} {\bibfnamefont {Y.}~\bibnamefont {Zhou}}\ and\ \bibinfo {author} {\bibfnamefont {M.~W.}\ \bibnamefont {Wu}},\ }\bibfield  {title} {\bibinfo {title} {Optical response of graphene under intense terahertz fields},\ }\href {https://doi.org/10.1103/PhysRevB.83.245436} {\bibfield  {journal} {\bibinfo  {journal} {Phys. Rev. B}\ }\textbf {\bibinfo {volume} {83}},\ \bibinfo {pages} {245436} (\bibinfo {year} {2011})}\BibitemShut {NoStop}%
\bibitem [{\citenamefont {Rudner}\ and\ \citenamefont {Lindner}(2020{\natexlab{b}})}]{Rudner2020_b}%
  \BibitemOpen
  \bibfield  {author} {\bibinfo {author} {\bibfnamefont {M.~S.}\ \bibnamefont {Rudner}}\ and\ \bibinfo {author} {\bibfnamefont {N.~H.}\ \bibnamefont {Lindner}},\ }\href@noop {} {\bibinfo {title} {{The Floquet Engineer's Handbook}}} (\bibinfo {year} {2020}{\natexlab{b}}),\ \Eprint {https://arxiv.org/abs/2003.08252} {arXiv:2003.08252 [cond-mat.mes-hall]} \BibitemShut {NoStop}%
\bibitem [{\citenamefont {B\"uttiker}(1985)}]{Buttiker1985}%
  \BibitemOpen
  \bibfield  {author} {\bibinfo {author} {\bibfnamefont {M.}~\bibnamefont {B\"uttiker}},\ }\bibfield  {title} {\bibinfo {title} {Small normal-metal loop coupled to an electron reservoir},\ }\href {https://doi.org/10.1103/PhysRevB.32.1846} {\bibfield  {journal} {\bibinfo  {journal} {Phys. Rev. B}\ }\textbf {\bibinfo {volume} {32}},\ \bibinfo {pages} {1846} (\bibinfo {year} {1985})}\BibitemShut {NoStop}%
\bibitem [{\citenamefont {Schnell}\ \emph {et~al.}(2023)\citenamefont {Schnell}, \citenamefont {Wu}, \citenamefont {Widera},\ and\ \citenamefont {Eckardt}}]{Schnell2023a}%
  \BibitemOpen
  \bibfield  {author} {\bibinfo {author} {\bibfnamefont {A.}~\bibnamefont {Schnell}}, \bibinfo {author} {\bibfnamefont {L.-N.}\ \bibnamefont {Wu}}, \bibinfo {author} {\bibfnamefont {A.}~\bibnamefont {Widera}},\ and\ \bibinfo {author} {\bibfnamefont {A.}~\bibnamefont {Eckardt}},\ }\bibfield  {title} {\bibinfo {title} {{Floquet-heating-induced Bose condensation in a scarlike mode of an open driven optical-lattice system}},\ }\href {https://doi.org/10.1103/PhysRevA.107.L021301} {\bibfield  {journal} {\bibinfo  {journal} {Phys. Rev. A}\ }\textbf {\bibinfo {volume} {107}},\ \bibinfo {pages} {L021301} (\bibinfo {year} {2023})}\BibitemShut {NoStop}%
\bibitem [{\citenamefont {Schnell}\ \emph {et~al.}(2024)\citenamefont {Schnell}, \citenamefont {Weitenberg},\ and\ \citenamefont {Eckardt}}]{Schnell2024}%
  \BibitemOpen
  \bibfield  {author} {\bibinfo {author} {\bibfnamefont {A.}~\bibnamefont {Schnell}}, \bibinfo {author} {\bibfnamefont {C.}~\bibnamefont {Weitenberg}},\ and\ \bibinfo {author} {\bibfnamefont {A.}~\bibnamefont {Eckardt}},\ }\bibfield  {title} {\bibinfo {title} {{Dissipative preparation of a Floquet topological insulator in an optical lattice via bath engineering}},\ }\href {https://doi.org/10.21468/SciPostPhys.17.2.052} {\bibfield  {journal} {\bibinfo  {journal} {SciPost Phys.}\ }\textbf {\bibinfo {volume} {17}},\ \bibinfo {pages} {052} (\bibinfo {year} {2024})}\BibitemShut {NoStop}%
\bibitem [{\citenamefont {Kitaev}(2006)}]{Kitaev2006}%
  \BibitemOpen
  \bibfield  {author} {\bibinfo {author} {\bibfnamefont {A.}~\bibnamefont {Kitaev}},\ }\bibfield  {title} {\bibinfo {title} {Anyons in an exactly solved model and beyond},\ }\href {https://doi.org/https://doi.org/10.1016/j.aop.2005.10.005} {\bibfield  {journal} {\bibinfo  {journal} {Annals of Physics}\ }\textbf {\bibinfo {volume} {321}},\ \bibinfo {pages} {2} (\bibinfo {year} {2006})},\ \bibinfo {note} {january Special Issue}\BibitemShut {NoStop}%
\bibitem [{\citenamefont {Bianco}\ and\ \citenamefont {Resta}(2011)}]{Bianco2011}%
  \BibitemOpen
  \bibfield  {author} {\bibinfo {author} {\bibfnamefont {R.}~\bibnamefont {Bianco}}\ and\ \bibinfo {author} {\bibfnamefont {R.}~\bibnamefont {Resta}},\ }\bibfield  {title} {\bibinfo {title} {Mapping topological order in coordinate space},\ }\href {https://doi.org/10.1103/PhysRevB.84.241106} {\bibfield  {journal} {\bibinfo  {journal} {Phys. Rev. B}\ }\textbf {\bibinfo {volume} {84}},\ \bibinfo {pages} {241106} (\bibinfo {year} {2011})}\BibitemShut {NoStop}%
\bibitem [{\citenamefont {Tran}\ \emph {et~al.}(2015)\citenamefont {Tran}, \citenamefont {Dauphin}, \citenamefont {Goldman},\ and\ \citenamefont {Gaspard}}]{Tran2015}%
  \BibitemOpen
  \bibfield  {author} {\bibinfo {author} {\bibfnamefont {D.-T.}\ \bibnamefont {Tran}}, \bibinfo {author} {\bibfnamefont {A.}~\bibnamefont {Dauphin}}, \bibinfo {author} {\bibfnamefont {N.}~\bibnamefont {Goldman}},\ and\ \bibinfo {author} {\bibfnamefont {P.}~\bibnamefont {Gaspard}},\ }\bibfield  {title} {\bibinfo {title} {{Topological Hofstadter insulators in a two-dimensional quasicrystal}},\ }\href {https://doi.org/10.1103/PhysRevB.91.085125} {\bibfield  {journal} {\bibinfo  {journal} {Phys. Rev. B}\ }\textbf {\bibinfo {volume} {91}},\ \bibinfo {pages} {085125} (\bibinfo {year} {2015})}\BibitemShut {NoStop}%
\bibitem [{\citenamefont {Privitera}\ and\ \citenamefont {Santoro}(2016)}]{Privitera2016}%
  \BibitemOpen
  \bibfield  {author} {\bibinfo {author} {\bibfnamefont {L.}~\bibnamefont {Privitera}}\ and\ \bibinfo {author} {\bibfnamefont {G.~E.}\ \bibnamefont {Santoro}},\ }\bibfield  {title} {\bibinfo {title} {{Quantum annealing and nonequilibrium dynamics of Floquet Chern insulators}},\ }\href {https://doi.org/10.1103/PhysRevB.93.241406} {\bibfield  {journal} {\bibinfo  {journal} {Phys. Rev. B}\ }\textbf {\bibinfo {volume} {93}},\ \bibinfo {pages} {241406} (\bibinfo {year} {2016})}\BibitemShut {NoStop}%
\bibitem [{\citenamefont {Caio}\ \emph {et~al.}(2019)\citenamefont {Caio}, \citenamefont {M{\"o}ller}, \citenamefont {Cooper},\ and\ \citenamefont {Bhaseen}}]{Caio2019}%
  \BibitemOpen
  \bibfield  {author} {\bibinfo {author} {\bibfnamefont {M.~D.}\ \bibnamefont {Caio}}, \bibinfo {author} {\bibfnamefont {G.}~\bibnamefont {M{\"o}ller}}, \bibinfo {author} {\bibfnamefont {N.~R.}\ \bibnamefont {Cooper}},\ and\ \bibinfo {author} {\bibfnamefont {M.~J.}\ \bibnamefont {Bhaseen}},\ }\bibfield  {title} {\bibinfo {title} {{Topological marker currents in Chern insulators}},\ }\href {https://doi.org/10.1038/s41567-018-0390-7} {\bibfield  {journal} {\bibinfo  {journal} {Nature Physics}\ }\textbf {\bibinfo {volume} {15}},\ \bibinfo {pages} {257} (\bibinfo {year} {2019})}\BibitemShut {NoStop}%
\bibitem [{\citenamefont {{Ul\ifmmode \check{c}\else \v{c}\fi{}akar, Lara and Mravlje, Jernej and Rejec, Toma\ifmmode \check{z}\else \v{z}\fi{}}}(2020)}]{Ulcakar2020}%
  \BibitemOpen
  \bibfield  {author} {\bibinfo {author} {\bibnamefont {{Ul\ifmmode \check{c}\else \v{c}\fi{}akar, Lara and Mravlje, Jernej and Rejec, Toma\ifmmode \check{z}\else \v{z}\fi{}}}},\ }\bibfield  {title} {\bibinfo {title} {{Kibble-Zurek Behavior in Disordered Chern Insulators}},\ }\href {https://doi.org/10.1103/PhysRevLett.125.216601} {\bibfield  {journal} {\bibinfo  {journal} {Phys. Rev. Lett.}\ }\textbf {\bibinfo {volume} {125}},\ \bibinfo {pages} {216601} (\bibinfo {year} {2020})}\BibitemShut {NoStop}%
\bibitem [{\citenamefont {Umer}\ \emph {et~al.}(2020)\citenamefont {Umer}, \citenamefont {Bomantara},\ and\ \citenamefont {Gong}}]{Umer2020}%
  \BibitemOpen
  \bibfield  {author} {\bibinfo {author} {\bibfnamefont {M.}~\bibnamefont {Umer}}, \bibinfo {author} {\bibfnamefont {R.~W.}\ \bibnamefont {Bomantara}},\ and\ \bibinfo {author} {\bibfnamefont {J.}~\bibnamefont {Gong}},\ }\bibfield  {title} {\bibinfo {title} {{Counterpropagating edge states in Floquet topological insulating phases}},\ }\href {https://doi.org/10.1103/PhysRevB.101.235438} {\bibfield  {journal} {\bibinfo  {journal} {Phys. Rev. B}\ }\textbf {\bibinfo {volume} {101}},\ \bibinfo {pages} {235438} (\bibinfo {year} {2020})}\BibitemShut {NoStop}%
\bibitem [{\citenamefont {Shi}\ \emph {et~al.}(2022)\citenamefont {Shi}, \citenamefont {Chen}, \citenamefont {Zhang},\ and\ \citenamefont {Zhang}}]{Shi2022}%
  \BibitemOpen
  \bibfield  {author} {\bibinfo {author} {\bibfnamefont {K.-Y.}\ \bibnamefont {Shi}}, \bibinfo {author} {\bibfnamefont {R.-Q.}\ \bibnamefont {Chen}}, \bibinfo {author} {\bibfnamefont {S.}~\bibnamefont {Zhang}},\ and\ \bibinfo {author} {\bibfnamefont {W.}~\bibnamefont {Zhang}},\ }\bibfield  {title} {\bibinfo {title} {{Topological invariants of Floquet topological phases under periodical driving}},\ }\href {https://doi.org/10.1103/PhysRevA.106.053301} {\bibfield  {journal} {\bibinfo  {journal} {Phys. Rev. A}\ }\textbf {\bibinfo {volume} {106}},\ \bibinfo {pages} {053301} (\bibinfo {year} {2022})}\BibitemShut {NoStop}%
\bibitem [{\citenamefont {Zhang}\ \emph {et~al.}(2023)\citenamefont {Zhang}, \citenamefont {Yi}, \citenamefont {Zhang}, \citenamefont {Jiao}, \citenamefont {Shi}, \citenamefont {Yuan}, \citenamefont {Zhang}, \citenamefont {Liu}, \citenamefont {Chen},\ and\ \citenamefont {Pan}}]{Zhang2023_d}%
  \BibitemOpen
  \bibfield  {author} {\bibinfo {author} {\bibfnamefont {J.-Y.}\ \bibnamefont {Zhang}}, \bibinfo {author} {\bibfnamefont {C.-R.}\ \bibnamefont {Yi}}, \bibinfo {author} {\bibfnamefont {L.}~\bibnamefont {Zhang}}, \bibinfo {author} {\bibfnamefont {R.-H.}\ \bibnamefont {Jiao}}, \bibinfo {author} {\bibfnamefont {K.-Y.}\ \bibnamefont {Shi}}, \bibinfo {author} {\bibfnamefont {H.}~\bibnamefont {Yuan}}, \bibinfo {author} {\bibfnamefont {W.}~\bibnamefont {Zhang}}, \bibinfo {author} {\bibfnamefont {X.-J.}\ \bibnamefont {Liu}}, \bibinfo {author} {\bibfnamefont {S.}~\bibnamefont {Chen}},\ and\ \bibinfo {author} {\bibfnamefont {J.-W.}\ \bibnamefont {Pan}},\ }\bibfield  {title} {\bibinfo {title} {{Tuning Anomalous Floquet Topological Bands with Ultracold Atoms}},\ }\href {https://doi.org/10.1103/PhysRevLett.130.043201} {\bibfield  {journal} {\bibinfo  {journal} {Phys. Rev. Lett.}\ }\textbf {\bibinfo {volume} {130}},\ \bibinfo {pages} {043201} (\bibinfo {year} {2023})}\BibitemShut {NoStop}%
\bibitem [{\citenamefont {Gross}\ \emph {et~al.}(2012)\citenamefont {Gross}, \citenamefont {Nesme}, \citenamefont {Vogts},\ and\ \citenamefont {Werner}}]{Gross2012}%
  \BibitemOpen
  \bibfield  {author} {\bibinfo {author} {\bibfnamefont {D.}~\bibnamefont {Gross}}, \bibinfo {author} {\bibfnamefont {V.}~\bibnamefont {Nesme}}, \bibinfo {author} {\bibfnamefont {H.}~\bibnamefont {Vogts}},\ and\ \bibinfo {author} {\bibfnamefont {R.~F.}\ \bibnamefont {Werner}},\ }\bibfield  {title} {\bibinfo {title} {Index theory of one dimensional quantum walks and cellular automata},\ }\href {https://doi.org/10.1007/s00220-012-1423-1} {\bibfield  {journal} {\bibinfo  {journal} {Communications in Mathematical Physics}\ }\textbf {\bibinfo {volume} {310}},\ \bibinfo {pages} {419} (\bibinfo {year} {2012})}\BibitemShut {NoStop}%
\bibitem [{\citenamefont {Roy}\ and\ \citenamefont {Harper}(2016)}]{Roy2016}%
  \BibitemOpen
  \bibfield  {author} {\bibinfo {author} {\bibfnamefont {R.}~\bibnamefont {Roy}}\ and\ \bibinfo {author} {\bibfnamefont {F.}~\bibnamefont {Harper}},\ }\bibfield  {title} {\bibinfo {title} {{Abelian Floquet symmetry-protected topological phases in one dimension}},\ }\href {https://doi.org/10.1103/PhysRevB.94.125105} {\bibfield  {journal} {\bibinfo  {journal} {Phys. Rev. B}\ }\textbf {\bibinfo {volume} {94}},\ \bibinfo {pages} {125105} (\bibinfo {year} {2016})}\BibitemShut {NoStop}%
\bibitem [{\citenamefont {Potter}\ \emph {et~al.}(2016)\citenamefont {Potter}, \citenamefont {Morimoto},\ and\ \citenamefont {Vishwanath}}]{Potter2016}%
  \BibitemOpen
  \bibfield  {author} {\bibinfo {author} {\bibfnamefont {A.~C.}\ \bibnamefont {Potter}}, \bibinfo {author} {\bibfnamefont {T.}~\bibnamefont {Morimoto}},\ and\ \bibinfo {author} {\bibfnamefont {A.}~\bibnamefont {Vishwanath}},\ }\bibfield  {title} {\bibinfo {title} {{Classification of Interacting Topological Floquet Phases in One Dimension}},\ }\href {https://doi.org/10.1103/PhysRevX.6.041001} {\bibfield  {journal} {\bibinfo  {journal} {Phys. Rev. X}\ }\textbf {\bibinfo {volume} {6}},\ \bibinfo {pages} {041001} (\bibinfo {year} {2016})}\BibitemShut {NoStop}%
\bibitem [{\citenamefont {Po}\ \emph {et~al.}(2016)\citenamefont {Po}, \citenamefont {Fidkowski}, \citenamefont {Morimoto}, \citenamefont {Potter},\ and\ \citenamefont {Vishwanath}}]{Po2016}%
  \BibitemOpen
  \bibfield  {author} {\bibinfo {author} {\bibfnamefont {H.~C.}\ \bibnamefont {Po}}, \bibinfo {author} {\bibfnamefont {L.}~\bibnamefont {Fidkowski}}, \bibinfo {author} {\bibfnamefont {T.}~\bibnamefont {Morimoto}}, \bibinfo {author} {\bibfnamefont {A.~C.}\ \bibnamefont {Potter}},\ and\ \bibinfo {author} {\bibfnamefont {A.}~\bibnamefont {Vishwanath}},\ }\bibfield  {title} {\bibinfo {title} {{Chiral Floquet Phases of Many-Body Localized Bosons}},\ }\href {https://doi.org/10.1103/PhysRevX.6.041070} {\bibfield  {journal} {\bibinfo  {journal} {Phys. Rev. X}\ }\textbf {\bibinfo {volume} {6}},\ \bibinfo {pages} {041070} (\bibinfo {year} {2016})}\BibitemShut {NoStop}%
\bibitem [{\citenamefont {Else}\ and\ \citenamefont {Nayak}(2016)}]{Else2016}%
  \BibitemOpen
  \bibfield  {author} {\bibinfo {author} {\bibfnamefont {D.~V.}\ \bibnamefont {Else}}\ and\ \bibinfo {author} {\bibfnamefont {C.}~\bibnamefont {Nayak}},\ }\bibfield  {title} {\bibinfo {title} {{Classification of topological phases in periodically driven interacting systems}},\ }\href {https://doi.org/10.1103/PhysRevB.93.201103} {\bibfield  {journal} {\bibinfo  {journal} {Phys. Rev. B}\ }\textbf {\bibinfo {volume} {93}},\ \bibinfo {pages} {201103} (\bibinfo {year} {2016})}\BibitemShut {NoStop}%
\bibitem [{\citenamefont {Khemani}\ \emph {et~al.}(2016)\citenamefont {Khemani}, \citenamefont {Lazarides}, \citenamefont {Moessner},\ and\ \citenamefont {Sondhi}}]{Khemani2016}%
  \BibitemOpen
  \bibfield  {author} {\bibinfo {author} {\bibfnamefont {V.}~\bibnamefont {Khemani}}, \bibinfo {author} {\bibfnamefont {A.}~\bibnamefont {Lazarides}}, \bibinfo {author} {\bibfnamefont {R.}~\bibnamefont {Moessner}},\ and\ \bibinfo {author} {\bibfnamefont {S.~L.}\ \bibnamefont {Sondhi}},\ }\bibfield  {title} {\bibinfo {title} {{Phase Structure of Driven Quantum Systems}},\ }\href {https://doi.org/10.1103/PhysRevLett.116.250401} {\bibfield  {journal} {\bibinfo  {journal} {Phys. Rev. Lett.}\ }\textbf {\bibinfo {volume} {116}},\ \bibinfo {pages} {250401} (\bibinfo {year} {2016})}\BibitemShut {NoStop}%
\bibitem [{\citenamefont {von Keyserlingk}\ and\ \citenamefont {Sondhi}(2016)}]{vonKeyserlingk2016}%
  \BibitemOpen
  \bibfield  {author} {\bibinfo {author} {\bibfnamefont {C.~W.}\ \bibnamefont {von Keyserlingk}}\ and\ \bibinfo {author} {\bibfnamefont {S.~L.}\ \bibnamefont {Sondhi}},\ }\bibfield  {title} {\bibinfo {title} {{Phase structure of one-dimensional interacting Floquet systems. I. Abelian symmetry-protected topological phases}},\ }\href {https://doi.org/10.1103/PhysRevB.93.245145} {\bibfield  {journal} {\bibinfo  {journal} {Phys. Rev. B}\ }\textbf {\bibinfo {volume} {93}},\ \bibinfo {pages} {245145} (\bibinfo {year} {2016})}\BibitemShut {NoStop}%
\bibitem [{\citenamefont {Harper}\ and\ \citenamefont {Roy}(2017)}]{Harper2017}%
  \BibitemOpen
  \bibfield  {author} {\bibinfo {author} {\bibfnamefont {F.}~\bibnamefont {Harper}}\ and\ \bibinfo {author} {\bibfnamefont {R.}~\bibnamefont {Roy}},\ }\bibfield  {title} {\bibinfo {title} {{Floquet Topological Order in Interacting Systems of Bosons and Fermions}},\ }\href {https://doi.org/10.1103/PhysRevLett.118.115301} {\bibfield  {journal} {\bibinfo  {journal} {Phys. Rev. Lett.}\ }\textbf {\bibinfo {volume} {118}},\ \bibinfo {pages} {115301} (\bibinfo {year} {2017})}\BibitemShut {NoStop}%
\bibitem [{\citenamefont {Moessner}\ and\ \citenamefont {Sondhi}(2017)}]{Moessner2017}%
  \BibitemOpen
  \bibfield  {author} {\bibinfo {author} {\bibfnamefont {R.}~\bibnamefont {Moessner}}\ and\ \bibinfo {author} {\bibfnamefont {S.~L.}\ \bibnamefont {Sondhi}},\ }\bibfield  {title} {\bibinfo {title} {{Equilibration and order in quantum Floquet matter}},\ }\href {https://doi.org/10.1038/nphys4106} {\bibfield  {journal} {\bibinfo  {journal} {Nature Physics}\ }\textbf {\bibinfo {volume} {13}},\ \bibinfo {pages} {424} (\bibinfo {year} {2017})}\BibitemShut {NoStop}%
\bibitem [{\citenamefont {Fidkowski}\ \emph {et~al.}(2019)\citenamefont {Fidkowski}, \citenamefont {Po}, \citenamefont {Potter},\ and\ \citenamefont {Vishwanath}}]{Fidkowski2019}%
  \BibitemOpen
  \bibfield  {author} {\bibinfo {author} {\bibfnamefont {L.}~\bibnamefont {Fidkowski}}, \bibinfo {author} {\bibfnamefont {H.~C.}\ \bibnamefont {Po}}, \bibinfo {author} {\bibfnamefont {A.~C.}\ \bibnamefont {Potter}},\ and\ \bibinfo {author} {\bibfnamefont {A.}~\bibnamefont {Vishwanath}},\ }\bibfield  {title} {\bibinfo {title} {{Interacting invariants for Floquet phases of fermions in two dimensions}},\ }\href {https://doi.org/10.1103/PhysRevB.99.085115} {\bibfield  {journal} {\bibinfo  {journal} {Phys. Rev. B}\ }\textbf {\bibinfo {volume} {99}},\ \bibinfo {pages} {085115} (\bibinfo {year} {2019})}\BibitemShut {NoStop}%
\bibitem [{\citenamefont {Gong}\ \emph {et~al.}(2020)\citenamefont {Gong}, \citenamefont {S\"underhauf}, \citenamefont {Schuch},\ and\ \citenamefont {Cirac}}]{Gong2020}%
  \BibitemOpen
  \bibfield  {author} {\bibinfo {author} {\bibfnamefont {Z.}~\bibnamefont {Gong}}, \bibinfo {author} {\bibfnamefont {C.}~\bibnamefont {S\"underhauf}}, \bibinfo {author} {\bibfnamefont {N.}~\bibnamefont {Schuch}},\ and\ \bibinfo {author} {\bibfnamefont {J.~I.}\ \bibnamefont {Cirac}},\ }\bibfield  {title} {\bibinfo {title} {{Classification of Matrix-Product Unitaries with Symmetries}},\ }\href {https://doi.org/10.1103/PhysRevLett.124.100402} {\bibfield  {journal} {\bibinfo  {journal} {Phys. Rev. Lett.}\ }\textbf {\bibinfo {volume} {124}},\ \bibinfo {pages} {100402} (\bibinfo {year} {2020})}\BibitemShut {NoStop}%
\bibitem [{\citenamefont {Zhang}\ and\ \citenamefont {Levin}(2023)}]{Zhang2023_c}%
  \BibitemOpen
  \bibfield  {author} {\bibinfo {author} {\bibfnamefont {C.}~\bibnamefont {Zhang}}\ and\ \bibinfo {author} {\bibfnamefont {M.}~\bibnamefont {Levin}},\ }\bibfield  {title} {\bibinfo {title} {{Bulk-Boundary Correspondence for Interacting Floquet Systems in Two Dimensions}},\ }\href {https://doi.org/10.1103/PhysRevX.13.031038} {\bibfield  {journal} {\bibinfo  {journal} {Phys. Rev. X}\ }\textbf {\bibinfo {volume} {13}},\ \bibinfo {pages} {031038} (\bibinfo {year} {2023})}\BibitemShut {NoStop}%
\bibitem [{\citenamefont {Tsuji}\ \emph {et~al.}(2008)\citenamefont {Tsuji}, \citenamefont {Oka},\ and\ \citenamefont {Aoki}}]{Tsuji2008}%
  \BibitemOpen
  \bibfield  {author} {\bibinfo {author} {\bibfnamefont {N.}~\bibnamefont {Tsuji}}, \bibinfo {author} {\bibfnamefont {T.}~\bibnamefont {Oka}},\ and\ \bibinfo {author} {\bibfnamefont {H.}~\bibnamefont {Aoki}},\ }\bibfield  {title} {\bibinfo {title} {{Correlated electron systems periodically driven out of equilibrium: $\text{Floquet}+\text{DMFT}$ formalism}},\ }\href {https://doi.org/10.1103/PhysRevB.78.235124} {\bibfield  {journal} {\bibinfo  {journal} {Phys. Rev. B}\ }\textbf {\bibinfo {volume} {78}},\ \bibinfo {pages} {235124} (\bibinfo {year} {2008})}\BibitemShut {NoStop}%
\bibitem [{\citenamefont {Aoki}\ \emph {et~al.}(2014)\citenamefont {Aoki}, \citenamefont {Tsuji}, \citenamefont {Eckstein}, \citenamefont {Kollar}, \citenamefont {Oka},\ and\ \citenamefont {Werner}}]{Aoki2014}%
  \BibitemOpen
  \bibfield  {author} {\bibinfo {author} {\bibfnamefont {H.}~\bibnamefont {Aoki}}, \bibinfo {author} {\bibfnamefont {N.}~\bibnamefont {Tsuji}}, \bibinfo {author} {\bibfnamefont {M.}~\bibnamefont {Eckstein}}, \bibinfo {author} {\bibfnamefont {M.}~\bibnamefont {Kollar}}, \bibinfo {author} {\bibfnamefont {T.}~\bibnamefont {Oka}},\ and\ \bibinfo {author} {\bibfnamefont {P.}~\bibnamefont {Werner}},\ }\bibfield  {title} {\bibinfo {title} {Nonequilibrium dynamical mean-field theory and its applications},\ }\href {https://doi.org/10.1103/RevModPhys.86.779} {\bibfield  {journal} {\bibinfo  {journal} {Rev. Mod. Phys.}\ }\textbf {\bibinfo {volume} {86}},\ \bibinfo {pages} {779} (\bibinfo {year} {2014})}\BibitemShut {NoStop}%
\bibitem [{\citenamefont {Uhrig}\ \emph {et~al.}(2019)\citenamefont {Uhrig}, \citenamefont {Kalthoff},\ and\ \citenamefont {Freericks}}]{Uhrig2019}%
  \BibitemOpen
  \bibfield  {author} {\bibinfo {author} {\bibfnamefont {G.~S.}\ \bibnamefont {Uhrig}}, \bibinfo {author} {\bibfnamefont {M.~H.}\ \bibnamefont {Kalthoff}},\ and\ \bibinfo {author} {\bibfnamefont {J.~K.}\ \bibnamefont {Freericks}},\ }\bibfield  {title} {\bibinfo {title} {{Positivity of the Spectral Densities of Retarded Floquet Green Functions}},\ }\href {https://doi.org/10.1103/PhysRevLett.122.130604} {\bibfield  {journal} {\bibinfo  {journal} {Phys. Rev. Lett.}\ }\textbf {\bibinfo {volume} {122}},\ \bibinfo {pages} {130604} (\bibinfo {year} {2019})}\BibitemShut {NoStop}%
\bibitem [{\citenamefont {Puviani}\ and\ \citenamefont {Manghi}(2016)}]{Puviani2016}%
  \BibitemOpen
  \bibfield  {author} {\bibinfo {author} {\bibfnamefont {M.}~\bibnamefont {Puviani}}\ and\ \bibinfo {author} {\bibfnamefont {F.}~\bibnamefont {Manghi}},\ }\bibfield  {title} {\bibinfo {title} {{Periodically driven interacting electrons in one dimension: Many-body Floquet approach}},\ }\href {https://doi.org/10.1103/PhysRevB.94.161111} {\bibfield  {journal} {\bibinfo  {journal} {Phys. Rev. B}\ }\textbf {\bibinfo {volume} {94}},\ \bibinfo {pages} {161111} (\bibinfo {year} {2016})}\BibitemShut {NoStop}%
\bibitem [{\citenamefont {Kalinowski}\ \emph {et~al.}(2023)\citenamefont {Kalinowski}, \citenamefont {Maskara},\ and\ \citenamefont {Lukin}}]{Kalinowski2023}%
  \BibitemOpen
  \bibfield  {author} {\bibinfo {author} {\bibfnamefont {M.}~\bibnamefont {Kalinowski}}, \bibinfo {author} {\bibfnamefont {N.}~\bibnamefont {Maskara}},\ and\ \bibinfo {author} {\bibfnamefont {M.~D.}\ \bibnamefont {Lukin}},\ }\bibfield  {title} {\bibinfo {title} {{Non-Abelian Floquet Spin Liquids in a Digital Rydberg Simulator}},\ }\href {https://doi.org/10.1103/PhysRevX.13.031008} {\bibfield  {journal} {\bibinfo  {journal} {Phys. Rev. X}\ }\textbf {\bibinfo {volume} {13}},\ \bibinfo {pages} {031008} (\bibinfo {year} {2023})}\BibitemShut {NoStop}%
\bibitem [{\citenamefont {Sun}\ \emph {et~al.}(2023)\citenamefont {Sun}, \citenamefont {Goldman}, \citenamefont {Aidelsburger},\ and\ \citenamefont {Bukov}}]{Sun2023}%
  \BibitemOpen
  \bibfield  {author} {\bibinfo {author} {\bibfnamefont {B.-Y.}\ \bibnamefont {Sun}}, \bibinfo {author} {\bibfnamefont {N.}~\bibnamefont {Goldman}}, \bibinfo {author} {\bibfnamefont {M.}~\bibnamefont {Aidelsburger}},\ and\ \bibinfo {author} {\bibfnamefont {M.}~\bibnamefont {Bukov}},\ }\bibfield  {title} {\bibinfo {title} {{Engineering and Probing Non-Abelian Chiral Spin Liquids Using Periodically Driven Ultracold Atoms}},\ }\href {https://doi.org/10.1103/PRXQuantum.4.020329} {\bibfield  {journal} {\bibinfo  {journal} {PRX Quantum}\ }\textbf {\bibinfo {volume} {4}},\ \bibinfo {pages} {020329} (\bibinfo {year} {2023})}\BibitemShut {NoStop}%
\bibitem [{\citenamefont {Nishad}\ \emph {et~al.}(2023)\citenamefont {Nishad}, \citenamefont {Keselman}, \citenamefont {Lahaye}, \citenamefont {Browaeys},\ and\ \citenamefont {Tsesses}}]{Nishad2023}%
  \BibitemOpen
  \bibfield  {author} {\bibinfo {author} {\bibfnamefont {N.}~\bibnamefont {Nishad}}, \bibinfo {author} {\bibfnamefont {A.}~\bibnamefont {Keselman}}, \bibinfo {author} {\bibfnamefont {T.}~\bibnamefont {Lahaye}}, \bibinfo {author} {\bibfnamefont {A.}~\bibnamefont {Browaeys}},\ and\ \bibinfo {author} {\bibfnamefont {S.}~\bibnamefont {Tsesses}},\ }\bibfield  {title} {\bibinfo {title} {{Quantum simulation of generic spin-exchange models in Floquet-engineered Rydberg-atom arrays}},\ }\href {https://doi.org/10.1103/PhysRevA.108.053318} {\bibfield  {journal} {\bibinfo  {journal} {Phys. Rev. A}\ }\textbf {\bibinfo {volume} {108}},\ \bibinfo {pages} {053318} (\bibinfo {year} {2023})}\BibitemShut {NoStop}%
\bibitem [{\citenamefont {Mambrini}\ and\ \citenamefont {Poilblanc}(2024)}]{Mambrini2024}%
  \BibitemOpen
  \bibfield  {author} {\bibinfo {author} {\bibfnamefont {M.}~\bibnamefont {Mambrini}}\ and\ \bibinfo {author} {\bibfnamefont {D.}~\bibnamefont {Poilblanc}},\ }\bibfield  {title} {\bibinfo {title} {{Quantum state preparation of topological chiral spin liquids via Floquet engineering}},\ }\href {https://doi.org/10.21468/SciPostPhys.17.1.011} {\bibfield  {journal} {\bibinfo  {journal} {SciPost Phys.}\ }\textbf {\bibinfo {volume} {17}},\ \bibinfo {pages} {011} (\bibinfo {year} {2024})}\BibitemShut {NoStop}%
\bibitem [{\citenamefont {Poilblanc}\ \emph {et~al.}(2024)\citenamefont {Poilblanc}, \citenamefont {Mambrini},\ and\ \citenamefont {Goldman}}]{Poliblanc2024}%
  \BibitemOpen
  \bibfield  {author} {\bibinfo {author} {\bibfnamefont {D.}~\bibnamefont {Poilblanc}}, \bibinfo {author} {\bibfnamefont {M.}~\bibnamefont {Mambrini}},\ and\ \bibinfo {author} {\bibfnamefont {N.}~\bibnamefont {Goldman}},\ }\bibfield  {title} {\bibinfo {title} {{Floquet dynamical chiral spin liquid at finite frequency}},\ }\href {https://doi.org/10.1103/PhysRevB.110.224408} {\bibfield  {journal} {\bibinfo  {journal} {Phys. Rev. B}\ }\textbf {\bibinfo {volume} {110}},\ \bibinfo {pages} {224408} (\bibinfo {year} {2024})}\BibitemShut {NoStop}%
\bibitem [{\citenamefont {Zhang}\ \emph {et~al.}(2024{\natexlab{b}})\citenamefont {Zhang}, \citenamefont {Mei}, \citenamefont {Xiao},\ and\ \citenamefont {Jia}}]{Zhang2024}%
  \BibitemOpen
  \bibfield  {author} {\bibinfo {author} {\bibfnamefont {J.-H.}\ \bibnamefont {Zhang}}, \bibinfo {author} {\bibfnamefont {F.}~\bibnamefont {Mei}}, \bibinfo {author} {\bibfnamefont {L.}~\bibnamefont {Xiao}},\ and\ \bibinfo {author} {\bibfnamefont {S.}~\bibnamefont {Jia}},\ }\bibfield  {title} {\bibinfo {title} {{Dynamical Detection of Topological Spectral Density}},\ }\href {https://doi.org/10.1103/PhysRevLett.132.036603} {\bibfield  {journal} {\bibinfo  {journal} {Phys. Rev. Lett.}\ }\textbf {\bibinfo {volume} {132}},\ \bibinfo {pages} {036603} (\bibinfo {year} {2024}{\natexlab{b}})}\BibitemShut {NoStop}%
\bibitem [{\citenamefont {Wang}\ \emph {et~al.}(2019)\citenamefont {Wang}, \citenamefont {Ronca},\ and\ \citenamefont {Sentef}}]{Wang2019}%
  \BibitemOpen
  \bibfield  {author} {\bibinfo {author} {\bibfnamefont {X.}~\bibnamefont {Wang}}, \bibinfo {author} {\bibfnamefont {E.}~\bibnamefont {Ronca}},\ and\ \bibinfo {author} {\bibfnamefont {M.~A.}\ \bibnamefont {Sentef}},\ }\bibfield  {title} {\bibinfo {title} {{Cavity quantum electrodynamical Chern insulator: Towards light-induced quantized anomalous Hall effect in graphene}},\ }\href {https://doi.org/10.1103/PhysRevB.99.235156} {\bibfield  {journal} {\bibinfo  {journal} {Phys. Rev. B}\ }\textbf {\bibinfo {volume} {99}},\ \bibinfo {pages} {235156} (\bibinfo {year} {2019})}\BibitemShut {NoStop}%
\bibitem [{\citenamefont {Sentef}\ \emph {et~al.}(2020)\citenamefont {Sentef}, \citenamefont {Li}, \citenamefont {K\"unzel},\ and\ \citenamefont {Eckstein}}]{Sentef2020}%
  \BibitemOpen
  \bibfield  {author} {\bibinfo {author} {\bibfnamefont {M.~A.}\ \bibnamefont {Sentef}}, \bibinfo {author} {\bibfnamefont {J.}~\bibnamefont {Li}}, \bibinfo {author} {\bibfnamefont {F.}~\bibnamefont {K\"unzel}},\ and\ \bibinfo {author} {\bibfnamefont {M.}~\bibnamefont {Eckstein}},\ }\bibfield  {title} {\bibinfo {title} {{Quantum to classical crossover of Floquet engineering in correlated quantum systems}},\ }\href {https://doi.org/10.1103/PhysRevResearch.2.033033} {\bibfield  {journal} {\bibinfo  {journal} {Phys. Rev. Res.}\ }\textbf {\bibinfo {volume} {2}},\ \bibinfo {pages} {033033} (\bibinfo {year} {2020})}\BibitemShut {NoStop}%
\bibitem [{\citenamefont {Li}\ \emph {et~al.}(2020)\citenamefont {Li}, \citenamefont {Golez}, \citenamefont {Mazza}, \citenamefont {Millis}, \citenamefont {Georges},\ and\ \citenamefont {Eckstein}}]{Li2020}%
  \BibitemOpen
  \bibfield  {author} {\bibinfo {author} {\bibfnamefont {J.}~\bibnamefont {Li}}, \bibinfo {author} {\bibfnamefont {D.}~\bibnamefont {Golez}}, \bibinfo {author} {\bibfnamefont {G.}~\bibnamefont {Mazza}}, \bibinfo {author} {\bibfnamefont {A.~J.}\ \bibnamefont {Millis}}, \bibinfo {author} {\bibfnamefont {A.}~\bibnamefont {Georges}},\ and\ \bibinfo {author} {\bibfnamefont {M.}~\bibnamefont {Eckstein}},\ }\bibfield  {title} {\bibinfo {title} {{Electromagnetic coupling in tight-binding models for strongly correlated light and matter}},\ }\href {https://doi.org/10.1103/PhysRevB.101.205140} {\bibfield  {journal} {\bibinfo  {journal} {Phys. Rev. B}\ }\textbf {\bibinfo {volume} {101}},\ \bibinfo {pages} {205140} (\bibinfo {year} {2020})}\BibitemShut {NoStop}%
\bibitem [{\citenamefont {Dmytruk}\ and\ \citenamefont {Schir{\`o}}(2022)}]{Dmytruk2022}%
  \BibitemOpen
  \bibfield  {author} {\bibinfo {author} {\bibfnamefont {O.}~\bibnamefont {Dmytruk}}\ and\ \bibinfo {author} {\bibfnamefont {M.}~\bibnamefont {Schir{\`o}}},\ }\bibfield  {title} {\bibinfo {title} {Controlling topological phases of matter with quantum light},\ }\href {https://doi.org/10.1038/s42005-022-01049-0} {\bibfield  {journal} {\bibinfo  {journal} {Communications Physics}\ }\textbf {\bibinfo {volume} {5}},\ \bibinfo {pages} {271} (\bibinfo {year} {2022})}\BibitemShut {NoStop}%
\bibitem [{\citenamefont {Eckhardt}\ \emph {et~al.}(2022)\citenamefont {Eckhardt}, \citenamefont {Passetti}, \citenamefont {Othman}, \citenamefont {Karrasch}, \citenamefont {Cavaliere}, \citenamefont {Sentef},\ and\ \citenamefont {Kennes}}]{Eckhardt2022}%
  \BibitemOpen
  \bibfield  {author} {\bibinfo {author} {\bibfnamefont {C.~J.}\ \bibnamefont {Eckhardt}}, \bibinfo {author} {\bibfnamefont {G.}~\bibnamefont {Passetti}}, \bibinfo {author} {\bibfnamefont {M.}~\bibnamefont {Othman}}, \bibinfo {author} {\bibfnamefont {C.}~\bibnamefont {Karrasch}}, \bibinfo {author} {\bibfnamefont {F.}~\bibnamefont {Cavaliere}}, \bibinfo {author} {\bibfnamefont {M.~A.}\ \bibnamefont {Sentef}},\ and\ \bibinfo {author} {\bibfnamefont {D.~M.}\ \bibnamefont {Kennes}},\ }\bibfield  {title} {\bibinfo {title} {{Quantum Floquet engineering with an exactly solvable tight-binding chain in a cavity}},\ }\href {https://doi.org/10.1038/s42005-022-00880-9} {\bibfield  {journal} {\bibinfo  {journal} {Communications Physics}\ }\textbf {\bibinfo {volume} {5}},\ \bibinfo {pages} {122} (\bibinfo {year} {2022})}\BibitemShut {NoStop}%
\bibitem [{\citenamefont {Appugliese}\ \emph {et~al.}(2022)\citenamefont {Appugliese}, \citenamefont {Enkner}, \citenamefont {Paravicini-Bagliani}, \citenamefont {Beck}, \citenamefont {Reichl}, \citenamefont {Wegscheider}, \citenamefont {Scalari}, \citenamefont {Ciuti},\ and\ \citenamefont {Faist}}]{Appugliese2022}%
  \BibitemOpen
  \bibfield  {author} {\bibinfo {author} {\bibfnamefont {F.}~\bibnamefont {Appugliese}}, \bibinfo {author} {\bibfnamefont {J.}~\bibnamefont {Enkner}}, \bibinfo {author} {\bibfnamefont {G.~L.}\ \bibnamefont {Paravicini-Bagliani}}, \bibinfo {author} {\bibfnamefont {M.}~\bibnamefont {Beck}}, \bibinfo {author} {\bibfnamefont {C.}~\bibnamefont {Reichl}}, \bibinfo {author} {\bibfnamefont {W.}~\bibnamefont {Wegscheider}}, \bibinfo {author} {\bibfnamefont {G.}~\bibnamefont {Scalari}}, \bibinfo {author} {\bibfnamefont {C.}~\bibnamefont {Ciuti}},\ and\ \bibinfo {author} {\bibfnamefont {J.}~\bibnamefont {Faist}},\ }\bibfield  {title} {\bibinfo {title} {{Breakdown of topological protection by cavity vacuum fields in the integer quantum Hall effect}},\ }\href {https://doi.org/10.1126/science.abl5818} {\bibfield  {journal} {\bibinfo  {journal} {Science}\ }\textbf {\bibinfo {volume} {375}},\ \bibinfo {pages} {1030} (\bibinfo {year} {2022})},\ \Eprint
  {https://arxiv.org/abs/https://www.science.org/doi/pdf/10.1126/science.abl5818} {https://www.science.org/doi/pdf/10.1126/science.abl5818} \BibitemShut {NoStop}%
\bibitem [{\citenamefont {Nguyen}\ \emph {et~al.}(2023)\citenamefont {Nguyen}, \citenamefont {Arwas}, \citenamefont {Lin}, \citenamefont {Yao},\ and\ \citenamefont {Ciuti}}]{Nguyen2023}%
  \BibitemOpen
  \bibfield  {author} {\bibinfo {author} {\bibfnamefont {D.-P.}\ \bibnamefont {Nguyen}}, \bibinfo {author} {\bibfnamefont {G.}~\bibnamefont {Arwas}}, \bibinfo {author} {\bibfnamefont {Z.}~\bibnamefont {Lin}}, \bibinfo {author} {\bibfnamefont {W.}~\bibnamefont {Yao}},\ and\ \bibinfo {author} {\bibfnamefont {C.}~\bibnamefont {Ciuti}},\ }\bibfield  {title} {\bibinfo {title} {{Electron-Photon Chern Number in Cavity-Embedded 2D Moir\'e Materials}},\ }\href {https://doi.org/10.1103/PhysRevLett.131.176602} {\bibfield  {journal} {\bibinfo  {journal} {Phys. Rev. Lett.}\ }\textbf {\bibinfo {volume} {131}},\ \bibinfo {pages} {176602} (\bibinfo {year} {2023})}\BibitemShut {NoStop}%
\bibitem [{\citenamefont {Masuki}\ and\ \citenamefont {Ashida}(2023)}]{Masuki2023}%
  \BibitemOpen
  \bibfield  {author} {\bibinfo {author} {\bibfnamefont {K.}~\bibnamefont {Masuki}}\ and\ \bibinfo {author} {\bibfnamefont {Y.}~\bibnamefont {Ashida}},\ }\bibfield  {title} {\bibinfo {title} {Berry phase and topology in ultrastrongly coupled quantum light-matter systems},\ }\href {https://doi.org/10.1103/PhysRevB.107.195104} {\bibfield  {journal} {\bibinfo  {journal} {Phys. Rev. B}\ }\textbf {\bibinfo {volume} {107}},\ \bibinfo {pages} {195104} (\bibinfo {year} {2023})}\BibitemShut {NoStop}%
\bibitem [{\citenamefont {Passetti}\ \emph {et~al.}(2023)\citenamefont {Passetti}, \citenamefont {Eckhardt}, \citenamefont {Sentef},\ and\ \citenamefont {Kennes}}]{Pasetti2023}%
  \BibitemOpen
  \bibfield  {author} {\bibinfo {author} {\bibfnamefont {G.}~\bibnamefont {Passetti}}, \bibinfo {author} {\bibfnamefont {C.~J.}\ \bibnamefont {Eckhardt}}, \bibinfo {author} {\bibfnamefont {M.~A.}\ \bibnamefont {Sentef}},\ and\ \bibinfo {author} {\bibfnamefont {D.~M.}\ \bibnamefont {Kennes}},\ }\bibfield  {title} {\bibinfo {title} {Cavity light-matter entanglement through quantum fluctuations},\ }\href {https://doi.org/10.1103/PhysRevLett.131.023601} {\bibfield  {journal} {\bibinfo  {journal} {Phys. Rev. Lett.}\ }\textbf {\bibinfo {volume} {131}},\ \bibinfo {pages} {023601} (\bibinfo {year} {2023})}\BibitemShut {NoStop}%
\bibitem [{\citenamefont {P\'erez-Gonz\'alez}\ \emph {et~al.}(2025)\citenamefont {P\'erez-Gonz\'alez}, \citenamefont {Platero},\ and\ \citenamefont {Gomez-Le\'on}}]{PerezGonzalez2025}%
  \BibitemOpen
  \bibfield  {author} {\bibinfo {author} {\bibfnamefont {B.}~\bibnamefont {P\'erez-Gonz\'alez}}, \bibinfo {author} {\bibfnamefont {G.}~\bibnamefont {Platero}},\ and\ \bibinfo {author} {\bibfnamefont {A.}~\bibnamefont {Gomez-Le\'on}},\ }\bibfield  {title} {\bibinfo {title} {{Light-matter correlations in Quantum Floquet engineering of cavity quantum materials}},\ }\href {https://doi.org/10.22331/q-2025-02-17-1633} {\bibfield  {journal} {\bibinfo  {journal} {Quantum}\ }\textbf {\bibinfo {volume} {9}},\ \bibinfo {pages} {1633} (\bibinfo {year} {2025})}\BibitemShut {NoStop}%
\bibitem [{\citenamefont {{Beatriz Pérez-González and Gloria Platero and Álvaro Gómez-León}}(2023)}]{PerezGonzalez2023}%
  \BibitemOpen
  \bibfield  {author} {\bibinfo {author} {\bibnamefont {{Beatriz Pérez-González and Gloria Platero and Álvaro Gómez-León}}},\ }\href {https://arxiv.org/abs/2312.10141} {\bibinfo {title} {{Many-body origin of anomalous Floquet phases in cavity-QED materials}}} (\bibinfo {year} {2023}),\ \Eprint {https://arxiv.org/abs/2312.10141} {arXiv:2312.10141 [quant-ph]} \BibitemShut {NoStop}%
\bibitem [{\citenamefont {Wagner}\ \emph {et~al.}(2023)\citenamefont {Wagner}, \citenamefont {Crippa}, \citenamefont {Amaricci}, \citenamefont {Hansmann}, \citenamefont {Klett}, \citenamefont {K{\"o}nig}, \citenamefont {Sch{\"a}fer}, \citenamefont {Sante}, \citenamefont {Cano}, \citenamefont {Millis}, \citenamefont {Georges},\ and\ \citenamefont {Sangiovanni}}]{Wagner2023}%
  \BibitemOpen
  \bibfield  {author} {\bibinfo {author} {\bibfnamefont {N.}~\bibnamefont {Wagner}}, \bibinfo {author} {\bibfnamefont {L.}~\bibnamefont {Crippa}}, \bibinfo {author} {\bibfnamefont {A.}~\bibnamefont {Amaricci}}, \bibinfo {author} {\bibfnamefont {P.}~\bibnamefont {Hansmann}}, \bibinfo {author} {\bibfnamefont {M.}~\bibnamefont {Klett}}, \bibinfo {author} {\bibfnamefont {E.~J.}\ \bibnamefont {K{\"o}nig}}, \bibinfo {author} {\bibfnamefont {T.}~\bibnamefont {Sch{\"a}fer}}, \bibinfo {author} {\bibfnamefont {D.~D.}\ \bibnamefont {Sante}}, \bibinfo {author} {\bibfnamefont {J.}~\bibnamefont {Cano}}, \bibinfo {author} {\bibfnamefont {A.~J.}\ \bibnamefont {Millis}}, \bibinfo {author} {\bibfnamefont {A.}~\bibnamefont {Georges}},\ and\ \bibinfo {author} {\bibfnamefont {G.}~\bibnamefont {Sangiovanni}},\ }\bibfield  {title} {\bibinfo {title} {Mott insulators with boundary zeros},\ }\href {https://doi.org/10.1038/s41467-023-42773-7} {\bibfield  {journal} {\bibinfo  {journal} {Nature Communications}\ }\textbf {\bibinfo
  {volume} {14}},\ \bibinfo {pages} {7531} (\bibinfo {year} {2023})}\BibitemShut {NoStop}%
\bibitem [{\citenamefont {Fainshtein}\ \emph {et~al.}(1978)\citenamefont {Fainshtein}, \citenamefont {Manakov},\ and\ \citenamefont {Rapoport}}]{Fainshtein1978}%
  \BibitemOpen
  \bibfield  {author} {\bibinfo {author} {\bibfnamefont {A.~G.}\ \bibnamefont {Fainshtein}}, \bibinfo {author} {\bibfnamefont {N.~L.}\ \bibnamefont {Manakov}},\ and\ \bibinfo {author} {\bibfnamefont {L.~P.}\ \bibnamefont {Rapoport}},\ }\bibfield  {title} {\bibinfo {title} {Some general properties of quasi-energetic spectra of quantum systems in classical monochromatic fields},\ }\href {https://doi.org/10.1088/0022-3700/11/14/020} {\bibfield  {journal} {\bibinfo  {journal} {Journal of Physics B: Atomic and Molecular Physics}\ }\textbf {\bibinfo {volume} {11}},\ \bibinfo {pages} {2561} (\bibinfo {year} {1978})}\BibitemShut {NoStop}%
\bibitem [{\citenamefont {Reynoso}\ and\ \citenamefont {Frustaglia}(2013)}]{Reynoso2013}%
  \BibitemOpen
  \bibfield  {author} {\bibinfo {author} {\bibfnamefont {A.~A.}\ \bibnamefont {Reynoso}}\ and\ \bibinfo {author} {\bibfnamefont {D.}~\bibnamefont {Frustaglia}},\ }\bibfield  {title} {\bibinfo {title} {{Unpaired Floquet Majorana fermions without magnetic fields}},\ }\href {https://doi.org/10.1103/PhysRevB.87.115420} {\bibfield  {journal} {\bibinfo  {journal} {Phys. Rev. B}\ }\textbf {\bibinfo {volume} {87}},\ \bibinfo {pages} {115420} (\bibinfo {year} {2013})}\BibitemShut {NoStop}%
\bibitem [{\citenamefont {Rudin}(1986)}]{Rudin1986}%
  \BibitemOpen
  \bibfield  {author} {\bibinfo {author} {\bibfnamefont {W.}~\bibnamefont {Rudin}},\ }\href@noop {} {\emph {\bibinfo {title} {Real and Complex Analysis}}},\ \bibinfo {edition} {3rd}\ ed.\ (\bibinfo  {publisher} {McGraw-Hill},\ \bibinfo {address} {New York},\ \bibinfo {year} {1986})\BibitemShut {NoStop}%
\bibitem [{\citenamefont {Royden}\ and\ \citenamefont {Fitzpatrick}(2010)}]{Royden2010}%
  \BibitemOpen
  \bibfield  {author} {\bibinfo {author} {\bibfnamefont {H.}~\bibnamefont {Royden}}\ and\ \bibinfo {author} {\bibfnamefont {P.}~\bibnamefont {Fitzpatrick}},\ }\href@noop {} {\emph {\bibinfo {title} {Real Analysis}}},\ \bibinfo {edition} {4th}\ ed.\ (\bibinfo  {publisher} {Pearson},\ \bibinfo {year} {2010})\BibitemShut {NoStop}%
\bibitem [{\citenamefont {Matsyshyn}\ \emph {et~al.}(2023)\citenamefont {Matsyshyn}, \citenamefont {Song}, \citenamefont {Villadiego},\ and\ \citenamefont {Shi}}]{Matsyshyn2023}%
  \BibitemOpen
  \bibfield  {author} {\bibinfo {author} {\bibfnamefont {O.}~\bibnamefont {Matsyshyn}}, \bibinfo {author} {\bibfnamefont {J.~C.~W.}\ \bibnamefont {Song}}, \bibinfo {author} {\bibfnamefont {I.~S.}\ \bibnamefont {Villadiego}},\ and\ \bibinfo {author} {\bibfnamefont {L.-k.}\ \bibnamefont {Shi}},\ }\bibfield  {title} {\bibinfo {title} {{Fermi-Dirac staircase occupation of Floquet bands and current rectification inside the optical gap of metals: An exact approach}},\ }\href {https://doi.org/10.1103/PhysRevB.107.195135} {\bibfield  {journal} {\bibinfo  {journal} {Phys. Rev. B}\ }\textbf {\bibinfo {volume} {107}},\ \bibinfo {pages} {195135} (\bibinfo {year} {2023})}\BibitemShut {NoStop}%
\bibitem [{\citenamefont {Shi}\ \emph {et~al.}(2024)\citenamefont {Shi}, \citenamefont {Matsyshyn}, \citenamefont {Song},\ and\ \citenamefont {Villadiego}}]{Shi2024}%
  \BibitemOpen
  \bibfield  {author} {\bibinfo {author} {\bibfnamefont {L.-k.}\ \bibnamefont {Shi}}, \bibinfo {author} {\bibfnamefont {O.}~\bibnamefont {Matsyshyn}}, \bibinfo {author} {\bibfnamefont {J.~C.~W.}\ \bibnamefont {Song}},\ and\ \bibinfo {author} {\bibfnamefont {I.~S.}\ \bibnamefont {Villadiego}},\ }\bibfield  {title} {\bibinfo {title} {{Floquet Fermi Liquid}},\ }\href {https://doi.org/10.1103/PhysRevLett.132.146402} {\bibfield  {journal} {\bibinfo  {journal} {Phys. Rev. Lett.}\ }\textbf {\bibinfo {volume} {132}},\ \bibinfo {pages} {146402} (\bibinfo {year} {2024})}\BibitemShut {NoStop}%
\bibitem [{\citenamefont {Kumari}\ \emph {et~al.}(2024)\citenamefont {Kumari}, \citenamefont {Seradjeh},\ and\ \citenamefont {Kundu}}]{Kumari2024}%
  \BibitemOpen
  \bibfield  {author} {\bibinfo {author} {\bibfnamefont {R.}~\bibnamefont {Kumari}}, \bibinfo {author} {\bibfnamefont {B.}~\bibnamefont {Seradjeh}},\ and\ \bibinfo {author} {\bibfnamefont {A.}~\bibnamefont {Kundu}},\ }\bibfield  {title} {\bibinfo {title} {{Josephson-Current Signatures of Unpaired Floquet Majorana Fermions}},\ }\href {https://doi.org/10.1103/PhysRevLett.133.196601} {\bibfield  {journal} {\bibinfo  {journal} {Phys. Rev. Lett.}\ }\textbf {\bibinfo {volume} {133}},\ \bibinfo {pages} {196601} (\bibinfo {year} {2024})}\BibitemShut {NoStop}%
\end{thebibliography}
\end{document}